%% file: PhD_Template.tex
\documentclass[twoside,12pt]{scrreprt}
\usepackage{amsfonts}
\usepackage{booktabs}        
\usepackage{graphicx}        
\usepackage{amsmath}         
\usepackage{amsfonts}
\usepackage{amssymb}
\usepackage{textcomp}        
\usepackage{fancyref}        
\usepackage{float}           
\usepackage{subfigure}
\usepackage{hyperref}
\usepackage{fancybox} 
\usepackage{picins} 
\usepackage{caption} 
\usepackage{draftcopy}
\usepackage{eso-pic}
\usepackage{units}

\usepackage[british]{babel}
\usepackage[utf8]{inputenc}
\usepackage[T1]{fontenc}
\addtocounter{footnote}{1}
\usepackage[a4paper,width=160mm,top=25mm,bottom=25mm,bindingoffset=6mm]{geometry}
\usepackage{type1cm}
\usepackage{booktabs}
\usepackage{tabto}
\usepackage[numbers,sort&compress]{natbib}

\clearpage

\renewcommand{\Im}{{\text{Im}}}
\renewcommand{\Re}{{\text{Re}}}

\newcommand{\e}{{\text{e}}}
\newcommand{\ii}{\text{i}}

\def\ds{\mathrm{dS}}
\def\dsd{\mathrm{dS_{D}}}
\def\x{\mathbf{x}}
\def\k{\mathbf{k}}
\def\lambdam{\lambda_{\mathrm{m}}}
\def\mds{m_{\mathrm{ds}}}
\def\rnd{\partial}
\def\wwp{\mathrm{W}_{\kappa,\mu}}
\def\wwm{\mathrm{W}_{\kappa,-\mu}}
\def\wmp{\mathrm{M}_{\kappa,\mu}}
\def\wmm{\mathrm{M}_{\kappa,-\mu}}
\def\w{\mathrm{W}}
\def\m{\mathrm{M}}

\def\I{\mathrm{I}}
\def\sem{\mathrm{sem}}
\def\comment#1{}
\def\nn{\nonumber}

\newcommand{\chap}{chapter }

\newcommand{\bra}[1]{\langle #1|}
\newcommand{\ket}[1]{|#1\rangle}

\renewcommand{\O}{{\mathcal{O}}}
\DeclareMathOperator\sgn{sgn}

\usepackage{slashed}
\usepackage{color}
\usepackage{bbm}

\usepackage[toc]{glossaries} 

\usepackage{glossary-mcols}

\usepackage{epigraph}

\makeglossaries

\input{./glossary.tex}

\begin{document}
\pagenumbering{roman}
\begin{titlepage}
\begin{center}
\thispagestyle{empty}
\begin{minipage}{0.32\textwidth}
 \centering
 \includegraphics[width=1\textwidth]{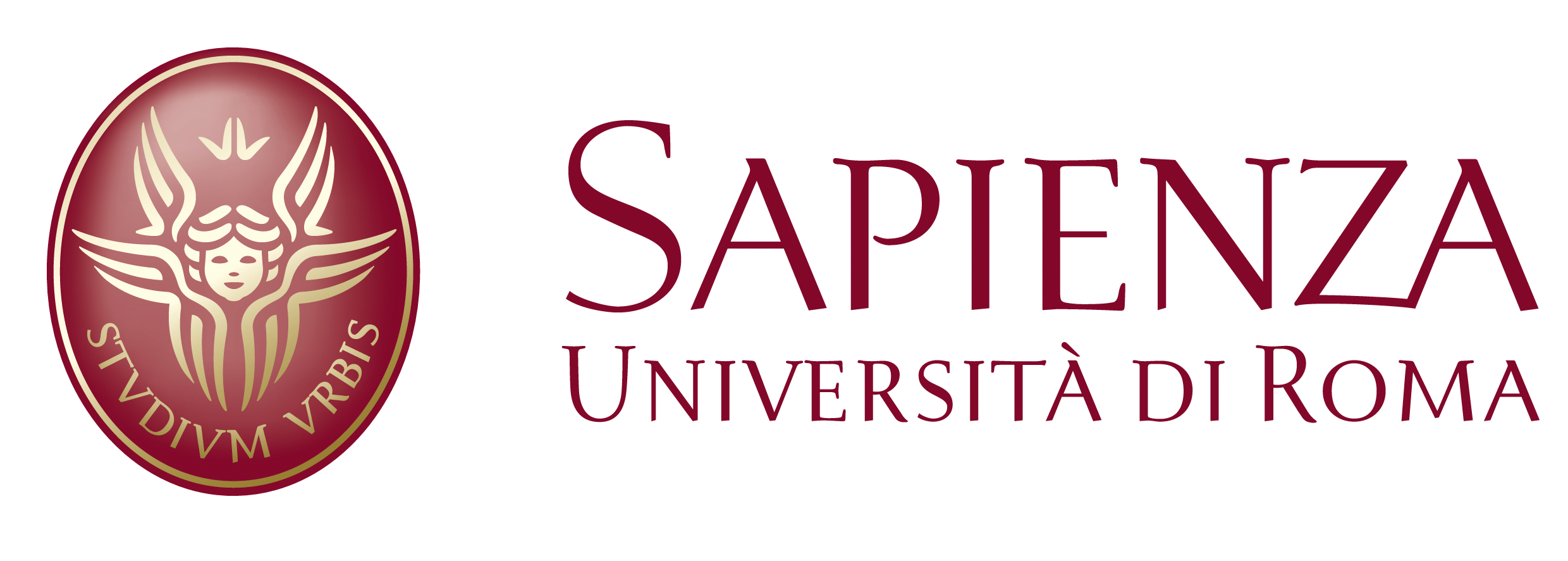}
\end{minipage}
\begin{minipage}{0.32\textwidth}
  \centering
 \includegraphics[width=0.45\textwidth]{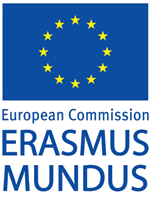}
\end{minipage}
\begin{minipage}{0.32\textwidth}
  \centering
 \includegraphics[width=0.6\textwidth]{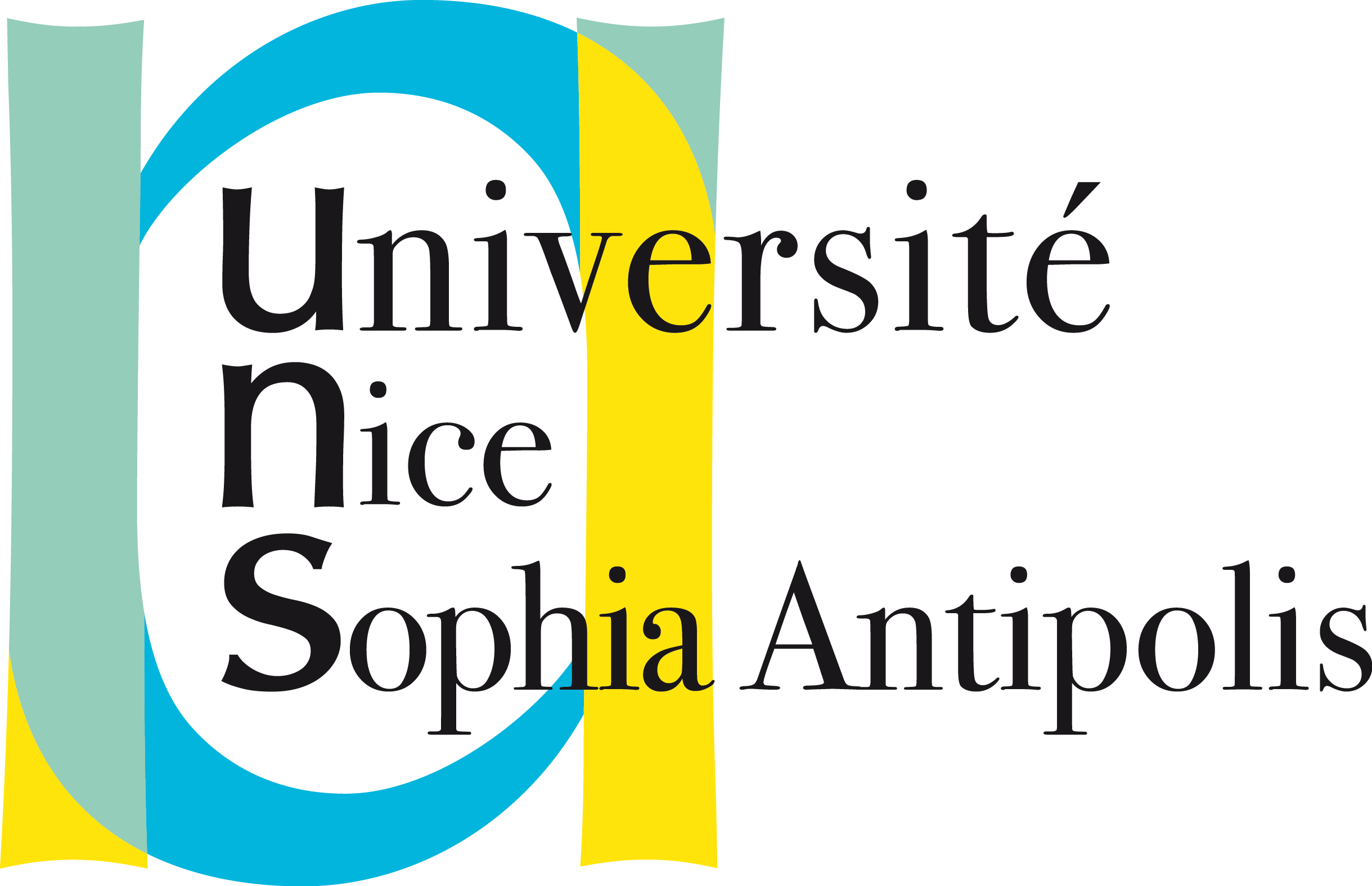}
\end{minipage}

\vfill

\noindent\rule[1ex]{16cm}{1pt}\\
\vspace{0.3cm} 
 {\Huge
\textbf{On early and late phases of acceleration of the expansion of the Universe}} \\
\vspace{0.3cm}
\noindent\rule[1ex]{16cm}{1pt}
\par
{\Large \textbf{International Relativistic Astrophysics PhD Thesis}}\\[1.2cm]
\end{center}

 \vfill
 \Large
 PhD Student \hfill Supervisor\\
 Cl\'ement Stahl \hfill   Remo Ruffini \\[0.5cm]

\begin{center}
             \noindent\rule[1ex]{0.5\textwidth}{1pt} \\
            Academic Year 2013-2016
\end{center}
 

\end{titlepage}

\newpage
\thispagestyle{empty}
\cleardoublepage
\thispagestyle{empty}
\phantomsection
\addcontentsline{toc}{chapter}{Preface}
\input{./Preface.tex}
\newpage
\phantomsection
\addcontentsline{toc}{chapter}{Contents}
\tableofcontents
\newpage\null\thispagestyle{empty}\newpage
\pagenumbering{arabic}
\part{Introduction to the Universe} 
\label{part:introcosmo}
 \input{cours/L1.tex}
   \input{cours/L3.tex}
    \input{cours/L4.tex}
    \input{cours/L5.tex}
\input{cours/infla.tex}
 \chapter{Pair creation in de Sitter space}
 \label{chap:early}
\input{semi/intro.tex}
 \input{semi/basis.tex}
 \section{Semiclassical estimates}
 \label{sec:smestim}
  \input{semi/proceed7.tex}
 \section{Bosons in $dS_3$: Pair creation and current}
 \label{sec:Bogbos}
 \input{current/boson.tex}
 \section{Fermions in $dS_2$: Pair creation and current}
 \label{sec:fermdsdeux}
 \input{current/Manuscript5.tex}
 \chapter{Backreaction of the created pairs}
 \label{chap:back}
 \input{back/backelectro.tex}
  \input{back/backgrav.tex}

 \input{cours/DE.tex}
 \chapter{Inhomogeneous cosmology}
 \label{chap:LTB}
 \input{LTB/LTB.tex}
\chapter{Interacting dark energy}
\label{chap:IDE}
 \input{IDE/IDE.tex}
 \input{./concl.tex}

 \appendix
\chapter{Appendix for \chap \ref{chap:introcosmo}} 
 \input{cours/appendix.tex}
 \label{ap:intro}
 \chapter{Appendix for \chap \ref{chap:inflation}}
  \label{ap:infl}
 \input{cours/appendixinfl.tex}
 \chapter{Appendix for \chap \ref{chap:early}} 
 \input{current/appendixbess.tex}
 \label{ap:bess}
 \chapter{Appendix for \chap \ref{chap:LTB}}
 \label{ap:LTB}
 \input{LTB/LTBappendix.tex}
\phantomsection
\addcontentsline{toc}{chapter}{List of Figures}
\listoffigures
\begingroup
\let\clearpage\relax
\phantomsection
\addcontentsline{toc}{chapter}{List of Tables}
\listoftables
\endgroup

\newpage
 \pagebreak
 \pagestyle{plain}

\newpage

\printglossary[style=mcolindex]
\input{./post.tex}

\end{document}

%% file: glossary.tex
\newglossaryentry{general relativity}{name={general relativity},description={(or GR)}}

\newglossaryentry{Slipher}{name={Slipher},description={\nopostdesc}}

\newglossaryentry{nebulae}{name={nebulae},description={\nopostdesc}}

\newglossaryentry{Hubble}{name={Hubble},description={\nopostdesc}}

\newglossaryentry{Hubble law}{name={Hubble law},description={\nopostdesc}}

\newglossaryentry{Hubble constant}{name={Hubble constant},description={\nopostdesc}}

\newglossaryentry{Hubble radius}{name={Hubble radius},description={\nopostdesc}}

\newglossaryentry{Hubble flow}{name={Hubble flow},description={\nopostdesc}}

\newglossaryentry{Hubble diagram}{name={Hubble diagram},description={\nopostdesc}}

\newglossaryentry{Einstein}{name={Einstein},description={\nopostdesc}}

\newglossaryentry{Einstein equation}{name={Einstein equation},description={\nopostdesc}}

\newglossaryentry{Einstein universe}{name={Einstein universe},description={\nopostdesc}}

\newglossaryentry{Einstein tensor}{name={Einstein tensor},description={\nopostdesc}}

\newacronym{EdS}{EdS}{Einstein de Sitter}

\newglossaryentry{equivalence principle}{name={equivalence principle},description={\nopostdesc}}

\newglossaryentry{symmetry}{name={symmetry},description={\nopostdesc}}

\newglossaryentry{de Sitter}{name={de Sitter},description={\nopostdesc}}

\newglossaryentry{dS}{name={de Sitter space},description={or (dS), it is also used when the number of spatial dimensions is precised, leading to the acronyms $\ds_2$, $\ds_3$, $\ds_4$, $\ds_D$}
}

\newglossaryentry{de Sitter metric}{name={de Sitter metric},description={\nopostdesc}}

\newglossaryentry{cosmological constant}{name={cosmological constant},description={\nopostdesc}}

\newglossaryentry{cosmological constant problem}{name={cosmological constant problem},description={\nopostdesc}}

\newglossaryentry{coincidence problem}{name={coincidence problem},description={\nopostdesc}}

\newglossaryentry{SM}{name={standard cosmological model},description={or the $\Lambda$CDM (lambda cold dark matter) model or the vanilla model or the FLRW model or the standard model or the big-bang model}}

\newglossaryentry{FLRW metric}{name={FLRW metric},description={Friedmann-Lemaître-Robertson-Walker metric}}

\newglossaryentry{SMP}{name={standard model},description={of particle physics}}

\newglossaryentry{manifold}{name={manifold},description={\nopostdesc}}

\newglossaryentry{cosmological principle}{name={cosmological principle},description={\nopostdesc}}

\newglossaryentry{Copernican principle}{name={Copernican principle},description={\nopostdesc}}

\newglossaryentry{Heisenberg uncertainty principle}{name={Heisenberg uncertainty principle},description={\nopostdesc}}

\newglossaryentry{anthropic principle}{name={anthropic principle},description={\nopostdesc}}

\newacronym{CMB}{CMB}{Cosmic Microwave Background}

\newglossaryentry{the Great Debate}{name={the Great Debate},description={\nopostdesc}}

\newglossaryentry{Friedmann}{name={Friedmann},description={\nopostdesc}}

\newglossaryentry{Friedmann equation}{name={Friedmann equation},description={\nopostdesc}}

\newglossaryentry{lui}{name={Lemaître},description={\nopostdesc}}

\newglossaryentry{dark energy}{name={dark energy},description={\nopostdesc}}

\newglossaryentry{dark matter}{name={dark matter},description={\nopostdesc}}

\newglossaryentry{topology}{name={topology},description={\nopostdesc}}

\newglossaryentry{algebric topology}{name={algebric topology},description={\nopostdesc}}

\newglossaryentry{lb}{name={Löbell topology},description={\nopostdesc}}

\newglossaryentry{locality}{name={locality},description={\nopostdesc}}

\newglossaryentry{vacuum}{name={vacuum},description={\nopostdesc}}

\newglossaryentry{vacuum fluctuations}{name={vacuum fluctuations},description={or quantum fluctuations}}

\newglossaryentry{Bunch-Davies vacuum}{name={Bunch-Davies vacuum},description={\nopostdesc}}

\newglossaryentry{vacuum energy}{name={vacuum energy},description={\nopostdesc}}

\newglossaryentry{holonomy group}{name={holonomy group},description={\nopostdesc}}

\newglossaryentry{BBN}{name={Big-Bang nucleosynthesis},description={\nopostdesc}}

\newglossaryentry{Humason}{name={Humason},description={\nopostdesc}}

\newglossaryentry{satellite}{name={satellite},description={\nopostdesc}}

\newglossaryentry{SNe Ia}{name={supernovae Ia},description={\nopostdesc}}

\newglossaryentry{supernovae II-P}{name={supernovae II-P},description={\nopostdesc}}

\newglossaryentry{supernovae}{name={supernovae},description={\nopostdesc}}

\newglossaryentry{scale factor}{name={scale factor},description={\nopostdesc}}

\newglossaryentry{Ricci tensor}{name={Ricci tensor},description={\nopostdesc}}

\newglossaryentry{Ricci scalar}{name={Ricci scalar},description={\nopostdesc}}

\newglossaryentry{Riemann tensor}{name={Riemann tensor},description={\nopostdesc}}

\newglossaryentry{Ricci-Weyl problem}{name={Ricci-Weyl problem},description={\nopostdesc}}

\newglossaryentry{Weyl tensor}{name={Weyl tensor},description={\nopostdesc}}

\newglossaryentry{Weyl basis}{name={Weyl basis},description={for spinor}}

\newglossaryentry{Christoffel symbol}{name={Christoffel symbol},description={\nopostdesc}}

\newglossaryentry{spin connection}{name={spin connection},description={see also Christoffel symbol}}

\newglossaryentry{tensor}{name={tensor},description={\nopostdesc}}

\newglossaryentry{energy momentum tensor}{name={energy momentum tensor},description={\nopostdesc}}

\newglossaryentry{metric tensor}{name={metric tensor},description={\nopostdesc}}

\newglossaryentry{redshift}{name={redshift},description={\nopostdesc}}

\newglossaryentry{photometric redshift}{name={photometric redshift},description={\nopostdesc}}

\newglossaryentry{redshift drift}{name={redshift drift},description={\nopostdesc}}

\newglossaryentry{spectroscopic redshift}{name={spectroscopic redshift},description={\nopostdesc}}

\newglossaryentry{spectrum}{name={spectrum},description={\nopostdesc}}

\newglossaryentry{power spectrum}{name={power spectrum},description={\nopostdesc}}

\newglossaryentry{number of pairs per momentum}{name={number of pairs per momentum},description={\nopostdesc}}

\newglossaryentry{physical number of pairs}{name={physical number of pairs},description={\nopostdesc}}

\newacronym{VIMOS}{VIMOS}{VIsible MultiObject Spectrograph}

\newacronym{KMOS}{KMOS}{K-band Multi Object Spectrograph}

\newacronym{MOSFIRE}{MOSFIRE}{Multi-Object Spectrometer for Infra-Red Exploration}

\newacronym{SDSS}{SDSS}{Sloan Digital Sky Survey}

\newacronym{MUSE}{MUSE}{Multi Unit Spectroscopic Explorer}

\newacronym{SINFONI}{SINFONI}{Spectrograph for INtegral Field Observations in the Near Infrared}

\newglossaryentry{Lyman break}{name={Lyman break},description={\nopostdesc}}

\newglossaryentry{standard candle}{name={standard candle},description={\nopostdesc}}

\newglossaryentry{standard ruler}{name={standard ruler},description={\nopostdesc}}

\newacronym{BAO}{BAO}{Baryon Acoustic Oscillation}

\newglossaryentry{angular distance}{name={angular distance},description={\nopostdesc}}

\newglossaryentry{parallax}{name={parallax},description={\nopostdesc}}

\newglossaryentry{luminosity distance}{name={luminosity distance},description={\nopostdesc}}

\newglossaryentry{Gamma Ray Bursts}{name={Gamma Ray Bursts},description={\nopostdesc}}

\newglossaryentry{distance duality relation}{name={distance duality relation},description={a consequence of Etherington reciprocity theorem}}

\newglossaryentry{covariant}{name={covariant},description={\nopostdesc}}

\newglossaryentry{covariant derivative}{name={covariant derivative},description={\nopostdesc}}

\newglossaryentry{contravariant}{name={contravariant},description={\nopostdesc}}

\newglossaryentry{curved spacetime}{name={curved spacetime},description={\nopostdesc}}

\newglossaryentry{Schwartz theorem}{name={Schwartz theorem},description={\nopostdesc}}

\newglossaryentry{Bianchi identity}{name={Bianchi identity},description={\nopostdesc}}

\newglossaryentry{Bianchi spacetimes}{name={Bianchi spacetimes},description={\nopostdesc}}

\newglossaryentry{perfect fluid}{name={perfect fluid},description={\nopostdesc}}

\newglossaryentry{neutron star}{name={neutron star},description={\nopostdesc}}

\newglossaryentry{barotropic index}{name={barotropic index},description={\nopostdesc}}

\newglossaryentry{radiation}{name={radiation},description={relativistic species present in the universe}}

\newglossaryentry{Hawking radiation}{name={Hawking radiation},description={\nopostdesc}}

\newglossaryentry{big-bang}{name={big-bang},description={\nopostdesc}}

\newglossaryentry{recombination}{name={recombination},description={see also CMB}}

\newglossaryentry{reheating}{name={reheating},description={\nopostdesc}}

\newglossaryentry{S}{name={structure},description={referring to the large scale structure of the universe}}

\newglossaryentry{structure constant}{name={structure constant},description={\nopostdesc}}

\newglossaryentry{inflation}{name={inflation},description={\nopostdesc}}

\newglossaryentry{eternal inflation}{name={eternal inflation},description={\nopostdesc}}

\newglossaryentry{warm inflation}{name={warm inflation},description={\nopostdesc}}

\newglossaryentry{axion inflation}{name={axion inflation},description={\nopostdesc}}

\newglossaryentry{magnetogenesis}{name={magnetogenesis},description={\nopostdesc}}

\newglossaryentry{perturbation}{name={perturbation},description={\nopostdesc}}

\newglossaryentry{Bardeen potential}{name={Bardeen potential},description={\nopostdesc}}

\newglossaryentry{MS}{name={Mukhanov-Sasaki variable},description={\nopostdesc}}

\newglossaryentry{slow roll}{name={slow roll},description={\nopostdesc}}

\newglossaryentry{particle creation}{name={particle creation},description={see also Schwinger effect and pair}}

\newglossaryentry{particle physics}{name={particle physics},description={\nopostdesc}}

\newglossaryentry{Schwinger effect}{name={Schwinger effect},description={see also particle creation}}

\newacronym{XFEL}{XFEL}{European X-Ray Free-Electron Laser}

\newacronym{HIBEF}{HIBEF}{Helmholtz International Beamline for Extreme Fields}

\newacronym{ELIA}{ELIA}{Extreme Light Infrastructure}

\newacronym{HiPer}{HiPer}{High Power Laser Energy Research}

\newacronym{XCELS}{XCELS}{Exawatt Center for Extreme Light Studies}

\newglossaryentry{adiabatic subtraction}{name={adiabatic subtraction},description={\nopostdesc}}

\newglossaryentry{isometry}{name={isometry},description={\nopostdesc}}

\newglossaryentry{tetrad}{name={tetrad},description={\nopostdesc}}

\newglossaryentry{semiclassical scattering method}{name={semiclassical scattering method},description={\nopostdesc}}

\newglossaryentry{conformal coupling}{name={conformal coupling},description={\nopostdesc}}

\newglossaryentry{Bogoliubov transformation}{name={Bogoliubov transformation},description={\nopostdesc}}

\newacronym{WKB}{WKB}{Wentzel Kramers Brillouin}

\newglossaryentry{Wronskian condition}{name={Wronskian condition},description={\nopostdesc}}

\newacronym{QED}{QED}{Quantum electrodynamics}

\newacronym{IRHC}{IR-HC}{infrared hyperconductivity}

\newglossaryentry{tachyon}{name={tachyon},description={\nopostdesc}}

\newglossaryentry{baryogenesis}{name={baryogenesis},description={\nopostdesc}}

\newglossaryentry{baryon}{name={baryon},description={\nopostdesc}}

\newglossaryentry{pair}{name={pair},description={of particle/antiparticle}}

\newglossaryentry{pair production rate}{name={pair production rate},description={\nopostdesc}}

\newglossaryentry{induced current}{name={induced current},description={\nopostdesc}}

\newglossaryentry{Breitenlohner-Freedman stability bound}{name={Breitenlohner-Freedman stability bound},description={\nopostdesc}}

\newglossaryentry{Maxwell equation}{name={Maxwell equation},description={\nopostdesc}}

\newglossaryentry{backreaction}{name={backreaction},description={\nopostdesc}}

\newglossaryentry{dynamical backreaction}{name={dynamical backreaction},description={\nopostdesc}}

\newglossaryentry{semiclassical}{name={semiclassical},description={\nopostdesc}}

\newglossaryentry{reproducibility}{name={reproducibility},description={\nopostdesc}}

\newglossaryentry{fine-tuning}{name={fine-tuning},description={\nopostdesc}}

\newglossaryentry{doomsday argument}{name={doomsday argument},description={\nopostdesc}}

\newglossaryentry{supersymmetry}{name={supersymmetry},description={\nopostdesc}}

\newglossaryentry{Lovelock theorem}{name={Lovelock theorem},description={\nopostdesc}}

\newglossaryentry{quintessence}{name={quintessence},description={\nopostdesc}}

\newglossaryentry{Killing vector}{name={Killing vector},description={\nopostdesc}}

\newacronym{LTB}{LTB}{Lemaître-Tolman-Bondi}

\newacronym{LRS}{LRS}{Local Rotational Symmetry}

\newglossaryentry{fractal}{name={fractal},description={\nopostdesc}}

\newacronym{QFC}{QFC}{Quantum Field Cosmology}

%% file: Preface.tex
\chapter*{Preface}
\epigraph{A man gazing at the stars is proverbially at the mercy of the puddles in the road}{A.~Smith, 1863 \cite{smith1863dreamthorp}}
\section*{Cosmology: a voyage to the depths of our cosmos}
My journey to the tree of knowledge started in my infancy. I will not lie and say that I always wanted to be astrophysicist or even physicist but I was always fascinated by the extremes, what is the hottest place on earth, the furthest point from our house and when I started to study general relativity, particle physics and cosmology, I felt the question to be answered by those disciplines were the most extreme ones the human being could come up with and I became soon addicted to those questions. On my climb to this tree, I had the chance to have wonderful people on my side that I will name in some details in the acknowledgment section.

This manuscript is the final product of three years of research about some selected topics of modern cosmology. I choose to link the questions I was trying to answer during this period under the same banner ``On early and late phases of acceleration of the expansion of the Universe''. This title can be read at several level which are reflected through this manuscript: on the first level, indeed our universe is expanding. This first point is now a matter of fact for the undergraduate students taking a general relativity or a cosmology class but it is worth contemplating this statement a little bit more. Our universe is expanding means that we are able to talk about the larger scales: we, humble humans being managed to develop tools and concept in order to discover facts happening at the border of our universe. In the part \ref{part:introcosmo}, I gathered with more or less mathematical details some knowledge, we possess about our cosmos. A thesis is a nice opportunity to include whatever I want with some scientific material, the tone and the writing of this thesis reflect my personal approach to science with a given percentage of mathematics, philosophy and of course physics. I also decided to try to sprinkle this thesis with a number of quotes, which might or might not be linked with the main text. They could reflect my state of mind while working on a every-day life, be openly provocative, let the reader to meditate or be just there for free.
\section*{Why bother about reading this thesis?}
I tried to write the part \ref{part:introcosmo} of this thesis in a textbook-like style so that already at a undergraduate level, the reader could learn something out of it. Furthermore, the part \ref{part:introcosmo} can be read independently but I tried to give enough motivation for the reader to go beyond the general knowledge of part \ref{part:introcosmo} and continue to discover the story told in this thesis. One common denominator in the presentation of my research works is the quest for simplicity and logical articulations. I tried to dislocate the mathematical calculations I performed to go to the essence of their realization. It is one of the guideline of this thesis which is specifically present in section \ref{sec:Newtcos} \ref{sec:dista} \ref{app:fried} \ref{sec:massless} \ref{sec:smestim}. Unlike research papers, I started with a oversimple calculation compared to the final result to present in order to grasp, the main steps, ideas and mathematical tricks of the calculations I performed. I hope this presentation reflects in some sense the researcher every-day life, when he tries to grok concepts for his next works.

The motivation to open this manuscript and start to read it could be nothing but the thirst to discover a point of view on the state of knowledge in cosmology in 2016. This thesis also put also into more context the research work I performed regardless of their nature. But maybe even more importantly my work are about the ``acceleration of the expansion of the universe'' which is, I believe, general enough to be called a burning topic of cosmology and physics and hence is one more motivation not to continue reading this thesis. More technical reasons are given in the manuscript but in short, while an expanding universe is a natural thing to expect in the context of general relativity. It has been an astonishing surprise when we measured that our universe is nowadays in an accelerated expansion. On the top of that it was so also in the past therefore the ``natural scenario'' of a decelerating expanding universe has been only a transitory phenomenon in the cosmic history. Even for this general presentation of my work, I will still define what I mean by ``natural'': the idea is that gravity described by general relativity is attractive so tends attenuate the effect of the expansion of the universe. The origin of the problem points already to a possible solution, maybe we are not understanding gravity on the larger scales, this idea will be developed through this thesis in connection with quantum phenomenons in chapter \ref{chap:early} and \ref{chap:back}. Other attempts to explain accelerated expansion phases developed in this thesis are about the role of the large scale (possibly fractal) structures (chapter \ref{chap:LTB}) and of microscopical interactions between cosmological constituents of our universe (chapter \ref{chap:IDE}).
\phantomsection
\addcontentsline{toc}{chapter}{List of publication}
\section*{List of publication}
On a pure factual point of view, this manuscript puts the following articles into context:
\begin{itemize}
\item E.~Bavarsad, C~Stahl, and S.-S~Xue, Scalar current of created pairs by Schwinger
mechanism in de Sitter spacetime, Phys.~Rev., vol.~D \textbf{94}, 2016. \cite{Bavarsad:2016cxh}
\item C.~Stahl and E.~Strobel, Semiclassical fermion pair creation in de Sitter spacetime, proceeding of the second Cesar Lattes meeting, 2015. \cite{Stahl:2015cra}
\item C.~Stahl, E.~Strobel, and S.-S.~Xue, Fermionic current and Schwinger effect in de
Sitter spacetime, Phys.~Rev., vol.~D \textbf{93}, 2016. \cite{Stahl:2015gaa}
\item C.~Stahl, E.~Strobel, and S.-S.~Xue, Pair creation in the early universe, proceeding of MG14, 2016. \cite{Stahl:2016qjs}
\item C.~Stahl and S.-S.~Xue, Schwinger effect and backreaction in de Sitter spacetime, Phys.~Lett., vol. B\textbf{760}, 2016. \cite{Stahl:2016geq}
\item C.~Stahl, Inhomogeneous matter distribution and supernovae, Int.~J.~Mod.~Phys.,
vol.~D\textbf{25}, 2016. \cite{Stahl:2016vcl}
\item D.~Bégué, C.~Stahl, and S.-S.~Xue, A model of interacting dark energy and supernovae, 2017. \cite{withD}

\end{itemize}
Beside, I hope this final product will not reduce to its factual definition. Regarding the article \cite{Bavarsad:2016cxh,Stahl:2015cra,Stahl:2015gaa,Stahl:2016qjs}, I reorganized all their content in order to show the connections between them and most of this material can be found in chapter \ref{chap:early}. For the other chapters, their corresponding publications have been indicated.
\phantomsection
\addcontentsline{toc}{chapter}{Acknowledgements}
\chapter*{Acknowledgements}
\epigraph{Buon pomodoro}{Anonymous}
\paragraph*{}
Mes premières pensées vont naturellement à ma famille et plus particulièrement mes parents qui m'ont guidés et encouragés dans les voies que je choisissais. Peut-être par pudeur, peut-être pour éviter les jaloux ou les oublis, je vais éviter d'énumerer toute ma nombreuse famille qui compte et sur qui j'ai toujours pu compter.
\paragraph*{}
Je remercie mon directeur de thèse qui a su créer un grand institut où tellement de personnes interessantes ont eu la chance de pouvoir travailler.
Eckhard Strobel who was my first collaborator and friend in Italy. Ich habe zu viel von dir gelernt. Charly et Eugenia, che bello tempo abbiamo passato qui, dedicated anche al vostro pequeño non ancora nato ma già famoso! The Pescara team was full when Fernanda and Hendrik were in town, our soup days and lasagna nights were a blast and will be remembered for long, with all those discussions about physics, life, dreaming...one time we've even sung! More than thanking you, I would like to wish you that the wind will always blow in your sails!
\paragraph*{}
I am also thankful to the professors I met to have accompany me throughout all this PhD. Prof.~Belinski for always interesting discussions, accurate opinions about modern science and for accepting to recommend me. Prof.~Xue who helped me to find a topic of research and was of good advices about the physical interpretation of the results. Prof.~Bavarsad with whom we started a fruitful collaboration. Prof.~Vereshchagin: very calm and rigorous when doing science. Mais aussi Pascal qui est plein de bons conseils et d'enthousiasme.
\paragraph*{}
Many PhD students were of great company in all the parts of the globe: Maxime (je dois te remercier pour tellement de choses que je vais me contenter de te remercier pour l'ice tea que tu m'as payé un jour plus brumeux que les autres à Trastavere), Disha, Tais, Damien (et toute ta tribu), Ivan, Xiaofeng (hope to eat strange stuffs and watch chinese opera with you one day in the future), Soroush, Rahim, Rashid, Gabriel (for bringing back some life in Pescara), Wang Yu, William, Juan David, Jose, Diego, Laura, Daniel, Alexander, Karen, Vahagn (carton!), Gabriel (viva matemática), Daria, David, Andreas, Mile, Cristina, Carlos, Eduardo, Jonas, Grasiele, Iarley, Kuantay (professor!), Bakytzhan, Yerlan, Giovanni (next trip soon!), Pavel, Marco, Milos, Camilo, Gerardo (alternatif!), Marina.
\paragraph*{}
Oltre i ricercatori, ch'è a Pescara un paio di personne che voglio ringraziare: un padrone di casa amichevole e compiacente, molti amici (Matteo, Francesca, Fabrizio, Marco, Giovanna, Andrea) e segretari efficaci e simpatici: Cinzia, Maria (sì al referendum), Gabi, Marzio, Cristina, Francesca, Silvia e naturalmente Federica: sei una persona meravigliosa, piena di qualita e ti auguro di trovare la felicità e di essere sempre radiosa. 
\paragraph*{}
Je dois aussi remercier tous mes amis des échecs: Manu, Oli, Pat', Louise, Alex, pour la partie alsace, P-L (goeter), Vruno, Mike, Tibo (je suis charni!), François (la serrure!), Manouk, Marc, Vince, Youssef (à quand le prochain match en 25 parties à 5h du mat'?), Maxime (et Sophie et Marco!), bébé Xavier, Jose (un jour on retournera à Lille!), Marianne, JB, pour la partie Paris.
\paragraph*{}
Bien sûr une petite pensée pour les amis de la fac: Marty (qui a contribué à cette thèse en doublant le nombre d'acronymes), JB, Anaïs, Guillaume, Yann, Satya, Paul T, Valentin, Bart, Marie-Co, Maud, Denis, Sophie ainsi que pour les professeurs qui m'ont guidés, Prof.~Contaldi, Parentani, Mathieu (qui m'a aussi prouvé qu'il pouvait être là dans les moments durs), Julien, Stefano et tous les autres.
\paragraph*{}
Et une dernière couche avec les amis d'Alsace: Ludan, Pit, Mytch, Maty, Léa, Soro, Rogar, Eric, Nico, Vince, Roger, Bebert, Mutchler, Clotaire, Louis.
\paragraph*{}
And finally, my last thoughts go to you the reader. I guess the first people to read the final version will be the members of my PhD defense commission: Jean Audouze, Paolo De Bernardis, Massimo Della Valle and Nikolaos Mavromatos, together with their substitutes Antonio Capone, Manuel Malheiro and Piero Rosati. But for also all the following readers, I wish you to find inside this thesis inspiration for all your future projects and hope you will conduct them well.
\newpage
\section*{Conventions}
\epigraph{It strikes me that mathematical writing is similar to using a language. To be understood you have to follow some grammatical rules. However, in our case, nobody has taken the trouble of writing down the grammar; we get it as a baby does from parents, by imitation of others. Some mathematicians have a good ear; some not (and some prefer the
slangy expressions such as “iff”). That’s life.}{Jean-Pierre Serre, letter to David Gross \cite{Serre}}
The mathematical conventions for this thesis will be mainly recalled when used if there is any ambiguity. However some general notations can be discussed here.
\paragraph*{}
The sign $\equiv$ denotes that we define a quantity.
The thesis deals with de Sitter spacetime that will be abbreviated $\ds$. For D-dimensional de Sitter spacetime, the following notation will be used $\ds_D$ in particular $\ds_2$ $\ds_3$ and $\ds_4$. We will work at some point in a $D$ dimensional spacetime, the corresponding number of spatial dimensions is defined as $d$ and for standard Lorentzian spacetimes we have $D= 1+d$.
\paragraph*{}
A vector will be written in bolt, as for instance in equation (\ref{eq:newtosecondlaw}), whereas the spacetime $D-$vectors and the scalars will not be discriminated. In most of the problems considered, a spherical symmetry exists, so that the quantities depend only on the magnitude of a given vector. We define the magnitude of a vector as $k=\sqrt{\textbf{k}^2}$ for $d \geq 2$. Sometimes we will use directly the magnitude without explicit reference to the vector itself. As a consequence vectorial equations will be directly transformed into scalar ones.
\paragraph*{}
The convention for the signature of the spacetime metric is (+ - - -). Regarding indices, we use the greek indices for spacetime indices and letters $i,j, k$ for general spatial indices. The Einstein summation convention that repeated indices are summed is also very often employed. We label the spatial dimension with arabic-persian numerals, e.g.~$1,2..$, and that letters, e.g.~$x,y..$, are reserved for Fourier space.
We will assume also in most of the cases $\hbar=c=1$, sometimes we will however restore them for clarity.
\paragraph*{}
The super or subscript denotes a present day measure of a given quantity, example $H_0$.
A quantity into squared brackets means that we consider the dimension of this quantity, for instance [k]= dimension of $k$. Beside two quantities into squared bracket denotes the commutator of those two quantities, we define it as: $[a,b]=ab-ba$. The anticommutator is defined as $\{a,b\}=ab+ba$. There exists a corresponding definition with Einstein notation, that we will be introduced in a footnote in after equation (\ref{riemann}). A quantity around chevrons eg. $<x>$ denotes an average quantity.
\paragraph*{}
A dot denotes a derivative, with respect to cosmic time or more generally with respect to $t$, eg. $\dot{a} = \frac{da}{dt}$. A prime denotes a derivative with respect to the radial coordinate $r$ in chapter \ref{chap:LTB}, that is $R'(r,t)=\frac{\partial R}{\partial r}$. However in all the other chapters, the prime is reserved for a derivative with respect to the conformal time defined in (\ref{time}). With Einstein notations, partial derivative operators are indicated in short by $\partial_{t} \equiv \frac{\partial}{\partial t}$.
\paragraph*{}
The limits or asymptotic expansion are usually presented with the following notation: $f(z) \underset{|z|\rightarrow\infty}{\sim}$ which corresponds well to a physicist view. Sometimes the notation $\lim_{z\rightarrow\infty} f(z)$ will be also used in more mathematical contexts. In chapter \ref{chap:early} and \ref{chap:back}, we introduced notations which might not be familiar to a physicist working in the field as we consider an electric field during inflation. Those notations are summed up in equations (\ref{lambda}), (\ref{kappa}) and (\ref{eq:mu}). For chapter \ref{chap:early} and \ref{chap:back}, the knowledge of those new notations should be enough to understand any equation because all the rest of the notations are pretty canonical or the terms used in a given equation are defined in its local neighborhood.
\paragraph*{}
A dagger, such as in equation (\ref{eq:dagger}), denotes the double action of transposing and conjugating an object, that is $\psi^ {\dagger} \equiv \text{ }^T\psi^*$. The identity matrix will be written $1$. The factorial is defined for $n \in \mathbb{N}, n! \equiv \prod_{i=1}^{n} i$. Its generalization to the complex plane is the Gamma function that we will represent as usual with $\Gamma$.

%% file: cours/L1.tex
\vspace{2cm}
\epigraph{Si vous avez jamais passé la nuit à la belle étoile, vous savez qu’à l’heure où nous dormons, un monde mystérieux s’éveille dans la solitude et le silence. Alors les sources chantent bien plus clair, les étangs allument des petites flammes. Tous les esprits de la montagne vont et viennent librement ; et il y a dans l’air des frôlements, des bruits imperceptibles, comme si l’on entendait les branches grandir, l’herbe pousser. Le jour, c’est la vie des êtres ; mais la nuit, c’est la vie des choses. Quand on n’en a pas l’habitude, ça fait peur...}{Les étoiles. Récit d'un berger provençal.
Alphonse Daudet, 1869 \cite{daudet1869lettres}}
\vspace{1cm}
\begin{center}
\includegraphics[width=0.8\textwidth]{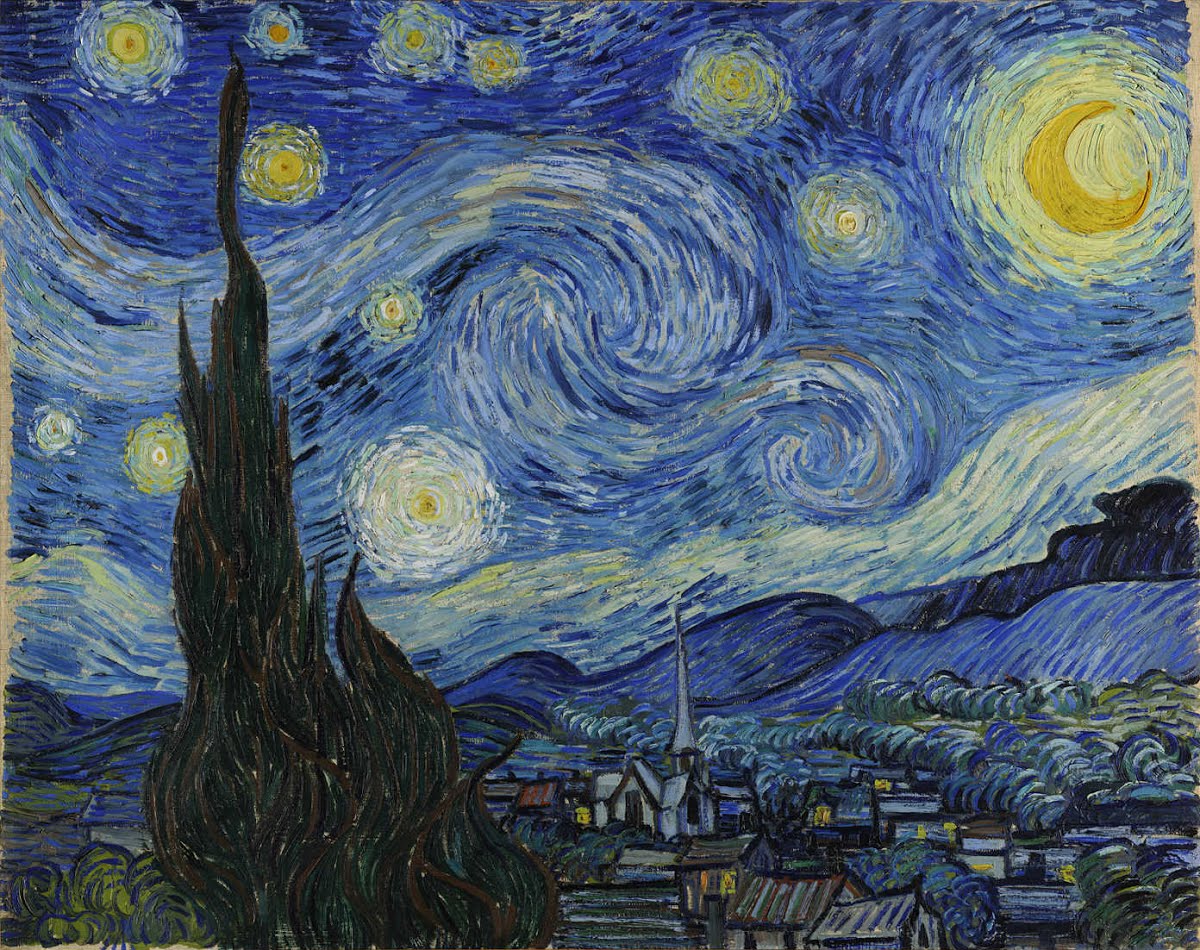}
\\
La nuit étoilée, Vincent Van Gogh, 1889
\end{center}
\chapter{An object of study being 14 billion years old}
\label{chap:introcosmo}
\epigraph{Aber selbst Gedanken, so substanzlos sie scheinen, brauchen einen Stützpunkt, sonst beginnen sie zu rotieren und sinnlos um sich selbst zu kreisen; auch sie ertragen nicht das Nichts.}{S.~Zweig, Schachnovelle, 1943 \cite{zweig2013schachnovelle}}
\paragraph*{}
\textit{This chapter aims at giving an overview of the current state of knowledge in cosmology with an emphasis on the standard model, its assumptions and its tremendous successes but also the questions it raises and its limits. After defining scientific cosmology, introducing a couple of required tools and concepts to study the universe as a whole, we will describe the cosmic history of universe mainly with a bird's eye view. As science is made by human beings, this chapter will be plagued by historical digressions in order to, I hope, give a more colorful picture of the fascinating quest to unveil the mysteries of our universe.  Many more material can be found in standard cosmology textbooks such as \cite{peebles1993principles,Dodelson:2003ft,Mukhanov,Peter,lyth2009primordial}}
\section{Historical overview and Foundations of Cosmology}
\subsection{Cosmology as a science}
\paragraph{}
Cosmology is a science that studies the Universe as a whole. It seeks to understand the nature of the Universe, as well as its origin, evolution, fate but also its structure and its laws. It is only recently that cosmology became a science. Before that, numerous (if not all) civilizations offered their idea about the organization of the World and its origin and end. Several philosophers also discussed this matter, but none of this was supported by a scientific method. With the advent of \gls{general relativity} (GR), cosmology soon became a science, and predictions that could be tested started to be made.
\paragraph{}
One could argue that cosmology is not a science, because it lacks one essential feature: the \gls{reproducibility} of experiment. If one formulates a model of the Universe, there is no way to create the corresponding Universe in the lab to see how it evolves! Cosmology is only a science in the sense that it is based on observational facts, but also on fundamental principles which are then confronted to observation. To escape from the lack of \gls{reproducibility}, statistical tools based on some ergodic principle are heavily used.
\subsection{First observational facts}
At the beginning of the twentieth century, astronomers had be looking at the sky for centuries and were believing that our Universe was entirely composed of our galaxy, the Milky way. One of the first to identify extra-galactic light was \gls{Slipher} in 1914. He studied \gls{nebulae} in the sky, and more precisely he studied the spectral shift of these objects (usually now called \gls{redshift}). His results were astonishing: these objects were moving very fast (hundreds of km.s$^{-1}$), but also mostly going \emph{away} from us.  He concluded that these objects were outside our galaxy : \gls{Slipher} just discovered that the Universe is gigantic \cite{Slipher:1917zz}. His results were later confirmed by Edwin \gls{Hubble}, in 1924. But as we will see later, \gls{Hubble} discovered another surprising fact: the Universe is expanding. Indeed, at this time, the Universe was believed to be static and eternal. To correctly describe this observed expansion, we need a tool that was brand new at that time: \textit{\gls{general relativity}}.
\subsection{General relativity for cosmology}
\label{sec:GRforC}
\paragraph{}
\Gls{general relativity} is currently the best theory to describe gravitational processes and was proposed by Albert \gls{Einstein} in 1915 \cite{Einstein:1915by,Einstein:1915ca,Einstein:1916vd}. This amazing fact is never emphasis enough: during the past century, generations of physics tried their best to falsify \glslink{general relativity}{GR} and failed. Despite all the ingenious tests, smart physicists invented to challenge \glslink{general relativity}{GR}, the theory always passed all the tests. Most of the tests are conducted in the solar system, for instance the founding principle of \glslink{general relativity}{GR}, the \gls{equivalence principle} is tested measuring the fractional differences of masses different freely falling bodies. They are parametrized by the so-called $\eta$ parameter: $\eta \equiv \frac{m_G-m_I}{m_G}$, where $m_G$ is the mass which goes into the gravitational force in Newton gravity and $m_I$ is the inertial mass which goes into Newton second law. They are postulated to be equal in Newton theory. The current observational bound from torsion balance experiments is $\eta < 2 \times 10 ^{-13}$ \cite{Wagner:2012ui}. More solar system tests can be found in \cite{Joyce:2014kja}. While in the solar system, the constraints are drastic, on the larger scales, \gls{general relativity} is way much less constrain and most of the test are mixing gravitational properties and matter properties of our universe. Hence there is room on cosmological scale to consider different theories than \gls{general relativity}. 
\paragraph{} A technical toolbox for computing the most important quantities of \glslink{general relativity}{GR} will be given in Sec.~\ref{sec:Einst} and here we will assume the existence of \glspl{Einstein equation} governing gravitational effects on large scales and discuss cosmological consequences.
\paragraph{Question:} Which solution to \glspl{Einstein equation}, best describes our Universe? \\ More modestly we could ask, which solution is simple enough to do calculations and is at least appreciatively correct\footnote{The word correct is defined here as in agreement with the observational data.}. To do so of course, one assumes the existence of astrophysical and cosmological data. \\
The answer to this question is not easy for several reasons. First the answer is time dependent in the sense that the cosmological model depends on the amount of observational data. Second as already discussed, unlike other sciences, we live in the object we are trying to describe! So we cannot rerun the Universe and reproduce an observational fact. Third we occupy a given spacetime position that we have not chosen. The observational data that we receive are located on our past light cone and hence are located on a part of a 3 dimensional hypersurface whereas we are trying to make statement about a 4-dimensional \gls{manifold}. Hence it might be that several spacetimes are compatible with the observational data but worth the interpretation depends on the spacetime considered to interpret the data.
\paragraph{Observable universe vs Universe} From the previous considerations, it is good to distinguish the \emph{observable universe} for which, by definition, we have data, from the \emph{Universe} which contains zones we do not have access to (in the sense that our past or future light cone will never intersect with the ones of those zones). Making statements about the Universe based on data on the observable universe needs of course extra-hypothesis, philosophical premises and other assumptions. Also careless cosmologists (including the author of this thesis) sometimes do not bother precising if they consider the observable universe of the Universe. Or they do not explicit which sort of extra-hypothesis they make to go from one to another. The context most of the time helps but sometimes the confusion is real. We will detail in Sec.~\ref{sec:cosmprin} the main working hypothesis of cosmology. But before that, after all those cautious remarks, we will give elements of answer to the question asked above and to do so, we review the birth of some solutions to \glspl{Einstein equation}, a toolbox to derive \glspl{Einstein equation} for a given metric is given in Sec.~\ref{sec:Einst}.
\paragraph{}
 The first cosmological solution was proposed by \gls{Einstein} himself in 1917 \cite{Einstein:1917ce}. One of the difficulty is that his theory (which relates the content of the Universe to its geometry) is highly non-linear. Doing some \gls{symmetry} assumptions, the problem can be greatly simplified. The most standard hypothesis is that our Universe is isotropic and homogeneous (see Sec.~\ref{sec:cosmprin} and Sec.~\ref{sec:groupth} for a more formal presentation of \gls{Einstein universe} and many more other universes). \gls{Einstein universe} followed the prevailing philosophical view at that time that our universe was static. To obtain a static solution to his equations, he introduced a \gls{cosmological constant} to enforce a steady-state universe. \gls{Einstein} could have predicted the expansion of the Universe which is nowadays a known observational fact (we will dedicate Sec.~\ref{sec:expofu} to it). \gls{Einstein} called this ``his biggest blunder''. However, the observation of the expansion didn't reject the possibility of the existence of a \gls{cosmological constant}, only the fact that its value must not cancel the expansion. The \glslink{SM}{standard model} of cosmology in fact consider a \gls{cosmological constant} to account for the acceleration of the expansion of the universe.
\paragraph{}
Willem \gls{de Sitter} also proposed a model of Universe in 1917 \cite{deSitter:1917zz}, which is a solution of \glspl{Einstein equation} of \gls{general relativity}. It is a homogeneous, isotropic and matter-less Universe, but filled with a \gls{cosmological constant}. \glslink{dS}{de Sitter} Universes are usually considered for \gls{inflation} scenarii, it will be the central object in the part \ref{part:early} devoted to early universe physics, we will introduce it in great details in Sec.~\ref{sec:strongintrodS}.
\paragraph{}
The derivation of the general cosmological solutions comes back to Alexander \gls{Friedmann} and Georges \gls{lui} in 1922 \cite{friedmann1922125,Friedmann:1924bb} and 1927 \cite{Lemaitre:1927zz} respectively.
\newpage
\shadowbox{\begin{minipage}{0.95\textwidth}
\section*{Alexander Friedmann (I/II)}
\paragraph*{}
As it is poetically expressed in \cite{magueijo2004faster}, it is kind of a running joke in the physics community that some Russian scientists claims the paternity of scientific discoveries over the western world. With the advent of internet, the harmonization of the scientific language and the fact that discoveries are nowadays much less the masterpiece of one unique person, the question of ``who was the first'' is gradually becoming an obsolete one, only bothering old jaundiced scientists eager for fame. However in the twenties, it was kind of a big deal and for once the ``russian'' camp was right: one of their fellow indeed was the first man deriving correctly the general equations governing an expanding universe. 
\paragraph*{} \piccaption*{Alexander Friedmann}
\parpic{\includegraphics[width=0.35\textwidth]{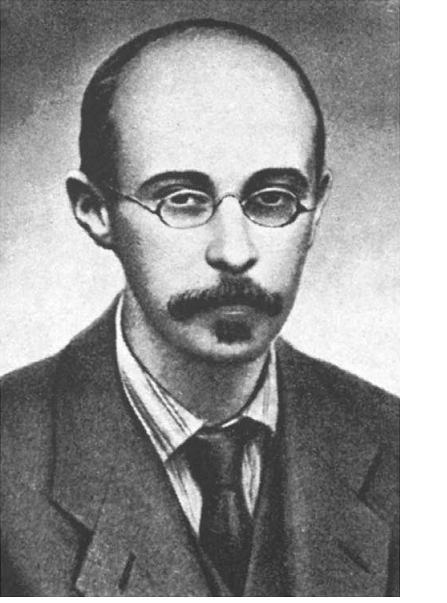}} Alexander \gls{Friedmann} (1888-1925) was one of those bright minds, workaholic with a vast field of interest: meteorology, aeronautics, fluids dynamics, mechanics and of course cosmology, who deserves to be better known. It is his supreme mathematical abilities conjugated to an inexhaustible energy which allow him to perform sharp research in so many topics in the twenties. While reading his biography \cite{biofried}, I felt like this man lived several lives. Born in an artist family as his father was ballet dancer and his mother pianist, he soon showed an outstanding ability for mathematics and physics. In 1905, in high school, he was very active politically, but for his entrance at the university, probably more interested in mathematics, he decided to focus on his study. In annual reports that he had to write to justify a grant that he was receiving from the university, one can note the incredible number of books the young Friedmann was studying. At the time of his graduation, in 1914, while everybody was seeing him working calmly in an office on physical and mathematical problems, he surprised everybody by engaging in the army. He revealed himself as a brave soldier trying to devote all his skills for his country, developing ballistics for bombs targeting and often going to fly to empirically test his theories. In a letter to a friend, he ironically described his daily experiences on February 5, 1915: ``My life is fairly quiet apart from such happenings as a shrapnel exploding at a distance of twenty steps, the explosion of a detonator of an Austrian bomb at a distance of half a step, when I got off nearly shot free, and a fall on my face and head resulting in nicking my upper lip and suffering some headaches. But you one gets used to all of this, of course, especially seeing things around which are thousand times more awful.''
\end{minipage}}
\newpage
\shadowbox{\begin{minipage}{0.95\textwidth}
\section*{Alexander Friedmann (II/II)}
\paragraph*{} Quiet, \gls{Friedmann} was more known for his acts than for his temperament which might be explained by the horrors he saw on the front. Due to the revolution, he came back in 1917 to more peaceful activities with his usual energy. On his cruise rhythm, he was both involved in administrative tasks as he founded many new research institutes, he would at the same time take at least three full-time jobs of teacher and, of course, was, on the top of that, performing cutting-edge research. In 1922, when he turned his working power to the newly born field of \gls{general relativity}, he started by writing monographs in Russian so that the next generations don't miss the opportunity to learn the crucial developments happening. He also, in the meantime, tried to play around with the newly born \glspl{Einstein equation}, in particular applying them to the whole universe. However unlike \gls{Einstein} and \gls{de Sitter}, he did not assume at the first place that the universe was in a steady-state, static, and you can imagine his surprise when he discovered that the universe might be expanding. However, trusting his mathematics, he submitted his result to the most popular journal of physics at that time: reference \cite{friedmann1922125}. It is at that time that \gls{Einstein} enters into play with the bad role for once. Still mired into his idea of a \glslink{Einstein universe}{static universe}, he tried to show that in addition to be philosophically dubious, Friedmann's solution was mathematically incorrect\footnote{Merely \gls{Einstein} wanted to show that the conservation of the \gls{energy momentum tensor} implied a static universe.}. He submitted two months after Friedmann article a nasty note\footnote{entitled Remark on the work of A. Friedmann (1922) ``On the curvature of space''} to the same journal mentioning a mathematical mistake in Friedmann's work. Having his scientific work criticized is never a good feeling, but having his scientific work heavily criticized by the great \gls{Einstein} himself, might have been kind of a stressful moment in Friedmann's life. After redoing and redoing his computations, Friedmann was formal: his calculation was correct and \gls{Einstein} was wrong. Due to the busy travel schedule of \gls{Einstein}, it is eventually almost one year after Friedmann's submission that \gls{Einstein} retracted his concerns in a new note send to the same journal. It was finally an underdog who emerged victorious of his controversy with \gls{Einstein} and made the universe expand!
\paragraph{} In 1925, Friedmann managed to find some time to break the world record (at that time) of balloon ascension by going to 7400 meters high (the previous record was 6400 m). During this flight Friedmann reports: ``While we were fussing with breathing of oxygen, an accident occurred. A deafening explosion was heard in complete silence, we looked up and saw that the balloon was all covered by smoke.  A quick thought occurred: we are on fire, so that chances for saving ourselves are very small. Then the smoke dispersed and we found that our 'oxygen  trunk' fractured. This is what happened: at high altitudes when the pressure is low, the oxygen trunk expanded and broke up and moist gas escaped and cooled in the process, so that the moisture condensed in the form of a cloud which we took for smoke''. After putting back their feet on solid ground, they improvised a little conference for the peasants of the small town of Okoroki who were intrigued by the strange crew who just landed. However, if Friedmann survived the ambush in the balloon trip, the Grim Reaper found a way to get him some months later: he contracted the typhus on the way back from the honeymoon of his second marriage in Crimea. The achievements of Friedmann are remarkable and as Hubble that we will discuss later, he is one of human who radically changed our way to see our surrounding, on an equal footing with the great Giordano Bruno, Galileo or Copernicus.\end{minipage}}
\newpage
\subsection{Building a Universe and the cosmological principle}
\label{sec:cosmprin}
\paragraph*{Question:} How to obtain a scientific cosmological model? \\ The current model of our Universe, called \emph{\gls{SM}}
relies on several hypotheses:
\subparagraph{Hypothesis 1:} Gravitation is described by \gls{general relativity}. \\
\\
We already discussed in Sec.~\ref{sec:GRforC} some elements of \glslink{general relativity}{GR}. The adoption of \glslink{general relativity}{GR} implies two corollary. One must specify the geometrical content of the universe and the matter content. It is the purpose of the \gls{cosmological principle} and of the next hypothesis respectively.
\subparagraph{Hypothesis 2:} Our universe is filled with \gls{dark energy}, \gls{dark matter}, \gls{baryon}ic matter and \gls{radiation}. \\
\\
We will detail in Sec.~\ref{sec:compoo}, the exact percentage of each fluid and their properties.
\subparagraph{Hypothesis 3:} The \gls{topology} of our universe is trivial. \\
\\
We will extend a bit the discussion here as this point is sometimes neglected in textbooks about cosmology. On the contrary to mathematics, physics is based on many, sometimes unwritten, ``physical'' rules such as \gls{locality}, separation of scales, microcausality, well defined initial value problems, no negative energies. Dropping some of those assumptions has to be done with great, great care, for instance if the energy is not bounded from below, the \gls{vacuum} would be unstable leading to a catastrophe for our every-day life: no matter would be stable! The \gls{topology} in mathematics is the classification of the non equivalent, different, shapes of a space. Applied to cosmology it consists in knowing the shape of a 1+3 D \gls{manifold} and is a often neglected question as all physical theories (\textit{eg.~}\glslink{general relativity}{GR}) are relying on \gls{locality} and microcausality\footnote{We will not dive here into the complex topic of the foundation of quantum mechanics} so do not say anything about the global properties and the global shape of our surroundings. As usually in cosmology, one always tries to do as simple as possible, hence the choice of Hypothesis 3. But we will list a couple of points to consider more elaborate shapes of space. More can be found about \gls{topology} for cosmology in the two reviews \cite{LachiezeRey:1995kj,Levin:2001fg}. \\
\\
If one consider the trivial \gls{topology}, as in the \glslink{SM}{standard model} of cosmology (hypothesis 3), we will see that three possibilities exists for the geometry of our spacetime: flat, hyperbolic or spherical. In the first two, it implies that the total space will be infinite. Whether our space is finite has been and will be a long lasting debate both on the physical and on the philosophical side of the question. Some are disturbed by an infinite space as many times in physics the presence of an infinite signal that something went wrong in a given theory, examples could be the \gls{big-bang} singularity, the need for renormalization in quantum field theory, the ultraviolet catastrophe before quantum mechanics... The Mach principle, stating that local physical laws are determined by the \glslink{S}{large scale structure} of the universe \cite{Hawking:1973uf} also motivates the existence of a finite universe. A last argument  \cite{Zeldovich:1984vk} for a finite universe comes from quantum cosmology as the probability of creating a universe of volume $V$ with a \gls{cosmological constant} $\Lambda$ out of the \gls{vacuum} behaves as $\mathcal{P} \propto \exp \left(- \frac{\Lambda V}{8 \pi \ell_P^2} \right)$ \cite{Peter}. \\
\\
In Appendix \ref{ap:topo}, the mathematical basis to study the different topologies of our universe are sketched. Possessing now a classification of the different spacetimes, a fair question is ``can it be tested with existing data sets?'' In non trivial topologies, one could observe multiple images for a given source as several geodesics link observer and the source. However the fact that the source evolves and hence the two different images would correspond to different phases of the life of an object leads to difficulties to identify it as a unique object. So far no strong constraints on the \gls{topology} of the universe were found by looking in galaxy survey or astrophysical objects \cite{Peter}. The \acrlong{CMB} (see Sec.~\ref{sec:status} for an introduction to its role in cosmology) which was emitted at the same instant in the whole universe offers a non local information and therefore is royal to study the \gls{topology} of the universe. Changing \gls{topology}, changes the boundary conditions of all physical quantities ; the way they change is given by the \gls{holonomy group} (see Appendix \ref{ap:topo}). The main signature of \gls{topology} is the existence of correlations in the anisotropies of the \acrfull{CMB}. The signature of a non-trivial \gls{topology} would be the presence of circles in the sky but despite a vigorous search, the circles were not found in the anisotropies of the \acrshort{CMB} favoring hypothesis 3 \cite{Ade:2013vbw}.
 \subsubsection{Principles...}
\paragraph{} To the three hypothesis discussed before, one usually adds a \gls{symmetry} assumption on the large scale matter distribution of the universe: \textit{the \gls{cosmological principle}}, which is a \gls{symmetry} assumption.
\subparagraph{Definition: (Cosmological principle)} On large scales ($ \mathcal{O} (100)$ Mpc), our universe is spatially  homogeneous and isotropic.
\subparagraph{Definition: (Copernican principle)} We do not occupy a preferred spot in the universe.
\paragraph{} 
The observations of the \acrshort{CMB} \cite{Ade:2015xua} and galaxies surveys \cite{Tegmark:2003ud,Seljak:2004xh,Abbott:2005bi} advocate for an universe statistically isotropic around us\footnote{As a bonus, we note the following: the quantum properties of space-time suggest that spacetime itself could present a foamy structure. A recent paper \cite{tamburini2011no} based on quasars observations towards different areas of the sky, show the absence of any directional dependence of quantum gravity effects, then extending the isotropy observation to the smaller scale. Obviously one of the assumption of this result is a parametrization of quantum gravity effect, see equation 1 of \cite{tamburini2011no}.}. Within these observations, usually one adds a \emph{principle} to determine which class of cosmological solutions describes our Universe. Above, we presented two principles which are related as the \gls{cosmological principle} can be deduced from a weaker principle: the \gls{Copernican principle} together with an hypothesis of isotropy. The \gls{cosmological principle} is very ambitious because it conjectures on the geometry of the (possibly) infinite Universe whereas the Copernican\footnote{While Copernicus (1473-1543) has most of the credit for the introduction of the heliocentrism, Aristarchus of Samos (310-230 B.C.) also introduced a heliocentric model.} principle applies only to the observable universe. Even though ``principle'' is an appealing wording, it reflects nothing but a confession of lack of knowledge on the matter distribution on large scale. Indeed, homogeneity cannot be directly observed in the galaxy distribution or in the \acrshort{CMB} since we observe on the past lightcone and not on spatial hypersurfaces (see figure \ref{fig:horizpro} for a graphical illustration). An interesting study might be to consider this lack of knowledge in the context of the information theory and try to quantify how much information one can extract with idealized measurements and if this permits to draw any cosmological conclusion. Note \cite{Celerier:2011zh} that \glslink{SM}{FLRW models} can only be called Copernican if one considers our spatial position. Considering \glslink{SM}{FLRW models} from a fully relativistic point of view: that is considering the observer in the four-dimensional spacetime, the model cannot be called Copernican. Indeed, whereas our position in space is not special, our temporal location is. Within the  \glslink{SM}{$\Lambda$CDM model}, this caveat is called the \gls{coincidence problem}, we will come back to it in Sec.~\ref{sec:coincp}. We will discuss again the \gls{Copernican principle} in the context of inhomogeneous spacetimes in Sec.~\ref{sec:whyinhomo}.
\paragraph{}
The hypothesis presented previously gave the frame for our cosmological model which is sustained by three main observational pillars:\\
\\
-\textbf{The Universe is expanding}: It is the first discovered pillar and the simplest to understand mathematically. It allows also many philosophical discussions on the nature of our universe. As it will be important for the work I carried then about inhomogeneous cosmologies, we will discuss it in Sec.~\ref{sec:expofu} \\
\\
-\textbf{The \gls{BBN}}: With the help of nuclear and \gls{particle physics}, one can predict the relative abundance of atoms in the primordial universe. ($\sim$75\% H, 25\% He and a very small amount of lithium) This is a testable and tested on 10 orders of magnitudes prediction of cosmology. We will discuss it more in Sec.~\ref{sec:status}, see also the recent review \cite{Cyburt:2015mya}. \\
\\
-\textbf{The Cosmic Microwave Background}: The \acrlong{CMB} is the first light emitted by the Universe and studying its properties tells us about the primordial universe. Many information can be found in two recent reviews \cite{Kamionkowski:2015yta,Guzzetti:2016mkm} with a focus on the burning question of the primordial gravitational waves.
\section{The expansion of the Universe}
\label{sec:expofu}
\begin{figure}[h]
\begin{center}
\includegraphics[width=0.7\textwidth]{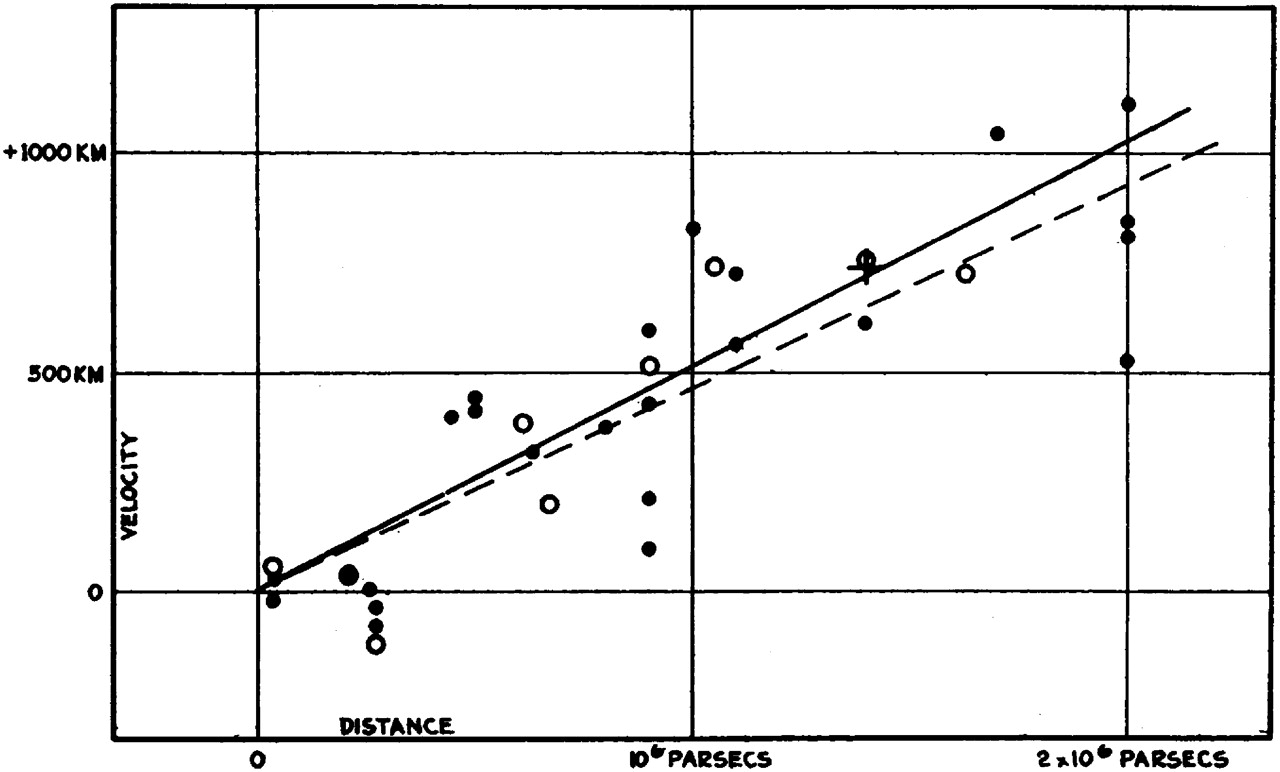}
\caption[Historical plot of Hubble's law]{The first \gls{Hubble diagram} illustrating \gls{Hubble law} from 1929 \cite{Hubble}. The radial velocity were determined with \gls{redshift} mesurements and are corrected from solar motion. Distances are estimated from involved stars and mean luminosities of \gls{nebulae} in a cluster. The black disks and full line correspond to the solution for solar motion using \gls{nebulae} individually; the circle and broken line represent the solution combining the \gls{nebulae} into groups; the cross represents the mean velocity corresponding to the mean distance of 22 \gls{nebulae} whose distance could not be estimated individually. It is amazing that \gls{Hubble} guessed a linear law between the distance and the velocity, moreover that the scale only is from 0 to 2 Mpc. In 1931, \gls{Hubble} and \gls{Humason} extended the measure up to 30 Mpc and found the same conclusion. This plot is to be compared with current measures, for instance figure \ref{fig:courbe}, where the distance ranges between 0 and 4200 Mpc!}\label{plotHubble}
\end{center}
\end{figure} 
\paragraph{}
During the twenties, \gls{the Great Debate} or Shapley-Curtis controversy was dividing the scientific community. The topic was to know whether our universe was only composed of our galaxy: the Milky Way or, was much more vast than our galaxy, following the ideas of Giordano Bruno and the concept of island universes of Immanuel Kant \cite{kant}. Vesto \gls{Slipher} gave the first blow using  several cepheid stars that he identified in some \gls{nebulae} (such as Andromeda nebula), to prove that these \gls{nebulae} were much too distant to be a part of our galaxy, and were in fact other galaxies: \gls{Slipher} discovered that the Universe was gigantic \cite{Slipher:1917zz}! In 1929, Edwin \gls{Hubble} discovered a linear relation between galaxies' spectral shifts $z$ and their distances (See Fig. \ref{plotHubble}).
The Doppler shift (\gls{redshift}) of an object is defined as:
\begin{equation}
\label{defredshift}
z \equiv \frac{\lambda_{obs}}{\lambda_{ref}}-1,
\end{equation}
 where $\lambda_{obs}$ is the observed wavelength and $\lambda_{ref}$ some reference wavelength. Using the standard formula for Doppler shifting, these \glspl{redshift} can be interpreted as the recession speed of the galaxies. This linear relation is now called \gls{Hubble law}\footnote{There is some controversy about who discovered this law. Indeed, Georges \gls{lui} published a paper in 1927 where he proposed that the Universe is expanding and suggested a value of this expansion rate. But it was \gls{Hubble} who brought an experimental confirmation of this hypothesis.} :
\begin{equation}
\label{hubble}
v= H_0 d,
\end{equation}
where $v$ is the recession speed of the galaxy, $d$ its comoving distance\footnote{In Sec.~\ref{sec:hubll} a derivation of \gls{Hubble law} together with definition of the different distances is given.} and $H_0$ is the proportionality coefficient, now called \gls{Hubble constant}. \gls{Hubble} estimated the value of this constant to be around $500$ km.s$^{-1}$.Mpc$^{-1}$. This value is now known to be overestimated due to an incorrect calibration. The first serious measurement of $H_0$ was done by Sandage in 1958 and gave a value of $H_0=75$ km.s$^{-1}$.Mpc$^{-1}$ \cite{hubleserisou}. Finally the current measurement with the \gls{satellite} Planck are $H_0=67.8 \pm 0.9$ km.s$^{-1}$.Mpc$^{-1}$ \cite{Ade:2015xua}. We will store the different numerical values useful for this thesis in table \ref{table:cosmnum}. The value deduced from Planck \gls{satellite} is rather low and in direct tension \cite{Verde:2013wza} with local measurements such as \glslink{SNe Ia}{supernovae measurements}. Some even argue that one should look for physics beyond \gls{general relativity} to explain the discrepancies \cite{Bernal:2016gxb}.
\newpage
\shadowbox{\begin{minipage}{0.95\textwidth}
\section*{Edwin \gls{Hubble}, the fame and galaxies chaser (I/II)}
\paragraph*{}
\piccaption*{Edwin \gls{Hubble}}
\parpic{\includegraphics[width=0.35\textwidth]{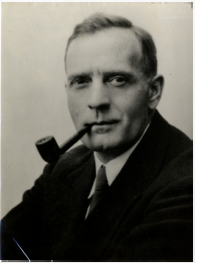}}
It is often said that accessing modern research for a person who did not study at the first place is a tremendous task\footnote{See \cite{blogpot} for a present-day view on that topic} and the vast majority of the research community followed a classical path: university studies with major in physics, graduation and so on. However, a few famous scientists are exception to this view. Edwin \gls{Hubble} (1889-1953) was one of those\footnote{Another example is John Moffat (1932- ) who stopped school at the age of 16 to become a painter. He left his native city Copenhagen to move to Paris where he lived the Boheme life, without any income. After some time of struggling, he decided to move back to Copenhagen and started studying physics. Impressively, he soon got a very good level and was starting \gls{general relativity} problems within a year. He started exchanging letters with \gls{Einstein} himself when he was around 20 year old. Moffat tells that the first letter was in german so he had to go to ask a barber of Copenhagen to translate it for him. When the local press started reporting the correspondence between this unknown young danish and \gls{Einstein}, many doors open for Moffat: the famous Niels Bohr gave him some attention. However, to start the first step of the research voyage: studying a topic during PhD, one usually needs the previous academic diploma: a master, diploma that Moffat did not possess. The solution came from Cambridge where all the rules are more elusive and ``to the peers assent'', and he finally graduated in 1958 under the supervision of Abdus Salam and Fred Hoyle. He is nowadays emeritus professor  at the university of Toronto and during his career, he touched many open and technical questions of gravity and cosmology, always with a bit of an offbeat view. João Magueijo reports also that they worked together on a theory of varying speed of light and that he was impress by many aspects of him, one being that he was a very conservative man, almost venerating Einstein theory of gravity, eventhough he was trying to challenge it on a everyday basis.}. Respecting his father wish to study law, he attended the university of Chicago and despite a B- average, he got a scholarship to study from 1910 to 1913 in Oxford. At that time, he reported that after saving a woman in distress after due to an unfortunate fell into a canal, he had, a few days later, to fight in a pistol duel with the jealous husband who was accusing them of flirting and wanted to defend the honor of the woman. Eventually \gls{Hubble} shot on purpose aside and so did his german contender. In trinity college, Oxford, he started to speak with an british english accent that he would maintain for the rest of his life, liberally adding to his sentences ``jolly good'', ``chap'', ``quite all right''. Some of his fellow scholar were joking that he would sometimes forget his accent in the middle of a sentence, so that he might take a bahhth in the bathtub. Back in America, he started wearing knickers, an Oxford cape and walked with a cane. While many report that he was mainly known in university for being a good athlete: baseball, football, basketball, tennis, swimming, boxing\footnote{While he was an undergraduate at the university of Chicago, he said that he was approached by a sport promoter who wanted to train him to fight the heavyweight champion of the world Jack Johnson. However the tales of his athletic prowess were often exaggerated in the literature, especially by \gls{Hubble} himself. He should also have boxed equal with the french national champion Georges Carpentier.}, one of his biographer Gale E.~Christianson defends that he never excelled as an athlete in college. At his father death, he started teaching spanish, mathematics and physics that he found some time to study between his others activities. Finally at the age of 25, he decided that he wanted to become a professional astronomer.
\end{minipage}}
\newpage
\shadowbox{\begin{minipage}{0.95\textwidth}
\section*{Edwin \gls{Hubble}, the fame and galaxies chaser (II/II)}
\paragraph*{} 
After rushing to finish his PhD about a photographic investigation of faint \gls{nebulae}, he got a position in the observatory of the mount Wilson, but first engaged to the US army who just started war again germany in 1917. Coming back victorious with the title of major, he finally started his job of astronomer in California, 20 km above Los Angeles. Edwin and his wife Grace socializes with many Hollywood stars and were often seen in company of Aldous and Maria Huxley, who were sharing their disdain for lower classes and were also anglophile. The Hubbles were also both racists saying mexicans were a ``mongrel race'' and using adjectives such as ``darky'' or ``pickaninny''. For professional, reasons, Edwin hired a publicist and finished once on the cover of the Time. When \gls{Einstein} came to visit the observatory of mount Wilson, \gls{Hubble} followed him everywhere so that the photographers could take many pictures together with him. One time, he got caught switching napkin rings in order to take a rival's spot at the head of the table. He also made up good heroic stories of him fighting a bear, saving two women from drowning... On his good sides, he loved animals, wanted zoo abolished and was a damn good astronomer, maybe the best of all time.

\paragraph{}
\piccaption*{a galaxy} \label{fig:galac}
\parpic{\includegraphics[width=0.4\textwidth]{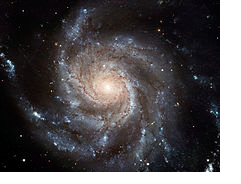}}
 Like Moffat, the fact that his background was not only scientific gave him greed and alternative view of the art. Helped by Milton \gls{Humason}, another school drop-out, originally in charge of the mules bringing the equipment for the observatory up in the mountain, \gls{Hubble} soon became at the edge of \gls{nebulae} observations. Hubble's telescope was so famous that some of his Hollywood social acquaintances, such as Charly Chaplin were begging to look into it. With practical skills, unbeatable flair and enthusiasm rather than solid academic knowledge, \gls{Hubble} tried a very unusual technique for the time: he installed his telescope inside a building and rotate it in order to exactly cancel the earth rotation. He could then, without looking into the telescope, target to fixed directions and thus use photographic plates with increased exposure time. You have to imagine that astronomy before the numeric era consisted in spending long lonely cold nights changing photographic plates of exposure time of order 45 minutes and then to develop and interpret them. It is after many sleepless nights at the top of the mount Wilson that \gls{Hubble} was able in 1923 to clearly make the statement that \gls{nebulae} were outside the milky wave, our galaxy. Actually other galaxies are not so small, some are even the size of the moon in the sky but it is \gls{Hubble}'s trick with the photographic plates which made them available to human's eyes. In 1929 he went further and showed that other galaxies were receding from us, pioneering the paradigm of the expending universe. Those two observations, of course, confirmed later by newer and more accurate observations\footnote{for instance the Pinwheel galaxy in the figure, from the Hubble telescope (ESA and NASA)} initiates changes comparable to Galileo, Copernicus: we were not the only galaxy in the universe. Actually, the Hubble Space telescope (2013) reported more than 225 billion galaxies. \\
Biographic elements based on \cite{magueijo2004faster,christianson1996edwin,teresi}
\end{minipage}}
\newpage
\epigraph{Gatsby, pale as death, with his hands plunged like weights in his coat pockets, was standing in a puddle of water glaring tragically into my eyes.}{F.~Scott Fitzgerald, The Great Gatsby, 1925 \cite{fitzgerald2003great}}
To finish this section on basics about the expansion of the Universe, we will discuss some popular misconceptions about the expansion of the universe.
\subsection*{In what is the Universe expanding ?} This question has been confusing generations of cosmologists because the answer is quite disturbing. A easy way to picture the expansion is to see galaxies as chocolate chips lying on a cake-universe which is expanding because of cooking. In this image, the cake is expanding in the oven but in actuality, the Universe do not need any other structure to expand in. \textit{The Universe is just getting bigger while remaining all that is.} For a pure homogeneous and isotropic Universe, one could represent the 4-dimensional curved Universe in a 5-dimensional flat space-time but there is not need for it and when small \glspl{perturbation} of this homogeneity are taken into account, this picture is no longer true. In short, as the universe is all, it expands into nothing.
 
\subsection*{Expansion is not only Doppler shift!} 
\begin{figure}[h]
\begin{center}
\includegraphics[scale=0.2]{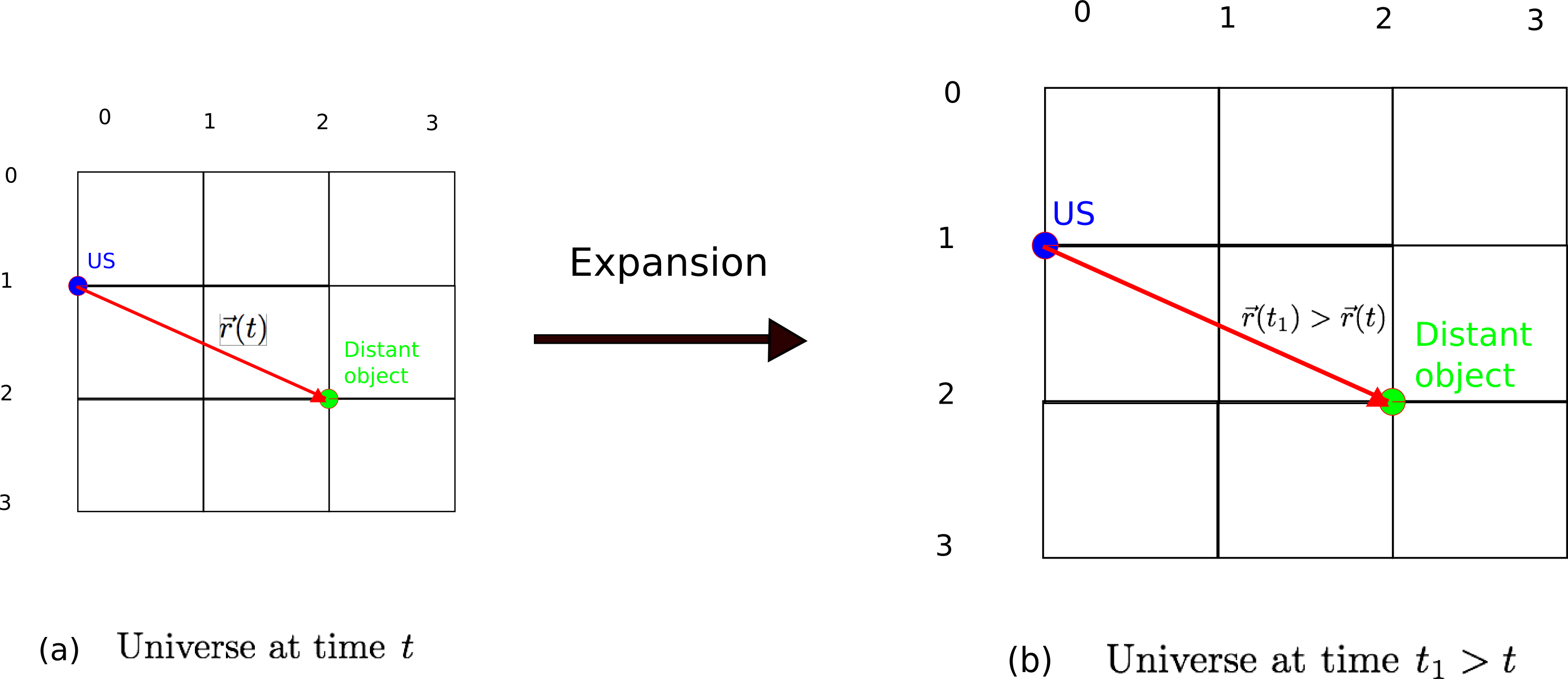}
\caption[Figure illustrating the expansion of the universe.]{Figure illustrating the expansion of the universe. The observer is not moving but it is the grid which is strechted. The distance evolves with the \gls{scale factor}: $r(t_1)= \frac{a(t)}{a(t_1)} r(t)$.}\label{expansion}
\end{center}
\end{figure}

One main difference between Doppler shift and \gls{redshift} is the fact that in the Doppler shift of an object (the siren of an ambulance for instance), there is a privileged direction: the one in which the object is moving. In the case of the \gls{redshift}, there is no privileged point in the Universe (\gls{Copernican principle}) and hence, the Universe expands equally about all of them: there is no center for expansion. See Fig.\ref{expansion} for an illustration that expansion do not imply any motion for the observers.
\subsection*{Can matter recede from us faster than light?} The answer is amazingly yes! For a spatial separation large enough, two observers on a time-like hypersurface can recede faster than light. No violation of special relativity is implied since this is not a local velocity and no information is transferred. As an example, the matter that emitted the \acrshort{CMB} was moving from us at a speed of $61c$ when it did so. It is however a point of view rather simplistic and Newtonian to understand the expansion of the universe as motions. 
\section{Newtonian Cosmology:}
\label{sec:Newtcos}
In this section, we will establish the equations governing an expanding universe using Newtonian physics. Let us consider a toy model universe which is:
\begin{itemize}
\item homogeneous
\item isotropic
\item filled with relativistic pressureless matter called ``dust'' in the literature\footnote{not to be confused with interstellar medium}
\end{itemize}
The point of studying such a model is that since it is very simple, the resulting equations will be simple. But it turns out that they are almost equivalent to the ones one obtained using \gls{general relativity} \cite{Wells:2014fia}. The key quantity to describe an expanding universe is the \emph{\gls{scale factor}} $a(t)$ which tells how distances are stretched (see figure \ref{expansion}). The comoving radius $\chi$ is the size of an object if no expansion of the space would be present.
\subsection*{Energy density conservation:}
Let us consider a sphere of radius R(t) and $\chi$ its comoving radius (\textit{cf.}~(\ref{eq:comv})):
\begin{equation}
R(t)=a(t) \chi.
\end{equation}
The matter is static with the expansion (no proper movement), so that the mass $M$ in the sphere is fixed. Therefore the matter density $\rho(t)=\frac{M}{\frac{4\pi}{3}R^{3}(t)}=\rho_{0}\left( \frac{R_{0}}{R(t)} \right)^{3}=\rho_{0}\left( \frac{a_{0}}{a(t)} \right)^{3}$. Remember that:
\begin{equation}
\label{eq:scaling}
\rho_{\text{matter}} \propto a^{-3}.
\end{equation}
As we will see in Sec.~\ref{sec:compoo}, each constituent of the Universe we model has a different evolution with $a(t)$ (\textit{cf.} table \ref{tab:flu}), (\ref{eq:scaling}) gives the one of matter which energy density is simply diluted with the third power of the \gls{scale factor} due to the three spatial dimensions. Deriving (\ref{eq:scaling}) with respect to time and using the definition of the \gls{Hubble constant} in term of the \gls{scale factor} given in (\ref{eq:H}) gives \textit{the density continuity equation of a ``dust'' Universe}:
\begin{equation}
\label{eq:cons}
\dot{\rho}(t)+3H(t)\rho(t)=0.
\end{equation}
It is possible to show that this equation is equivalent to the continuity equation in fluid dynamics:
\begin{equation}
\frac{\partial \rho}{\partial t}+\nabla \cdot (\rho\textbf{ v})=0.
\end{equation}

\subsection*{Acceleration equation}
\underline{Remark}: Since gravity is attractive, we expect to have a universe expanding at a decelerated rate \textit{ie.}~$\ddot{R}<0$.  Indeed, the presence of matter stifles the expansion of the universe. Let us consider a comoving test mass $m$ on the surface of the sphere. Let us christen the acceleration of the mass $\vec{\gamma}$. Newton's second law stipulates: 
\begin{equation}
\label{eq:newtosecondlaw}
\textbf{F}=-\frac{mMG}{R^{2}(t)} \textbf{u}_r,
\end{equation}
giving, due to the spherical symmetry:
\begin{equation}
\ddot{R}=-\frac{MG}{R^{2}}.
\end{equation}
This is the acceleration equation we were seeking. In fact it is usually written:
\begin{equation}
\frac{\ddot{R}}{R}=\frac{\ddot{a}}{a}=-\frac{4 \pi G}{3} \frac{M}{R^3}.
\label{eq:NFried2}
\end{equation}
Multiplying both sides by $2 \dot{R}$ and integrating yields to:
\begin{equation}
\frac{\dot{R}^2}{2}-\frac{G M}{R}=\text{cst}.
\label{EnergyConservation}
\end{equation}
This equation is similar to a mechanical energy conservation, with the first term looking like a kinetic energy per unit mass and the second a potential energy per unit mass. But most importantly, this equation reveals two very different situations:
\begin{itemize}
\item if ``cst'' $> 0$, the kinetic term dominates so the expansion will never end (loose system),
\item if ``cst'' $< 0$, the potential term dominates so the expansion will stop at some point and the Universe will collapse (bound system).
\end{itemize}
But this ``cst'' is not very intuitive. A little reformulation will make it more striking:
\begin{equation}
\eqref{EnergyConservation} \Rightarrow R^2 \left(\frac{H^2}{2}-\frac{4 \pi G}{3} \rho \right) = R_{0}^2 \left(\frac{H_{0}^2}{2}-\frac{4 \pi G}{3} \rho_{0} \right),
\label{AlmostFriedmann}
\end{equation}
where we have used the fact that since ``cst'' is constant, it is equal to its value today.\\
This way we see that the previous distinction reveals \textit{a natural scale for matter density} namely\footnote{when the right hand side vanishes}:
\begin{equation}
\label{eq:rhocont}
\rho_c \equiv \frac{3 H_0^2}{8 \pi G},
\end{equation}
so that, remarking from (\ref{AlmostFriedmann}) and (\ref{eq:rhocont}), that $\rho_c - \rho= \text{cst}$:
\begin{itemize}
\item if $\rho <  \rho_c$ the expansion will never end (loose system),
\item if $\rho >  \rho_c$ the expansion will stop at some point and the Universe will collapse (bound system).
\end{itemize}
Now, what is the dimension of this ``cst''? $[cst]=[R_0^2 H_0^2]=[\text{velocity}]^2$. But the only fundamental constant that has the dimension of a velocity is the speed of light c. So let us set by convention\footnote{the change of sign is simply to recover completely the usual notations}:
\begin{equation}
k c^2 = R_{0}^2 \left(\frac{4 \pi G}{3} \rho_{0} - \frac{H_{0}^2}{2}\right),
\end{equation}
with $k$ dimensionless.
\eqref{AlmostFriedmann} then becomes \textit{the first \gls{Friedmann equation}}:
\begin{equation}
\label{eq:NFried}
H^{2}=\frac{4\pi G}{3}\rho-\frac{kc^{2}}{R^{2}}
\end{equation}
This equation is the most important equation of cosmology. It is the equation of motion of the \gls{scale factor} $a(t)$. In other words it relates the expansion of the Universe $\{H(t)\}$ with its content $\{\rho\}$ and... a mysterious term with a parameter ``$k$''. Within Newtonian cosmology, how can we interpret this parameter ``$k$''? No clue so far! Note that we can reformulate this \gls{Friedmann equation} by defining: 
\begin{equation}
\label{eq:omeg}
\Omega(t) \equiv \frac{\rho(t)}{\rho_{c}}.
\end{equation}
Thus:
\begin{equation}
\label{eq:friedref}
H^{2}=H_{0}^{2} \left[\Omega(t)-\frac{kc^{2}}{R^{2}}\right].
\end{equation}
\underline{Remark:} The whole idea of cosmology is to consider a certain number of possible constituents in the Universe (like matter, \gls{dark matter}, \gls{radiation} or \gls{dark energy}) and to model the evolution in time of the proportions of each of them. The proportions are precisely the $\Omega(t)$'s, hence its importance. They will be discussed in Sec.~\ref{sec:compoo}. In this model we considered a full matter Universe ($\Omega=\Omega_{\text{Matter}}$) and we have:
\begin{equation}
\Omega(t)=\Omega_{0}\left( \frac{R_{0}}{R(t)}\right)^{3}=\Omega_{0}\left( \frac{R_{0}}{R(t)}\right)^{3}.
\end{equation}
That is:
\begin{equation}
\Omega_{\text{Matter}}(t) \propto a^{-3}(t).
\end{equation}
\subsection*{A toy model in the toy model: k=0}
Setting $k=0$ can be understood as a ``cosmological'' equipartition of the energy. This model is called \acrfull{EdS} which was introduced by \gls{Einstein} and \gls{de Sitter} together in a two pages paper \cite{Einstein:1932zz} once they accepted that the universe was expanding. It was the prevailing view of the universe until the end of the 20th century. The \gls{Friedmann equation} can be solved: we can deduce \glslink{scale factor}{$a(t)$} out of it. Indeed, we then have $\rho=\rho_{c}$ and $H^{2}=H_{0}^{2}\Omega_{M}$. We set in the previous equation $\Omega_{M,0}=1$. These equations can be integrated and we get:
\begin{equation}
\label{eq:matterbeha}
a(t) \propto t^{2/3},
\end{equation}
$H(t)=\frac{2}{3}t^{-1}$ so with the value of $H_0$ in table \ref{table:cosmnum}, we can get the age of the Universe:
\begin{equation}
\label{eq:wrong}
t_{\text{BB}}= \frac{2}{3H_0}=9 \text{ Gyr}.
\end{equation}
This is a huge problem since astronomers have seen stars aged of 12-14 Gyr! How can we  reconcile this experimental fact with the theory ? This time the problem comes from the drastic assumptions we have made in this toy model: only dust, Newtonian equation of motion. This wrong prediction will be rectified in equation (\ref{eq:tbbt}) and is an invitation to go further...!

%% file: cours/L3.tex
\section{FLRW Metric}
From here, we will assume that the reader knows some elements of general relativity, some definition will be given in Sec.~\ref{sec:Einst} but the purpose will be more a booster shot than a pedagogical introduction. The most general metric which satisfies the \gls{cosmological principle} is \cite{krasinski2007introduction,Peter}:
\begin{equation}
\label{FLRW1}
\mathrm{d}s^2=\mathrm{d}t^2 - R(t)^2 \left(\frac{\mathrm{d}r^2}{1-kr^2}+r^2\mathrm{d} \theta^2+r^2 \sin^2(\theta)\mathrm{d}\varphi^2 \right),
\end{equation} 
or equivalently :
\begin{equation}
\label{FLRW2}
\mathrm{d}s^2=\mathrm{d}t^2 - R(t)^2 \left(\mathrm{d}\chi^2 + f_k(\chi)^2 \left(\mathrm{d} \theta^2+\sin^2(\theta) \mathrm{d}\varphi^2 \right) \right),
\end{equation} 
where $k$, the measure of the curvature of space, it can take the values $-1,0$ or $+1$ and is directly related to the mysterious quantity of Sec.~\ref{sec:Newtcos} and can be interpreted only in a full \glslink{general relativity}{GR} treatment.  $t$ is the cosmic time and $R(t)$ represents the \gls{scale factor}. Observe that it is only a function of time and multiply the spatial part of the metric in order to encode a spatial expansion of the universe. (\ref{FLRW1}) and (\ref{FLRW2}) are two expressions of the \gls{FLRW metric}. We see that the assumption of isotropy and homogeneity lead to great simplifications for the line element which is now only dependent on a free time function $a(t)$ and one number $k$.
\paragraph{}
In (\ref{FLRW2}), if we look in a fixed direction (meaning $\theta$ and $\varphi$ are constants), the line element reduces to:
\begin{equation}
\mathrm{d}\ell=R(t)\mathrm{d}\chi.
\end{equation} 
We rewrite it now in terms of the \gls{scale factor} :
\begin{equation}
\label{comovline}
\mathrm{d}\ell=a(t)\mathrm{d}\chi,
\end{equation} 
with $R(t)= \frac{a(t)}{a_0} R_0$. Notice that in (\ref{comovline}), $\chi$ has now the dimension of a length. We can then rewrite \ref{FLRW2} :
\begin{equation}
\mathrm{d}s^2=\mathrm{d}t^2 - a(t)^2 \left(\mathrm{d}\chi^2 + g_k(\chi)^2 \left(\mathrm{d} \theta^2+\sin^2(\theta) \mathrm{d}\varphi^2 \right) \right),
\label{eq:hopla}
\end{equation} 
where $g_k(r)= R_0 f_k(\frac{\chi}{R_0}$).
\paragraph{About units and dimensions}
When writing \ref{comovline}, $\mathrm{d}\ell$ is the physical radial line element, and has the dimension of length. On the right side of the equation however, the dimension of length can be absorbed either in $\chi$ or $a(t)$. The choice is purely conventional but leads to slightly different formulations.
\begin{itemize}
\item If we chose the \gls{scale factor} $a(t)$ to be dimensionless, then $\chi$ corresponds to the comoving coordinate. We can then set $a(t_0) =a_0 = 1$.
\item If we choose the \gls{scale factor} to have the dimension of length, then $k$ can be normalized to $-1,0,+1$. This is what we used before.
\end{itemize}
\subsubsection*{Discussion on homogeneity (warming up for chapter \ref{chap:LTB})}
\paragraph*{} The assumption of homogeneity and isotropy are valid up to some scale called the isotropic and homogeneity scale. On an epistemological point of view, it is unsatisfactory to have a model for a universe which is homogeneous but does not have any built-in prescription for its scale of homogeneity.  A common order of magnitude for the homogeneity scale is hundred of Megaparsec but observations of larger and larger \glslink{S}{structures} are being reported in the past decades (Ref.~\cite{Balazs:2015xsa} for a recent example). Many people see them as Black Swan events but they could be more fundamental. It is furthermore interesting to note that the \gls{Hubble law} is starting to be true when the proper velocity of the galaxy can be neglected \emph{ie.} around 10 Mpc whereas the homogeneity scale is known to be much above ($\mathcal{O}(100)$ Mpc).

\paragraph*{} It has also been questioned whether the homogeneity assumption should be applied to \gls{Einstein tensor} (\ref{einsteinTenseur}) or the metric itself \cite{1963SvA.....6..699S}. It is indeed known that in the general case $<G_{\mu \nu}(g_{\mu \nu})> \neq G_{\mu \nu}(<g_{\mu \nu}>)$, where $<...>$ denotes spatial average. This idea is really hard to implement because \glspl{Einstein equation} are not \gls{tensor} equations after the averaging procedure (changing from \gls{covariant} to \gls{contravariant} indices alters the equations). Then only \glspl{tensor} of rank 0 and scalars would have well behaved average \cite{Buchert:1999er}. The consequences of this approach are still unclear \cite{Zalaletdinov:2008ts}.
\paragraph*{} Describing our Universe with a \glslink{FLRW metric}{FLRW spacetime} does not tackle the so-called \emph{\gls{Ricci-Weyl problem}}. This usual FLRW geometry is characterized by a vanishing \gls{Weyl tensor} and a non-zero \gls{Ricci tensor} whereas in reality, it is believed that light is traveling mostly in \gls{vacuum} where the \gls{Ricci tensor} vanishes and the \gls{Weyl tensor} is non-zero (see eg. Ref.~\cite{Fleury:2013sna} and references therein for more details). 
\section{Cosmography}
\label{sec:dista}
As the universe is expanding the distance between two objects can be expressed in different manners, which all agree on small scales. A good old earth bound observers, for instance, look back in time as looking further away simply because the speed of light is constant. Mainly without the dynamics of the universe (\gls{Friedmann equation}), we will give in this section a list of many cosmological distances and their relation one to each other. The notation are based on \cite{Hogg:1999ad} and for the every day life calculations of the cosmologists, many codes to compute the distance exist online\footnote{For instance \url{http://cosmocalc.icrar.org/}, \url{https://ned.ipac.caltech.edu/help/cosmology_calc.html}, \url{http://home.fnal.gov/~gnedin/cc/},\url{http://www.astro.ucla.edu/~wright/CosmoCalc.html}, \url{http://www.kempner.net/cosmic.php}, \url{http://www.icosmos.co.uk/}.}. A good understanding of those distance is important also to challenge the \glslink{SM}{standard model} of cosmology, as we will do in chapter \ref{chap:LTB} where we will compute the \gls{luminosity distance} in an inhomogeneous spacetime.
\subsection*{Hubble's law}
\label{sec:hubll}
\paragraph{}
We are going to see that from the \gls{FLRW metric}, we can obtain the \gls{Hubble law}. In other words, assuming that the Universe is homogeneous and isotropic, we can deduce the fact that it is expanding!
\paragraph{}
Consider an observer at point $O$ and an object at point $P$. We are going to write the distance traveled by a photon between $P$ and $0$. Looking in one fixed direction, the physical distance between $O$ and $P$ is the time of flight of a null-like geodesic ($\mathrm{d}s^2 =0$) and given by (\ref{comovline}) so we have, after integrating between $O$ and $P$:
\begin{equation}
r(t)= a(t) \chi.
\label{eq:hubbagain}
\end{equation}
Taking the derivative with respect to cosmic time of (\ref{eq:hubbagain}), we find:
\begin{equation}
\label{eq:hubbleagain2}
\dot{r}(t)=H(t)r(t),
\end{equation}
where $H$ is the \gls{Hubble constant} linked to the \gls{scale factor} by: 
 \begin{equation}
 H(t) \equiv \frac{\dot{a}}{a}.
 \label{eq:H}
 \end{equation}
We see by expanding (\ref{eq:hubbleagain2}) around the present time that we recover the \gls{Hubble law} (\ref{hubble}).
\subsection*{Redshift}
\label{sec:redshift}
In this section, following the definition of the \gls{redshift} given in equation (\ref{defredshift}), we derive it in term of geometric quantities, such a calculation is important as if one wants to propose a new metric aiming at describing cosmic observables, one needs to generalize such a calculation to the new proposed metric. This is what we have done for a inhomegeneous spacetime \emph{cf.}~(\ref{SNR}). 
To link the \gls{redshift} to the \gls{scale factor}, we consider again O and P related by (\ref{comovline}). If we integrate $\frac{\mathrm{d}t}{a(t)}$ between the time of the emission and the time of the observation, we get the comoving distance $\chi$ between the observation point (which we take to be the origin) and the source:
\begin{equation}
\int_{t_{\text{em}}}^{t_{\text{obs}}} \frac{\mathrm{d}t}{a(t)} = \int_0^{\chi} \mathrm{d}\chi =\chi.
\end{equation}
Now consider two consecutive light wave-fronts, emitted with a small time difference $\delta t_{em}$ and observed with a difference $\delta t_{obs}$
They must travel the same comoving distance, so we can write : 
\begin{equation*}
\int_{t_{\text{em}}}^{t_{\text{obs}}} \frac{\mathrm{d}t}{a(t)} = \int_{t_{\text{em}}+\delta t_{\text{em}}}^{t_{\text{obs}}+\delta t_{\text{obs}}} \frac{\mathrm{d}t}{a(t)} .
\end{equation*}
Now call $F$ a primitive of $a^{-1}$. Then 
\begin{eqnarray*}
  \int_{t_{\text{em}}}^{t_{\text{obs}}} \frac{\mathrm{d}t}{a(t)} - \int_{t_{\text{em}}+\delta t_{\text{em}}}^{t_{\text{obs}}+\delta t_{\text{obs}}} \frac{\mathrm{d}t}{a(t)}  & = & 0 \nonumber \\
   & = &  F(t_{\text{obs}}+ \delta t_{\text{obs}}) - F(t_{\text{em}} + \delta t_{\text{em}}) - F(t_{\text{obs}}) + F(t_{\text{em}}). \nonumber \\
\end{eqnarray*}
Since $\delta t$ is small we can Taylor expand at first order :
\begin{equation*}
0 = F'(t_{\text{obs}})\delta t_{\text{obs}} - F'(t_{\text{em}}) \delta t_{\text{em}}  = \frac{\delta t_{\text{obs}}}{a(t_{\text{obs}})} - \frac{\delta t_{\text{em}}}{a(t_{\text{em}})},
\end{equation*}
which finally gives :
\begin{equation}
\frac{\delta t_{\text{obs}}}{a(t_{\text{obs}})} = \frac{\delta t_{\text{em}}}{a(t_{\text{em}})}.
\end{equation}
Note now that $\delta t = \frac{1}{\nu} = \lambda$, where $\nu$ is the frequency of the light and $\lambda$ its corresponding wavelength, we find then:
which gives : 
\begin{equation}
1+z = \frac{a(t_{\text{obs}})}{a(t_{\text{em}})} = \frac{1}{a(t)}.
\label{eq:scalered}
\end{equation}
In the last step we used the fact that $a_0$ can be set to $1$. 
So, provided we solve the \gls{Friedmann equation} to get \glslink{scale factor}{$a(t)$}, the \gls{redshift} is a measure of cosmic time! But the \gls{redshift}, is also directly related to observations: two main techniques to determine the \gls{redshift} exist.
\subsubsection*{Spectroscopic redshift} 
\begin{figure}[h]
\begin{center}
\includegraphics[width=0.7\textwidth]{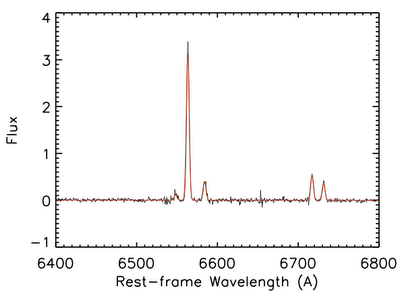}
\caption[Example of emission lines in a galaxy spectrum used to measure redshifts]{An example of emission lines in a galaxy \gls{spectrum}. Credit: Jeyhan Kartaltepe}
\label{fig:spectro}
\end{center}
\end{figure}

This technique consists in studying the \textit{\gls{spectrum}} of a galaxy: it shows how the light of a galaxy is distributed in terms of frequency or wavelength. With such details, it is possible to observe emission lines due to typical atoms or molecules in a galaxy, see figure \ref{fig:spectro}. 
By knowing from different independent experiments on earth, where the emission line or absorption lines are situated at rest, it is possible to obtain the \gls{redshift} with (\ref{defredshift}). Note that, as the effect of the \gls{redshift} is independent of the wavelength, the difficulty lies in identifying to which atom or molecule the line belongs and as a consequence two lines are usually enough to obtain a result. Historically, the \gls{spectrum} was placed on transparent table and adjusted with a wheel to match the known \gls{spectrum} and obtain the \gls{redshift}. Nowadays, the method are less archaic and the instruments \acrshort{VIMOS} \cite{Lilly:2006va,2016arXiv160505503S} and \acrshort{KMOS} \cite{Wisnioski:2014xwa} on the Very Large Telescope or \acrshort{MOSFIRE} \cite{Kriek:2014ita} on the Keck observatory are among the most modern instruments to determine a \gls{spectroscopic redshift}. However, obtaining a \gls{spectrum} is a difficult task: for most of the spectrographs, the light coming from the object to observe is diffracted through a slit or a circle. The incidence angle is then related to the wavelength of the photon. For \acrfull{SDSS} \cite{Kessler:2009ys}, for instance, in order to iterate that procedure, one needs to change the plaque with the holes, which usually takes time: one has to choose during an observation for which objects, the \gls{spectrum} is wanted. For \acrshort{MOSFIRE} \cite{Kriek:2014ita}, the replacement of the plaque is done automatically and takes (only) a couple of minutes. The most modern versions like \acrshort{MUSE} \cite{Karman:2014sqa} and \acrshort{SINFONI} \cite{2012A&A...539A..91C} are able to obtain spectra for any pixel of a 2D map of the sky but the data analysis has been reported to be painful. The \gls{spectrum} is the only quantity to precisely determine the \gls{redshift} but another technique acting as a proxy for the spectroscopic determination exists: the \gls{photometric redshift}.
\subsubsection*{Photometric redshift}
\begin{figure}[h]
\begin{center}
\includegraphics[width=0.65\textwidth]{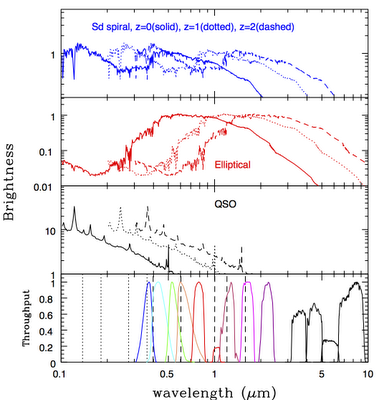}
\caption[Example of a spectral energy distribution used to measure redshifts]{Examples of spectral energy distribution, one can see the evolution due to redshift for various galaxies, the bottom figure represents the different filter bands used to determine the spectral energy density, from \url{http://candels-collaboration.blogspot.it/2012/08/how-far-away-is-this-galaxy.html}.}
\label{fig:SED}
\end{center}
\end{figure}

\begin{figure}[h]
\begin{center}
\includegraphics[width=0.65\textwidth]{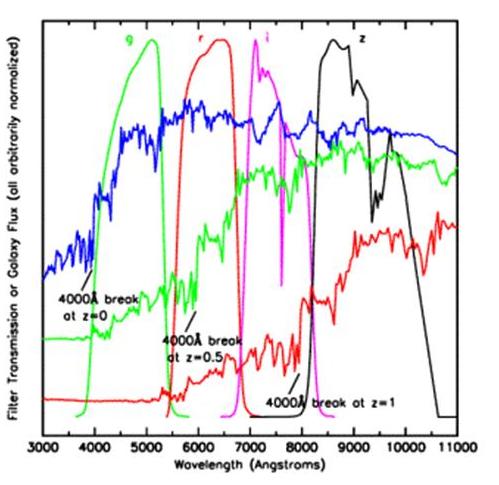}
\caption[Illustration of the 4000 \AA~break]{Spectral energy density of three galaxies. Illustration of the 4000 \AA~break. Credit: DES collaboration \cite{Abbott:2005bi}}
\label{fig:4000}
\end{center}
\end{figure}

For a \gls{photometric redshift}, one considers the \textit{spectral energy distribution}, which is again the light of a galaxy but taken with given filters and with less much details, it is somehow averaged, integrated, with a given binning. The border between spectroscopic and \gls{photometric redshift} is now becoming fuzzy with the advent of technologies for filter: some extra-fine filters are now designed to see specific lines, for instance the $H\alpha$ lines. Even if the spectral lines are not present, it is possible to determine the \gls{redshift} by fitting to existing well-known galaxy and stellar population models (for instance \cite{2003MNRAS.344.1000B}) and identifying known features in a galaxy such as the \gls{Lyman break} or the 4000 \AA ngström break.  Physically, the \gls{Lyman break} is due to the fact that the radiation above the Lyman limit of 912 \AA~is almost totally absorbed by the neutral gas around a star forming region of a given galaxy. Of course those effects can be displaced due to the expansion of the universe so for instance for $z=3$, the \gls{Lyman break} appears at 3600 \AA. See figure \ref{fig:SED}, for a complete illustration of a spectral energy density distribution.
 The 4000 \AA ngstrom break happens because, in a given galaxy, some metals\footnote{In astronomy, a metal is defined as any element but hydrogen or helium.} in the atmosphere of old stars absorb radiation from younger stars at 4000 \AA, see figure \ref{fig:4000} for a graphical illustration. Depending on the stellar population of a galaxy, the break can evolve: as the opacity increases with decreasing stellar temperature, the 4000 \AA ngstrom break gets larger with older ages, and it is largest for old and metal-rich stellar populations. The metallicity is of minor influence for populations with ages less than 1 Gyr \cite{2003MNRAS.344.1000B}. A second event reinforces the 4000 \AA ngstrom break: the Balmer break at 3646 \AA, it is the end of the Balmer serie and is the strongest in A-type main-sequence stars.
\paragraph*{} In big galaxy surveys, such as the dark energy survey \cite{Abbott:2005bi} or \acrshort{SDSS} \cite{Kessler:2009ys}, the \gls{redshift} is measured only with photometric techniques for several objects at the same time. As this \gls{redshift} determination relies on specific models of galaxy which include many parameters difficult to constrain, the uncertainties on the \gls{redshift} determination are bigger and can even reach $\delta z =0.5$ for a single object. Those failures are known as \textit{catastrophic outliers} but happen only in 1 \% of the cases. Statistically the photometric method for \gls{redshift} determination is robust: for more than 10 filter bands, one finds $\frac{\delta z}{z} \sim 1 \%$. To finish, the \gls{photometric redshift} is usually the first step before dedicating telescope time to produce the \gls{spectrum} of the object to observe.
\subsubsection*{Redshift drift}
An important effect for cosmology is the \gls{redshift drift} \cite{Sandage}. Let us assume we observe a cosmological object, typically a galaxy at two different times: $t_0$ and $t_0+\delta t_0$. The time difference will induce a \gls{redshift} difference of:
\begin{equation}
1+z+\delta z = \frac{a(t_0+\delta t_0)}{a(t+\delta t)}.
\end{equation} 
Expanding to first order in $\delta t$ and $\delta t_0$ and using the fact that light travel on null geodesics, that is $\frac{\delta t}{a(t)}= \delta t_0$, we find:
\begin{equation}
\frac{\delta z}{\delta t_0}=H_0 \left((1+z)-\frac{H(t)}{H_0} \right).
\end{equation}
At small redshift, assuming the \glslink{SM}{standard model} of cosmology, that the value of $H_0$ in table \ref{table:cosmnum}, we find $\frac{\delta z}{\delta t} \sim 10^{-8}$/century which is small and has not been detected so far. However unlike other type of measurements this formula does not depend on the evolution or the properties of a source so it gives direct information about $H(t)$. On the contrary the observations with the \glslink{luminosity distance}{luminosity} or \gls{angular distance} are sometimes dependent on the assumption of the type of source, \textit{cf.}~the discussion about \glspl{standard candle} in the section about the \gls{luminosity distance}. A detection of the \gls{redshift drift} would be an independent novel approach to constrain the history of the universe. In inhomogeneous models (see chapter \ref{chap:LTB}) the prediction for the \gls{redshift drift} is very different and a measurement of the \gls{redshift drift} would be of tremendous importance to discriminate those models. See \cite{Steinmetz:2008gp} for the experimental prospects using laser combs and to, in term, measure the changing expansion of the universe in real time. This would be also a mine of information to characterize \gls{dark energy} that we will discuss in chapter \ref{chap:DE}.
\subsection*{Hubble's radius}
\paragraph{}
Because the \gls{Hubble constant} $H_0$ has the dimension of the inverse of a time, we can construct a distance from it, called the \emph{Hubble distance} or \emph{\gls{Hubble radius}} $D_H$:
\begin{equation}
D_H(t) \equiv \frac{1}{H(t)}.
\label{eq:Hdist}
\end{equation}
This \gls{Hubble radius} is a characteristic scale of the Universe. We can think of it as the distance at which the relativistic kinematics start to be non-negligible. The \gls{Hubble radius} will be used through out this thesis for qualitative reasoning.
\subsection*{Curvature radius of the Universe today}
\paragraph{}
Using the \gls{Friedmann equation} (\ref{eq:NFried}):
\begin{equation}
\label{eq:FRiedebut}
H(t)^2 = \frac{8 \pi G}{3}\rho(t)  - \frac{k}{R(t)^2} = \left(\frac{\dot{R}}{R}\right)^2 = \left(\frac{\dot{a}}{a}\right)^2,
\end{equation}
$\rho$ is not only the matter density: we must take into account all \glspl{radiation}, neutrinos, the \gls{dark matter}, the \gls{cosmological constant}, etc..
We can rewrite this \gls{Friedmann equation} in terms of a critical density $\rho_c = \frac{3 H_0^2}{8 \pi G}$.
Defining
\begin{eqnarray}
 \Omega_k(t) & \equiv & \frac{-k}{H_0^2 R(t)^2}, \nonumber\\
 &=& -k\left( \frac{D_H}{R(t)}\right)^2, \label{eq:oK}
\end{eqnarray} 
and using (\ref{eq:omeg}), the \gls{Friedmann equation} (\ref{eq:FRiedebut}) can then be written: 
\begin{equation}
H(t)^2 = H_0^2 \left[\Omega(t) + \Omega_k(t)\right].
\end{equation}
The interesting thing is that now, at $t=t_0$ :
\begin{equation}
\Omega(t_0) + \Omega_k(t_0) =1,
\end{equation}
$\Omega_k(t_0) = -k\left(\frac{D_H}{R_0}\right)^2$ is called the curvature parameter. With the parametrization, the \gls{Friedmann equation} (\ref{eq:FRiedebut}) is nothing but a constraint equation.
From this we can get the curvature radius of the universe today
\begin{equation}
R_0= D_H \sqrt{\frac{-k}{1-\Omega(t_0)}}.
\label{eq:curvatureradius}
\end{equation} 
\subsection*{Comoving distance}
We already used many times the comoving distance which is the distance between two objects moving through the \gls{Hubble flow}: they do not have any proper or peculiar velocity. Applied to one earth bound observer, it is sometimes called the line of sight and has already been be simply inferred in equation (\ref{comovline}) from the spatial part of metric (\ref{eq:hopla}):
\begin{equation}
\label{eq:comv}
D_C \equiv \chi = \int \frac{\text{d} t}{a(t)}.
\end{equation}
Using the changes of variables $\frac{\text{d}a}{\text{d} t}=a H$ (equation (\ref{eq:H})) and $a = \frac{1}{1+z}$ (equation (\ref{eq:scalered})), one can find:
\begin{equation}
\label{eq:comovingC}
\chi = \int \frac{dz}{H(z)}.
\end{equation}
While this distance is of theoretical importance, when it comes to discuss the notion of horizon in the cosmological spacetimes, for instance related to \gls{inflation}, it cannot be measured directly and two relativistic effects need to be taken into account in order to obtain a ready-to-measure distance: the curvature of space and the expansion of the universe.
\subsection*{Transverse comoving distance}
\paragraph{}
Consider two galaxies at the same \gls{redshift}. In order to account for the curvature of space, their transverse comoving distance is given by:
\begin{equation}
D_M = R_0 \left\{ \begin{array}{ccc} 
\sin \frac{\chi}{R_0} &\text{for } k=+1,   \\ 
\frac{\chi}{R_0} &\text{for } k=0, \\
\sinh \frac{\chi}{R_0} &\text{for } k=-1,
\end{array}  \right.
\label{eq:comovingdist}
\end{equation}
where $R_0$ is the curvature radius of the universe defined in (\ref{eq:curvatureradius}).
\subsection*{Angular distance}
\begin{figure}[h]
\begin{center}
\includegraphics[width=0.8\textwidth]{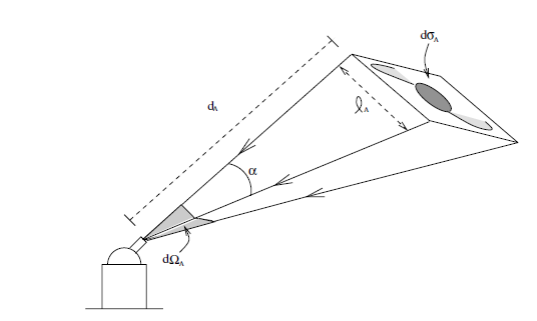}
\caption[Definition of the angular distance]{Angular distance defined as the ratio between the intrisic cross-sectional area of the source and the observed solid angle (Credit: \cite{Ribeiro:2004nk})}
\label{fig:distancegoet}
\end{center}
\end{figure}

The \gls{angular distance} is used to evaluate an object (eg. galaxy) with physical size $(\text{d} \sigma)$, it is the generalization to cosmological purposes of the astrophysical \gls{parallax}. The \gls{angular distance} is defined by :
\begin{equation}
D_A \equiv \frac{\text{d} \sigma}{\text{d} \Omega},
\end{equation}
where $\text{d} \Omega$ is the angle sustained in the sky by the object. See figure \ref{fig:distancegoet} for a graphical representation. 
This distance is sometimes called the diameter distance and is the \textit{physical} equivalent to $D_M$ as it takes into account the expansion of the universe.
We have the relation \begin{equation}
D_A= \frac{a(t)}{a_0} D_M = \frac{D_M}{1+z}.
\end{equation}
To obtain one measurement of the \gls{angular distance}, one needs to know the physical size $(\text{d} \sigma)$ of the object under consideration. An object of physical size known or measured by an independent experiment is called a \gls{standard ruler}. The canonical example of a \gls{standard ruler} is the \acrfull{BAO} scale. It corresponds to the maximal distance that a sound wave could travel in the early universe. It can be measured both the \acrshort{CMB} experiments and in galaxy surveys.
\subsection*{Luminosity distance}
The flux emitted by any luminous object is defined by strict analogy with the ``every day life''\footnote{$F=\frac{I}{4 \pi D_F^2}$, I being the intensity and D the distance.}. Let us take a candle (= a source) with intrinsic (and bolometric) luminosity L. The flux received at a distance d is :
\begin{equation}
F=\frac{L}{4 \pi d^2}.
\end{equation}
Therefore, we define the \gls{luminosity distance} $D_L$ by :
\begin{equation}
D_L \equiv \sqrt{\frac{L}{4 \pi F}}.
\end{equation}
Observe that, if one manages to find a family of astrophysical objects with the same intrinsic luminosity, measuring their flux is equivalent to measure their \gls{luminosity distance}. Such a family is christen \textit{\glspl{standard candle}}. The classical example of \gls{standard candle} is the \glslink{SNe Ia}{supernovae Ia} which will be used in chapter \ref{chap:LTB} and \ref{chap:IDE}. A \glslink{SNe Ia}{supernova Ia} is observed when a white dwarf\footnote{The nature of the progenitor is still not clear to the whole community \cite{2012NewAR..56..122W}.} reaches the Chandrasekhar mass ($1.4 M_{\odot}$), by accreting matter from a companion star, and runaway thermonuclear explosion. A priori, this phenomenon has no reason to depend on local environment, as the Chandrasekhar mass is known and after standardizing type Ia SNe via Phillips relation \cite{1993ApJ...413L.105P}, the intrinsic luminosity is fixed thus leading to the concept of \glspl{standard candle}. 

Other promising possibilities are \gls{supernovae II-P} which would be an independent way of probing the \glslink{SM}{standard cosmology} \cite{2007ecf..book...95D,2015MNRAS.448.2312B}, see also Sec.~5.3 of \cite{CristinaPhD}\footnote{available on demand} for an complete overview of this field. \gls{Gamma Ray Bursts} are also awaited for cosmology but so far the systematics are too high to make any strong statement \cite{2013IJMPD..2230028A,izzo2015new}.
\paragraph{Relation with other distances} We now derive a relation with the radial distance. To do that, we will have to understand how $L$ and $F$ evolve in the \gls{Hubble flow}, that is how it evolves for an expanding universe. The definition of $L$ is $L_S \equiv \frac{\Delta E_1}{\Delta t_1}$. One simplifying assumption is that the source is monochromatic, hence we have $\delta E_1= h \nu_1$. The observed luminosity is $L_0=\frac{\Delta E_0}{\Delta t_0}$ (where $\Delta E_0 = h \nu_0$).
\paragraph{} With (\ref{defredshift}), taking $\Delta t$ as the unit of time, $\nu_1 \Delta t_1=\nu_0 \Delta t_0$, it is possible to derive the relation:
\begin{equation}
L_0= \frac{L_S}{(1+z)^2}.
\end{equation}
The measured flux is $F=\frac{L_0}{\Sigma}$, where $\Sigma$ is the the surface of the sphere centered on a source S. Using (\ref{eq:hopla}), this surface is given by : $\Sigma= 4 \pi a_0^2 g_{k}^2 (\chi_0)$, thus $D_L=\sqrt{\frac{L_S}{4 \pi F}}=(1+z) a_0 g_k(\chi_0)$. Taking $a_0=1$ and putting everything together, we get the final relation:
\begin{equation}
D_L = (1+z) D_M.
\end{equation}
Using (\ref{eq:comovingdist}) and (\ref{eq:comovingC}), we have $D_L = (1+z) D_M$ and $D_A = \frac{D_M}{1+z}$. This gives the relation:
\begin{equation}
D_L = (1+z)^2 D_A. 
\label{eq:Etherigg}
\end{equation}
This equation is known as the \gls{distance duality relation} and it is valid for any space time, given that the metric that describes it has no torsion and that light travels along null geodesics. It is also a consequence of \glslink{distance duality relation}{Etherington reciprocity theorem} \cite{Ether} and is very useful because all the quantities involved in (\ref{eq:Etherigg}) can be measured and hence the assumptions of the theorem can be challenged: finding modified theories gravity or viable alternatives to the \glslink{SM}{standard model} of cosmology \cite{Uzan:2006mf}. Note that, it is not a smoking gun for a departure from \glslink{general relativity}{GR} as it has been shown to be valid also for Nordstrom's, $f(R)$ gravity. We will make use of this theorem in the derivation of the \gls{luminosity distance} for inhomogeneous spacetimes in Sec.~\ref{distances}, equation \ref{Th}. Interestingly $D_A$ has been defined as a geometric quantity whereas $D_L$ was defined by the received flux but these quantities are always related!
\subsubsection*{Distance modulus}
\paragraph{} In astronomy, the apparent magnitude $m$ is a measure of the brightness of the source with respect to a reference source\footnote{this reference flux is the one of the star Vega} :
\begin{equation}
m \equiv -2.5 \log \left( \frac{\text{Flux}}{\text{reference Flux}} \right).
\end{equation}
\paragraph{Absolute Magnitude} By definition, the absolute magnitude $M$ is the one that the source would have if it were located at 10 pc from the earth. The distance modulus is then defined to be:
\begin{equation}
\mu_B \equiv m_B - M,
\label{eq:magni}
\end{equation} the subscript B meaning bolometric (ie. integrated over all the frequency).
In practice, the magnitude are measured in a limited band of wave-length using filters. To take this into account, one usually adds at term to (\ref{eq:magni}) called k-corrections \cite{1996PASP..108..190K,Nugent:2002si}. Finally understanding well the distances is important to test the \glslink{SM}{Friedmann model} at the so-called background level as we will do in chapter \ref{chap:LTB} and \ref{chap:IDE}, indeed as we will see the observations of \glslink{SNe Ia}{supernovae Ia} gave many observational data for $D_L$ (or $\mu_B$) and then different cosmological models can be challenged using these data.

%% file: cours/L4.tex
\section{Toward constructing Einstein Field Equation}
\label{sec:Einst}
In this section, we will give some essential elements of \gls{general relativity}.  We will focus on the quantites which will help us to derive the \gls{covariant} generalization of (\ref{eq:NFried}) and (\ref{eq:NFried2}). Various introductions to GR can be found in \cite{d1992introducing,dirac1996general,lichnerowicz1955elements,Peter,krasinski2007introduction}
\subsection*{Calculation toolbox:}
\gls{Einstein}'s coup was to realize that physics should not be frame dependent so he introduced the notion of \gls{tensor} to the physics community. Those objects are quantities from differential geometry that transform in a specific \emph{\gls{covariant}} way under coordinates or frame transformation. \gls{Einstein} postulated that the equation governing gravity should be \gls{covariant} and of second order and so he wrote a very general second order \gls{tensor} to describe gravitational effects. \gls{Lovelock theorem}, states actually that it was the only one, see Sec.~\ref{sec:DEsolution}. This is what we will try to do also in a first time. To do so, one has to realize that the existence of \gls{tensor} is bound to the existence of \textit{curved} spacetimes. In curved spacetime, the derivative $\partial_{\mu}$ of a given quantity is in the general case not a \gls{tensor} anymore and, for instance for $v^{\mu}$ a given vector, one has to define a new \gls{covariant derivative}:
\begin{equation}
\nabla_\mu v^\nu \equiv \partial_\mu v^\nu + \Gamma^\nu_{\mu \alpha} v^{\alpha}.
\label{eq:cov}
\end{equation}
The $\Gamma^\nu_{\mu \alpha}$ are called the \glspl{Christoffel symbol} and have been precisely introduced to account for the non-tensoricity of the first term. By considering a general coordinates transformation, it is possible to obtain an expression for $\Gamma^\nu_{\mu \alpha}$ in term of the \gls{metric tensor} $g_{\mu \nu}$ and its derivatives:
\begin{equation}
\Gamma^{\alpha}_{\mu \nu}=  \frac{1}{2} g^{\alpha \beta} \left( \partial_{\nu} g_{\mu \beta} + \partial_{\mu} g_{\nu \beta} - \partial_{\beta} g_{\mu \nu}  \right).
\label{eq:defCHR}
\end{equation}
By definition, the \glspl{Christoffel symbol} encode some curvature effect as they would vanish in a non-curved or flat spacetime, however they are not \gls{tensor}. Out of them, it is possible to construct a \gls{tensor} which quantifies how curved a spacetime is the \gls{Riemann tensor} $R^{\mu}_{\delta \alpha \beta}$
\begin{equation}
\label{eq:Riemann}
[\nabla_\alpha, \nabla_\beta] v^\mu \equiv R^{\mu}_{\delta \alpha \beta}v^\delta,
\end{equation}
for $v^{\mu}$ a given vector. Using (\ref{eq:cov}) and (\ref{eq:defCHR}), a brute force calculation of (\ref{eq:Riemann}) gives an expression for the \gls{Riemann tensor} in term of the \glspl{Christoffel symbol}:
\begin{equation}
R_{\alpha \beta \mu} ^{\delta} = \partial_{\beta} \Gamma^{\delta}_{\alpha \mu} - \partial_{\alpha} \Gamma^{\delta}_{\beta \mu} + \Gamma^{\nu}_{\alpha \mu} \Gamma^{\delta}_{\beta \nu} -\Gamma^{\nu}_{\beta \mu}\Gamma^{\delta}_{\alpha \nu}.
\label{riemann}
\end{equation} 
The \gls{Riemann tensor} verifies the following properties. It:
\begin{itemize}
\item vanishes for a flat space because of \gls{Schwartz theorem} to swap derivatives,
\item is antisymmetric under the permutation of the last two indices,
\item is symmetric under the permutation of the first two index with the last two ones,
\item verifies the \glslink{Bianchi identity}{Bianchi identities} : $R^{\mu}_{(\nu \alpha \beta)}=0$ and $\nabla_{[\lambda} R_{\mu \nu]\alpha \beta}=0$\footnote{ the brackets denote the symmetrisation of the indices and square brackets the anti-symmetrisation, for instance $A_{(\mu  \nu)}=A_{\mu  \nu}+A_{\nu  \mu}$ or $A_{[\mu  \nu]}=A_{\mu  \nu}-A_{\nu  \mu}$ .}.
\end{itemize}
One can now define the \gls{Ricci tensor}\footnote{There is less information in the \gls{Ricci tensor} than in the \gls{Riemann tensor} since the \gls{Riemann tensor} has a traceless part called the \gls{Weyl tensor}. The \gls{Weyl tensor} describes gravitational effects in the absence of matter, for instance for a Schwarzschild black hole, the \gls{Ricci tensor} is zero but not the \gls{Weyl tensor}.}:
\begin{equation}
\label{ricci}
R_{\alpha \beta} \equiv R^{\mu}_{\alpha \mu \beta},
\end{equation} which is symmetric from the properties of the \gls{Riemann tensor}. We finally define the \gls{Ricci scalar}:
\begin{equation}
\label{courbure}
R \equiv R^\mu_\mu.
\end{equation}
We are now in position to perform the first step announced previously: find a candidate general \gls{tensor} which would describe gravitational phenomenon. It is called \gls{Einstein tensor} and is defined as:
\begin{equation}
\label{einsteinTenseur}
G_{\mu \nu} \equiv R_{\mu \nu} - \frac{R}{2} g_{\mu \nu}.
\end{equation}
From the \gls{Bianchi identity}, one can show that:
\begin{equation}
\nabla^{\mu} G_{\mu \nu}=0.
\end{equation}
Observe here that replacing $R \rightarrow R- 2 \Lambda$, the \gls{Bianchi identity} would still be fulfilled giving the most general \gls{Einstein tensor}:
\begin{equation}
G_{\mu \nu} \equiv R_{\mu \nu} - \frac{R}{2} g_{\mu \nu} + \Lambda g_{\mu \nu},
\end{equation}
which value should be taken to be proportional to $g_{\mu \nu}$ will be discussed in Sec.~\ref{sec:CCP} in the context of the \gls{cosmological constant problem}. This is the historical way \gls{Einstein} introduced the \gls{cosmological constant} $\Lambda$ to enforce a \glslink{Einstein universe}{static universe}. The \glspl{Einstein equation} reads:
\begin{equation}
\label{eq:Eisteq}
G_{\mu \nu} = 8 \pi G T_{\mu \nu},
\end{equation}
where $T_{\mu \nu}$ is the energy momentum of the matter content of the universe. \Glspl{Einstein equation} tell how the matter ($T_{\mu \nu}$) evolves regarding a given geometry ($G_{\mu \nu}$) but at the same time, the matter tells to the geometry how to curve. Observe that sending the \gls{cosmological constant} term to the right-hand side of the \glspl{Einstein equation}, we find that it corresponds to an energy momentum:
\begin{equation}
\label{eq:defCC}
T_{\mu \nu}^{DE}= \Lambda g_{\mu \nu}.
\end{equation}
We will see in Sec.~\ref{sec:compoo} that in the \glslink{SM}{standard model} of cosmology this \gls{perfect fluid} accounts for 68 \% of the content of the universe  today! However, with such a definition, nothing has been said about the nature of such a fluid, a proposed explanation for the nature of the fluid will be discussed in Sec.~\ref{sec:CCP} but we will see that this attempt leads to one of the wrongest prediction of all time!

With all those geometrical definitions, it is possible now to derive the \glspl{Friedmann equation}. It is performed in Appendix \ref{ap:intro}, where the assumption of perfects fluids containing the universe is made in (\ref{eq:perfectf}). The final result is:
\begin{align}
& H^2 \equiv (\frac{\dot{a}}{a})^2=\frac{8\pi G}{3}\rho-\frac{k}{a^2}+\frac{\Lambda}{3}, \label{eq:Fried1}\\
& \frac{\ddot{a}}{a}=-\frac{4\pi G}{3}(\rho+3P)+\frac{\Lambda}{3}. \label{eq:Fried2}
\end{align}
If one compares with the Newtonian results (\ref{eq:NFried2}) and (\ref{eq:NFried}), the presence of the relativistic pressure and of the \gls{cosmological constant} are the only difference. It is only in cosmological context that the pressure self gravity has been measured: for the acceleration of the expansion of the universe \cite{Narimani:2014zha} or in \glslink{BBN}{big-bang nucleosynthesis} \cite{Rappaport:2007ct}. In compact objects such as \glspl{neutron star}, the pressure self gravity has not been measured yet because of uncertainties on the equation of state \cite{Schwab:2008ce,Kamiab:2011am} 
\paragraph*{} Another difference between (\ref{eq:Fried1}), (\ref{eq:Fried2}) and (\ref{eq:NFried2}), (\ref{eq:NFried}) is the content of the universe. In (\ref{eq:perfectf}), \glspl{perfect fluid} were assumed but nothing was said on their number so:
\begin{align}
& \rho(t) = \sum_i \rho_i(t), \\
& P(t) = \sum_i P_i(t),
\end{align}
where the sum over $i$ is made over the different possible fluids. The pressure of single fluid is related to the energy density by the \gls{barotropic index} $w$ which is assumed to be constant for the \glslink{SM}{standard model} of cosmology:
\begin{equation}
P_i(t)=w_i \rho_i.
\end{equation}
See that (\ref{equarho}), the generalization of (\ref{eq:cons}) becomes in this case:
\begin{equation}
\label{eq:conservatoten}
\sum_i \left[\dot{\rho}_i(t)+3 H (1+w_i) \rho_i \right]=0.
\end{equation}
One possibility is that all the terms of this sum vanish. This would correspond to non-interacting fluids: no energy transfer between them. We will study models where an interaction between \gls{dark energy} and \gls{dark matter} is postulated in chapter \ref{chap:IDE}.
\section{The present composition of our universe}
\label{sec:compoo}
\begin{figure}[h]
\begin{center}
\includegraphics[width=1\textwidth]{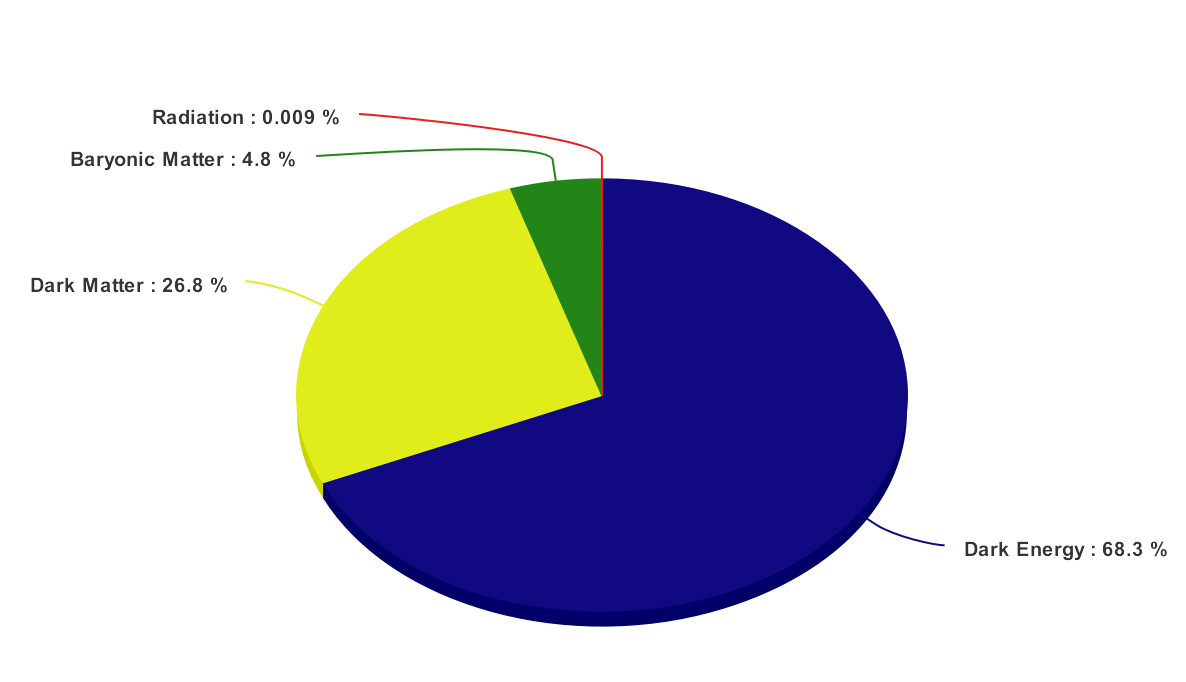}
\caption[The cosmic pie]{The current composition of our universe. The curvature is not displayed as discussed in the main text.}\label{fig:pie}
\end{center}
\end{figure}

In the case of single fluids, as in (\ref{eq:scaling}), it is possible to solve the conservation equation (\ref{equarho}), the solution reads:
\begin{equation}
\rho(t)= \rho_0 \left( \frac{a(t)}{a_0} \right)^{-3(1+w)}.
\end{equation}
It is possible to associate to the curvature a pressure and an energy density as $\rho_K(t)=\frac{-3K}{8\pi G a^2(t)}$. In the same manner, the \gls{dark energy}'s density is $\rho_{\Lambda}(t)=\frac{\Lambda}{8 \pi G}$. Their respective equation of state can be seen in table \ref{tab:flu}. The current \glslink{SM}{standard model} of cosmology, following Hypothesis 2 of Sec.~\ref{sec:cosmprin}, assumes that our universe is composed of \gls{dark energy}, \gls{dark matter}, \gls{baryon}ic matter, \gls{radiation} and ``curvature''. Table \ref{tab:flu} sums up the different cosmological solution for each fluid, and the behavior of the \gls{scale factor} for a universe composed of a single \gls{perfect fluid} which can be obtained by solving (\ref{eq:Fried1}) specified for a single fluid. 
\begin{table}[h]
\begin{tabular}{|c||c|c|c|}
\hline
fluid & equation of state parameter & $\rho(a)$ & $a(t)$ \\
\hline
cold dark matter & 0 & $\propto a^{-3}$ & $\propto t^{2/3}$ \\
baryonic matter & 0 & $\propto a^{-3}$ & $\propto t^{2/3}$ \\
radiation & 1/3 & $\propto a^{-4}$ & $\propto t^{1/2}$ \\
curvature & -1/3 & $\propto a^{-2}$ & $\propto t$ \\
cosmological constant & -1 & $\propto a^{0}$ & $\propto \exp(Ht)$ \\
\hline
\end{tabular}
\caption{Equation of state and cosmological behavior of different fluids}
\label{tab:flu}
\end{table}
Observing the third column of table \ref{tab:flu}, we can already infer a cosmic history, by looking at the dominating fluids through the cosmic history. It has been plotted in figure \ref{fig:coinc}. The \gls{radiation} is being diluted by the cosmic history by the fourth-power so it dilutes faster than matter. Such reasoning reveals that the universe was, deep in the past, in a hot dense phase dominated by \gls{radiation} and that then it undergoes a matter dominated phase and finally finished with a \gls{cosmological constant} dominated phase. A curvature dominated phase would be possible between the matter and \gls{cosmological constant} phase but we will see that the observation point toward a flat universe: the curvature being zero. Following the definition (\ref{eq:omeg}) and (\ref{eq:friedref}), applied to each fluids, we are now in position to discuss the composition of our universe depicted in figure \ref{fig:pie}.
\paragraph{Radiation} By \gls{radiation} in the previous paragraphs, we meant ``relativistic species present in the universe''. Mainly neutrinos and photons, most of it is in the form of the photons of the \acrshort{CMB}. \Gls{radiation} evolves as $\rho_{\text{rad}} \propto a^{-4}$, the 4 stands for 3+1: 3 for the spatial dilution plus 1 for the wavelength \gls{redshift}. \Gls{radiation} accounts\footnote{The value is not indicated directly in the Planck paper \cite{Ade:2015xua}  but can be deduced with the value of the \gls{redshift} at the equilibrium matter-radiation $z_{\text{eq}}$. At that redshift, we have $\Omega_M^0 (1+z_{\text{eq}})^3= \Omega_{\text{rad}}^0 (1+z_{\text{eq}})^4 $, so the knowledge of $\Omega_M^0$ and $z_{\text{eq}}$ gives $\Omega_{\text{rad}}^0$} for $\Omega_{\text{rad}}^0 = (9.16 \pm 0.24) \times 10^{-5}$ in the cosmic budget.
\paragraph{Baryonic Matter} In table \ref{tab:flu}, we tried to draw a clear line between \gls{baryon}ic and non-baryonic matter. The first one corresponds to ordinary matter composed of protons, neutrons and other \glspl{baryon} (made up of three quarks) together with electrons (which are not \gls{baryon} but lepton and much lighter) and other leptons. The ordinary matter amounts only for \cite{Ade:2015xua} $\Omega_b^0 = 0.0483 \pm 0.0005$. 
\paragraph{Dark Matter} The rest of the matter in the universe is in the form of \gls{dark matter} \cite{Bertone:2004pz,Bertone:2016nfn}. It has been postulated to facilitate the \glslink{S}{larger scale structure} formation and the galaxies formation and to account for many independent observational facts: rotation curves of galaxies, \acrshort{CMB} statistics. In the most simple version, it is cold in the sense that it decouples when it is non relativistic and dark in the sense that it does not interact electromagnetically. It amounts for \cite{Ade:2015xua} $\Omega_{\text{CDM}}^0 = 0.268 \pm 0.013$.
\paragraph{Curvature} All the observation constraints are consistent with a flat universe $K=0$. A curved universe would be challenging for cosmologists and would open the door for exotic cosmologies such as non trivial \gls{topology} as discussed in Sec.~\ref{sec:cosmprin}. The current \cite{Ade:2015xua} constraint on the curvature parameter is: $\Omega_K^0 = -0.005^{+0.016}_{-0.017}$.
\paragraph*{Dark energy} We will dedicate a whole chapter (chapter \ref{chap:DE}) to the last (dominating) constituent of our universe. Its introduction is motivated by the late time observation of the acceleration of the expansion of the universe and could be a manifestation of the \gls{cosmological constant}. The \gls{dark energy} current contribution is \cite{Ade:2015xua} $\Omega_{DE}^0 = 0.683 \pm 0.013$.
\paragraph{} All those parameters are reported in table \ref{table:cosmnum}. Armed with this view that the universe has been dominated by several fluids through its cosmic history, it is possible to integrate through its history (\ref{eq:Fried1}) to find a precise age for our universe. The answer is more satisfactory than the one found in Newtonian cosmology (\ref{eq:wrong}):
\begin{equation}
t_{\text{BB}} \equiv \frac{1}{H_0}\int_0^1 \frac{\text{d}a}{\sqrt{\sum_i \Omega_i^0 a^{-1-3 w_i}}} \sim 0.95 \frac{1}{H_0}=13.799 \pm 0.021 \text{ Gyr}.
\label{eq:tbbt}
\end{equation}
This calculation finally justifies the title of this chapter.

%% file: cours/L5.tex
\section{A brief cosmic history}
\label{sec:status}
\begin{figure}[h]
\begin{center}
\includegraphics[width=1\textwidth]{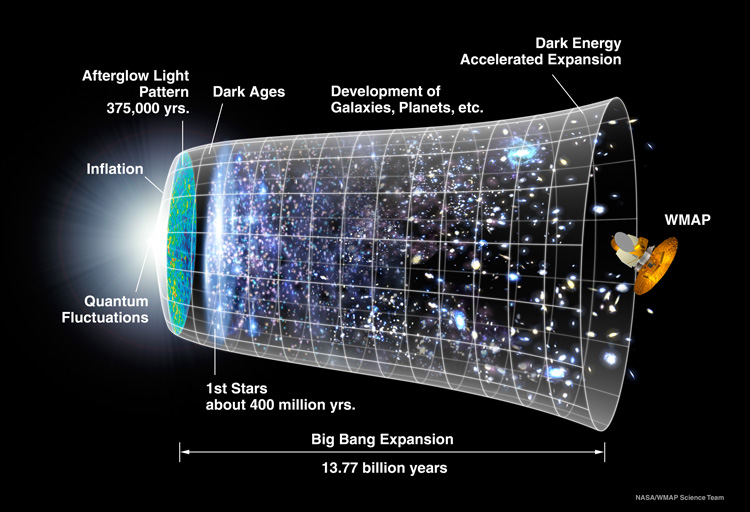}
\caption[The cosmological trumpet]{The sometimes called cosmological trumpet, a summary of the $\sim$ 14 billion years history of the universe}\label{fig:trompette}
\end{center}
\end{figure}

\begin{figure}[h]
\begin{center}
\includegraphics[width=0.7\textwidth]{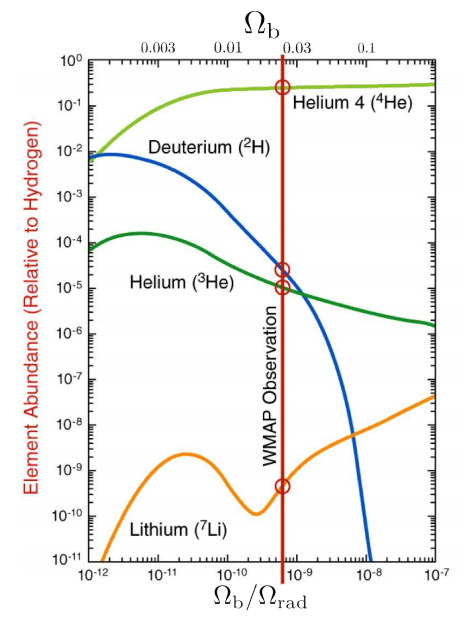}
\caption[Big-Bang Nucleosynthesis]{Light elements abundances (relative to hydrogen) as a function of the \gls{baryon} density. The \gls{satellite} WMAP measured the baryon density inducated with the vertical red line \cite{2013ApJS..208...19H}, hence giving a measure (red circle) of the different abundances. Credit: NASA/WMAP}\label{fig:BBN}
\end{center}
\end{figure}

\begin{figure}[h]
\begin{center}
\includegraphics[width=0.7\textwidth]{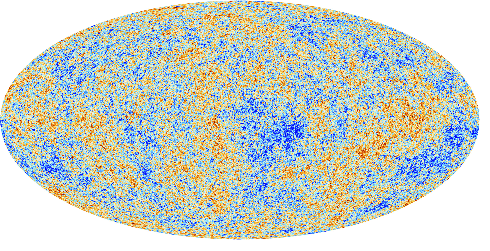}
\caption[Cosmic Microwave Background anisotropies]{Full sky map of the temperature anisotopies in the universe as seen by the \gls{satellite} Planck \cite{Ade:2013sjv}. The color indicates deviation to homogeneities which are of order $10^{-5}$, blue for colder spot and red for hotter spot. Credit: Planck Collaboration}\label{fig:CMB}
\end{center}
\end{figure}

So far we have discussed how to construct a cosmological models, the assumptions for our cosmological models and the observational pillars of cosmology. We extended the discussion of one of the observation pillar: the expansion of the universe with both observational aspects and the theoretical derivation of the \gls{Friedmann equation} performed in Appendix \ref{app:fried}. The discussion on the composition of the universe with different fluids stressed again that the universe has a thermal history. Deep in the past, it was dominated by \gls{radiation} getting hotter and hotter until a time of an infinite temperature christen the \gls{big-bang} \cite{Lemaitre:1931zzb,Lemaitre:1933gd}\footnote{It is Fred Hoyle, one of the most vocal critics of the \glslink{SM}{hot big-bang model} who invented this name on 28 March 1949 during the BBC radio’s Third Programme broadcast at 18:30 GMT. His purpose was to mock this hypothetical instant of the birth of the universe but the name stayed.}. Figure \ref{fig:trompette} shows a nice sum-up of the key eras of our universe. We will now discuss, with this picture in mind, some of the aspects of the cosmological history.
\paragraph{Inflation} The first period of the universe lasted around $10^{-32}$ seconds and is christen \gls{inflation}. It is a postulated phase of expansion and will be discussed with mathematical details in chapter \ref{chap:inflation}. Right after the \gls{inflation}, a period called \gls{reheating}\footnote{While the way \gls{reheating} should work is perfectly known, only few observations can constrain \gls{reheating} mechanisms \cite{Martin:2014nya}.} generated all the particles known in the \glslink{SM}{standard model} of cosmology and left our universe in a hot dense state where it could start to expand and cool down due to this expansion. As time flows, more and more particles started to decouple from the \gls{Hubble flow} \textit{ie.}~not follow the expansion of the universe anymore and live their own life.
\paragraph{Big-Bang Nucleosynthesis} Starting from $t \sim 0.01$ s to $t \sim 3$ min, nuclear fusion processes dominate and the light elements (H, D, He, Li and Be) are \glslink{BBN}{formed} \cite{Gamow:1946eb} \cite{Alpher:1948ve}\footnote{This article appeared in physical review letter on the 1st April 1948 and the name of Bethe as an author was included only for the purpose of joking and calling this paper the alphabetical paper or the $\alpha \beta \gamma$ paper} \cite{meneguzzi1971production}. From nuclear physics calculations, it possible to predict the relative abundance of the elements and to compare them to their observed value, for instance in the \acrshort{CMB} in figure \ref{fig:BBN}. We see that a total agreement for many different elements over 10 orders of magnitude. The rest of the elements of the Mendeleev table are only formed later by fusion processes inside stars and with \gls{supernovae} explosions \cite{Burbidge:1957vc}.
\paragraph{Cosmic Microwave Background} As the universe keeps cooling down, it reached the temperature when the electrons and the protons could form hydrogen atoms (the \gls{recombination}). Sensibly at the same time, $t \sim 375$ $000$ years, photons decoupled from the matter and could then travel almost unaffected to us: this is the celebrated light from the \acrshort{CMB}. We see it nowadays as the purest blackbody radiation ever measured with its peak temperature nowadays at $T_{\text{CMB}} = 2.725$ K. It is one of the key observation that our universe was in the past very homogeneous and isotropic. The \acrshort{CMB} was predicted by Gamov in 1948 \cite{gamow1948origin}, discovered by Penzias and Wilson in 1964  \cite{Penzias:1965wn} and interpreted in Princeton by Dicke \textit{et.~al.}~\cite{Dicke:1965zz}. The \acrshort{CMB} is however not exactly isotropic and its anisotropies can be distinguished into two classes: primordial which are the key prediction of \gls{inflation} and will be discussed in chapter \ref{chap:inflation} and secondary due to effects of \glslink{S}{structures} to the \acrshort{CMB} photons travel. For illustration, we propose in figure \ref{fig:CMB} the anisotropies of the \acrshort{CMB} as measured by the Planck \gls{satellite} (2013). The figure is the most striking proof that our universe was isotropic to one part out of 100 000. The tiny departure from homogeneity are one of the key prediction of \gls{inflation} and will be discussed in chapter \ref{chap:inflation}.
\paragraph{Structure formation} After the release of the \acrshort{CMB}, \glslink{S}{structures} such as galaxies and stars started to form, this part is very important in cosmology and is called structure formation (see for instance \cite{Peter,Ma:1995ey,Amendola:2012ys} and chapter 10 of \cite{Joyce:2014kja}). Some of the formalism of the cosmological \gls{perturbation} theory developed in Appendix \ref{ap:infl} can also be used to describe the formation of structure in the late time universe. Structure formation is a fruitful playground to test \gls{dark energy} models: the second focus of this thesis, discussed in chapter \ref{chap:DE}. 
\paragraph{} The two focus of this thesis are both related to accelerated phases of expansion which are usually awkwardly described by Einstein \gls{general relativity} as intuitively gravity is attractive leading to decelerated expansion of the universe. 
\section{Status of the big-bang theory, the cosmological problems and inflation} \label{sec:probinfl}
\begin{figure}[h]
\begin{center}
\includegraphics[width=1\textwidth]{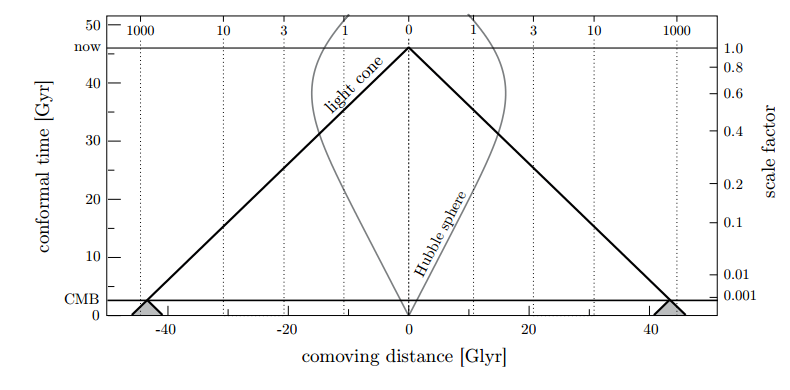}
\caption[Illustration of the horizon problem]{Spacetime diagram illustrating the horizon problem \cite{Misner:1967uu} in comoving coordinate. The dotted lines represent different worldlines of comoving observers and their associated current \gls{redshift}. We are situated in the center at \gls{redshift} 0 and can only make light-like observations on our past light-cone. On this diagram, the angles have been ignored so that the intersection of the light cone with the space-like line called \acrshort{CMB} is two diametrically opposed points in the sky. The time between the \acrshort{CMB} and the initial sigularity is too small to allow for a causal mechanism explaining the observed uniformity of the \acrshort{CMB}. Credit: \cite{Baumann:2014nda}}\label{fig:horizpro}
\end{center}
\end{figure}

\paragraph*{} One thing to be emphasized about the \glslink{SM}{big-bang model} presented in the previous sections is that it is a considerably good \glslink{SM}{standard model}. It can account for \textit{all} the observations made so far and modifying it requires to reinterpret many, many independent observation, which has never been achieved by any candidate. From $10^{-2}$ seconds to nowadays, the \glslink{SM}{big-bang theory} relies on known physics: \gls{general relativity}, nuclear physics, thermodynamics, \gls{particle physics}. However for the first instants, at high temperature, the model is less reliable as the physics laws were extrapolated to regions where there were not tested. The \glslink{SM}{standard model} of cosmology suffers also from some internal problems, some of them will motivate the work carried in this thesis. It is the study of those problems which motivates the introduction of \gls{inflation} that we will study from now and in chapter \ref{chap:inflation}.
\paragraph{The flatness problem}
In the sections describing the \gls{FLRW metric} and the \gls{Friedmann equation}, we were always very careful to keep the curvature term, which has been shown to be small. From equation (\ref{eq:oK}), we have:
\begin{equation}
\Omega_K(t) = \frac{1}{1+\frac{\sum\Omega_i}{\Omega_K}}.
\end{equation}
Thus evaluating this quantity in the early universe, when the \gls{radiation} is the dominant constituent ($\sum_i \Omega \sim \Omega_{\text{rad}})$,we find:
\begin{equation}
\Omega_K = \frac{\Omega_K^0}{\Omega^0_{\text{rad}}} \left(\frac{1}{1+z}\right)^2.
\end{equation}
The consequence is a universe close to flat must have been even flatter in the past. As an illustration, at the Planck time $t_{Pl}=10^{-43}$ s, the curvature would be
\begin{equation}
|\Omega_K(t_{Pl}) | < 10^{-62}.
\end{equation} 
Such a tiny value is the reason of the flatness problem \cite{flatness}. It can be reformulated by stating that the absolute curvature is an increasing function of the time: a flat universe is an unstable solution of the \gls{Friedmann equation}.
\paragraph{The horizon problem}
We have defined in equation (\ref{eq:comv}) the comoving distance. With it, one can find the maximal distance a (massless) particle can travel since the \gls{big-bang} \cite{Rindler:1956yx}:
\begin{equation}
\label{eq:horizp}
\chi = \int_0^t \frac{\text{d}t'}{a(t')} = \int_0^a \frac{\text{d} \ln a}{aH}.
\end{equation}
During the \gls{radiation} and matter dominated era (\textit{cf.}~table \ref{tab:flu}), the \gls{Hubble radius} (equation (\ref{eq:Hdist})) grows faster than the \gls{scale factor}, that is $\frac{d}{dt} \left( \frac{1}{a(t) H(t)} \right)>0$. The integral (\ref{eq:horizp}) is hence dominated by late-times contributions or in other words as time is flowing, more and more information enters the comoving \gls{Hubble radius}. In such cases, the conformal time (defined in equation (\ref{time})) between the \acrshort{CMB} and us is much larger than the time between the \acrshort{CMB} and the initial singularity, not letting causal processes to take place, see figure \ref{fig:horizpro} for a graphical illustration. 
However in its early stage the universe was more and more homogeneous, see figure \ref{fig:CMB}. A precise calculation implies that around 40 000 causally disconnected patches of size roughly one degree (size of the moon or a thumb at arm's length) have been observed in the \acrshort{CMB} with the same temperature, even if they were never in causal contact.

\begin{figure}[h]
\begin{center}
\includegraphics[width=0.8\textwidth]{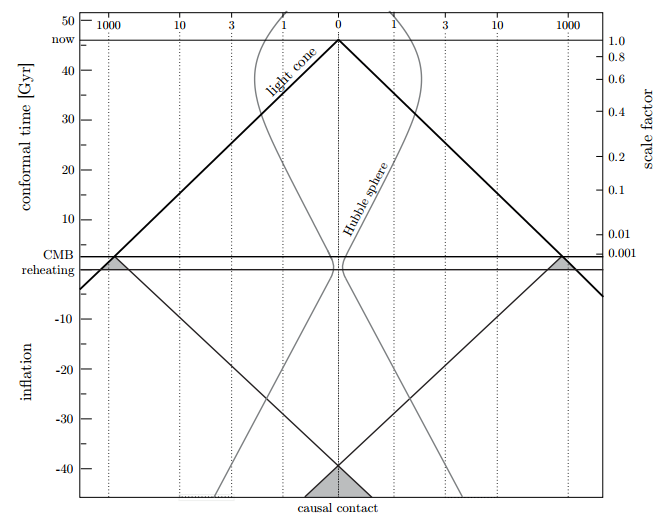}
\caption[Illustration of the solution to the horizon problem]{Spacetime diagram illustrating the inflationary solution to the horizon problem. The conformal time is extended to negative values and during that time the \gls{Hubble radius} diminishes: the universe is in accelerated expansion. If the \gls{inflation} lasts long enough, all point observed in the \acrshort{CMB} can be in causal contact. The \gls{big-bang} singularity is replaced by the \gls{reheating} where \gls{inflation} ends and all the particle of the \gls{SMP} are created in a hot dense state. Credit: \cite{Baumann:2014nda}}\label{fig:solhp}
\end{center}
\end{figure}

The horizon and the flatness problems are problems in the sense in order for the universe to be the way it is, one requires a huge amount of \gls{fine-tuning} of the parameters governing the dymamics of the universe in the early universe. Whether this is problematic is an interesting philosophical debate, an argument for this debate is given in section \ref{sec:coincp}.
\subparagraph{The monopole problem} In the very early universe, by phases transitions, a huge amount of topological defects such as magnetic monopoles could have been produced and would have soon dominate the energy budget in the universe preventing any \glslink{S}{structure} to form \cite{Guth:1979bh,Zeldovich:1978wj,Preskill:1979zi}. This problem has a historic importance as it inspires the pioneers of \gls{inflation} but it could not be the only motivation for \gls{inflation} as it is nothing but the non-detection of a hypothetical particle.
\subparagraph{The origin of structure problem} \glslink{FLRW metric}{Friedmann solutions} describe homogeneous and isotropic spacetimes, which is a fairly good approximation, as can be seen in figure \ref{fig:CMB} for instance. We see in this figure also that small inhomogeneities exists. Before \gls{inflation}, no mechanism was able to predict them, so they were simply \textit{postulated}. It is deeply unsatisfactory for a physicist, thermal mechanism have been tried but in a \gls{radiation} dominated era, they were not efficient enough to form \glslink{S}{structure}.
\paragraph{Inflation solving the problems:}
The idea of \gls{inflation} \cite{Guth:1980zm} is to postulate that the comoving \gls{Hubble radius} is diminishing, even though the physical size of the universe is increasing. It implies that the integral (\ref{eq:horizp}) is dominated by early time contributions, it extends the conformal time to negative values and gives enough time for a causal process to occur, see figure \ref{fig:solhp}.
We thus want:
\begin{equation}
\frac{d}{dt} \left( \frac{1}{a(t) H(t)} \right)= -\frac{1}{a(t)} \left( \frac{\dot{H}}{H^2}+1 \right)<0.
\end{equation}
In this case, objects/information/\glspl{perturbation} are exiting the horizon. The mechanism allows therefore for a smoothing of the universe describing the large-scale homogeneity of our universe and solving the horizon problem.
\paragraph{}
We define the \gls{slow roll} parameter as:
\begin{equation}
\label{eq:srp}
\epsilon \equiv \frac{\dot{H}}{H^2},
\end{equation}
to achieve an \gls{inflation} phase, we need to have $\epsilon <1$, the \gls{slow roll} condition is $\epsilon \ll 1$ and the limit $\epsilon=0$ corresponds to \glslink{dS}{de Sitter space} which will be studied in chapter \ref{chap:early}, the space would, in this case, grow exponentially, see equation (\ref{eq:dssf}).
To support such an expansion, a \gls{perfect fluid} in a flat \gls{FLRW metric} would need to have an unconventional behavior: negative pressure. In this case the \glspl{Friedmann equation} (\ref{eq:Fried1}) and (\ref{eq:Fried2}) read:
\begin{align}
& H^2=\frac{\rho}{3 M_{\text{Pl}}}, \label{eq:F1infl}\\
& 6 M_{\text{Pl}}(\dot{H}+H^2)=-(\rho +3P). \label{eq:R2infl}
\end{align}
With (\ref{eq:srp}), (\ref{eq:F1infl}) and (\ref{eq:R2infl}), we find:
\begin{equation}
\label{eq:epress}
\epsilon=\frac{3}{2} \left( 1+ \frac{P}{\rho} \right).
\end{equation}
So the condition for \gls{inflation} to happen corresponds to $P< - \frac{1}{3} \rho$ violating the strong energy condition \cite{Baumann:2014nda}. Observe that considering the \gls{cosmological constant} as a fluid, it would unleash a period of \gls{inflation} as $P_{\Lambda} = -\rho_{\Lambda}$, this case correspond to \glslink{dS}{de Sitter space}, is unrealistic because \gls{inflation} would never stop but is a useful approximation to study physical processes deep in the \gls{inflation} regime.
\paragraph{}
In a phase of accelerated expansion, the curvature will become a decreasing function of the time and thus giving a mechanism for its small value at \gls{reheating}. Any existing curvature would be diluted if \gls{inflation} lasts long enough. More precisely it should last at least as much conformal time than the time from \gls{reheating} to us. Even more precisely, if we define:
\begin{equation}
\frac{|\Omega_K(t_f)|}{|\Omega_K(t_i)|}=\left(\frac{a_f}{a_i} \right)^{-2} \equiv e^{-2N},
\end{equation}
where the subscripts $i$ and $f$ stand for the initial time and final time of the postulated phase of accelerated expansion. To solve the flatness problem we need $\Omega_K(t_i) \sim \mathcal{O}(1)$ and $|\Omega_K(t_f)| \sim 10^{-62}$. This would happen if $N > 71$. $N$ is the so-called number of e-fold, it characterizes how long \gls{inflation} lasted.

An accelerated period of expansion would also wash out any topological defect and actually any classical hair (classical inhomogeneities) leading to an almost homogeneous and isotropic universe. After a long enough period of \gls{inflation}, the only non-trivial physical phenomenon is \glslink{vacuum fluctuations}{quantum fluctuations} stretched to large scales due to the acceleration of the expansion. Those fluctuations correspond to the small \glspl{perturbation} seen in the \acrshort{CMB} and will eventually collapse into galaxies, clusters of galaxies. Thus providing a satisfactory mechanism for the origin of structures and solving the origin of structure problem.
\paragraph{} We will discuss in detail in chapter \ref{chap:inflation}, how to realize such a period of accelerated expansion and how the \gls{vacuum fluctuations} are produced. To finish, the nature of \gls{dark matter} and \gls{dark energy} is also to be characterized in the realm of modern cosmology. Two problems for \gls{dark energy} are the \gls{cosmological constant problem} and the \gls{coincidence problem}, we will discuss them in the chapter \ref{chap:DE}.

%% file: cours/infla.tex
\part{Early Universe physics}
  \label{part:early}
  \vspace{2cm}
\textit{Homer has disappeared into a wall in the living room.}\\
\textbf{Lisa:} Well, where's my dad? \\
\textbf{Frink:} Well, it should be obvious to even the most dimwitted individual who holds an advanced degree in hyperbolic topology, n'gee, that Homer Simpson has stumbled into...[the lights go off] the third dimension. \\
\textbf{Lisa:} \textit{flips the light switch back.} Sorry. \\
\textbf{Frink:} \textit{drawing on a blackboard.} Here is an ordinary square.... \\
\textbf{Wiggum:} Whoa, whoa--slow down, egghead!\\
\textbf{Frink:} ... but suppose we extend the square beyond the two dimensions of our universe, along the hypothetical z-axis, there. \\
\textbf{Everyone:} \textit{gasps} \\
\textbf{Frink:} This forms a three-dimensional object known as a ``cube'', or a ``Frinkahedron'' in honor of its discoverer, n'hey, n'hey. \\
\textbf{Homer's voice:} Help me! Are you helping me, or are you going on and on? \\
\textbf{Frink:} Oh, right. And, of course, within, we find the doomed individual. \\
\textbf{Chief Wiggum:} Enough of your borax, Pointdexter! A man's life's at stake. We need action! \\
\textit{Fires gun at portal} \\
\textbf{Chief Wiggum:} Take that, you lousy dimension! \\
  \vspace{1cm}
\begin{center}
\includegraphics[width=0.8\textwidth]{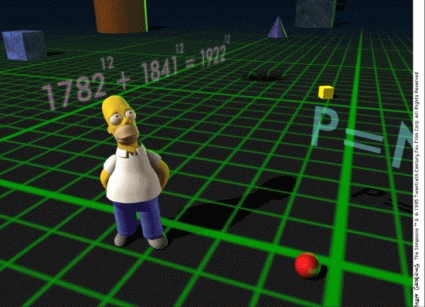} \\
The Simpsons, Treehouse of Horror VI, Matt Groening, 1995, 
\end{center}
\chapter{Inflation}
\label{chap:inflation}
\epigraph{Tout a sans doute déjà été dit, mais comme personne n'écoute il faut recommencer.}{J.~Sfar, Le chat du Rabbin, 2006}
\textit{In this chapter, we introduce inflation as a way to solve some problems of the \glslink{SM}{standard model} of cosmology discussed in the previous part. The stress is to show how particles in an inflationary universe are created. We will also encounter some formalism useful for the rest of the thesis. More systematic introductions on inflation can be found in \cite{Mukhanov,Peter,lyth2009primordial,Baumann:2014nda}. For later use, a link to Schwinger effect is presented in \cite{Martin:2007bw}.}
\paragraph*{}
We finished the previous chapter by picturing some internal problems of the \glslink{SM}{standard model} of cosmology and indicating that a period of accelerated expansion would solve some of them. Different ways to achieve a period of accelerated expansion exist. The canonical one is a play involving two actors accounting for the dynamical evolution of the universe: a \glslink{slow roll}{slowly rolling} scalar field (the inflaton) and the \gls{metric tensor} describing the gravitational degrees of freedom. This simplest model seems, as often in cosmology, to be the most fashionable now: it made a couple of predictions which were all confirmed: (1) primordial \glspl{perturbation} are adiabatic, (2) the \gls{power spectrum} of primordial \gls{perturbation} is nearly scale invariance but not exactly (3) the distribution of primordial \glspl{perturbation} should be nearly Gaussian, (4) there should be primordial \glspl{perturbation} which are outside the \gls{Hubble radius} (with wavelength greater than the observable universe) at the period of the decoupling of the \acrshort{CMB}. The success of those predictions suggests that \gls{inflation} is an important block of the \glslink{SM}{standard model} of cosmology and should be studied further. To do so, we will present two calculations. One will be basic: the study of a massless field in a \glslink{dS}{de Sitter spacetime} will be performed in Sec.~\ref{sec:massless} We will see that already in this case \gls{vacuum fluctuations} would create small \glspl{perturbation} stretched on large scale. This calculation is also the first step to understand in order to calculate more involved processes of \gls{particle creation} in \gls{dS}, as we will discuss in chapter \ref{chap:early}. Finally, in Sec.~\ref{sec:quasi}, we will describe a second calculation where, we will work in quasi \gls{dS} with non-standard fluids, again we will be able to compute the amount of quantum \glspl{perturbation} leading to \gls{particle creation}.
\section{Massless field in de Sitter}
\label{sec:massless}
In this section, we study a massless scalar field $\varphi$ in \gls{dS}. In this space the wavelength of any fluctuations increases exponentially until being of the order $H^{-1}$, at that time it is said to be frozen: its amplitude do not evolve anymore and become constant: \glslink{particle creation}{particles} were created out the \gls{vacuum}. The action, we consider is the following:
\begin{equation}
\label{eq:actsimpl}
S=\int \mathrm{d}^{4}x \sqrt{-g} (\frac{1}{2}\partial_{\mu}\varphi \partial^{\mu}\varphi).
\end{equation}
Generalizations of this action will be studied in (\ref{eq:actmore}), (\ref{eq:actmoremore}) and (\ref{action:sca}). The \gls{de Sitter metric} is:
\begin{equation}
\label{g}
\mathrm{d}s^{2}=  \mathrm{d}t^{2}-a^{2}(t) \mathrm{d} \textbf{x}^2,
\end{equation}
where an expression for $a(t)$ is given in equation (\ref{eq:dssf}). More details about \gls{de Sitter metric} will be given in Sec.~\ref{sec:strongintrodS}. Aiming at studying \gls{vacuum fluctuations}, we develop the scalar field on a basis of creation and annihilation operators:
\begin{equation}
\label{eq:decompo}
\varphi(\textbf{x},t)= \int \frac{\text{d}^3 \textbf{k}}{(2 \pi)^3} \left[ \varphi_k(t) e^{\textbf{k}.\textbf{x}} \hat{a}_{\textbf{k}}+\varphi_k^*(t) e^{-\textbf{k}.\textbf{x}} \hat{a}^{\dagger}_{\textbf{k}} \right].
\end{equation}
The creation and annihilation operators satisfy the following commutation relations:
\begin{equation} \label{ECTR a}\begin{split}
&[\hat{a}_{\textbf{k}},\hat{a}_{\textbf{k}'}]=[\hat{a}_{\textbf{k}}^{\dagger},\hat{a}_{\textbf{k}'}^{\dagger}]=0, \\
&[\hat{a}_{\textbf{k}},\hat{a}_{\textbf{k}'}^{\dagger}]= \delta^{(3)}( \textbf{k}-\textbf{k}'), \\
\end{split}
\end{equation}
where $\delta^{(3)}$ corresponds to the 3-Dirac distribution. It will be used throughout the rest of the manuscript in various dimensions. In one dimension, it will be written simply $\delta$. Varying the action with respect to $\varphi$, we get the Klein-Gordon equation for the mode functions $\varphi_k(t)$:
\begin{equation}
\label{eq:KGct}
\ddot{\varphi_k}+3H \dot{\varphi_k}+\frac{k^2}{a^2} \varphi_k=0.
\end{equation}
Two regimes are possible: for $k \gg a(t) H(t)$, the modes are inside the comoving \gls{Hubble radius} and the equation corresponds to an harmonic oscillator with a frequency getting redshifted. Conversely, for modes outside the \gls{Hubble radius}, $k \gg a(t) H(t)$, one can neglect the gradient term and a constant term is a solution to this equation. The mode is said to be amplified in the sub-Hubble region and then frozen when it is super-Hubble. A similar reasoning will be applied in Sec.~\ref{sec:backEM}.
\paragraph{} To solve (\ref{eq:KGct}), we will consider the conformal time defined in (\ref{time}). Considering the auxiliary field $v(t) \equiv a(\tau) \varphi$, the Klein-Gordon equation (\ref{eq:KGct}) become:
\begin{equation}
\label{eq:KGmassless}
v_k''+\left(k^2-\frac{2}{\tau^2} \right)v_k=0.
\end{equation}
We will later study generalizations of this equation in (\ref{eq:geneapp}), (\ref{eq:hoho}) and (\ref{vphieq}). A general solution is:
\begin{equation}
v_k(\tau)=\left[ A(k) H_{3/2}^{(1)}(-k\tau)+B(k) H_{3/2}^{(2)}(-k \tau) \right] \sqrt{-\tau},
\end{equation}
where $H_{3/2}^{(1,2)}$ are the Hankel function of first and second order, in chapter \ref{chap:early}, due to the presence of an electric field, the Hankel function will generalized to the Whittaker function described in Appendix \ref{app:aw}. For this calculation, we can simply use $H_{3/2}^{(2)}(z) = \left[H_{3/2}^{(1)}(z)\right]^*=-\frac{2}{\pi z} e^{-i z} \left(1+\frac{1}{i z} \right),$ to obtain an expression for $v_k(\tau)$ in term of the usual functions. To construct then the Fock representation of the Hilbert space, one needs to define the \gls{vacuum} state ; all the other states will be obtained by applying $a^{\dagger}$'s. We define it such that $\forall \textbf{k}, a_{\textbf{k}} \ket{0}=0$. Determining the \gls{vacuum} of the field theory will determine the constants $A(k)$ and $B(k)$. In all this thesis, we will use the same prescription known as the \textit{\gls{Bunch-Davies vacuum}} \cite{Bunch:1978yq,book:Parker}. The prescription is that in the remote past, the field behaves as a plane wave of positive frequency:
\begin{equation}
v_k \underset{k \tau \rightarrow -\infty}{\sim} \frac{e^{- i k \tau}}{\sqrt{k}}.
\end{equation} 
In the Sec.~\ref{sec:quasi}, we will propose two other ways of seing the \gls{Bunch-Davies vacuum}, one as being the minimal energy allowed by \gls{Heisenberg uncertainty principle}, another being the minimal energy allowed to an harmonic oscillator. The full solution for the Klein-Gordon equation reads then:
\begin{equation}
\varphi_k(\tau) =\frac{H \tau}{\sqrt{2k}} \left( 1+ \frac{1}{i k \tau} \right) e^{-i k \tau}.
\end{equation}
We have seen that in the asymptotic past, the solution behaves as plane wave, in the asymptotic future, it has as announced a constant amplitude:
\begin{equation}
\label{eq:afr}
|\varphi_k(\tau)| \underset{k \tau \rightarrow 0}{\sim} \frac{H}{\sqrt{2k^3}}. 
\end{equation}
A customary way of representing the amount of \glslink{particle creation}{particles} created from the \gls{vacuum} is the \gls{power spectrum} which is related to the two point correlator. We define the correlation function as:
\begin{equation}
\label{eq:corrl}
\xi_v \equiv \bra{0} \hat{v}(\textbf{x},\tau) \hat{v}(\textbf{x'},\eta) \ket{0}.
\end{equation}
With the \gls{symmetry} of the problem, the correlation function depends only on the distance $|\textbf{x'}-\textbf{x}|$ and it is possible to define the \gls{power spectrum}:
\begin{equation}
\label{eq:PS}
\xi_v \equiv \int \frac{dk}{k} \frac{\sin kr}{kr} P_v(k).
\end{equation}
Using (\ref{eq:decompo}),(\ref{ECTR a}),(\ref{eq:afr}), (\ref{eq:corrl})  (\ref{eq:PS}), we can find in the asymptotic future:
\begin{equation}
\label{eq:psmassles}
P_{\varphi}(k) \equiv \left( \frac{H}{2\pi} \right)^2
\end{equation}
The \gls{power spectrum} is not a function of $k$: it is said to be scale invariant.
\section{General perturbation theory in quasi-de Sitter space:}
\label{sec:quasi}
\paragraph{} We will in this part generalize the result of the previous part in two ways, first we will not consider \gls{dS} anymore in which \gls{inflation} never ends but quasi \gls{dS}, second, we will allow for a general fluid. We will only underline the most important steps and relegate some details in Appendix \ref{ap:infl}. \\
We will consider that a scalar field $\varphi$ drives \gls{inflation} and will consider small \glspl{perturbation} $\varphi=\varphi_{0}(\tau)+\delta\varphi(\tau,\textbf{x})$. Those inflaton \glspl{perturbation} will induce \glspl{perturbation} of the metric since the inflaton dynamics and the geometry are coupled with \glspl{Einstein equation}. To describe those metric \glspl{perturbation}, we will use the formalism SVT.
\subsection*{Decomposition SVT (Scalar-Vector-Tensor)}
The most general metric describing small \glspl{perturbation} on a flat \glslink{FLRW metric}{FLRW background} reads:
\begin{equation}
\label{g}
\mathrm{d}s^{2}=a^{2}(\tau)\left[(1+2\phi)\mathrm{d}\tau^{2}+B_{i}\mathrm{d}x^{i}\mathrm{d}\tau-E_{ij} \mathrm{d}x^{i}\mathrm{d}x^{j} \right].
\end{equation}
Now it is interesting to decompose those \glspl{perturbation} into their scalar, vector and \gls{tensor} part \cite{Bardeen:1980kt}. They will indeed evolve independently because of the choice of background metric. \\
Any vector field can be decomposed into the divergence of a scalar and a divergence free vector:
\begin{equation}
B_i = \partial_i B + \bar{B}_i, \text{  with } \partial^i\bar{B}_i=0.
\end{equation}
In the same way, any \gls{tensor} field can be decomposed into:
\begin{equation}
E_{ij}=(1-2\psi) \delta_{ij}-2\partial_i \partial_j E+2\partial_{(i}\bar{E}_{j)}, \text{ with } \partial_i \bar{E}^{ij}=0 \text{ and } \bar{E}^i_i=0.
\end{equation} 
During \gls{inflation}, it can be shown that the vector \glspl{perturbation} are exponentially suppressed \cite{Mukhanov:1990me,Peter} and are usually ignored, see however \cite{Gomez:2013xza}. \Gls{tensor} \glspl{perturbation} (or in other words gravity waves) and scalar \glspl{perturbation} are the two main focus. In this thesis, we will only present the case for scalar \glspl{perturbation} which reads:
\begin{equation}
\label{g}
\mathrm{d}s^{2}=a^{2}(\tau)\left[(1+2\phi)\mathrm{d}\tau^{2}+2 \partial_i B \mathrm{d}x^{i}\mathrm{d}\tau-\left((1-2\psi)\delta_{ij}-2 \partial_i \partial_j E \right)\mathrm{d}x^{i}\mathrm{d}x^{j} \right],
\end{equation}
where the small \glspl{perturbation} of the metric are encoded in 4 functions: $\phi$, $\psi$ called the \glslink{Bardeen potential}{Bardeen gravitational potentials} and $E$ and $B$ which encodes anisotropic shear. \Gls{general relativity} is invariant by any diffeomorphism reflecting the arbitrary of the coordinates hence the above approach is redundant. For \gls{general relativity} \glspl{perturbation}, it is possible to describe the gravity sector with one simple gauge invariant quantity: one \gls{Bardeen potential} given by:
\begin{equation}
\label{eq:Bardeen}
\Phi(\tau,\textbf{x}) \equiv \phi+\frac{1}{a(\eta)}\left[a(B-E')\right]'.
\end{equation}
In the same way, it is possible to study the \glslink{vacuum fluctuations}{fluctuations} of the scalar field in a gauge invariant way:
\begin{equation}
\overline{\delta \varphi}(\tau,\textbf{x}) \equiv \delta \varphi - \varphi'_0 (B-E').
\end{equation}
$\Phi$ and $\overline{\delta \varphi}$ are related to each other by the \glspl{Einstein equation} and to unambiguously describe the \glspl{perturbation} of the coupled inflaton-gravity system, we introduce the \glslink{MS}{Mukhanov-Sasaki variable} as a combination of the previous quantities. It will be the canonical variable to quantize:
\begin{equation}
\label{eq:MSS}
v(\tau,\textbf{x}) \equiv a(\tau) \left[ \overline{\delta \varphi}(\tau,\textbf{x}) + \varphi'_0(\tau) \frac{\Phi(\tau,\textbf{x})}{\mathcal{H(\tau)}} \right],
\end{equation}
where $\mathcal{H} \equiv \frac{a'}{a}$ is the conformal Hubble rate. We will now propose an action for $v(\tau,\textbf{x})$ and will carry out a computation similar but more elaborated than the one done in Sec.~\ref{sec:massless}.
To obtain an equation for the linear \glspl{perturbation}, one needs to expand the action $S(g_{\mu \nu},\varphi)$ for gravity and for the scalar field up to second order \cite{Sasaki:1986hm,Mukhanov:1988jd,Mukhanov:1990me}:
\begin{equation}
\label{eq:actmoremore}
S=\frac{1}{2} \int \mathrm{d}\tau \mathrm{d}^{3}\textbf{x} \left[ (v')^{2} + c_{s} \delta^{ij}\partial_i v \partial_j v + \frac{z''}{z} v^{2} \right].
\end{equation}
The Lorentz \gls{symmetry} is broken by the time dependence of the background so it is possible that the \glspl{perturbation} have a non-trivial speed, hence we introduced the speed of the sound $c_s$ defined in (\ref{eq:cs}). In the case of the massless scalar field in \gls{dS} of Sec.~\ref{sec:massless}, this was not allowed and the \glslink{particle creation}{particles} created propagated at $c=1$. $z$ is not the redshift and is defined as: $z=\frac{a}{c_s} \frac{2 \epsilon}{3}$, see also (\ref{eq:defz}). Varying (\ref{eq:actmoremore}) with respect to $v$ gives:
\begin{equation}
v'' -\left( c_{s}\partial_i \partial^i + \frac{z''}{z} \right) v=0,
\label{eq:eomag}
\end{equation}
$v$ will now be promoted and will become a quantum operator evolving on a classical geometry.
\paragraph{} We define the conjugate momentum as: $\pi \equiv \frac{\partial \mathcal{L}}{\partial v'}=v'$ Then the variables $v$ and $\pi$ become operators satisfying the equal time commutation relations:
\begin{equation} \label{ECTR v} \begin{split}
&[\hat{v}(\textbf{x}),\hat{v}(\textbf{y})]=[\hat{\pi}(\textbf{x}),\hat{\pi}(\textbf{y})]=0, \\
&[\hat{v}(\textbf{x}),\hat{\pi}(\textbf{y})]=[\hat{v}(\textbf{x}),\hat{v'}(\textbf{y})]=i \delta( \textbf{x}-\textbf{y}). \\
\end{split}
\end{equation}
After going to the Fourier domain, a general solution for $\hat{v}$ can be written:
\begin{equation}
\label{ergebniss}
\hat{v}(\eta,\textbf{x}) =\frac{1}{\sqrt{2}}\int \frac{\mathrm{d}^{3}\textbf{k}}{(2\pi)^{\frac{3}{2}}}\left( v_{\textbf{k}}^{*}(\tau) e^{i \textbf{k}.\textbf{x}}\hat{a}_{\textbf{k}}+v_{\textbf{k}}(\tau) e^{-i \textbf{\textbf{k}}.\textbf{x}}\hat{a}_{\textbf{k}}^{\dagger} \right).
\end{equation}
The temporal modes satisfy a Schrödinger-like equation:
\begin{equation}
\label{eq:hoho}
v_{\textbf{k}}''+\omega_{k}^{2}(\tau) v_{\textbf{k}}=0,
\end{equation}
where the time dependent effective frequency is:
\begin{equation}
\omega_{k}^{2} \equiv c_{s}^{2}k^{2}-\frac{z''}{z}.
\end{equation}
The creation and annihilation operators satisfy the commutation relatition (\ref{ECTR a}). Using (\ref{ECTR a}) with (\ref{ergebniss}) and (\ref{ECTR v}), integrating and using the reality condition ($v_{k}^{*}=v_{-k}$), we get the \glslink{Wronskian condition}{normalization of the Wronskian}: 
\begin{equation}
v_{\textbf{k}}^{*}v'_{\textbf{k}}-v_{\textbf{k}}v'^{*}_{\textbf{k}}=2\ii.
\label{eq:wronsk}
\end{equation}
\paragraph{} We define as in Sec.~\ref{sec:massless} the \gls{vacuum} as $\forall \textbf{k}, a_{\textbf{k}} \ket{0}=0$. To find the initial conditions for $v$, we force the field to have the minimal energy allowed by \glslink{vacuum fluctuations}{quantum fluctuations}, another demonstration based on the \gls{Heisenberg uncertainty principle} is given in Appendix \ref{ap:infl}, they are all equivalent to the \gls{Bunch-Davies vacuum} of the field theory. We will here use the ansatz $v_{\textbf{k}}=r_{\textbf{k}} \exp{i \alpha_{\textbf{k}}}$, multiplying by $v'$, integrating the equation of the harmonic oscillator (\ref{eq:hoho}) and using (\ref{eq:wronsk}), one can find the energy of the harmonic oscillator:
\begin{equation} E_{\textbf{k}} =\frac{1}{2} \left( \mid v'_{\textbf{k}}\mid ^{2}+\omega_{\textbf{k}}^{2} \mid v_{\textbf{k}} \mid^{2} \right)=\frac{1}{2}\left( r_{\textbf{k}}^{'2}+\frac{1}{r_{\textbf{k}}^{2}}+\omega_{k}^{2} r_{\textbf{k}}^{2} \right).
\label{eq:Heinsenb}
\end{equation} Then, we find that the minimum of $E_{\textbf{k}}$ is at $r_{\textbf{k}}(\eta_{i})=\omega_{k}^{-1/2}$ and hence requiring the field to carry this minimal energy, the initial conditions become:
\begin{equation}
\label{CI}
\begin{split}
&v_{\textbf{k}}(\tau_{i})= \frac{1}{\sqrt{\omega(\tau_{i},k)}},\\
& v'_{\textbf{k}}(\tau_{i})= \ii \sqrt{\omega_{k}(\tau_{i},k}).\\
\end{split}
\end{equation}
Note that we assumed $\omega_{k}^{2}$ to be positive such that we consider only positive frequency modes at the beginning of the \gls{inflation}. The negative frequency modes are not in the \gls{Hubble radius} when \gls{inflation} begins so they will be stretched to huge unobservable scales.
\paragraph{} One can then obtain an expression for the \gls{power spectrum} at the end of \gls{inflation} \cite{Mukhanov}:
\begin{equation}
\label{Final power spectrum}
P_{\Phi}(k)=\frac{32}{9}\left( \frac{H^2}{M_{\text{Pl}}^2 c_{s}\epsilon}\right)_{c_{s}k \sim H a}.
\end{equation}
The numerical factor depends on the approximation done to integrate (\ref{eq:hoho}) and on the convention for the \gls{power spectrum} but all the other factors are important. As in (\ref{eq:PSAP}), the correction to (\ref{eq:psmassles}) are dependent on the \gls{slow roll} parameter and the de Sitter solution is singular. As the quantities in (\ref{Final power spectrum}) have to be taken at horizon crossing, a time dependence of $H$ or $c_s$ implies a scale dependence of the \gls{power spectrum}. In the case of (\ref{eq:psmassles}), the \gls{power spectrum} was scale invariant as $H$ was constant. Departure from scale invariance are quantified by the spectral tilt:
\begin{equation}
n_S-1 \equiv \frac{\text{d} \ln P_{\Phi}}{\text{d} k}.
\end{equation} 
This definition is motivated by the fact that one can, at some scales, approximate the \gls{power spectrum} by a power law, namely $P_{\Phi}^{2}(k) \sim k^{n_{s} -1}$. The physical interpretation of non flat \gls{power spectrum} is the need for a graceful exit for the \gls{inflation} and the determination of the exact value of the spectral tilt is an exciting experimental challenge. Currently, the observations are $n_{s} = 0.968 \pm 0.006$ \cite{Ade:2015xua}, on a theoretical point of view, the inspection of the different \gls{inflation} scenarii allows: $0.92<n_{s}<0.97$. In such a case of a red-tilted \gls{power spectrum}, the modes which cross the horizon earlier will have a larger $P_{\Phi}$ than those which enter later. 
\section{Conclusion}
In this chapter, we have described with mathematical details  a way to implement a phase of accelerated expansion in the early universe. The bottom line, we will use in chapter \ref{chap:early} is the creation of \glslink{particle creation}{particles} out of the \gls{vacuum} which are stretched to large scale due to the expansion. We presented two text-book ways of calculating the amount of \glslink{particle creation}{particles} created: the final results were presented in equations (\ref{eq:psmassles}) and (\ref{Final power spectrum}). The goal of chapter \ref{chap:early} is to generalize the calculation in Sec.~\ref{sec:massless}, valid deep in the inflation regime, to the case when a constant electric field is present as well. The calculation of Sec.~\ref{sec:quasi} has been presented to give an idea of a more realistic case which can be tested with observation as was briefly discussed with the spectral index. It is known that a pure \gls{dS} cannot fully describe \gls{inflation} because \gls{inflation} never ends in \gls{dS}.
Beyond the simple cases, we presented in this chapter, the literature on \gls{inflation} is very rich, as many ways exists to induce an accelerated period of expansion. A classification of the models of \gls{inflation} has been proposed in \cite{Martin:2013tda}. The detection of the celebrated B-modes is awaited for the close future \cite{Kamionkowski:2015yta,Guzzetti:2016mkm} and would be the crowning of inflationary scenarios.

%% file: semi/intro.tex
\epigraph{We are all agreed that your theory is crazy. The question that divides us is whether it is crazy enough to have a chance of being correct. My own feeling is that it is not crazy enough.}{N.~Bohr talking to W.~Pauli, 1958, \cite{dyson1997innovation}}

\textit{In this chapter, we present our investigation on pair production in \gls{dS} in the presence of an electric field. In the previous chapter, we have seen the need for inflation and its implementation as a test field in de Sitter and quasi de Sitter space. Our goal is to show how the standard results of inflation generalize in the presence of an electric field. This chapter is adapted from the following publications \cite{Bavarsad:2016cxh,Stahl:2015cra,Stahl:2015gaa,Stahl:2016qjs} }

\subsection{Introduction}
\paragraph{} After this introductory chapter on inflation and the physics of the very early universe, we turn to the core research work of this thesis. As we saw previously, in the main models of \gls{inflation}, the key physical phenomenon is the amplification of \gls{vacuum fluctuations}{quantum fluctuations} of an inflaton to large scales due to an accelerated expansion. In other words, the accelerated expansion renders the quantum virtual particles real. A legitimate question to ask is: is there other ways to creating \glslink{particle creation}{particles} which would also be stretched to large scales in the early universe? A possibility for \glslink{particle creation}{particle} creation, very famous in flat spacetime, is the \gls{Schwinger effect} which was first discovered by Sauter in 1931 \cite{Sauter1931} and then develloped by Heisenberg and his student Euler \cite{Heisenberg1936} and finally Schwinger \cite{Schwinger1951}, see \cite{Gelis:2015kya} for a modern review.  The \gls{Schwinger effect} is a process where \glslink{particle creation}{particles}\footnote{As in this process is CPT invariant, the particles are always created by pairs of particle/antiparticle, in this thesis, we will use either \gls{pair} or \gls{particle creation} to describe the annihilation of quanta out of the vacuum.} are also created from the \gls{vacuum} under the influence of a critical or overcritical electric field. It is a nonperturbative (in the coupling) effect of quantum field theory which, despite tremendous efforts on the experimental point of view, has never been detected so far. The main reason is that the number of particles created is exponentially damped before a critical value for the electric field $E_{\mathrm{critical}}\simeq10^{18}$V/m. New laser facilities\footnote{\acrfull{XFEL} \cite{XFEL}, see also \acrfull{HIBEF}, \cite{HIBEF}; \acrfull{ELIA} \cite{ELIA}, \acrfull{HiPer} \cite{HiPer}, \acrfull{XCELS} \cite{XCELS}} are planned to be operational in the next ten years and might approach this critical electric field. Textbooks introductions on \acrshort{QED} and Schwinger effect can be found in \cite{book:Parker,Peskin:1995ev,greiner2008quantum,itzykson2012quantum,Toms:2012bra}.
\paragraph{}  During \glslink{inflation}{inflationary} \gls{magnetogenesis} strong electric fields are also produced \cite{Durrer:2013pga}, which provides motivation for considering \glslink{Schwinger effect}{Schwinger pair production} in this context. It is moreover an interesting framework for the study of false vacuum decay and bubble nucleation\cite{Garriga1994,Froeb2014}. Constraints on \gls{magnetogenesis} scenarios were put in \cite{Kobayashi:2014zza} via the \gls{backreaction} of the produced \glspl{pair}. In some models of \gls{reheating}, it might give clues on the open problem of \gls{baryogenesis}. Via the AdS-CFT correspondence\footnote{Anti-de-Sitter/Conformal Field Theory correspondence or gauge/gravity duality, a conjecture that to any conformal field theory in curved could correspond a field theory on the boundary of an anti-de-Sitter spacetime.}, it has also been  used as a playing field to test the ER=EPR conjecture\footnote{it is conjectured that a pair of entangled particles (EPR stand for Einstein-Podolsky-Rosen paradox related to entagled particles) could be related to a Einstein-Rosen bridge, sometimes also called wormhole.} \cite{Fischler2014}. Additionally it might help to better understand renormalization schemes in \gls{curved spacetime} \cite{book:Parker, Landete2014} and the relations of these schemes to each other. Recently a unified thermal picture of the \gls{Schwinger effect} in both $\ds$ and anti-de Sitter spacetime and \gls{Hawking radiation} near Reissner-Nordström black holes has been proposed \cite{kim2015}. 
\paragraph{}
The study of the \gls{Schwinger effect} in $\ds$ has been under focus exactly at the time of this PhD: it was studied in depth for various types of particles and spacetime dimension \cite{Froeb2014,Kobayashi:2014zza,Fischler2014,Stahl:2015gaa,Hayashinaka:2016qqn,Hayashinaka:2016dnt,Yokoyama:2015wws}.
In \cite{Froeb2014}, \cite{Bavarsad:2016cxh} and \cite{Kobayashi:2014zza}, the authors computed the \gls{Schwinger effect} for a charged scalar test field in $\ds_{2}$, $\ds_{3}$ and $\ds_{4}$, respectively. In \cite{Stahl:2015gaa} and \cite{Hayashinaka:2016qqn}, the generalization to $\ds_{2}$ and $\ds_{4}$, respectively, for fermionic particles was performed aiming at checking if the known equivalence in flat spacetime between boson and fermion for a constant electric field still holds. The answer was that there was a difference between boson and fermion. To see that, it was necessary to compute the \gls{induced current} which turns out, as also noted in \cite{Froeb2014,Kobayashi:2014zza}, to be the right quantity to describe the \gls{Schwinger effect} in \gls{curved spacetime}. Indeed, it is not plagued by the need of the notion of particle in the adiabatic future which allows one to explore a broader parameter space.
\paragraph{}
But to cure infinities arising from momentum integration, this current needs to be renormalized.
The \gls{adiabatic subtraction} is the most used method. The Pauli-Villars method was implemented in \cite{Froeb2014} and can be shown to agree with the \gls{adiabatic subtraction}. In \cite{Hayashinaka:2016dnt}, the point-splitting method was shown to agree with the \gls{adiabatic subtraction} in $\ds_{4}$ for bosons. In \cite{Kobayashi:2014zza,Stahl:2015gaa,Hayashinaka:2016qqn}, an \gls{adiabatic subtraction} method was used to regularize the current.
\paragraph{}
We will present in the next sections, the results on the \gls{Schwinger effect} in \glslink{dS}{de Sitter spacetime} ($\ds$) based on \cite{Stahl:2015cra,Stahl:2015gaa,Bavarsad:2016cxh} and compare with the relevant results of the authors listed in the previous paragraph. First in Sec.~\ref{sec:strongintrodS}, we will introduce some aspects of \gls{dS} that is one of the central object we are manipulating. Then in the next sections we will describe the different techniques we used to calculate the \gls{pair} production under the influence of an electric and a gravitational field. First, in Sec.~\ref{sec:setup}, we will introduce the actors of the play: electric and gravitational fields, bosons, fermions, and their evolution as time is flowing. Second we will present a \gls{semiclassical}  computation to obtain our first results for \gls{pair} production under a gravitational and an electric field. We will introduce with more detail this technique in Sec.~\ref{sec:smestim}. Third and fourth, in Secs.~\ref{sec:Bogbos} and \ref{sec:fermdsdeux}, we will study on two specific cases: the creation of bosons in 1+2 D and fermions in 1+1 D and we will present more advanced techniques which allow to explore a broader range of the parameter space and grasp more physical effects. Fifth, in Sec.~\ref{sec:IRHC} we will focus more precisely on an effect which seems to be a common prediction of \gls{pair} production in \gls{dS} together with an electric field: \acrfull{IRHC} and discuss its possible relation to conformality and \glslink{tachyon}{tachyonic excitations}. We will conclude in Sec.~\ref{sec:cclpairs} and will propose a perspective related to the multiverse proposal and \gls{baryogenesis}.

%% file: semi/basis.tex
\subsection{About de Sitter space}
\label{sec:strongintrodS}
\gls{dS} \cite{deSitter:1917zz} is a space of tremendous interest for the study of quantum effects in the presence of gravity. Quantum field theory has been shown to be a fantastic tool to investigate quantum process when the gravitational effect are turned off, that is in minkowski spacetime \cite{Peskin:1995ev}. Taking the point of view that in the presence of gravity those effects should be generalized, one needs to investigate simple spacetime with a high level of \gls{symmetry} such as \gls{dS}.

\begin{figure}[h]
\begin{center}
\includegraphics[width=0.5\textwidth]{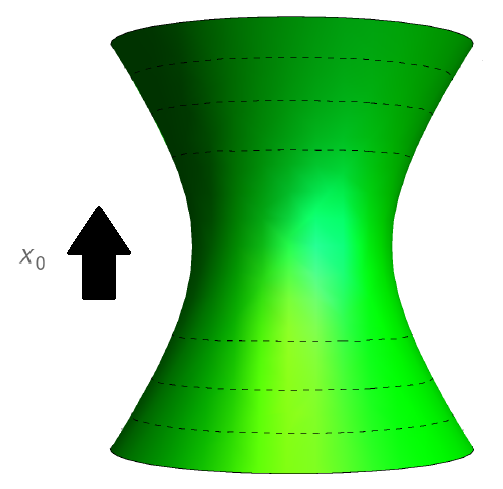}
\caption[A graphical representation of the de Sitter space]{Graphical representation of the hyperboloid given by equation (\ref{eq:desphere}), the dotted lines represent an extremal volume: a $D-1$ sphere of radius $X_0^2+H^{-2}$.}
\label{fig:pictds}
\end{center}
\end{figure}

A $D$ dimensional de Sitter space ($\dsd$) may be realized by considering a $D$ dimensional hypersurface described by:
\begin{equation}
\label{eq:desphere}
-X_0^2 +\sum_{i=1}^{D} X_i^2 = H^{-2},
\end{equation} embedded in a 1+D dimensional flat Minkowski spacetime of coordinates $\{X_i\}$. $H^{-1}$ is the de Sitter radius and is introduced in this way to correspond to the \gls{Hubble radius} given in equation (\ref{eq:Hdist}). A graphical representation is given in figure \ref{fig:pictds}.
  The \gls{isometry} group (see Sec.~\ref{sec:groupth} for a discussion) is O(D,1) for $\ds$ and is manifest in (\ref{eq:desphere}). Roughly speaking, $\ds$ is nothing but the Lorentzian generalization of a D-sphere. The line element would in this case reads:
\begin{equation}
\label{eq:lelconstr}
ds^2=dX_0^2 -dX_i dX^i,
\end{equation}
where we stress that the sum is over $ i \in (1,D)$ and the coordinates are related by (\ref{eq:desphere}). It is possible to find intrinsic coordinates varying between $- \infty$ and $+ \infty$ sometimes dubbed planar coordinates (t, $x^i$) and to show (\textit{eg.}~\cite{Birrell1984} p. 130) that the line element (\ref{eq:lelconstr}) takes the form:
\begin{equation}
ds^2= dt^2 -e^{2Ht} dx_i dx^i,
\end{equation}
where here the sum over i is for $i \in (1, D-1) \equiv (1,d)$. These coordinates only cover one half of the de Sitter manifold as it can be seen in figure \ref{fig:pictds}.

Generalizing the \glspl{Einstein equation} (\ref{eq:Eisteq}) to $D$ dimension gives:
\begin{equation}\label{eq:einstein}
R^{\mu\nu}-\frac{1}{2}Rg^{\mu\nu}+\Lambda g^{\mu\nu}=-8\pi G_{D}T^{\mu\nu}_{\sem},
\end{equation}
$G_D$ is defined after equation (\ref{einstein}). We find that the \gls{de Sitter metric} is a \gls{vacuum} solution ($T_{\mu \nu}$=0) of (\ref{eq:einstein}) if:
\begin{equation}
\Lambda = \frac{(D-2)(D-1)}{2} H^2.
\end{equation} The only non trivial component the \glspl{Einstein equation} is obtained by taking the trace of (\ref{eq:einstein}):
\begin{equation}
R=\frac{2D}{D-2} \Lambda >0.
\end{equation}
 The fact that the \gls{Ricci scalar} is constant and positive reveals that \gls{dS} is maximally \glslink{symmetry}{symmetric} and its local structure has a positive scalar curvature. For more about \gls{dS}, see \textit{eg.} \cite{deSitter:1917zz,Hawking:1973uf,Birrell1984,book:Parker,Barata:2016ves}
 \section{Generalities}
\label{sec:setup}
In the next sections, we will present the basic computations of the \gls{Schwinger effect} in \gls{dS}. We will first present the case for scalar particles in $D$ dimensional spacetime, it will be similar to Sec.~\ref{sec:massless}, where in (\ref{eq:actsimpl}), the action for a massless scalar field has been considered. Then we will discuss how it can be generalized to fermions in $2D$ and $4D$. The starting point of our studies is the action for \acrshort{QED} coupled to a complex scalar field $\varphi(x)$ or a spinor field $\psi(x)$:
We consider the action of scalar and spinor \acrfull{QED} in $D$ dimension,
\begin{align}\label{action:sca}
& S=\int{\text{d}^D}x\sqrt{-g}\Big\{-\frac{1}{\kappa}R +\big(\rnd_{\mu}+ieA_{\mu}\big)\varphi\big(\rnd^{\mu}-ieA^{\mu}\big)
\varphi^{\ast}-\big(m^{2}+\xi R\big)\varphi\varphi^{\ast}-\frac{1}{4}F_{\mu\nu}F^{\mu\nu}\Big\},\\
\label{action:fer}
& S=\int \text{d}^D x \sqrt{-g} \Big\{  -\frac{1}{\kappa}R +\frac{i}{2} (\bar{\psi} \gamma^{\mu} \nabla_{\mu} \psi -\nabla_{\mu} \bar{\psi}\gamma^{\mu}\psi) -m \bar{\psi} \psi  -\frac{1}{4} F_{\mu \nu}F^{\mu \nu}\Big\}, 
\end{align}
 where $m$ is the mass of the field and $e$ is the gauge coupling. The scalar density of weight 1/2 which generalizes the flat spacetime Lagrangian is $S:=\int \text{d}^D x \mathcal{L}$. The $\dsd$ metric $g_{\mu\nu}$ has been introduced in Sec.~\ref{sec:strongintrodS}. $g$ is the determinant of the metric, $R$ is the scalar curvature defined in equation \ref{courbure}, and $\xi$ is the dimensionless \gls{conformal coupling} introduced to make the theory more general and arises naturally in string \gls{inflation} framework \cite{Stahl:2016qjs,Kachru:2003sx}. The framework in which we will be working in the next section is such that the bosonic/fermionic particles see two backgrounds: gravitational through \gls{dS} and electromagnetic. We will now describe some of the properties of those two backgrounds. While studying those two backgrounds we will also define the quantities appearing in (\ref{action:sca}) and (\ref{action:fer}) which were not defined yet.
 \subsection{Electromagnetic side of the theory:}
 \label{sec:em}
  The field strength is defined in the usual way:
\begin{align}
F_{\mu \nu}\equiv \partial_{\mu} A_{\nu}-\partial_{\nu} A_{\mu}.
\end{align}
 The vector potential describing
a constant electric field background is
\begin{equation}\label{vector}
A_{\mu}(\tau)=-\frac{E}{H^{2}\tau}\delta^1_{\mu},
\end{equation}
where $E$ is constant. This is due to the fact that a comoving observer, with a proper velocity $u^{\mu}$, would measure an electric field of
\begin{equation}
E_{\mu} = u^{\nu} F_{\mu \nu}= a E\delta^{1}_{\mu},
 \end{equation}
which leads to a constant field strength since $E_{\mu}E^{\mu}=E^2$. Then, the only nonzero components of the electromagnetic field strength tensor are
\begin{align}\label{tensor}
F_{01}&=-F_{10} \nn\\
&=\rnd_{0}A_{1}-\rnd_{1}A_{0}=a^{2}(\tau)E.
\end{align}
\subsection{Gravitational side of the theory}
\label{sec:ds}
We have seen in Sec.~\ref{sec:strongintrodS} that the $\dsd$ metric is
\begin{align}\label{line}
ds^{2}=dt^{2}-e^{2Ht}d\x^{2}, &&t\in(-\infty,+\infty), &&\x\in\mathbb{R}^{d}.
\end{align}
It corresponds to the line element of the flat \gls{FLRW metric} (\ref{FLRW2}) with the scale factor:
\begin{equation}
\label{eq:dssf}
a(t)=\exp(Ht),
\end{equation}
$t$ is the cosmological time, and $H$ is the \gls{Hubble constant} defined in (\ref{eq:H}). Considering the transformation defined by the conformal time $\tau$,
\begin{align}\label{time}
\tau&=-\frac{1}{H}e^{-Ht}, & \tau&\in(-\infty,0),
\end{align}
the line element~(\ref{line}) reads
\begin{align}\label{metricds}
ds^{2}&=a^{2}(\tau)\big(d\tau^{2}-d\x^{2}\big), & a(\tau)&=-\frac{1}{\tau H},
\end{align}
revealing that this portion of $\ds$ is conformal to a portion of Minkowski spacetime. Observe that in order to investigate the other patch of the \gls{dS} one needs to consider $\tau\in(0,+\infty)$ but the limit $\tau \rightarrow - \infty$ and $\tau \rightarrow + \infty$ approach the same null hypersurface while $\tau \rightarrow 0^-$ and $\tau \rightarrow 0^+$ approach null hypersurface of opposite ends of the de Sitter pseudo sphere. In order to study \gls{particle creation}, we will only consider that $\tau\in(-\infty,0)$ as indicated in equation (\ref{time}).

We also introduce the \gls{tetrad} field $e^a_{\mu}(x)$ with the help of the Minkowski metric $\eta_{ab}$ 
\begin{align}
g_{\mu \nu}(x) \equiv e^a_{\mu}(x) e^b_{\nu}(x) \eta_{ab}. \label{eq:tetraddefinition}
\end{align}
The \gls{covariant derivative} for fermions is given by
\begin{align}
  \nabla_{\mu} \equiv \left(\partial_{\mu}-\frac{i}{4}\omega^{ab}_\mu\sigma_{ab}\right) +i e A_{\mu}(x) ,
\end{align}
where we defined the commutator of the gamma matrices 
\begin{align}
 \sigma_{ab}& \equiv \frac{i}{2}[\gamma_a,\gamma_b]
\end{align}
as well as the \gls{spin connection}
\begin{align}
\begin{split}
\omega_{\mu}^{ab}&:= \frac{1}{4}\left[e^{b\alpha}(x) \partial_{\mu}e^a_{\alpha}(x)-e^{a\alpha}(x) \partial_{\mu}e^b_{\alpha}(x)+ e^{a \alpha}(x) \partial_{\alpha}e^b_{\mu}(x)-e^{b\alpha}(x) \partial_{\alpha} e^{a \mu}(x)\right.\\ & \hspace{2cm}\left.+e^{b\nu}(x)e^{a \lambda}(x)e_{c \mu}(x)\partial_{\lambda} e^c_{\nu}(x)-e^{a\nu}(x)e^{b \lambda}(x)e_{c \mu}(x)\partial_{\lambda} e^c_{\nu}(x) \right]. \label{eq:spinconnection}
\end{split}
\end{align}
The gamma matrices in \gls{curved spacetime} $\underline{\gamma}^\mu$ are related to the usual ones in Minkowski spacetime via 
 \begin{align}
\gamma^a \equiv \underline{\gamma}^{\mu} e^a_{\mu}. 
\end{align} 
The canonical momentum is given by 
\begin{align}
 \pi(x) \equiv \frac{\partial \mathcal{L}}{\partial(\partial_0\psi(x))}=\sqrt{-g(x)} \,i \overline{\psi}(x)\underline{\gamma}^0. \label{eq:mometum}
\end{align}
In \gls{curved spacetime}, the Hermitian adjoint is given by \cite{book:Parker,Parker1980,Pollock2010}\footnote{This definition using the gamma matrix $\gamma^0$ of flat spacetime ensures that $ \overline{\psi}(x) \psi(x)$ transforms as a scalar, is real and that the probability current $j^{\mu}= \overline{\psi}(x) \underline{\gamma}^{\mu} \psi(x)$ is conserved.}
\begin{align}
 \overline{\psi}(x) \equiv \psi^\dagger(x)\gamma^0.
\end{align}
Following from (\ref{eq:tetraddefinition}) and (\ref{metricds}) the \glspl{tetrad} for $\ds$ are given by 
\begin{align}
 e^a_{\mu}(x)=a(\eta)\delta_{\mu}^a. \label{eq:tetrad}
\end{align}
\subsubsection{Tetrads in $2D$}
The non-zero components of the \gls{spin connection} (\ref{eq:spinconnection}) are found to be 
  \begin{align}
 \label{eq:connection}
 \omega_{1}^{01}(x)=-\omega_{1}^{10}(x)=\frac{a'(\tau)}{2a(\tau)}, 
 \end{align}
We choose to work in the \gls{Weyl basis}
\begin{align}
 \gamma^0=\begin{pmatrix}
           0&1\\
           1&0
          \end{pmatrix},&&
 \gamma^{1}=\begin{pmatrix}
           0&1\\
           -1&0
          \end{pmatrix}.
          \end{align}
          \subsubsection{Tetrads in $4D$}
The non-zero components of the \gls{spin connection} (\ref{eq:spinconnection}) can be shown to be 
  \begin{align}
 \label{connection}
 \omega_{1}^{01}=\omega_{2}^{02}=\omega_{3}^{30}=-\omega_{1}^{10}=-\omega_{2}^{20}=\omega_{3}^{30}=\frac{a'(\tau)}{2a(\tau)}.
 \end{align}
In $4D$, it will be more convinient to work with the Dirac representation of the gamma matrices, \emph{i.e.}
\begin{align}
 \gamma^{j}=\begin{pmatrix}
             0 &\sigma^j\\
             -\sigma^j & 0
            \end{pmatrix},
&&
  \gamma^{0}=\begin{pmatrix}
             I_2 &0\\
             0 & -I_2
            \end{pmatrix},
&& \text{ where }&&
   \sigma^1=\begin{pmatrix}
             0 & 1\\
             1 & 0
            \end{pmatrix},
&&
   \sigma^2=\begin{pmatrix}
             0 & -i\\
             i & 0
            \end{pmatrix},
            &&
   \sigma^3=\begin{pmatrix}
             1 & 0\\
             0 & -1
            \end{pmatrix}.
\end{align}
\subsection{\label{sec:KG}The Klein-Gordon equation in $\dsd$}
We derive the equation of motion for the scalar field $\varphi$ by using the Euler-Lagrange equation for the Lagrangian coming from the action~(\ref{action:sca}). We then obtain the Klein-Gordon equation
\begin{equation}\label{kgeq}
\frac{1}{\sqrt{-g}}\rnd_{\mu}\big(\sqrt{-g}g^{\mu\nu}\rnd_{\nu}\varphi\big)+2ieg^{\mu\nu}A_{\nu}\rnd_{\mu}\varphi
-e^{2}g^{\mu\nu}A_{\mu}A_{\nu}\varphi+\mds^{2}\varphi=0,
\end{equation}
where we defined
\begin{equation}\label{mds}
\mds^{2}\equiv m^{2}+\xi R.
\end{equation}
In $\dsd$, the \gls{Ricci scalar} reads:
\begin{equation}
\label{eq:RicciS}
R=D(D-1) H^2.
\end{equation}
After substituting explicit expressions of the the vector potential and the $\dsd$ metric given in Eqs.~(\ref{vector}) and (\ref{metricds}), respectively, Eq.~(\ref{kgeq}) takes the form
\begin{equation}\label{phieq}
\bigg[\rnd_{0}^{2}-\delta^{ij}\rnd_{i}\rnd_{j}+(D-2)H a(\tau)\rnd_{0}-\frac{2ieE}{H} a(\tau)\rnd_{1}
+\Big(\frac{e^{2}E^{2}}{H^{2}}+\mds^{2}\Big) a^{2}(\tau)\bigg]\varphi(x)=0.
\end{equation}
If we define
\begin{equation}\label{varphi}
\tilde{\varphi}(x) \equiv a^{\frac{D-2}{2}}(\tau)\varphi(x),
\end{equation}
it can be shown that Eq.~(\ref{phieq}) leads to
\begin{equation}\label{vphieq}
\bigg[\rnd_{0}^{2}-\delta^{ij}\rnd_{i}\rnd_{j}+\frac{2ieE}{\tau H^{2}}\rnd_{1}+\frac{1}{\tau^{2}}
\Big(\frac{e^{2}E^{2}}{H^{4}}+\frac{\mds^{2}}{H^{2}}
+\frac{1-d^{2}}{4}\Big)\bigg]\tilde{\varphi}(x)=0.
\end{equation}
This equation is nothing but the generalization of (\ref{eq:KGmassless}), for a massive field in $D$ dimensions coupled to a constant electric field. Based on the invariance of Eq.~(\ref{vphieq}) under translations along the spatial directions, let
\begin{equation}\label{fpm}
\tilde{\varphi}(\tau,\x)=e^{\pm i\k\cdot\x}f^{\pm}(\tau),
\end{equation}
where the superscript $\pm$ denotes the positive and negative frequency solutions, respectively. Substituting~(\ref{fpm}) into Eq.~(\ref{vphieq}) leads to
\begin{equation}\label{feq}
\frac{d^{2}}{dz_{\pm}^{2}}f^{\pm}(z_{\pm})+\Big(-\frac{1}{4}+\frac{\kappa}{z_{\pm}}
+\frac{1/4-\gamma^{2}}{z_{\pm}^{2}}\Big)f^{\pm}(z_{\pm})=0,
\end{equation}
where the variables $z_{+}$ and $z_{-}$ are defined by
\begin{align}\label{zpm}
z_{+}& \equiv +2ik\tau, & z_{-}& \equiv e^{i\pi}z_{+}=-2ik\tau,
\end{align}
with $k=|\k|=\sqrt{k^2}$. In terms of the dimensionless parameters
\begin{align}
\gamma&:=\frac{\mds}{H}, & 
\lambda&:=\frac{eE}{H^{2}}, & \label{lambda}
\rho&:=+\big(\gamma^{2}+\lambda^{2}\big)^{\frac{1}{2}}, &
r&:=\frac{k_{x}}{k},
\end{align}
the coefficients $\kappa$ and $\mu$ read
\begin{eqnarray}
\kappa&=&i\lambda r, \label{kappa} \\
\mu^{2}&=&\frac{d^{2}}{4}-\rho^{2}. \label{eq:mu}
\end{eqnarray}
In Secs.~\ref{sec:sctscalar} and~\ref{sec:pcb} we consider that it will be possible to neglect the ``$\frac{d^{2}}{4}$'' in \ref{eq:mu}, so that $\mu^{2}<0$, and then the coefficient $\mu$
becomes purely imaginary: in this case, we use
the convention $\mu=+i|\mu|$. Equation~(\ref{feq}) is the Whittaker differential equation, and its most general solution in
terms of the conventions of \cite{Olver2010}, can be written as
\begin{equation}\label{general}
f^{\pm}(z_{\pm})=C_{1}\w_{\kappa,\pm\mu}(z_{\pm})+C_{2}\m_{\kappa,\pm\mu}(z_{\pm}),
\end{equation}
where $C_{1}$ and $C_{2}$ are arbitrary constant coefficients. From Eqs.~(\ref{varphi}) (\ref{fpm}) (\ref{general}) the
corresponding solutions of Eq.~(\ref{phieq}) for positive and negative frequency solutions are
\begin{align}
\label{u}
U(x)&=a^{\frac{2-D}{2}}(\tau)e^{+i\k\cdot\x}\Big(C_{1}\wwp(z_{+})+C_{2}\wmp(z_{+})\Big), \\
\label{v}
V(x)&=a^{\frac{2-D}{2}}(\tau)e^{-i\k\cdot\x}\Big(C_{1}\wwm(z_{-})+C_{2}\wmm(z_{-})\Big),
\end{align}
where the choice of the sign for the $\mu$ parameter will follow without ambiguity from the discussion in Sec.~\ref{sec:mode}.
\subsection*{\label{sec:mode}Mode functions}
We need mode functions that determine the creation and annihilation operators and hence the \gls{vacuum} state of the quantum field
theory. This \gls{vacuum} will be determined by
specifying the asymptotic form of the mode functions \cite{Birrell1984,book:Parker}. In order to determine the mode functions
at early times, which is approached as $t\rightarrow-\infty$,
we impose that the functions $f^{\pm}(z_{\pm})$, given by Eq.~(\ref{general}), asymptotically take the form
$f^{\pm}(z_{\pm})\sim e^{\mp ik\tau}$ as $\tau\rightarrow-\infty$.
A comparison with the Minkowski spacetime mode functions shows that the functions $f^{+}(z_{+})$ and $f^{-}(z_{-})$ are positive
and negative frequency mode functions, respectively.
By the virtue of asymptotically expansions of the Whittaker functions $\wwp(z)$ and $\wmp(z)$ as $|z|\rightarrow\infty$, see
Eqs.~(\ref{win}) and~(\ref{min}) respectively,
the normalized positive and negative frequency mode functions are \cite{Kim:2016dmm}, respectively:
\begin{align}
\label{uin}
U_{in}(x)&=(2k)^{-\frac{1}{2}}e^{\frac{i\pi}{2}\kappa}a^{\frac{2-D}{2}}(\tau)e^{+i\k\cdot\x}\wwp(z_{+}), \\
\label{vin}
V_{in}(x)&=(2k)^{-\frac{1}{2}}e^{-\frac{i\pi}{2}\kappa}a^{\frac{2-D}{2}}(\tau)e^{-i\k\cdot\x}\wwm(z_{-}).
\end{align}
A similar discussion is possible in the asymptotic future ($t \rightarrow \infty$). The desired asymptotic form is
$f^{\pm}(z_{\pm})\sim e^{\mp i|\mu|t}$,
leading with Eqs.~(\ref{mout}) and~(\ref{wout}) to the mode functions \cite{Kim:2016dmm}
\begin{align}
\label{uout}
U_{out}(x)&=(4|\mu|k)^{-\frac{1}{2}}e^{\frac{i\pi}{2}\mu}a^{\frac{2-D}{2}}(\tau)e^{+i\k\cdot\x}\wmp(z_{+}),
\\ \label{vout}
V_{out}(x)&=(4|\mu|k)^{-\frac{1}{2}}e^{\frac{i\pi}{2}\mu}a^{\frac{2-D}{2}}(\tau)e^{-i\k\cdot\x}\wmm(z_{-}).
\end{align}
The subscripts $in/out$ denote that these mode functions have the desired asymptotic form at early/late times and the
corresponding \gls{vacuum} state is referred to as the in-vacuum and out-vacuum respectively.
\par
Since the orthonormality of the mode functions should be independent of time, there exists a conserved scalar product. Between
two scalar functions $u_{1}(x)$ and $u_{2}(x)$ it is defined in $D=1+d$ dimension by
\begin{equation}\label{produc}
\big(u_{1},u_{2}\big) \equiv i\int d^{d}\textbf{x}\sqrt{|g|}g^{0\nu}\Big(u_{1}^{\ast}\rnd_{\nu}u_{2}-u_{2}\rnd_{\nu}u_{1}^{\ast}\Big),
\end{equation}
where the integral is taken over a constant $x^{0}$ hypersurface \cite{Birrell1984,book:Parker}. If $u_{1}(x)$ and $u_{2}(x)$
are solutions of the field equation~(\ref{kgeq})
which vanish at spacial infinity, then $(u_{1},u_{2})$ is conserved \cite{book:Parker}. The mode
functions~(\ref{uin})-(\ref{vout}) will be
orthonormal with respect to the scalar product~(\ref{produc}) integrated over a constant $\tau$ hypersurface. Then, the following orthonormality relations are satisfied
\begin{align}\label{orthonorm}
\big(U_{in(out)\k},U_{in(out)\k'}\big)&=-\big(V_{in(out)\k},V_{in(out)\k'}\big)
=(2\pi)^{d}\delta^{(d)}(\k-\k'), \nn\\
\big(U_{in(out)\k},V_{in(out)\k'}\big)&=0.
\end{align}
 \subsection{The Dirac equation in $\ds_2$}
 The Dirac equation can be derived from the action (\ref{action:fer}) by varying with respect to the field $\overline{\psi}(x)$, which gives
 \begin{align}
\left(i\underline{\gamma}^{\mu} \nabla_{\mu} -m \right) \psi(x) =0. \label{eq:CurvedDiracGen}
 \end{align}
 Using the explicit form of the \glspl{tetrad} (\ref{eq:tetrad}) and the \gls{spin connection} (\ref{eq:connection}) the Dirac equation in $\ds_2$ can be derived from
(\ref{eq:CurvedDiracGen}) as
  \begin{equation}
\left[i \left(\underline{\gamma}^{\mu} \partial_{\mu}+\frac{1}{2} \frac{a'(\tau)}{a(\tau)} \underline{\gamma}^0 +i e A_{\mu}(x) \underline{\gamma}^{\mu} \right) -m \right] \psi(x)=0. \label{eq:CurvedDirac}
 \end{equation}
Using a momentum mode decomposition of the form\footnote{The exact form of the momentum decomposition is specified later in (\ref{eq:decomp2}).}
\begin{align}
\psi(x)\sim\e^{i k x_1}  \begin{pmatrix}
           \psi_1(\tau)\\
           \psi_2(\tau) 
          \end{pmatrix}. \label{eq:decomp1}
\end{align}
For fermions, unlike for bosons, we will keep the form of the vector potential as abstract as possible and put the explicit formula for constant electric field \ref{vector} only when required. We hence consider now a solely time dependent electric field $A_\mu(x)=(0,A(\tau))$ we find that (\ref{eq:CurvedDirac}) takes the form
\begin{align}
 i {\psi_1}'(\tau)+p(\tau)\psi_1(\tau)+\frac{1}{2} \frac{a'(\tau)}{a(\tau)}\psi_1(\tau)-m a(\tau)\psi_2(\tau)=&0,\label{eq:CoupledDirac1}\\
 i {\psi_2}'(\tau)-p(\tau)\psi_2(\tau)+\frac{1}{2} \frac{a'(\tau)}{a(\tau)}\psi_2(\tau)-m a(\tau)\psi_1(\tau)=&0 \label{eq:CoupledDirac2},
\end{align}
where
\begin{align}
 p(\tau) \equiv k+eA(\tau)
\end{align}
is the kinetical momentum. Decoupling these equations leads to
\begin{align}
 \psi_1''(\tau)+\left(\omega_k(\tau)^2-i\, p(\tau) \left[\frac{p'(\tau)}{p(\tau)}-\frac{a'(\tau)}{a(\tau)}\right]+\left[\frac{a''(\tau)}{2a(\tau)}-\frac{3 a'(\tau)^2}{4 a(\tau)^2}\right]\right)\psi_1(\tau)=&0, \label{eq:DecoupledDirac1}\\
 \psi_2''(\tau)+\left(\omega_k(\tau)^2+i\, p(\tau) \left[\frac{p'(\tau)}{p(\tau)}-\frac{a'(\tau)}{a(\tau)}\right]+\left[\frac{a''(\tau)}{2a(\tau)}-\frac{3 a'(\tau)^2}{4 a(\tau)^2}\right]\right)\psi_2(\tau)=&0
 ,\label{eq:DecoupledDirac2}
\end{align}
where we defined the the effective frequency
 \begin{align}
\omega_{k}(\tau)^2 := p(\tau)^2+m^2a(\tau)^2. \label{eq:omega}
\end{align}
See the corresponding equation in $\ds_4$ for a discussion about those equations. We now turn to the corresponding problem in $\ds_4$.
\subsection{The Dirac equation in $\ds_4$:}
Varying the action with respect to the spinor field gives the Dirac equation
 \begin{equation}
\left(i \gamma^{\mu} \nabla_{\mu} -m \right) \psi(\textbf{x}, \tau) =0.
 \end{equation}
Using Eqs.~(\ref{eq:tetrad}) and (\ref{connection}), this equation becomes
  \begin{equation}
  \label{eq:diracavantcahnge}
\left\{i \left(\gamma^{\mu} \partial_{\mu}+\frac{3}{2} a H \underline{\gamma^0} + i e A_{\mu} \gamma^{\mu} \right) -m \right\} \psi(\textbf{x}, \tau)=0.
 \end{equation}
One now considers the auxiliary field $ \Psi(\textbf{x}, \tau) = a^{3/2}(\tau) \psi(\textbf{x}, \tau)$ which can be thought of as the equivalent of the \glslink{MS}{Mukhanov-Sasaki variable} (\ref{eq:MSS}). With this substitution the Dirac equation takes the form
\begin{equation}
\left\{\gamma^{\mu}  ( i \partial_{\mu} - e A_{\mu})  - m \right\} \Psi(\textbf{x}, \tau)=0. \label{eq:Dirac}
 \end{equation}
 We will also decompose this field in momentum modes according to
\begin{align}
 \Psi(\textbf{x},\tau)\sim\e^{i\textbf{k}.\textbf{x}}\psi_{\textbf{k}}.\label{TF}
\end{align}
To solve the Dirac equation, for the purpose of the calculation of the \gls{pair production rate}, it is often useful to use the squared version of the Dirac equation because of its similarities to the Klein-Gordon equation, see e.g. \cite{Strobel:2014tha,Dumlu2011}. The squared Dirac equation can be found using
 \begin{align}
  \Psi(\textbf{x},\tau)&=\gamma^{\mu}\left[(i \partial_{\mu}-e A_{\mu}(\tau)) +m a(\tau)\right]\phi(\textbf{x},\tau) \label{eq:squared}
 \end{align}
 with
 \begin{align}
 \phi_{\textbf{k}}(\tau)&=\begin{pmatrix}\phi_1(\tau)\\  \phi_2(\tau) \end{pmatrix},&&  \phi_i(\tau)=\begin{pmatrix} \phi_i^+(\tau)\\  \phi_i^-(\tau)\end{pmatrix}.
 \end{align}  
in the Dirac equation. In the previous equation, $ \phi_{\textbf{k}}(\tau)$ is the Fourier transform (in the sense of (\ref{TF})) of $\phi(\textbf{x},\tau)$. Again, we will not specify the electric field until required but different from $\ds_2$, we consider it pointing in the $z$-direction:
\begin{equation}
\label{eq:efield}
A_{\mu}\equiv A(\tau)\delta^{3}_{\mu}.
\end{equation}
For such fields, the squared Dirac equation takes the form 
\begin{align}
&\left( \partial_{\tau}^2+\omega_{\textbf{k}}(\tau)^2-i ma'(\tau) \right)\phi_1^{\pm} \pm i e A'(\tau) \phi_2^{\pm}(\tau)=0, \\
&\left(\partial_{\tau}^2+\omega_{\textbf{k}}(\tau)^2+i ma'(\tau) \right)\phi_2^{\pm} \pm i e A'(\tau) \phi_1^{\pm}(\tau)=0,   \label{to solve}
\end{align}
where the effective pulsation and the kinetical momentum are defined as
\begin{align}
& \omega_{\textbf{k}}(\tau)^2\equiv p_z(\tau)^2+k_{\perp}^2+m^2 a(\tau)^2, \label{eq:omega}
& p_z(\tau)\equiv k_z+e A(\tau),
&& k_\perp^2\equiv k_x^2+k_y^2.
\end{align}
At this point, it is interesting to compare fermionic and bosonic results, it possible to do it in any dimension, but it is more straightforward in the Dirac basis, so we will restrict to the bosonic problem in $4D$. Putting $\xi=0$ and $d=3$ in (\ref{vphieq}) together with (\ref{fpm}) gives: 
\begin{equation}
\label{boson to solve}
\left(\partial_{\tau}^2+\omega_{\textbf{k},B}(\tau)^2 \right) f_{\textbf{k}} =0,
\end{equation}
with
\begin{align}
\omega_{\textbf{k},B}(\tau)^2=\omega_{\textbf{k}}(\tau)^2-\frac{2}{\tau^2}. \label{eq:omegaboson}
\end{align}
The equation of the bosonic problem (\ref{to solve}) can be understood as a harmonic oscillator with a time dependent pulsation. The two other terms in Eq.~(\ref{to solve}) are new and due to the fermionic nature of the particles considered. On the one hand, the mass term was already derived \emph{e.g}. in \cite{Landete2014} where no background electric field was considered. On the other hand, the electric term is present for instance in \cite{Kluger1992} in flat spacetime. Such that this equation is a generalization of the Dirac equation in \gls{curved spacetime} with background electric and gravitational fields.
\par The squared Dirac equation is also analogous to the Dirac equation for two-component fields in flat spacetime. In \cite{Strobel:2014tha} a method was used to \gls{semiclassical}ly compute the \gls{pair production rate} for these fields. It is possible to use the same method for the case studied here. Instead of looking for a solution of the squared Dirac-equation (\ref{to solve}) we will however use the ansatz 
\begin{align}
\label{newansatz}
 \psi_{\vec{k},\uparrow}(\tau)=\begin{pmatrix}
                      -i k_\perp \, \psi_2^+(\tau)\\
                       (k_x+i k_y)\, \psi_2^+(\tau)\\
                      -i k_\perp \, \psi_1^+(\tau)\\
                      -(k_x+i k_y)\, \psi_1^+(\tau)\\
                     \end{pmatrix}, &&
   \psi_{\vec{k},\downarrow}(\tau)=\begin{pmatrix}
                      (k_x-i k_y)\, \psi_2^-(\tau)\\
                      i k_\perp \, \psi_2^-(\tau)\\
                      -(k_x-i k_y) \, \psi_1^-(\tau)\\
                      i k_\perp\, \psi_1^-(\tau)\\
                     \end{pmatrix} .                  
\end{align}
This ansatz can be derived by finding the solution of the squared equation analogous to \cite{Strobel:2014tha} and then use (\ref{eq:squared}) to construct a solution for the Dirac equation (\ref{eq:Dirac}).
Observe that \(\psi_{\vec{k},\uparrow}(\tau)\) and \(\psi_{\vec{k},\downarrow}(\tau)\) are independent since
\begin{align}
\label{eq:dagger}
 \psi_{\textbf{k},\uparrow}(\tau)^\dagger\cdot \psi_{\textbf{k},\downarrow}(\tau)=0.
\end{align}
Putting (\ref{newansatz}) in the Dirac equation (\ref{eq:Dirac}) leads to the equations
\begin{align}
 i\, \psi_1^{'\pm}(\tau)+ m a(\tau)\, \psi_1^\pm(\tau)\pm (p_z(\tau)+i k_\perp)\,\psi_2^\pm(\tau)=0,\label{to solve 1}\\
 i\, \psi_2^{'\pm}(\tau)- m a(\tau)\, \psi_2^\pm(\tau)\pm (p_z(\tau)-i k_\perp)\,\psi_1^\pm(\tau)=0,\label{to solve 2}
\end{align}
 which we will resolve in Sec.~\ref{WKB}.

%% file: semi/proceed7.tex
\par We will in this section present a first technique to evaluate the \gls{particle creation} in \gls{dS} under the influence of an electric field. This technique does not require an analytic expression for the electric field but on that the electric field depends solely on time and has only one component as described for instance in (\ref{eq:efield}). The method is called the \gls{semiclassical scattering method} \cite{Brezin1970,Popov1971,Popov1972,Popov1973,Popov2001,Kleinert2008,Kleinert2013,Strobel:2013vza,StrobelPhD}. In Eckhard Strobel's PhD thesis \cite{StrobelPhD}\footnote{available on demand}, it was well explained that it relies ultimately on approximating integrals by the value of their residutes at the turning points. It is usually referred to as \acrshort{WKB} but it is more precise \cite{StrobelPhD,Blinne:2015zpa} to call it the \emph{\gls{semiclassical scattering method}}. The \gls{semiclassical scattering method} has also been successfully applied in flat spacetime problems for a two-components electrical field \cite{Strobel:2014tha}. We generalize here the \gls{semiclassical scattering method} to \gls{curved spacetime} examples. 
\par The \gls{semiclassical} expansion is considering cases where a \gls{vacuum} state for the created particles exits in the asymptotic future. That is true if the background fields are evolving slowly This is called the adiabatic condition and is a \gls{semiclassical} approximation. To explicit it, we will consider the case of bosons and positive frequencies, but the final result will hold in the general case. The mode equation~(\ref{feq}) can be rewritten
\begin{equation}\label{refeq}
\left(\partial_{\tau}^2+\omega_{\textbf{k},B}(\tau)^2 \right) f_{\textbf{k}} =0.
\end{equation}
where the momentum dependent frequency is
\begin{equation}\label{omega}
\omega_{\textbf{k},B}^2 \equiv k^{2}-\frac{2eEk_{x}}{H^{2}\tau}+\frac{1}{\tau^{2}}\Big(\frac{m^{2}}{H^{2}}+\frac{e^{2}E^{2}}{H^{4}}
+\frac{1-d^{2}}{4}+\xi D(D-1)\Big)=k^{2}-\frac{2eEk_{x}}{H^{2}\tau}+\frac{1}{\tau^{2}}\left(\frac{1}{4}-\mu^2\right).
\end{equation}
In this section, we are not interested in the role of the \gls{conformal coupling} $\xi$ so for the sake of simplicity, we will set it to zero: $\xi=0$. The adiabatic condition requires the effective frequency $\omega_{\textbf{k},B}$ to vary slowly in the asymptotic future, that is to satisfy the relations
\begin{align}\label{condition}
\frac{\omega_{\textbf{k},B}^{'2}}{\omega_{\textbf{k},B}^{4}}&\ll 1, & \frac{\omega''_{\textbf{k},B}}{\omega_{\textbf{k},B}^{3}}&\ll 1.
\end{align}
Explicit calculations in the asymptotic future ($ \tau \rightarrow 0$) gives:
\begin{equation}\label{omegao}
\frac{\omega_{\textbf{k},B}^{'2}}{\omega_{\textbf{k},B}^{4}}\sim\frac{1}{2}\frac{\omega^{''}_{\textbf{k},B}}{\omega_{\textbf{k},B}^{3}}
\sim\Big(\frac{1}{4}-\mu^2\Big)^{-1}=\Big(\lambdam^2+\lambda^2 +\frac{1-d^{2}}{4}\Big)^{-1},
\end{equation}
under the condition that
\begin{equation}\label{assume}
\frac{m^{2}}{H^{2}}+\frac{e^{2}E^{2}}{H^{4}} \gg \frac{d^2-1}{4},
\end{equation}
the adiabatic condition is satisfied. We add that in term of the definitions (\ref{lambda}) and (\ref{eq:mu}), this gives: $\rho \sim |\mu| \gg \frac{d^2-1}{4}$. 
The condition~(\ref{assume}) justifies our assumption about the range of parameters $m/H$ and
$eE/H^{2}$ which leads to $\mu^{2}< 0$.
In our investigation, the spacetime dimension is not too large, \textit{i.e.}, $d\sim 1$. Hence, the condition~(\ref{assume}) implies $\rho^{2}\gg 1.$
Observe that assuming~(\ref{assume}), implies also that if one does not want to have trivial flat spacetime results, one needs to assume as well:
\begin{equation}\label{lambdamgg}
\gamma^2=\left(\frac{m}{H}\right)^{2} \gg \frac{d^{2}-1}{4}.
\end{equation}
The proof of the previous statement is that if one assumes $\gamma\ll 1$, together with the \gls{semiclassical} condition~(\ref{assume}), it will be
equivalent to assume $\lambda\gg 1$.
In this limit the scalar field, the electromagnetic field and the \gls{dS} are conformally invariant, leading to flat spacetime results.
Under the \gls{semiclassical} approximation, two regimes can be discussed: strong electric field that we define as $\lambda\gg\max(1,\gamma)$ which will give results very close to the flat spacetime results and heavy scalar field that we define as $\gamma\gg\max(1,\lambda)$.
\par When the adiabatic condition is realized, it is also possible to derive a heuristic relation between the comoving momentum $k$ and the conformal time $\tau$ when most of the \glslink{particle creation}{particles} are created. To find this relation, one has to study when the violation of the adiabaticity is maximal, \textit{i.e.}~when the rate of change of the frequency $\omega_{\textbf{k},B}$ is extremal. 
 One can show that the maximum gives the estimate for the creation time \cite{Froeb2014} 
\begin{equation}
\label{estimate}
\tau \sim -\frac{|\mu|}{|k|}.
\end{equation}
Using this the \(k\)-integral can be changed into a time integral. Detailed justification of this procedure can be found in\cite{Gavrilov:1996pz,Kluger:1998bm,Anderson:2013ila}.
\par We will now turn to the \gls{semiclassical scattering method}, the strategy for the calculation will be the following:
\begin{itemize}
\item Reformulate the equation of motion in terms of an equation for the mode functions \(\alpha(\tau)\), \(\beta(\tau)\).
\item Perform a multiple integral iteration to compute $|\beta|^2$.
\item Calculate the resulting integrals with a \gls{semiclassical} saddle point approximation to derive the number of produced \glspl{pair} for each momentum mode \(\mathbf{k}\).
\end{itemize}
\subsection{Klein-Gordon field in $\dsd$}
\label{sec:sctscalar}
\subsubsection{Equations for the mode functions}
From the equation of motion (\ref{refeq}), it is
possible to do a \gls{Bogoliubov transformation} and reformulate
this equation in term of the Bogoliubov coefficients $\alpha_{\k,B}(\tau)$ and $\beta_{\k,B}(\tau)$. The starting point is to implement the \gls{Bogoliubov transformation} using an ansatz inspired by a \acrshort{WKB} expansion
\begin{align}
\label{wkbf}
f_{\k}(\tau)&=\frac{\alpha_{\k,B}(\tau)}{\sqrt{\omega_{\k,B}(\tau)}}e^{-iK_0(\tau)}
+\frac{\beta_{\k,B}(\tau)}{\sqrt{\omega_{\k,B}(\tau)}}e^{iK_0(\tau)}, \\
\label{wkbfd}
\dot{f}_{\k}(\tau)&=-i\omega_{\k,B}(\tau)\bigg[\frac{\alpha_{\k,B}(\tau)}{\sqrt{\omega_{\k,B}(\tau)}}e^{-iK_0(\tau)}
-\frac{\beta_{\k,B}(\tau)}{\sqrt{\omega_{\k,B}(\tau)}}e^{iK_0(\tau)}\bigg],
\end{align}
where
\begin{equation}\label{keq}
K_0(\tau) \equiv \int_{-\infty}^{\tau}\omega_{\k,B}(\tau')d\tau'.
\end{equation}
To preserve the commutation relation, it is necessary to impose the \gls{Wronskian condition} $|\alpha_{\k,B}(\tau)|^2-|\beta_{\k,B}(\tau)|^2=1$.
In this basis, the \gls{number of pairs per momentum} reads
\begin{equation}\label{spectrum}
n_{\k,B} \equiv \lim_{\tau\rightarrow0}\left|\beta_{\k,B}(\tau)\right|^2.
\end{equation}
The initial conditions for the mode functions are chosen such that at past infinity there are only negative frequency modes, \textit{i.e.}
\begin{align}\label{boundary}
\beta_{\k,B}(-\infty)=0, && \alpha_{\k,B}(-\infty)=1.
\end{align}
It is possible to find a first order coupled differential equation for the Bogoliubov coefficients
\begin{align}
\label{alphadot}
\alpha'_{\k,B}(\tau)&=\frac{\omega'_{\k,B}(\tau)}{2\omega_{\k,B}(\tau)}e^{2iK_0(\tau)}\beta_{\k,B}(\tau), \\
\label{betadot}
\beta'_{\k,B}(\tau)&=\frac{\omega'_{\k,B}(\tau)}{2\omega_{\k,B}(\tau)}e^{-2iK_0(\tau)}\alpha_{\k,B}(\tau).
\end{align}
\subsubsection{Multiple integral iteration}
\label{sec:miscalar}
Note that at this point, the equations derived are still exact. Aiming at finding the \gls{number of pairs per momentum} (\ref{spectrum}), it is possible
to integrate formally Eqs.~(\ref{alphadot}) and~(\ref{betadot}) by using the boundary condition~(\ref{boundary}). One finds \cite{Berry1982}
\begin{eqnarray}\label{multint}
\beta_{\k,B}(0)&=&\sum_{m=0}^\infty\int_{-\infty}^{0}d\tau_{0}
\frac{\omega'_{\k,B}(\tau_{0})}{2\omega_{\k,B}(\tau_{0})}e^{-2iK_0(\tau_{0})} \nn\\
&\times&\prod_{n=1}^{m}\int_{-\infty}^{\tau_{n-1}}dt_{n}\frac{\omega'_{\k,B}(t_{n})}{2\omega_{\k,B}(t_{n})}e^{2iK_0(t_{n})}
\int_{-\infty}^{t_{n}}d\tau_{n}\frac{\omega'_{\k,B}(\tau_{n})}{2\omega_{\k,B}(\tau_{n})}e^{-2iK_0(\tau_{n})}.
\end{eqnarray}
Each of these integrals can be calculated using a saddle point approximation. Those integrals are dominated by the regions around the
turning point, \textit{i.e.}, $\omega_{\k,B}(\tau_{p}^{\pm})=0$, where
the superscript $\pm$ denotes the two conjugate pairs in the complex plane of $\tau$. More precisely, by deforming the contour of
integration, we consider the singularities for the turning point for which
\begin{equation}\label{impart}
\Im[K(\tau_{p})]<0.
\end{equation}
From now on, the subscript $\pm$ will be dropped, and we will consider $\tau_{p}$ the turning point which corresponds to~(\ref{impart}).
Following \cite{Berry1982}, it is possible to describe the
behavior of $\omega_{\k,B}^2(\tau)$ near the turning point assuming first order singularity which is the case contemplating Eq.~(\ref{omega}),
\begin{equation}\label{model}
\omega_{\k,B}^2(\tau)\simeq A(\tau-\tau_{p}),
\end{equation}
with $A$ being a constant which can be calculated. One can find then an expression for $K(\tau)$ near the turning point
\begin{align}
\label{turneq1}
K_0(\tau)&\simeq K(\tau_{p})+\frac{2}{3}A(\tau-\tau_{p})^{\frac{3}{2}}, \\
\label{turneq2}
\frac{\omega '_{\k,B}(\tau)}{\omega_{\k,B}(\tau)}&\simeq\frac{1}{3\big(K_0(\tau)-K_0(\tau_{p})\big)}
\frac{dK_0(\tau)}{d\tau}.
\end{align}
Changing variables to $\xi_{n}=K_0(\tau_{p})-K_0(\tau_{n})$ and $\eta_{n}=K_0(\tau_{p})-K_0(\tau'_{n})$ one gets an approximate expression for
the integrals
\begin{equation}\label{semibeta}
\beta_{\k,B}(0)\simeq-2i\pi e^{-2iK(\tau_{p})}\sum_{m=0}^{\infty}\frac{(-1)^m}{6^{m+1}}I_{m},
\end{equation}
where
\begin{equation}\label{im}
I_{m}=\frac{1}{2i\pi}\int_{-\infty}^{\infty}d\xi_{0}\frac{e^{i\xi_{0}}}{\xi_{0}}\prod_{n=1}^{m}
\int_{-\infty}^{\xi_{n-1}}d\eta_{n}\frac{e^{-i\eta_{n}}}{\eta_{n}}\int_{\eta_{n}}^{\infty}
d\xi_{n}\frac{e^{i\xi_{n}}}{\xi_{n}}=\frac{\pi^{2m}}{(2m+1)!}.
\end{equation}
Using (\ref{spectrum}), the final result reads then
\begin{equation}\label{final}
n_{\k,B}=\left|e^{-2iK_0(\tau_{p})}\right|^2.
\end{equation}
\subsubsection{Explicit calculation of the number of pairs}
For the \gls{semiclassical} approximation to hold, one needs the notion of adiabatic \gls{vacuum} in the asymptotic future. Hence, the \gls{semiclassical}
approximation holds if the relation~(\ref{assume}) is satisfied.
The remaining step is to compute the integral~(\ref{keq}). The turning point is given by
\begin{equation}\label{taup}
\tau_{p}=\frac{1}{k}\Big[i\kappa-i\Big(\rho^{2}+\kappa^{2}\Big)^{\frac{1}{2}}\Big],
\end{equation}
where the coefficients $\rho$ and $\kappa$ have been defined in Eqs.~(\ref{lambda}) and~(\ref{kappa}), respectively. Then one can find
the imaginary part of $K(\tau)$
\begin{equation}\label{findim}
\Im[K_0(\tau_{p})]=-\pi\big(\rho-\lambda r\big)\theta\big(\lambda r\big),
\end{equation}
where the Heaviside step function $\theta$ is there to ensure that the condition~(\ref{impart}) holds. Eq.~(\ref{findim}) implies
that, e.g., a particle with charge $e>0$ is only created with a momentum $k_{x}>0$. We see that in the \gls{semiclassical} limit,
the upward tunneling is suppressed
and only the screening direction or downward tunneling stays. The number of \glspl{pair} in the \gls{semiclassical} limit is eventually given by
\begin{equation}
n_{\k,B}=\exp\Big[-2\pi\Big(\rho-\lambda r\Big)\Big]\theta(\lambda r).
\label{mainB}
\end{equation}
The \gls{pair production rate} is defined as
\begin{equation}\label{semidif}
\Gamma_B \equiv \frac{1}{\Delta V}\int\frac{d^{d}k}{(2\pi)^{d}}\,n_{\k},
\end{equation}
where $\Delta V=a^{D}(\tau)\Delta\tau$ is the slice of $D$-volume in the conformal time interval $\Delta \tau$. The procedure then is to transform the $k$-integral into a $\tau$-integral
by using an estimate for the time when most of the
\glslink{particle creation}{particles} are created, see Eq.~(\ref{estimate}). Using Eqs.~(\ref{formul}) (\ref{formula}) (\ref{area}), it is then possible to present
the final expression for the scalar \gls{pair production rate} in $\dsd$ under the influence of a constant electric field:
\begin{equation}\label{smifinal}
\Gamma_B=\frac{H^{D}}{(2\pi)^{d}}\rho^{d}|\lambda|^{\frac{1-d}{2}}e^{-2\pi(\rho-|\lambda|)}.
\end{equation}
The scalar \gls{pair production rate} is constant with respect to conformal time. It signals that \gls{pair} production in $\dsd$ from electric and gravitational fields exactly
balances the dilution from the expansion of the universe.
This implies that the population of scalars is always dominated by the \glslink{particle creation}{particles} created within a Hubble time \cite{Kobayashi:2014zza}. It is interesting to note that when one changes the space dimension, what
changes is the prefactor before the exponential.
The exponential factor is the classical trajectory
which is not a function of the dimension $d$. However,
the one loop integration depends on $d$ and hence the prefactor to the classical trajectory is a function of $d$.
\subsection{Dirac field in $\ds_2$}
\label{sec:ssmf}
\subsubsection{Equations for the mode functions}
We will now detail the calculation for fermions in $\ds_2$. The main novelty is the way the spinor sector is handled. For this reason, the corresponding calculation in $\ds_3$ present few interest because the spin sector in $2D$ is the same as in $3D$. Inspired by similarities between Eqs.~(11)-(12) of \cite{Strobel:2014tha} and (\ref{to solve 1})-(\ref{to solve 2}), we propose the following ansatz
\begin{align}\Psi_{1}(\tau)&=\frac{1}{\sqrt{2\omega_{k}(\tau)}}\left(\alpha(\tau)\sqrt{\omega_{k}(\tau)-p(\tau)}\,{\e^{-\frac{i}{2}K_0(\tau)}} +\beta(\tau)\sqrt{\omega_{k}(\tau)+p(\tau)}\,{\e^{\frac{i}{2}K_0(\tau)}}\right),\label{eq:ansatz1w} \\
 \Psi_{2}(\tau)&=\frac{1}{\sqrt{2\omega_{k}(\tau)}}\left(\alpha(\tau)\sqrt{\omega_{k}(\tau)+p(\tau)}\,{\e^{-\frac{i}{2}K_0(\tau)}} -\beta(\tau)\sqrt{\omega_{k}(\tau)-p(\tau)}\,{\e^{\frac{i}{2}K_0(\tau)}}\right), \label{eq:ansatz2w}
\end{align}
with the integral is defined slightly differently than in Sec.~\ref{sec:sctscalar}, this is done only to harmonize the notation but has no physical meaning.
 \begin{align}
 K_0(\tau)&  \equiv  2 \int_{-\infty}^{\tau} \omega_{k}(\tau) d \tau
 \label{eq:K_xyw}.
 \end{align}
 The Dirac equation (\ref{to solve 1})-(\ref{to solve 2}) leads to coupled differential equations for the mode functions by inserting the ansatz (\ref{eq:ansatz1w})-(\ref{eq:ansatz2w}). They read
\begin{align}
 \alpha'(\tau)&=\frac{\omega_{k}'(\tau)}{2 \omega_{k}(\tau)}G(\tau)\, \e^{i K_0(\tau)}\beta(\tau),\label{eq:alphafermw}\\
 \beta'(\tau)&=-\frac{\omega_{k}'(\tau)}{2 \omega_{k}(\tau)}G(\tau)\, \e^{-i K(\tau)}\alpha(\tau),\label{eq:betafermw}
\end{align}
with
\begin{align}
& G(\tau) \equiv \frac{p(\tau)}{ma(\tau)}-\frac{\omega_{k}(\tau)p'(\tau)}{ma(\tau)\omega_{k}'(\tau)}, 
\end{align}
which can be seen as fermionic corrections to the analog bosonic case. The initial conditions for the mode functions are chosen such that at past infinity there are only negative frequency modes, see equation (\ref{boundary}).
\subsubsection{Multiple integral iteration}
It is now possible to iteratively integrate  (\ref{eq:alphafermw}) and (\ref{eq:betafermw}), which leads to
\begin{equation}
 \begin{split}
 \beta(0)=&\sum_{m=0}^\infty (-1)^{m+1}\int_{-\infty}^0 d\eta_0\frac{\omega_k'(\eta_0)}{2\omega_k(\eta_0)}G(\eta_0)\e^{-i K(\eta_0)} \\&  \hspace{2cm}\times
\prod_{n=1}^m \int_{-\infty}^{\eta_{n-1}} d\tau_n\frac{\omega_k'(\tau_n)}{2\omega_k(\tau_n)}G(\tau_n)\e^{i K_0(\tau_n)} \int_{-\infty}^{\tau_n}d\eta_n\frac{\omega_k'(\eta_n)}{2\omega_k(\eta_n)} G(\eta_n)\e^{-i K_0(\eta_n)}. \\\label{eq:multintw}
 \end{split}
\end{equation}
One can use the fact these integrals are dominated by the classical turning points \cite{Berry1982}, which are given by 
\begin{align}
\label{eq:TPw}
 \omega_k(\tau_p^\pm)=0 .
\end{align}
It is possible to show that for one pair of simple turning points the \gls{number of pairs per momentum} in a \gls{semiclassical} saddlepoint approximation is given by (see \cite{Strobel:2014tha,Dumlu2011,Berry1982})
\begin{align}
 n_{k} = \lim_{\tau\rightarrow0} \left|\beta(\tau)\right|^2 = \left|\e^{-i K_0(\tau_p^{-})}\right|^2 \label{eq:MomentumSpectrumConstw}.
\end{align}
The detailed intermediate steps of the derivation can be found in Eqs.~(32)-(38) of \cite{Strobel:2014tha} and are very similar to the bosonic case depicted also in Sec.~\ref{sec:miscalar}. Observe that in \cite{Strobel:2014tha,Dumlu2011,Berry1982} the integration contour is closed in the upper imaginary half plane whereas, we (as in \cite{Stahl:2015cra}) close it in the lower imaginary half plane because of opposite convention for the phases in (\ref{eq:multintw}).\\
The above result represents the \gls{semiclassical} \gls{number of pairs per momentum} $k$ for general electric fields in $\ds_2$. In the following we will concentrate on the constant electric field, where more explicit computations can be performed.
\subsubsection{Explicit calculation of the number of pairs}
Using an explicit expression for $\omega_k$ (\ref{eq:omega}) together with (\ref{vector}) and (\ref{metricds}), it is possible to get a result for the \gls{number of pairs per momentum}. The starting point is to compute the turning points (\ref{eq:TPw}), which are given by
\begin{align}
\tau_p^{\pm}=-\frac{\lambda}{k}\pm\frac{i\gamma}{|k|},
\end{align}
where the absolute value of \(k\) was introduced such that \(\tau_p^-\) always denotes the turning point in the lower imaginary plane. We now find that the real part of \(\tau_p^{\pm}\) is only negative, \textit{i.e.}~inside the integration contour which is used for the approximation of (\ref{eq:multintw}), if \(k\) and \(\lambda\) have the same sign. In \cite{Froeb2014} this is called \gls{pair} production in ``screening'' direction because the produced \glspl{pair} would reduce the electric field if we would allow \gls{backreaction}. In the language of false vacuum decay this would be ``downward tunneling''. We find that as in flat spacetime in the \gls{semiclassical} limit no \glspl{pair} are produced in ``anti-screening'' direction which would be connected to ``upward'' tunneling. 
\par Only the imaginary part of $K(\tau_p^-)$ is contributing to (\ref{eq:MomentumSpectrumConstw}). It is given by
\begin{align}
\Im[K_0(\tau_p^-)]=-\pi\left(\sqrt{\gamma^2+\lambda^2}-|\lambda|\right)\theta(k \lambda).
\end{align}
Thus we find that the \gls{number of pairs per momentum} \(k\) (\ref{eq:MomentumSpectrumConstw}) in the \gls{semiclassical} limit is 
\begin{align}
n_{k}  \equiv \exp\left[-2\pi \left(\sqrt{\gamma^2+\lambda^2}-|\lambda|\right)\right]\theta(k \lambda).
\label{main2}
\end{align}
Observe that in the regime we are working, this result is the same than the bosonic case (\ref{mainB}), it is a common feature of \gls{semiclassical} techniques: the spin is washed out. At this point it is also insightful to discuss what \gls{semiclassical} means in this context. The number of produced \glspl{pair} is related to the effective action via \(n_k= \exp(-S)\). The \gls{semiclassical} limit is the limit of large action, \textit{i.e.}~$|\mu|-|\lambda| \gg 1$. A necessary condition for this to happen is $|\mu| \gg 1$. Two regimes can be discussed:
\paragraph*{Weak electric field:} For $\lambda \ll \gamma$, one finds $S \sim 2 \pi (\gamma - |\lambda|)$. The \glspl{pair} are mainly produced by gravitational (cosmological) \gls{pair} production. The first term is the usual Boltzmann factor for non-relativistic massive particles at the Gibbons-Hawking temperature, while the second term is the correction of the small electric field. The number of created \glslink{particle creation}{particles} gets suppressed by the electric field in this limit. 
\paragraph*{Strong electric field:}
For a strong electric field one finds $S= \pi m^2/(eE)$, which is the usual \gls{semiclassical} action for the \gls{Schwinger effect} in flat spacetime. The effect of curvature is negligible. 
\subsection{Dirac field in $\ds_4$}
\label{WKB}
\subsubsection{Equations for the mode functions}
We will now generalize this result to $\ds_4$, again the difficulty will be in handling the spinor sector of the theory, leading to more elaborate ansatz. To derive analogous equations in the fermionic case we start from the equations (\ref{to solve 1}) and (\ref{to solve 2}). For the \gls{semiclassical} treatment we make the following ansatz  
\begin{align}\psi^{\pm}_{1}(\tau)&=\frac{C_\pm}{\sqrt{\omega_{\textbf{k}}(\tau)}}\frac{\sqrt{p(\tau)}}{\sqrt{p_z(\tau)-i k_\perp}}\left(\alpha^\pm(\tau)[\omega_{\textbf{k}}(\tau)-m a(\tau)]{\e^{-\frac{i}{2}K(\tau)}} +\beta^\pm(\tau)[\omega_{\textbf{k}}(\tau)+m a(\tau)]{\e^{\frac{i}{2}K(\tau)}}\right),\label{eq:ansatz1} \\
 \psi^{\pm}_{2}(\tau)&=\frac{\mp C_\pm}{\sqrt{\omega_{\textbf{k}}(\tau)}}\frac{\sqrt{p(\tau)}}{\sqrt{p_z(\tau)+i k_\perp}}\left(\alpha^\pm(\tau)[\omega_{\textbf{k}}(\tau)+m a(\tau)]{\e^{-\frac{i}{2}K(\tau)}} +\beta^\pm(\tau)[\omega_{\textbf{k}}(\tau)-m a(\tau)]{\e^{\frac{i}{2}K(\tau)}}\right), \label{eq:ansatz2}
\end{align}
with the integrals
 \begin{align}
 K(\tau)& \equiv K_0(\tau)+K_1(\tau),\label{eq:K_s}\\
 K_1(\tau)& \equiv k_\perp\int_{-\infty}^{\tau} \frac{m a(\tau)p_z'(\tau)}{\omega_{\textbf{k}}(\tau)p(\tau)^2}d\tau \label{eq:K_xy}.
 \end{align}
 Using the ansatz (\ref{eq:ansatz1})-(\ref{eq:ansatz2}) in (\ref{to solve 1}) and (\ref{to solve 2}), we find that the mode functions are connected through coupled differential equations
\begin{align}
 \alpha'^\pm(\tau)&=-\frac{\omega_{\textbf{k}}'(\tau)}{2 \omega_{\textbf{k}}(\tau)}G_{\alpha}(\tau) \e^{i K(\tau)}\beta^\pm(\tau),\label{eq:alphaferm}\\
 \beta'^\pm(\tau)&=\frac{\omega_{\textbf{k}}'(\tau)}{2 \omega_{\textbf{k}}(\tau)}G_{\beta}(\tau) \e^{-i K(\tau)}\alpha^\pm(\tau),\label{eq:betaferm}
\end{align}
with
\begin{align}
& G_{\alpha} \equiv \frac{ma(\tau)}{p(\tau)}-\frac{\omega_{\textbf{k}}(\tau)ma'(\tau)-i k_\perp p_z'(\tau)}{p(\tau)\omega_{\textbf{k}}'(\tau)}, \\
&G_{\beta} \equiv \frac{ma(\tau)}{p(\tau)}-\frac{\omega_{\textbf{k}}(\tau)ma'(\tau)+i k_\perp p_z'(\tau)}{p(\tau)\omega_{\textbf{k}}'(\tau)},
\end{align}
representing the fermionic corrections to the analog bosonic case.
\subsubsection{Multiple integral iteration}
In this section we will perform the multiple integral iteration for the fermionic case. By iteratively using Eqs.~(\ref{eq:alphaferm}) and (\ref{eq:betaferm}) and the boundary conditions (\ref{boundary}) one finds
 \begin{equation}
 \begin{split}
& \beta^{\pm}(0)=\sum_{m=0}^\infty \int_{-\infty}^\infty d\eta_0\frac{\omega_k'(\eta_0)}{2\omega_k(\eta_0)}G_{\beta}(\eta_0)\e^{-i K(\eta_0)} \\
& \times \prod_{n=1}^m \int_{-\infty}^{\eta_{n-1}} d\tau_n\frac{\omega_k'(\tau_n)}{2\omega_k(\tau_n)}G_{\alpha}(\tau_n)\e^{i K(\tau_n)} \int_{-\infty}^{\tau_n}d\eta_n\frac{\omega_k'(\eta_n)}{2\omega_k(\eta_n)} G_{\beta}(\eta_n)\e^{-i K(\eta_n)}. \\\label{eq:multint}
 \end{split}
\end{equation}
Again the integration procedure is the same as described in Sec.~\ref{sec:sctscalar} and \ref{sec:ssmf}, and can be also found in the references provided there. The result is:
\begin{align}
 n_{\textbf{k},F}=\left|\e^{-i K(\tau_p^{-})}\right|^2 \label{eq:MomentumSpectrumConst}.
\end{align}
By a direct inspection of (\ref{final}), we find that, analogously to the case of two-component fields in flat spacetime, the difference between fermions and bosons is a factor of the form \(\exp(K_1(\tau_p^-))\).
\subsubsection{Explicit calculation of the number of pairs}
\label{sec:const}
To compute the \gls{pair production rate} we first have to compute the integrals \(K_0(\tau)\) and \(K_1(\tau)\) defined in (\ref{eq:K_xyw}) and (\ref{eq:K_xy}) respectively. The value of $\tau$ at the turning point (\ref{eq:TPw}) is found to be
\begin{equation}
\tau_p^{-}=\frac{-\lambda\frac{k_z}{k}-i\sqrt{\gamma^2+\lambda^2\left(1-\left(\frac{k_z}{k}\right)^2\right)}}{k}.\label{eq:TP}
\end{equation}
The imaginary parts of $K_0(\tau_p^-)$ and $K_1(\tau_p^-)$ are the only ones contributing to (\ref{eq:MomentumSpectrumConst}). One can show that
\begin{align}
&\Im[K_0(\tau_p^-)]=-\pi\left(\rho-\frac{k_z}{k} \lambda\right)\theta(k_z \lambda),\\
& \Im[K_1(\tau_p^-)]=0,
\end{align}
\par $K_1(\tau_p^-)$ was the only difference between bosons and fermions and is not contributing to the number of \glspl{pair} produced. Thus we find that the \gls{number of pairs per momentum} in the \gls{semiclassical} limit for both bosons and fermions is given by 
\begin{equation}
n_{\textbf{k}} = \exp\left[-2\pi \left(\rho-\frac{k_z}{k} \lambda\right)\right] \theta(k_z \lambda).
\label{main4}
\end{equation}
The definition of the \gls{pair production rate} is
\begin{equation}
\Gamma \equiv \frac{1}{(2 \pi)^3 V} \int d^3 \textbf{k} \,n_{\textbf{k}},
\end{equation}
where $ V= a(\tau)^4 d\tau$ is the unit four volume of the spacetime. Using (\ref{estimate}) the \(k\)-integral can be changed into a time integral. Going to spherical coordinates the $k_z$ integral can be performed. Putting everything together, one finds
\begin{equation}
\Gamma_F= \frac{H^4}{(2\pi)^3} \frac{\rho^3}{|\lambda|} \left(\e^{2 \pi |\lambda|}-1\right) \e^{-2 \pi \rho}.
\end{equation}
This is to be compared with \ref{smifinal}. We see that in the \gls{semiclassical} limit $\rho \gg 1$ and $\lambda \gg 1 $ the expressions are equal. As in the bosonic case the \gls{physical number of pairs} \(n\) at the time $\tau$ is given by
\begin{equation}
n \equiv \frac{1}{a(\tau)^3} \int_{-\infty}^{\tau} d\tau a(\tau)^4 \Gamma= \frac{\Gamma}{3H}.
\end{equation}
The fact that it is constant shows that the dilution from the expansion of the universe is exactly compensated by the \glslink{particle creation}{particles} created from \glslink{Schwinger effect}{Schwinger} and gravitational \gls{particle creation}. Hence in the \gls{semiclassical} limit, the population of fermions is always dominated by the \glslink{particle creation}{particles} created within a Hubble time. We will close this section on \gls{semiclassical} estimates by mentioning the flat spacetime limit, it could have been performed for all the case studied there it is ultimately the same calculation so we mention it only here as an illustrative example.
\par The \gls{vacuum} decay rate is defined for fermions as
\begin{equation}
\Upsilon=\log(1-|\beta_{\textbf{k}}|^2).
\end{equation}
The limit in which the \gls{Hubble constant} is negligible compared to the gravitational and electrical strength corresponds to the limit to flat spacetime. After some calculations which are analogous to the ones in \cite{Kobayashi:2014zza} one can find the Minkowski limit by taking $H\rightarrow0$, which gives
\begin{equation}
\label{M4f}
\begin{split}
& \lim_{H\rightarrow 0} \Gamma = \frac{(eE)^2}{(2 \pi)^3} \exp \left(-\frac{\pi m^2}{|eE|} \right), \\
& \lim_{H\rightarrow 0} \Upsilon =\sum_{i=1}^{\infty} \frac{1}{i^2} \frac{(eE)^2}{(2 \pi)^3} \exp \left(-\frac{i\pi m^2}{|eE|} \right).
\end{split}
\end{equation}
These are the familiar results for the \gls{Schwinger effect} in Minkowski spacetime \cite{Sauter1931,Schwinger1951,Heisenberg1936,Narozhnyi:1970uv,Nikishov:1970br}.
\subsection*{Concluding remarks}
\paragraph{} We have presented a first technique to obtain results for \gls{pair} production in \gls{dS} with an electric field. We apply it for boson in $D$ dimension, for fermions in $2D$ and $4D$, see equations (\ref{mainB}), (\ref{main2}) and (\ref{main4}) for those main results in each case. See equation (\ref{final}), (\ref{eq:MomentumSpectrumConstw}) and (\ref{eq:MomentumSpectrumConst}) to remark that the results are independent of the specific form of the field and can be used to compute the \gls{pair production rate} for general time dependent fields. An important point is the fact that, one finds no differences between the fermion and boson results. This equivalence between fermions and bosons in the \gls{semiclassical} limit occurs also for one-component fields in flat spacetime when there is only one pair of turning points (see e.g. \cite{Strobel:2014tha}). With our result, it is possible to obtain also a flat spacetime limit that we presented for the specific case of fermion $4D$ in (\ref{M4f}), it agrees with the usual expression for the \gls{Schwinger effect}.
\paragraph{} Several clues points toward the fact that these results are not the end of the story, first from examples in \gls{dS} and FLRW spacetime \cite{Landete2014,book:Parker}, it is known that in \gls{curved spacetime} differences for \gls{particle creation} exists for varying spin, so it is from the regime we considered that they did not appear. Second the discussion on the \gls{semiclassical} limit revealed that it restricts much the parameter space one can explore, indeed one has to assume (\ref{assume}), which leads to (\ref{lambdamgg}), in order to obtain non-trivial results. The \gls{semiclassical} limit was a condition to have an adiabatic future and a well defined notion of particle. However, as the time derivative is not a \gls{Killing vector}, this notion is not too relevant in \gls{curved spacetime}. We argue that the \gls{induced current} of and the \gls{energy momentum tensor} are quantities better suited to study the \gls{Schwinger effect} in $\ds$. First, those two quantities are not plagued by the absence of a clear notion of particle and so allow to explore a regime where the notion of adiabatic \gls{vacuum} and of particles does not necessarily exist. Second, their definition does not depend on the characteristic time of creation of the \glspl{pair} and the rough estimate (\ref{estimate}) can be avoided. Third, they are well suited to explore \gls{backreaction} effects as they are the quantities going on the right-hand side of \glslink{Maxwell equation}{Maxwell} and \glspl{Einstein equation}. 

%% file: current/boson.tex
\subsection{Pair production}
\label{sec:pcb}
We will now derive an alternative way of calculating (\ref{mainB}) with the use of a \gls{Bogoliubov transformation}, see also \cite{Kim:2016dmm}. For this calculation, we will continue to work in $D$ dimensions but later in Sec.~\ref{sec:curent}, we will restrict to $D=3$.
Similar methods have been used to compute the \gls{pair production rate} in time-dependent fields in flat spacetime for general D-dimensional fields \cite{Gavrilov:1996pz}, without an electric field for bosons in $\ds$ in \cite{Anderson:2013ila} and for the constant field in $\ds_2$ for bosons and fermions respectively in \cite{Froeb2014} and \cite{Haouat2013}. In \cite{Kluger:1998bm} the connection of this technique to kinetic theory was shown in the bosonic case.
\subsubsection*{Quantization of the theory and Bogoliubov transformation}
In Sec.~\ref{sec:KG}, two complete sets of orthonormal mode functions were obtained: $\{U_{in\k},V_{in\k}\}$ given by
Eqs.~(\ref{uin}) (\ref{vin}), and $\{U_{out\k},V_{out\k}\}$ given by Eqs.~(\ref{uout}) (\ref{vout}).
The scalar field operator $\varphi(x)$ is expanded in terms of the $\{U_{in\k},V_{in\k}\}$ set
\begin{equation}\label{phin}
\varphi(x)=\int\frac{d^{d}\textbf{k}}{(2\pi)^{d}}\Big[U_{in\k}(x)a_{in\k}+V_{in\k}(x)b^{\dag}_{in\k}\Big],
\end{equation}
where $a_{in\k}$ annihilates particles described by the mode function $U_{in\k}$, and $b^{\dag}_{in\k}$ creates antiparticles described by the mode
function $V_{in\k}$. The quantization of the theory is implemented by adopting the commutation relations
\begin{equation}\label{comin}
\big[a_{in\k},a_{in\k'}^{\dag}\big]=\big[b_{in\k},b_{in\k'}^{\dag}\big]=(2\pi)^{d}\delta^{(d)}(\k-\k').
\end{equation}
The \gls{vacuum} state is defined as
\begin{align}\label{vacin}
a_{in\k}|0\rangle_{in}=0, && \forall\,\k,
\end{align}
and then the construction of the Fock space can be done similarly to the Minkowski spacetime case. However, there is no $\vec{\rnd_{\tau}}$ \gls{Killing vector} to define positive frequency mode functions, and consequently a unique mode decomposition of the scalar field operator $\varphi(x)$ does not exist.
Therefore, $\varphi(x)$ can be expanded in terms of a second complete set of orthonormal mode functions in the form
\begin{equation}\label{phiout}
\varphi(x)=\int\frac{d^{d}\textbf{k}}{(2\pi)^{d}}\Big[U_{out\k}(x)a_{out\k}+V_{out\k}(x)b^{\dag}_{out\k}\Big],
\end{equation}
where $a_{out\k}$ annihilates particles described by the mode function $U_{out\k}$, and $b^{\dag}_{out\k}$ creates antiparticles described by the
mode function $V_{out\k}$. In this case, the commutation relations are
\begin{equation}\label{comout}
\big[a_{out\k},a_{out\k'}^{\dag}\big]=\big[b_{out\k},b_{out\k'}^{\dag}\big]=(2\pi)^{d}\delta^{(d)}(\k-\k').
\end{equation}
The decomposition of $\varphi(x)$ in Eq.~(\ref{phiout}) defines a new \gls{vacuum} state
\begin{align}\label{vacout}
a_{out\k}|0\rangle_{out}=0, && \forall\,\k,
\end{align}
and a new Fock space. Since both sets are complete, the orthonormal mode functions $U_{out\k}$ can be expanded in terms of the first complete
set of orthonormal mode functions. This is precisely a \gls{Bogoliubov transformation}:
\begin{equation}\label{bogolu}
U_{out\k}(x)=\int \frac{d^{d}\textbf{k'}}{(2\pi)^{d}}\Big[\alpha_{\k,\k'}U_{in\k'}(x)+\beta_{\k,\k'}V_{in\k'}(x)\Big].
\end{equation}
With the orthonormality relations~(\ref{orthonorm}) the Bogoliubov coefficients $\alpha_{\k,\k'}$ and $\beta_{\k,\k'}$ are
\begin{align}\label{bogolab}
\alpha_{\k,\k'}& \equiv \big(U_{out\k},U_{in\k'}\big), & \beta_{\k,\k'}& \equiv -\big(U_{out\k},V_{in\k'}\big),
\end{align}
and have to satisfy the additional relations
\begin{align}\label{bogol}
\int\frac{d^{d}\textbf{k}}{(2\pi)^{d}}\Big[\alpha^{\ast}_{\k,\k'}\alpha_{\k,\k''}-\beta_{\k,\k'}\beta^{\ast}_{\k,\k''}\Big]
&=(2\pi)^{d}\delta^{(d)}\big(\k'-\k''\big), \nn\\
\int\frac{d^{d}\textbf{k}}{(2\pi)^{d}}\Big[\alpha^{\ast}_{\k,\k'}\beta_{\k,\k''}-\beta_{\k,\k'}\alpha^{\ast}_{\k,\k''}\Big]&=0.
\end{align}
As a consequence of Eqs.~(\ref{phin}) (\ref{phiout}) (\ref{bogolu}) the late time annihilation operator $a_{out\k}$ is related to the early time
annihilation operator $a_{in\k}$ by a \gls{Bogoliubov transformation}
\begin{equation}\label{boga}
a_{out\k}=\int\frac{d^{d}\textbf{k'}}{(2\pi)^{d}}\Big[\alpha^{\ast}_{\k,\k'}a_{in\k'}-\beta^{\ast}_{\k,\k'}b^{\dag}_{in\k'}\Big].
\end{equation}
Using $a_{out\k}$ and the \gls{vacuum} state $|0\rangle_{in}$ we can calculate the expectation value of the particle number operator\footnote{One can
verify that $_{in}\langle 0|a^{\dag}_{out\k}a_{out\k}|0\rangle_{in}=\,_{out}\langle 0|a^{\dag}_{in\k}a_{in\k}|0\rangle_{out}$.}
\begin{equation}\label{nout}
_{in}\langle 0|N_{out\k}|0\rangle_{in}=\,_{in}\langle 0|a^{\dag}_{out\k}a_{out\k}|0\rangle_{in}
=\int\frac{d^{d}\textbf{k'}}{(2\pi)^{d}}\big|\beta_{\k,\k'}\big|^{2}.
\end{equation}
Therefore, if $\big|\beta_{\k,\k'}\big|^{2}\neq0$ then particles are created.
\subsubsection*{\label{sec:den}Number of pairs}
In order to obtain the \gls{physical number of pairs} an explicit expression of the Bogoliubov coefficients is needed.
To identify the Bogoliubov coefficients, the orthonormal mode functions, given by Eqs.~(\ref{uin})-(\ref{vout}), should be substituted into
Eq.~(\ref{bogolab}). We then obtain
\begin{align}
\label{alpha}
\alpha_{\k,\k'}&=(2\pi)^{d}\delta^{(d)}\big(\k-\k'\big)\alpha_{\k}, &
\alpha_{\k}&=(2|\mu|)^{\frac{1}{2}}\frac{\Gamma(-2\mu)}{\Gamma(\frac{1}{2}-\mu-\kappa)}
e^{\frac{i\pi}{2}(\kappa-\mu)}, \\
\label{beta}
\beta_{\k,\k'}&=(2\pi)^{d}\delta^{(d)}\big(\k+\k'\big)\beta_{\k}, &
\beta_{\k}&=-i(2|\mu|)^{\frac{1}{2}}\frac{\Gamma(-2\mu)}{\Gamma(\frac{1}{2}-\mu+\kappa)}
e^{\frac{i\pi}{2}(\kappa+\mu)},
\end{align}
observe that the bosonic normalization condition $|\alpha_{\k}|^{2}-|\beta_{\k}|^{2}=1$ is satisfied.
The \gls{number of pairs per momentum}, in the in-vacuum is given by Eq.~(\ref{nout}).
After a short calculation, Eq.~(\ref{beta}) results in
\begin{align}\label{betasq}
|\beta_{\k,\k'}|^{2}=\Big((2\pi)^{d}\delta^{(d)}(\k+\k')\Big)^{2}|\beta_{\k}|^{2}, &&
|\beta_{\k}|^{2}=\frac{e^{-2\pi|\mu|}+e^{2\pi i\kappa}}{2 \sinh(2\pi|\mu|)}.
\end{align}
For convenience we normalize the $d$-volume of $\dsd$ in a box with dimensions $L^{d}$.
Then, \gls{number of pairs per momentum} is
\begin{equation}\label{nv}
\frac{1}{L^{d}}\times\int\frac{d^{d}\textbf{k'}}{(2\pi)^{d}}\big|\beta_{\k,\k'}\big|^{2}=|\beta_{\k}|^{2}.
\end{equation}
Using (\ref{formul}) (\ref{area}) (\ref{areaint}), the number of produced \glspl{pair} per unit $d$-volume is
\begin{equation}\label{intk}
\int\frac{d^{d}\textbf{k}}{(2\pi)^{d}}|\beta_{\k}|^{2}
=\frac{1}{(4\pi)^{\frac{d}{2}}\sinh(2\pi|\mu|)}\Big(\frac{e^{-2\pi|\mu|}}{\Gamma(\frac{d}{2})}
+(\pi\lambda)^{1-\frac{d}{2}}\I_{\frac{d}{2}-1}(2\pi\lambda)\Big)
\int_{0}^{\infty}k^{d-1}\,dk,
\end{equation}
where $\I_{\nu}$ is the modified Bessel function, see Appendix~\ref{app:bessel}.
This integral is not finite, since it takes into account the total number of produced \glspl{pair} from the infinite past to the infinite future.
However, the \gls{pair production rate} is finite. Thus, we convert again the $k$-integral into a $\tau$-integral by using (\ref{estimate})\footnote{Here we also used the fact that $\rho \sim |\mu|$}
\begin{equation}\label{intt}
\int\frac{d^{d}\textbf{k}}{(2\pi)^{d}}|\beta_{\k}|^{2}=
\frac{1}{(4\pi)^{\frac{d}{2}}\sinh(2\pi\rho)}\Big(\frac{e^{-2\pi\rho}}{\Gamma(\frac{d}{2})}
+(\pi\lambda)^{1-\frac{d}{2}}\I_{\frac{d}{2}-1}(2\pi\lambda)\Big)
H^{D}\rho^{d}\int_{-\infty}^{0}\Omega^{D}(\tau)d\tau.
\end{equation}
The \gls{pair production rate} is then given by
\begin{equation}\label{rate}
\Gamma \equiv \frac{1}{\Delta V}\times\int\frac{d^{d}\textbf{k}}{(2\pi)^{d}}|\beta_{\k}|^{2}
=\frac{H^{D}\rho^{d}}{(4\pi)^{\frac{d}{2}}\sinh(2\pi\rho)}
\Big(\frac{e^{-2\pi\rho}}{\Gamma(\frac{d}{2})}+(\pi\lambda)^{1-\frac{d}{2}}
\I_{\frac{d}{2}-1}(2\pi\lambda)\Big),
\end{equation}
where
\begin{equation}\label{slice}
\Delta V=\Omega^{D}(\tau)\Delta\tau
\end{equation}
is the slice of $D$-volume in the conformal time interval $\Delta\tau$.
Using Eqs.~(\ref{cosh}) and~(\ref{sinh}), it can be checked that in the cases $D=2$ and $D=4$, Eq.~(\ref{rate}) agrees with \cite{Froeb2014} and \cite{Kobayashi:2014zza}. The \gls{pair production rate} (\ref{rate}) is independent of time, as a consequence the \gls{physical number of pairs} in the comoving frame at time $\tau$ reads
\begin{equation}\label{density}
n \equiv \Omega^{-d}(\tau)\int_{-\infty}^{\tau}d\tau'\Omega^{D}(\tau')\Gamma=\frac{\Gamma}{Hd}.
\end{equation}
It is constant with respect to time, therefore, the number of \glspl{pair} produced by the background electric and gravitational fields is exactly balanced
by the expansion of the spacetime.
\par
Provided that Eq.~(\ref{assume}) is satisfied, then the Bogoliubov coefficient~(\ref{betasq}) is approximated as 
\begin{equation}\label{betaapprox}
|\beta_{\k}|^{2}\simeq e^{-4\pi\rho}+e^{-2\pi(\rho-\lambda r)}.
\end{equation}
From the definitions in Eqs.~(\ref{lambda}), $|r|\leq 1$ implying $\rho\geq\lambda r$, so the first term in the right-hand side of Eq.~(\ref{betaapprox}) is smaller than the second one. Then, to leading order, we find
\begin{equation}\label{approxbeta}
|\beta_{\k}|^{2}\simeq\exp\Big[-2\pi\Big(\Big(\frac{\mds^{2}}{H^{2}}+
\frac{(eE)^{2}}{H^{4}}\Big)^{\frac{1}{2}}+\frac{eE}{H^{2}}\frac{k_{x}}{k}\Big)\Big].
\end{equation}
Therefore, under the \gls{semiclassical} condition~(\ref{assume}), $\beta_{\k}$ is nonzero for both $k_{x}>0$ and $k_{x}<0$ \cite{Garriga:1994bm}.
In the language of nucleation of bubbles, considering $E>0$ and taking the particle with charge $|e|$ to the right of the particle with charge
$-|e|$, the \glspl{pair} can nucleate in both the screening and the antiscreening orientations (corresponding to $k_{x}<0$ and $k_{x}>0$, respectively)
because of the gravitational effects \cite{Garriga1994}.
Hence, creating charges in the screening orientation tends to decrease the background electrical field while creating them in anti-screening
orientation tends to increase it.
Usually screening and antiscreening orientations are referred to as downward and upward tunneling \cite{Froeb2014}.
\par
The Minkowski spacetime limit is obtained in the limit $H\rightarrow0$. The \gls{pair production rate} (\ref{rate}) in this limit approaches
\begin{equation}\label{limit}
\lim_{H\rightarrow0}\Gamma=\frac{|eE|^{\frac{D}{2}}}{(2\pi)^{d}}e^{-\frac{\pi m^{2}}{|eE|}},
\end{equation}
which is the same result as the \glslink{Schwinger effect}{Schwinger pair production} rate in $D$ dimensional Minkowski spacetime \cite{Gavrilov:1996pz}.
In $\dsd$, the \gls{pair} production rate is higher than in flat spacetime, due to the gravitational \gls{pair} production contribution.
\subsection{\label{sec:curent}Induced current and conductivity in $D=3$ dimension}
In this section, we confine ourselves to the case of $\ds_{3}$ and compute the \gls{induced current} and the conductivity without imposing (\ref{assume}).
Whereas the number of pairs has no meaning when the adiabatic future does not exist, the current is well defined and is indeed the right quantity
to describe the \gls{Schwinger effect} as we discussed at the end of Sec.~\ref{sec:smestim}. We will study this question in Sec.~\ref{sec:backEM}.
It can be shown that the current operator of the charged scalar field
\begin{equation}\label{operator}
j^{\mu}(x)=\frac{ie}{2}g^{\mu\nu}\Big(\{(\rnd_{\nu}\phi+ieA_{\nu}\phi),\phi^{\ast}\}-\{(\rnd_{\nu}\phi^{\ast}
-ieA_{\nu}\phi^{\ast}),\phi\}\Big),
\end{equation}
is conserved, i.e., $\nabla_{\mu}j^{\mu}=0$ \cite{book:Parker}.
Using Eqs.~(\ref{phin})-(\ref{vacout}) it can be shown that in the in-vacuum and out-vacuum, $\langle j^{0}\rangle=0$.
However, in the in-vacuum state, the expectation of the spacelike component of the current operator is
\begin{align}\label{vev}
\langle j^{1}\rangle_{in}&=\,_{in}\langle0|j^{1}|0\rangle_{in} \nn\\
&=2e\Omega^{-3}(\tau)\int\frac{d^{2}\textbf{k}}{(2\pi)^{2}}\Big(k_{x}+eA_{1}(\tau)\Big)
\frac{e^{\kappa\pi i}}{2k}\big|\wwp(z_{+})\big|^{2}.
\end{align}
In order to compute the vacuum expectation value of the current operator~(\ref{operator}) we choose the in-vacuum state because this state is
Hadamard \cite{Garriga:1994bm,Froeb2014}.
Hence, the expectation value has a UV behavior similar to the flat spacetime one.
Substituting explicit expressions, the integral~(\ref{vev}) can be rewritten as
\begin{align}
\langle j^{1}\rangle_{in}&=\frac{e}{2\pi^{2}}H^{2}\Omega^{-1}(\tau) \nn\\
&\times\lim_{\Lambda\rightarrow\infty}\int_{-1}^{1}\frac{dr}{\sqrt{1-r^{2}}}
\int_{0}^{\Lambda}dp\big(rp-\lambda\big)e^{\lambda r\pi}\big|\w_{-i\lambda r,\gamma}(-2ip)\big|^{2}, \label{curent}
\end{align}
where $\Lambda=-K\tau$ and $K$ is an upper cutoff on the momentum $k$ introduced for convenience and that will be taken to infinity at the end of the
calculation. We also have introduced
\begin{equation}\label{pk}
p=-k\tau.
\end{equation}
Some details of computation of the integral~(\ref{curent}) are given in Appendix~\ref{app:int}. The final result is
\begin{align}\label{unreg}
&\langle j^{1}\rangle_{in}=\frac{e}{2\pi^{2}}H^{2}\Omega^{-1}(\tau)\bigg[-\frac{\pi}{2}\lambda
\lim_{\Lambda\rightarrow\infty}\Lambda \nn\\
&+\frac{\pi}{4}\lambda\gamma\cot(2\pi\gamma)+\frac{\gamma}{4\sin(2\pi\gamma)}\Big(3\I_{1}(2\pi\lambda)
-2\pi\lambda \I_{0}(2\pi\lambda)\Big)+\frac{i}{2\sin(2\pi\gamma)} \nn\\
&\times\int_{-1}^{1}\frac{dr}{\sqrt{1-r^{2}}}b_{r}\Big\{\big(e^{2\pi\lambda r}
+e^{-2\pi i\gamma}\big)\psi(\frac{1}{2}+i\lambda r-\gamma)-\big(e^{2\pi\lambda r}
+e^{2\pi i\gamma}\big)\psi(\frac{1}{2}+i\lambda r+\gamma)\Big\}\bigg],
\end{align}
where $\psi$ denotes the digamma function and the coefficient $b_{r}$ is defined as
\begin{equation}\label{br}
b_{r} \equiv -\frac{3}{2}\lambda^{2}r^{3}+\Big(\frac{1}{8}-\frac{\gamma^{2}}{2}+\lambda^{2}\Big)r.
\end{equation}
\subsubsection{\label{sec:ren}Adiabatic subtraction}
In order to remove the UV divergent term from the expression~(\ref{unreg}) we need to apply a renormalization scheme. Various methods exist to regularize and renormalize physical quantities. To name some of them, there are proper-time regularization, dimensional regularization, zeta-function regularization, Pauli-Villars subtraction, point splitting regularization (in particular by the Hadamar method) and adiabatic regularization (or subtraction). For this thesis, we choose the \gls{adiabatic subtraction} method. It is achieved by subtracting terms computed in the limit of
slowly varying backgrounds to obtain a finite expression.
The idea of slow varying backgrounds is implemented by introducing adiabatic orders which in our problem  will be nothing but counting time derivatives in a given quantity. 
Adiabatic regularization was first introduced by Parker to cure the UV divergence and the rapid oscillation of the particle number operator \cite{Parker1966}. Parker and Fulling generalized it to take care of the UV divergences of the \gls{energy momentum tensor} of scalar fields in homogeneous cosmological backgrounds \cite{Parker1974,Fulling1974,Fulling1974B}.
We will perform an adiabatic expansion of the mode functions up to the minimal order which makes the original expression~(\ref{unreg})
finite. To do so, we express the solution of the mode equation~(\ref{refeq}) as a \acrshort{WKB} type solution
\begin{equation}\label{fa}
f_{\mathrm{A}}(\tau)=\big(2W(\tau)\big)^{-\frac{1}{2}}\exp\Big[-i\int^{\tau}W(\tau')d\tau'\Big],
\end{equation}
where in order to fulfill Eq.~(\ref{refeq}), the function $W$ satisfies the equation
\begin{equation}\label{weq}
W^{2}(\tau)=\omega^{2}(\tau)+\frac{3}{4}\frac{W'^{2}}{W^{2}}-\frac{1}{2}\ddot{W''}{W}.
\end{equation}
Provided that the adiabatic condition~(\ref{condition}) holds, derivative terms in Eq.~(\ref{weq}) will be negligible compared to $\omega^{2}$
terms. The zeroth order of the adiabatic expansion is enough to remove the UV divergent term from~(\ref{unreg}), it is
\begin{equation}\label{wzero}
W^{(0)}(\tau)=\omega_{0}(\tau),
\end{equation}
where the superscript denotes the adiabatic order. The last term in $\omega^{2}$, see Eq.~(\ref{omega}), can be rewritten in the form
\begin{equation}
\frac{6}{\tau^{2}}\Big(\xi-\frac{1}{8}\Big)=6\Big(\xi-\frac{1}{8}\Big)\frac{\dot{\Omega}^{2}}{\Omega^{2}},
\label{last}
\end{equation}
revealing that this term is of adiabatic order 2. Therefore, $\omega_{0}$ in Eq.~(\ref{wzero}) is given by
\begin{equation}\label{omegaz}
\omega_{0}(\tau)=+\Big(k^{2}-\frac{2eE}{H^{2}\tau}k_{x}+\frac{m^{2}}{H^{2}\tau^{2}}+
\frac{e^{2}E^{2}}{H^{4}\tau^{2}}\Big)^{\frac{1}{2}}.
\end{equation}
With Eqs.~(\ref{varphi}) (\ref{fpm}) (\ref{fa}) (\ref{wzero}) (\ref{omegaz}), the zeroth order adiabatic expansion of the positive frequency
$U_{\mathrm{A}}$ and of the negative frequency $V_{\mathrm{A}}$ mode functions are
\begin{eqnarray}\label{uava}
U_{\mathrm{A};\k}(x)&=&\Omega^{-\frac{1}{2}}(\tau)\big(2\omega_{0}\big)^{-\frac{1}{2}}
\exp\Big[i\k\cdot\x-i\int^{\tau}\omega_{0}(\tau')d\tau'\Big], \nn\\
V_{\mathrm{A};-\k}(x)&=&\Omega^{-\frac{1}{2}}(\tau)\big(2\omega_{0}\big)^{-\frac{1}{2}}
\exp\Big[i\k\cdot\x+i\int^{\tau}\omega_{0}(\tau')d\tau'\Big].
\end{eqnarray}
We use this complete set of orthonormal mode functions to expand the charged scalar field operator, then with Eq.~(\ref{operator}) we find to the zeroth order adiabatic expansion of the current operator
\begin{equation}\label{ja}
\langle j^{1}\rangle_{\mathrm{A}}=-\frac{eH^{2}}{4\pi}\lambda\Omega^{-1}(\tau)
\lim_{\Lambda\rightarrow\infty}\Lambda.
\end{equation}
We emphasize that in the expression~(\ref{ja}) there is no finite term or $\Lambda$-independent contribution.
Applying the \gls{adiabatic subtraction} scheme:
\begin{eqnarray}\label{scheme}
\langle j^{1}\rangle_{\mathrm{reg}}&=&\langle j^{1}\rangle_{in}-\langle j^{1}\rangle_{\mathrm{A}} \nn\\
&=&\Omega^{-1}(\tau)J,
\end{eqnarray}
gives the regularized current as
\begin{align}\label{reg}
&J=\frac{eH^{2}}{8\pi^{2}}\frac{\gamma}{\sin(2\pi\gamma)}\bigg[\pi\lambda\cos(2\pi\gamma)
+3\I_{1}(2\pi\lambda)-2\pi\lambda\I_{0}(2\pi\lambda)+\frac{2i}{\gamma} \nn\\
&\times\int_{-1}^{1}\frac{b_{r}dr}{\sqrt{1-r^{2}}}\Big\{\big(e^{2\pi\lambda r}
+e^{-2\pi i\gamma}\big)\psi\big(\frac{1}{2}+i\lambda r-\gamma\big)
-\big(e^{2\pi\lambda r}+e^{2\pi i\gamma}\big)\psi\big(\frac{1}{2}+i\lambda r+\gamma\big)\Big\}\bigg].
\end{align}
With Eq.~(\ref{parity}), one can show that $J$ is an odd function under the transformation $\lambda\rightarrow-\lambda$, illustrating that if one inverts the electrical field sense, the particles move in the opposite direction.
\subsubsection{\label{sec:result}Regularized current and conductivity}
\begin{figure}[h]
\begin{center}
\includegraphics[width=0.7\textwidth]{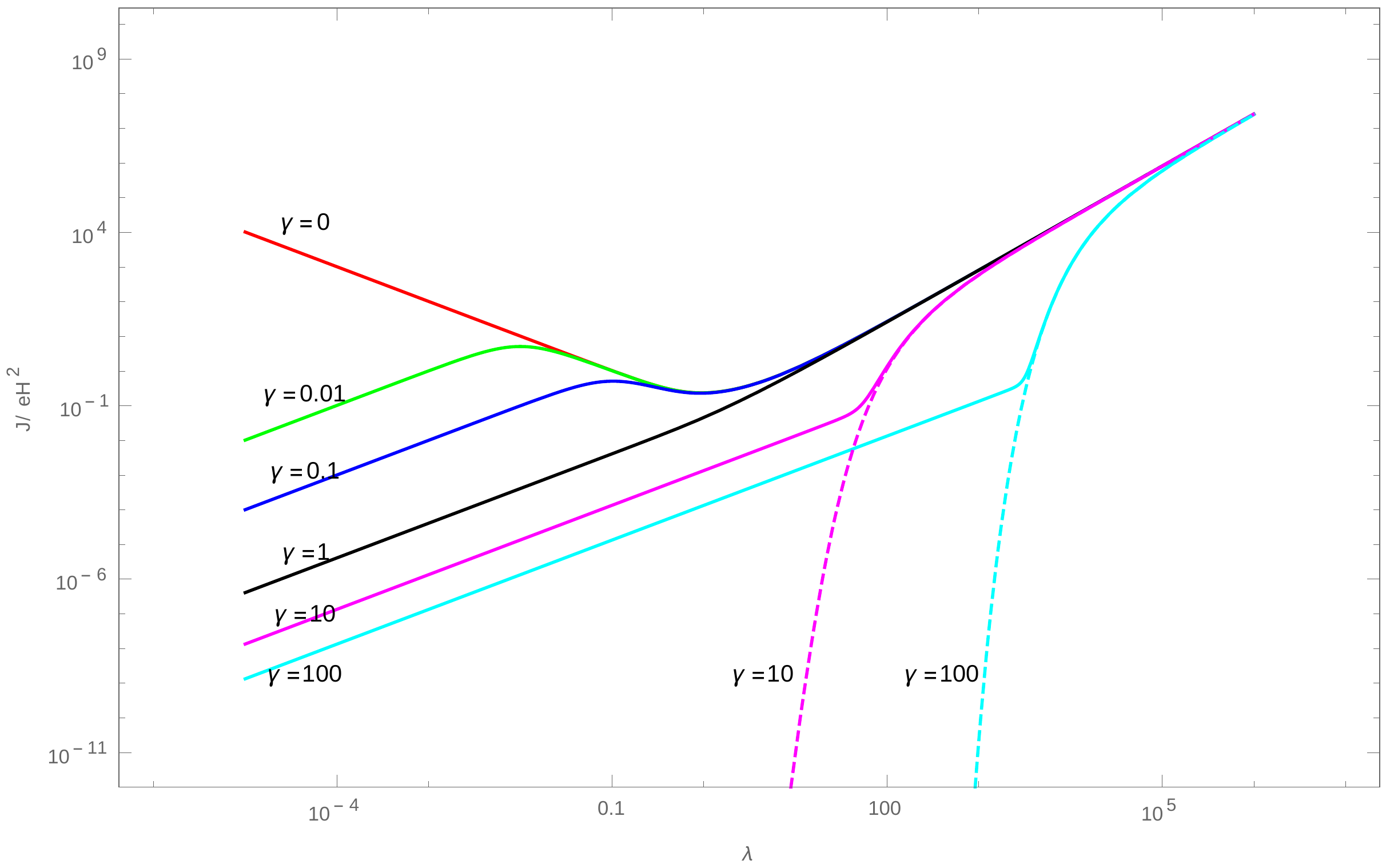}
\caption[Induced current in 1+2 D for bosons]{For different values of $\gamma$, the normalized quantum vacuum expectation value of the induced current $J/eH^{2}$ and the
semiclassical current $J_{\sem}/eH^{2}$ in $D=3$ dimension, are plotted as a function of $\lambda$ with solid and dashed lines, respectively.}
\label{fig1}
\end{center}
\end{figure}

\begin{figure}[h]
\begin{center}
\includegraphics[width=0.7\textwidth]{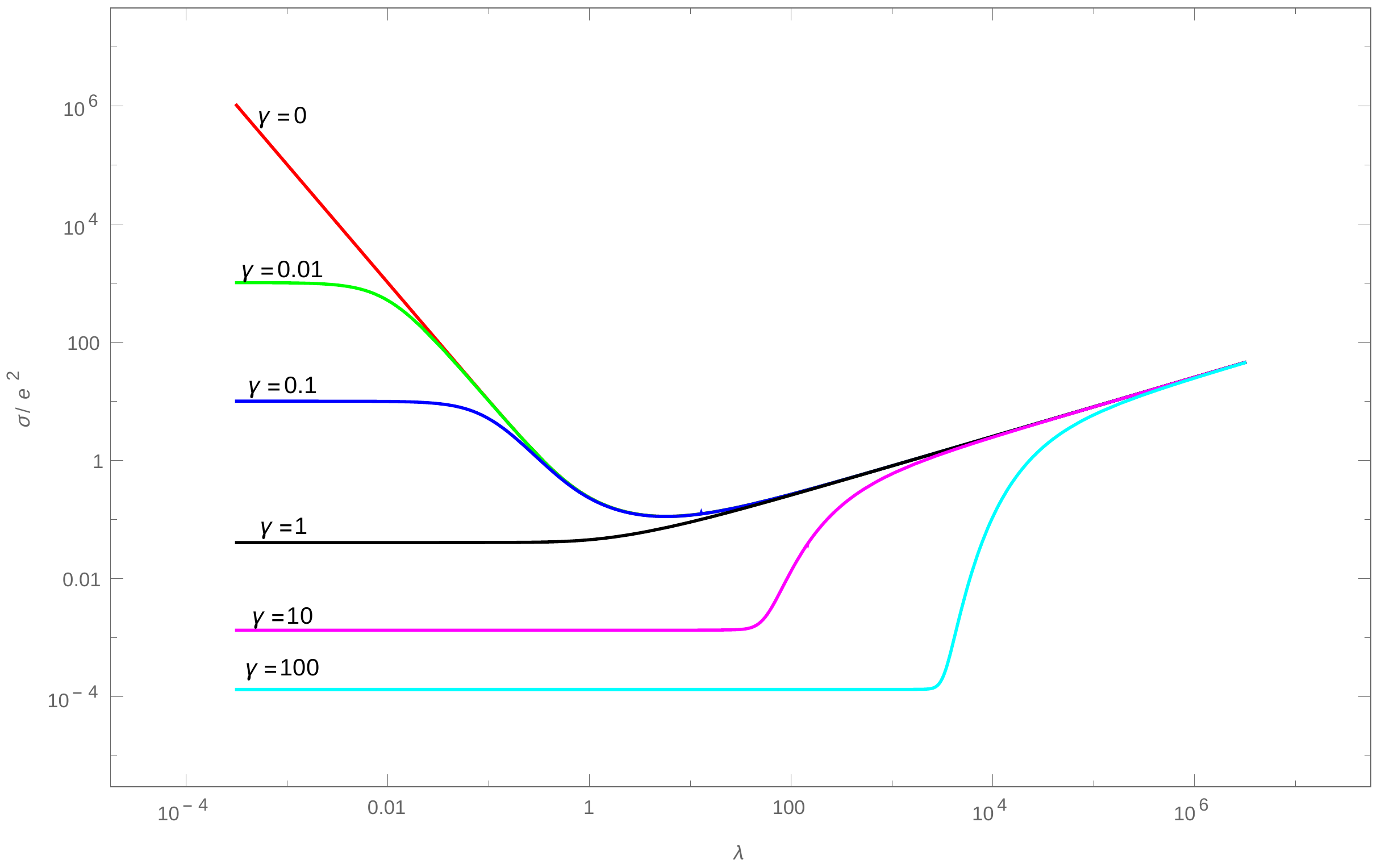}
\caption[Conductivity in 1+2 D for bosons]{For different values of $\gamma$, the normalized conductivity $\sigma/e^{2}$ is plotted as a function of $\lambda$.
The phenomenon of \acrlong{IRHC} appears for $\gamma<\sqrt{3/4}$.} \label{fig2}
\end{center}
\end{figure}

After computing the renormalized current, we consider the conductivity defined as
\begin{equation}\label{sigma}
\sigma \equiv \frac{J}{E}.
\end{equation}
We present a plot of the current~(\ref{reg}) and of the conductivity~(\ref{sigma}) in Figs.~\ref{fig1} and \ref{fig2}, respectively.
The general features of these figures are that in the strong electric field regime $\lambda\gg\max(1,\gamma)$ all the curves have
the same asymptotic behavior, and in the weak electric field regime $\lambda\ll\min(1,\gamma)$ the current and conductivity are
suppressed for increasing scalar field mass.
For the case of a massless minimally coupled scalar field, i.e., $\gamma=0$, for $\lambda\lesssim1$, the current and
conductivity are increasing as the electric field is decreasing.
This phenomenon was dubbed \acrfull{IRHC} in \cite{Froeb2014}.
In the following subsections, we analytically investigate the limiting behaviors of the current and the conductivity.
In this analysis, for simplicity, we use the sign conventions $\lambda=|\lambda|$ and $J=|J|$.
\paragraph{\label{sec:strong}Strong electric field regime}
Taking $\lambda\rightarrow\infty$ in the current expression~(\ref{reg}) with $\gamma$ fixed, the leading order terms are
\begin{align}\label{strong}
J&\simeq\frac{e^{2}}{4\pi^{2}}\frac{|eE|^{\frac{1}{2}}}{H}E, &
\sigma&\simeq\frac{e^{2}}{4\pi^{2}}\frac{|eE|^{\frac{1}{2}}}{H}.
\end{align}
The results~(\ref{strong}) analytically describe the behaviors of the current and conductivity shown by Figs.~\ref{fig1} and \ref{fig2},
respectively. As illustrated in the figures, in this limit, the current and conductivity become increasing functions of electric field $E$ and
independent of $\mds$.
In the cases of $\ds_{2}$ \cite{Froeb2014} and $\ds_{4}$ \cite{Kobayashi:2014zza}, the authors showed that the current responds as $E^{1}$ and
$E^{2}$, respectively, in this regime.
Indeed, in this limit the \gls{semiclassical} computation is a good approximation, and as we found in Secs.~\ref{sec:sctscalar} and \ref{sec:pcb} the current responds
as $E^{\frac{D}{2}}$ in this regime.
\paragraph{\label{sec:weak}Weak electric field regime}
The behavior of the current~(\ref{reg}) in the weak electric field regime $\lambda\ll\min(1,\gamma)$ is obtained by a series expansion
around $\lambda=0$ with $\gamma$ fixed. In the case of heavy particles, i.e., $\gamma\gg 1$, the leading order terms are
\begin{align}\label{wheavy}
J&\simeq\frac{e^{2}H}{24\pi\mds}E, & \sigma&\simeq\frac{e^{2}H}{24\pi\mds},
\end{align}
and in the case of light particles, i.e., $\gamma\ll 1$, the leading order terms are given by
\begin{align}\label{wlight}
J&\simeq\frac{e^{2}H^{2}}{\pi^{2}\mds^{2}}E, & \sigma&\simeq\frac{e^{2}H^{2}}{\pi^{2}\mds^{2}}.
\end{align}
The results given by Eqs.~(\ref{wheavy}) and~(\ref{wlight}) are in agreement with the curves shown in Figs.~\ref{fig1} and~\ref{fig2}.
As illustrated in Fig.~\ref{fig1} the current monotonically increases for increasing electric field $E$.
Figure \ref{fig2} shows that the conductivity is independent of the electric field $E$.
For both, the current and the conductivity, we see an inverse dependence on the scalar field mass parameter $\mds$.
In the case of $\ds_{2}$ the current responds as $J\propto mE\exp(-2\pi m/H)$ for heavy particles and
behaves as $J\propto E/m^{2}$ for light particles \cite{Froeb2014}.
In the case of $\ds_{4}$, for heavy and light particles, the current behaves as $J\propto E/m^{2}$ \cite{Kobayashi:2014zza}.
\paragraph{\label{sec:heavy}Heavy Scalar Field Regime}
The behavior of the current~(\ref{reg}) in the heavy scalar field regime $\gamma\gg\max(1,\lambda)$ is obtained by taking the limit $\gamma\rightarrow\infty$ with $\lambda$ held fixed. We then obtain the leading order terms as
\begin{align}\label{heavy}
J&\simeq\frac{e^{2}H}{24\pi\mds}E, & \sigma&\simeq\frac{e^{2}H}{24\pi\mds},
\end{align}
which are the same as the result~(\ref{wheavy}). It is a general feature regardless of the dimension of the dimension and the nature of the particle: the heavy scalar field limit is equivalent to the weak electric field limit.
\paragraph{\label{sec:hyper}Massless Minimally Coupled Scalar Field Case}
In the case of a massless minimally coupled scalar field, i.e., $\gamma=0$ we now examine the behavior of the current in two limiting regimes.
In the limit $\lambda\rightarrow\infty$, the current and conductivity are approximated by Eq.~(\ref{strong}), whereas in the limit $\lambda\rightarrow 0$, the leading order terms are
\begin{align}\label{hyper}
J&\simeq\frac{H^{4}}{\pi^{2}E}, & \sigma&\simeq\frac{H^{4}}{\pi^{2}E^{2}}.
\end{align}
The results~(\ref{hyper}) agree with the asymptotic behavior of the red curves corresponding to $\gamma=0$ in Figs.~\ref{fig1} and \ref{fig2}. In the regime $\lambda\ll 1$ the current and the conductivity are not bounded from above and increase as $\lambda$ decreases,
as illustrated in Figs.~\ref{fig1} and~\ref{fig2}, respectively.
This divergence signals that the framework used to derive this result is not valid anymore and \gls{backreaction} to the reservoir fields needs to be
taken into account. More about that regime will be given in Sec.~\ref{sec:IRHC} and about \gls{backreaction} in chapter \ref{chap:back}.
\begin{figure}
\begin{center}
\includegraphics[width=0.7\textwidth]{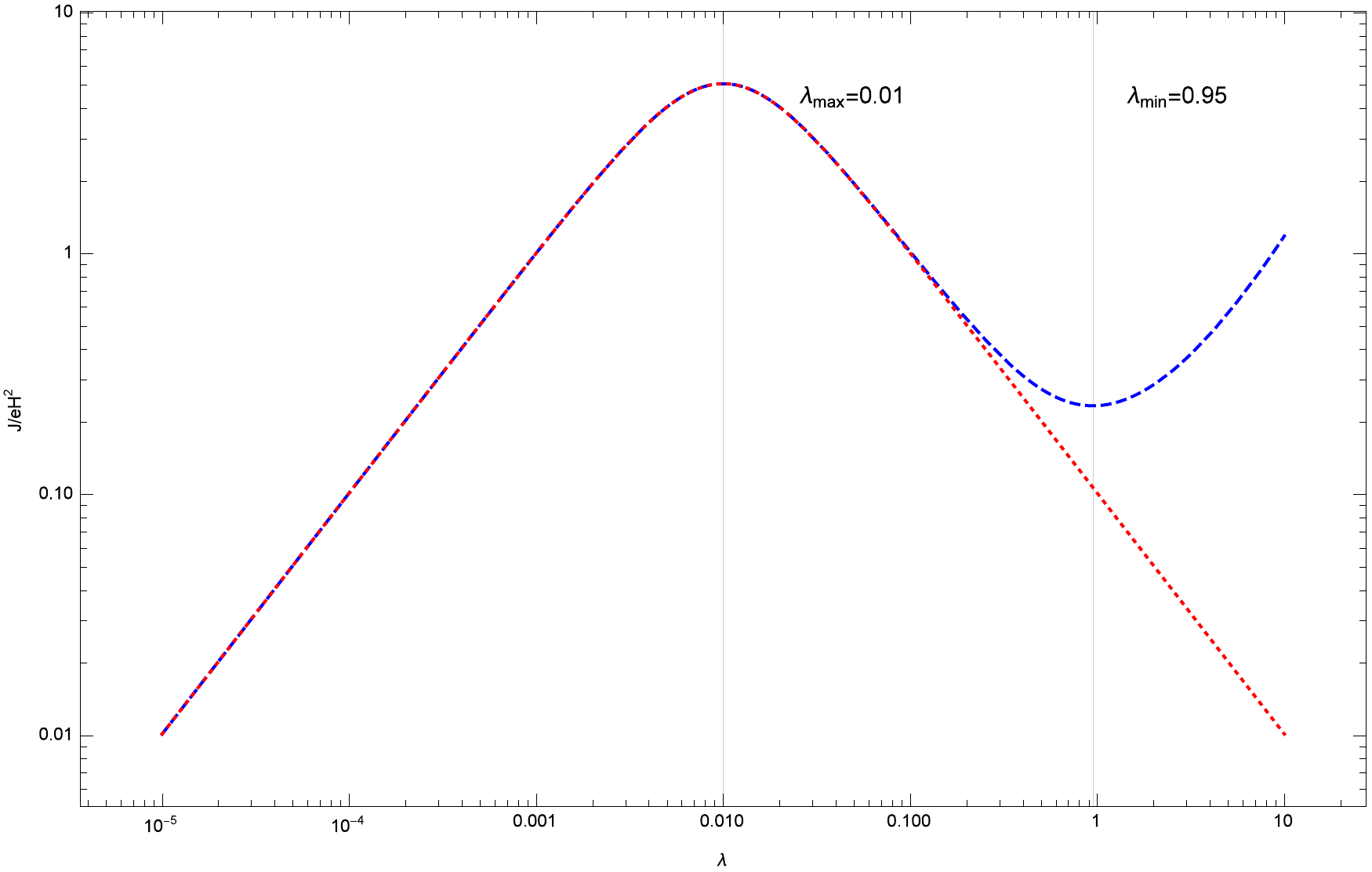}
\caption[Zoom on the infrared hyperconductivity regime]{Zoom on the \acrshort{IRHC} regime: the normalized current~(\ref{reg}) (in the blue dashed line) and its approximation~(\ref{light}) (in the red dotted line) are plotted as a function of $\lambda$ with $\gamma=0.01$.} \label{fig4}
\end{center}
\end{figure}
\paragraph{\label{sec:light}Light Scalar Field Case}
We now study the behavior of the current~(\ref{reg}) for a light scalar field case, i.e., $\gamma\ll 1$ and more specifically $\gamma<\sqrt{3/4}$.
In the regime $\lambda\gg 1$, the current and conductivity scale as indicated by Eq.~(\ref{strong}).
Numerical analyses show that in the regime $0\lesssim\lambda\lesssim 1$ the current and conductivity behave as
\begin{align}\label{light}
J&\simeq\frac{eH^{2}}{\pi^{2}}\Big(\frac{\lambda}{\lambda^{2}+\gamma^{2}}\Big), &
\sigma&\simeq\frac{e^{2}}{\pi^{2}}\Big(\frac{1}{\lambda^{2}+\gamma^{2}}\Big).
\end{align}
In Sec.~\ref{sec:IRHC} we will derive Eq.~(\ref{light}) analytically. In Fig.~\ref{fig4}, we plot the current~(\ref{light}) together with the current~(\ref{reg}), in the \acrshort{IRHC} regime.
This figure illustrates the quite good agreement between the numerical and analytical results.
The current has a local minimum in $\lambda_{\mathrm{min}}\simeq\sqrt{3/4}+\epsilon$, where $\epsilon$ is a small positive parameter, and a local
maximum in $\lambda_{\mathrm{max}}\simeq\gamma$.
In the interval $\lambda\in\big(\lambda_{\mathrm{max}},\lambda_{\mathrm{min}}\big)$ the phenomenon of \acrshort{IRHC} occurs; i.e., the current increases
for decreasing $\lambda$.
Beyond $\lambda_{\mathrm{max}}$, in the interval $(0,\lambda_{\mathrm{max}})$, the current has a linear response for $\gamma\neq 0$ which agrees
with Eq.~(\ref{wlight}).
From this and Sec.~\ref{sec:hyper}, we can conclude that for decreasing $\lambda$, if $\gamma$ becomes real, then there would be a period of
\acrshort{IRHC}; and if one keeps decreasing $\lambda$, it will be followed by a linear behavior for $\gamma\neq 0$ or continued unbounded for $\gamma=0$.

%% file: current/Manuscript5.tex
We will now investigate the analogous problem studied for bosons in Sec.~\ref{sec:Bogbos} but for fermions in 1+1 D. We will compute the \gls{pair production rate} with a \gls{Bogoliubov transformation} and compare it to the result of Sec.~\ref{sec:smestim}. Finally we will present a calculation of the expectation value of the \gls{induced current}. The main motivations to study fermions in the context of the \gls{Schwinger effect} in \gls{dS} are to strengthen and compare to boson results. It is for instance known from flat spacetime examples that the bosons and the fermions are created at the same rate (\textit{eg.}~\cite{Strobel:2014tha}) under the influence of a strong electric field. We will see that it is not the case in \gls{curved spacetime}.
\subsection*{Explicit solutions to (\ref{eq:CurvedDirac})}
\label{sec:solutions}
In two dimensions, in the specific case of a constant electric field, it possible to find a complete analytic solution to the problem of fermions seeing a de Sitter and a constant electric field background. More precisely, we construct here the positive and negative frequency solutions at past and future infinity for the constant field. Similarly to (\ref{zpm}), introducing the new time variable:
\begin{align}
 z \equiv 2i k\tau,
\end{align}
and an equation similar to (\ref{kappa}), but with the opposite sign convention\footnote{To match to the convention of \cite{Stahl:2015gaa}, we do not use exactly (\ref{kappa}) but the opposite sign convention, see equation (48) of \cite{Stahl:2015gaa}. If we would follow our convention (\ref{kappa}), the consequence would be a change of sign for $z$ and swapping $\psi_1$ and $\psi_2$, we will come back to this issue while talking about antiparticles.}, in (\ref{eq:DecoupledDirac1}) and (\ref{eq:DecoupledDirac2}), for the constant field (\ref{vector}), we get:
\begin{align}
 \psi_1''(z)+\left(\frac{1}{z^2}\left[\frac{1}{4}-\mu^2\right]+\frac{1}{z}\left[\kappa-\frac{1}{2}\right]-\frac{1}{4}\right)\psi_1(z)=&0, \label{eq:Whittaker1}\\
 \psi_2''(z)+\left(\frac{1}{z^2}\left[\frac{1}{4}-\mu^2\right]+\frac{1}{z}\left[\kappa+\frac{1}{2}\right]-\frac{1}{4}\right)\psi_2(z)=&0.
 \label{eq:Whittaker2}
\end{align}
Observe that in two dimension, the parameter $r= \frac{k_x}{k}$ is nothing but $r=\sgn(k)$.
Solutions of these equations are the Whittaker functions $W_{\kappa\pm\frac12,\mu}(z),\,M_{\kappa\pm\frac12,\mu}(z)$ \cite{Olver2010}. We can make the following ansatz for two independent solutions of the decoupled Dirac equation (\ref{eq:DecoupledDirac1})-(\ref{eq:DecoupledDirac2})
\begin{align}
 \psi^a(z)=\begin{pmatrix}
            C_1\,W_{\kappa-\frac12,\mu}(z)\\
            C_2\,W_{\kappa+\frac12,\mu}(z)
           \end{pmatrix},&&
 \psi^b(z)=\begin{pmatrix}
            C_3\,W_{-\kappa+\frac12,-\mu}(-z)\\
            C_4\,W_{-\kappa-\frac12,-\mu}(-z) \label{eq:Csolutions}
           \end{pmatrix}.
\end{align}
Since we solved the two decoupled equations separately we lost the information about the coupling. It can be recovered by using the solutions (\ref{eq:Csolutions}) in one of the coupled equations (\ref{eq:CoupledDirac1})-(\ref{eq:CoupledDirac2}). Using the identity (\ref{eq:pmonehalfIdentity}) 
we find
\begin{align}
  C_{\nicefrac{1}{4}}=-\sqrt{\mu^2-\kappa^2}\,C_{\nicefrac{2}{3}}=-i\gamma \,C_{\nicefrac{2}{3}}. \label{eq:C_23}
\end{align}
We can show with the help of (\ref{eq:Wconjugated}) that with this values for the constants $C_{\nicefrac{2}{3}}$  the solutions (\ref{eq:Csolutions}) are orthogonal
\begin{align}
 \psi^a(z)^\dagger\cdot\psi^b(z)=0. \label{eq:orthogonality}
\end{align}
The value of the remaining constants will be found by asking for normalization after quantizing.\\
\subsubsection{Construction of positive and negative frequency solutions}
We can now use the limit of the function $W_{\kappa,\mu}(z)$ for $|z|\rightarrow\infty$ given in (\ref{win}) to find the behavior of the solutions (\ref{eq:Csolutions}) at past infinity
\begin{align}
 \lim_{\tau\rightarrow-\infty}\psi^a(z)=\lim_{\tau\rightarrow-\infty}
 \begin{pmatrix}
  C_1\,z^{\kappa-\frac{1}{2}}\\
  C_2\,z^{\kappa+\frac{1}{2}}
 \end{pmatrix} \exp\left(-i k\tau\right),\\
 \lim_{\tau\rightarrow-\infty}\psi^b(z)=\lim_{\tau\rightarrow-\infty}
 \begin{pmatrix}
  C_3\,(-z)^{-\kappa+\frac{1}{2}}\\
  C_4\,(-z)^{-\kappa-\frac{1}{2}}
 \end{pmatrix} \exp\left(i k\tau\right).
\end{align}
Comparing this to the desired asymptotic behavior of the mode functions, we find that the positive frequency solution $\psi_\text{in}^+(\tau)$ is given by $\psi^a(z)$ for $k>0$ and by $\psi^b(z)$ for $k<0$. Thus the positive and negative frequency solutions in the asymptotic past  can be constructed as
\begin{align}
 \psi_\text{in}^+(z)=\begin{cases}\psi^a(z)& k>0\\
                        \psi^b(z)& k<0
              \end{cases},&&
\psi_\text{in}^-(z)=\begin{cases}\psi^b(z)& k>0\\
                        \psi^a(z)& k<0
              \end{cases}.&& \label{eq:2DFermSol}
\end{align}
Following from (\ref{eq:orthogonality}) also these solutions are orthogonal.\\
\subsubsection*{Quantization} We can now construct the spinor field operator by specifying the momentum decomposition (\ref{eq:decomp1})   
\begin{align}
 \psi(x)=\int \frac{dk}{2\pi}\e^{i k x_1}\left[b(k) \psi^+(\tau)+d^\dagger(-k)\psi^-(\tau)\right].\label{eq:decomp2}
\end{align}
We impose the canonical equal time anti-commutation relations
\begin{align}
 \left\{\psi_\alpha(x_1,\tau),\pi(x_1',\tau)\right\}=i\,\delta(x_1-x_1')\delta_{\alpha\beta}. 
\end{align}
Using the conjugate momentum (\ref{eq:mometum}) in $\ds$ we find that this is equivalent to
\begin{align}
\left\{\psi_\alpha(x_1,\tau),\psi_\beta^\dagger(x_1',\tau)\right\}=\frac{1}{a(\tau)}\delta(x_1-x_1')\delta_{\alpha\beta}. \label{eq:comm}
\end{align}
This holds true if the creation and annihilation operators follow the anti-commutation relations
\begin{align}
 \left\{b(k),b(k')^\dagger\right\}=\left\{d(k),d(k')^\dagger\right\}= 2\pi \,\delta(k-k'). \label{eq:comm2}
\end{align}
 and the mode functions fulfil the \gls{Wronskian condition}
\begin{align}
\psi^+(\tau)\psi^+(\tau)^\dagger+\psi^-(\tau)\psi^-(\tau)^\dagger=\frac{1}{a(\tau)}1.\label{eq:WronskianCondition}
\end{align}
This is true for the solutions (\ref{eq:2DFermSol}) if (see App.~\ref{sec:WronskianCondition})
\begin{align}
 C_2=C_3 =\sqrt{\frac{H}{2|k|}}\e^{\frac{\pi}{2}i\kappa\sgn(k)} \label{eq:C_14}.
\end{align}
\subsubsection{Final result for the solution to Dirac equation (\ref{eq:CurvedDirac})}
Accordingly we find that the positive and negative frequency solutions at asymptotic past infinity are given by
\begin{align}
& \psi_\text{in}^+(\tau)=\sqrt{\frac{H}{2|k|}}\begin{cases}
               \e^{\frac{\pi}{2}i\kappa}
               \begin{pmatrix}
                \frac{\gamma}{i}\,W_{\kappa-\frac12,\mu}(2i v)\\
                 \,W_{\kappa+\frac12,\mu}(2i v)
               \end{pmatrix}& k>0
               \\
               {\e^{-\frac{\pi}{2}i\kappa}}\begin{pmatrix}
                \,W_{-\kappa+\frac12,-\mu}(2i v)\\
                \frac{\gamma}{i}\,W_{-\kappa-\frac12,-\mu}(2i v)
               \end{pmatrix}&k<0
              \end{cases}, \label{eq:2DFermSol2p} \\
& \psi_\text{in}^-(\tau)=\sqrt{\frac{H}{2|k|}}\begin{cases}
               {\e^{\frac{\pi}{2}i\kappa}}
               \begin{pmatrix}
                \,W_{-\kappa+\frac12,-\mu}(-2i v)\\                
               \frac{\gamma}{i}\,W_{-\kappa-\frac12,-\mu}(-2i v)
               \end{pmatrix}& k>0\\
               {\e^{-\frac{\pi}{2}i\kappa}}\begin{pmatrix}
                \frac{\gamma}{i}\,W_{\kappa-\frac12,\mu}(-2i v)\\
                \,W_{\kappa+\frac12,\mu}(-2i v)
               \end{pmatrix}&k<0
              \end{cases}, \label{eq:2DFermSol2m}
\end{align}
where we introduced $v:=|k|\tau$.
In an analogous way we can construct the positive and negative frequency solutions at $\tau\rightarrow0$ as
\begin{align}
& \psi^+_\text{out}(\tau)=\frac{1}{2}\sqrt{\frac{H}{|k|}}
               \e^{\frac{\pi}{2}i r\mu }
               \begin{pmatrix}
                 \sqrt{\frac{\mu-\kappa}{\mu}}\,M_{\kappa-\frac12,\mu}(z)\\
                 \sqrt{\frac{\mu+\kappa}{\mu}}\,M_{\kappa+\frac12,\mu}(z)
               \end{pmatrix} 
, \label{eq:2DFermSoloutp} \\
 & \psi_\text{out}^-(\tau)=\frac{1}{2}\sqrt{\frac{H}{|k|}}           
               \e^{\frac{\pi}{2}i r\mu }
                \begin{pmatrix}
                 \sqrt{\frac{\mu+\kappa}{\mu}}\,M_{-\kappa+\frac12,-\mu}(-z)\\
                 -\sqrt{\frac{\mu-\kappa}{\mu}}\,M_{-\kappa-\frac12,-\mu}(-z)
                \end{pmatrix},
\label{eq:2DFermSoloutm}
\end{align}
where we introduced 
\begin{align}
 r:=\sgn(k).
\end{align}
Using the limit of the function $M_{\kappa,\mu}(z)$ for $z\rightarrow0$ given by (\ref{wout}) we find
\begin{align}
 \lim_{\tau\rightarrow0}\psi^+_\text{out}(\tau)&=\lim_{\tau\rightarrow0}\frac{1}{2}\sqrt{\frac{H}{|k|}}
               \e^{i\frac{\pi}{2}r\mu}
               \begin{pmatrix}
                 \sqrt{\frac{\mu-\kappa}{\mu}}\\
                 \sqrt{\frac{\mu+\kappa}{\mu}}
               \end{pmatrix}\left(2i |k|\tau\right)^{\mu+\frac{1}{2}}
,\\
 \lim_{\tau\rightarrow0} \psi_\text{out}^-(\tau)&=
 \lim_{\tau\rightarrow0}\frac{1}{2}\sqrt{\frac{H}{|k|}}\e^{i\frac{\pi}{2}r\mu}
                \begin{pmatrix}
                 \sqrt{\frac{\mu+\kappa}{\mu}}\\
                 -\sqrt{\frac{\mu-\kappa}{\mu}}
                \end{pmatrix}\left(-2i|k|\tau\right)^{-\mu+\frac{1}{2}}.
\end{align}
One can show that the modes (\ref{eq:2DFermSoloutp}) (\ref{eq:2DFermSoloutm}) have the right asymptotic behavior and follow the \gls{Wronskian condition} (\ref{eq:WronskianCondition}) by performing steps analogous to the ones found in Appendix \ref{sec:WronskianCondition}.\\
Observe that the positive and negative frequency also correspond to the particle and antiparticle solution. This can be seen by defining the charge conjugate spinor representing the antiparticle of $\psi(x)$ as:
\begin{equation}
\label{eq:charge}
\psi^c(x) \equiv i \sigma_2 \psi^*(x),
\end{equation}
with the Pauli matrix 
\begin{align}
 \sigma_2=\begin{pmatrix}
           0&-i\\
           i&0
          \end{pmatrix}.
\end{align}
This is a physical illustration of Feynman's picture that antiparticles are traveling backwards in time \cite{Feynman:1949hz}.
\subsection{Number of created pairs using a Bogoliubov transformation}
\label{sec:Bogoliubov}
In this section, we propose an alternative derivation of the result (\ref{main2}) based on a \gls{Bogoliubov transformation}, it is the analog method presented in Sec.~\ref{sec:Bogbos} for boson so we refer the reader there for an introduction to this method. 

To use the the method of Bogoliubov coefficients we use the fact that the positive frequency mode at past infinity is connected to the modes at $\tau\rightarrow0$ through
\begin{align}
 \psi_\text{in}^+(\tau)=\alpha_k \,\psi_\text{out}^+(\tau)+\beta_k\, \psi_\text{out}^-(\tau) \label{eq:inout},
\end{align}
where the Bogoliubov coefficients are normalized as
\begin{align} 
 |\alpha_k|^2+|\beta_k|^2=1.
\end{align}
The coefficients can now be found by putting the explicit form of the solutions (\ref{eq:2DFermSol2p}) (\ref{eq:2DFermSol2m}) and (\ref{eq:2DFermSoloutp}) (\ref{eq:2DFermSoloutm}) in  (\ref{eq:inout}) and using the connection between the Whittaker functions $W_{\kappa,\mu}(z)$ and $M_{\kappa,\mu}(z)$ given in (\ref{eq:WtoM}). Using (\ref{conectionw}) (\ref{conectionm}) this leads to 
\begin{align}
\alpha_k=\frac{\Gamma(-2\mu)}{\Gamma(-\mu- r \kappa)}\frac{\sqrt{2\mu}}{\sqrt{\mu+ r \kappa}}\e^{-\frac{\pi}{2}i(\mu- r \kappa)}\e^{\frac{\pi}{4}i(r-1)}
,&&
\beta_k=\frac{\Gamma(2\mu)}{\Gamma(\mu- r \kappa)}\frac{\sqrt{2\mu}}{\sqrt{\mu- r \kappa}}\e^{\frac{\pi}{2}i(\mu+ r \kappa)}\e^{\frac{\pi}{4}i(r+1)}.
\end{align}
We thus find the \gls{number of pairs per momentum} to be
\begin{align}
 n_k=|\beta_k|^2=\e^{-\pi(|\mu|-i r \kappa)}\frac{\sinh(\pi(|\mu|+i r \kappa))}{\sinh(2\pi |\mu|)}\label{eq:nk}.
\end{align}
This result can be shown to be equivalent to Eq.~(19) of  \cite{Haouat2013}, where it has been derived in an equivalent way. Comparing it to the bosonic result in $\ds_2$ of \cite{Froeb2014} (see Eq.~(2.18)) we find that the only difference is a $\sinh(\pi(|\mu|-  i r \kappa))$ instead of a $\cosh(\pi(|\mu|-i r \kappa))$.  However this difference vanishes in the relevant limit (\ref{assume}).\\
Comparing to the \gls{semiclassical} result given in (\ref{main2}) the most striking difference is that according to (\ref{eq:nk}) particles can also be created in ``anti-screening'' direction which corresponds to ``upward'' tunneling. This was already found in the bosonic case \cite{Froeb2014}. However in the limit, i.e.~$|\mu|\gg1, \lambda\gg 1$, the two results agree, since \gls{pair} production in ``anti-screening'' direction gets exponentially suppressed.
\subsubsection{Pair production rate from the number of pairs}
\label{sec:currentfromnk}
One can compute the \gls{pair production rate} from the \gls{number of pairs per momentum} $k$ by integrating
\begin{equation}
\Gamma \equiv \frac{1}{ V} \int \frac{d k}{2\pi}\, n_{k},
\end{equation}
where $ V = a(\tau)^2 d\tau$ is the unit two volume of the spacetime. Using (\ref{estimate}), the \(k\)-integral can be changed into a time integral and the \gls{pair production rate} can thus be estimated from (\ref{eq:nk}) by 
\begin{equation}
\Gamma \approx \frac{ |\mu|  H^2}{2\pi}  \frac{\cosh\left(2 \pi \lambda\right)-\e^{-2\pi|\mu|}}{\sinh(2 \pi |\mu|)}. \label{eq:paircreationrate}
\end{equation}
We can also compute the \gls{physical number of pairs} $n$ of produced \glspl{pair} at the time $\tau$ with the help of
\begin{equation}
n \equiv \frac{1}{a(\tau)} \int_{-\infty}^{\tau}\, d\eta a(\eta)^2 \Gamma= \frac{\Gamma}{H}.
\end{equation}
\subsubsection{Flat spacetime limit} Performing calculations in analogy with the ones performed in \cite{Kobayashi:2014zza} one finds that in the limit of flat spacetime, i.e. $H\rightarrow0$, one recovers the familiar results for the \gls{Schwinger effect} in Minkowski spacetime (see e.g.~\cite{Gavrilov:1996pz})
\begin{equation}
\label{M4}
 \lim_{H\rightarrow 0} \Gamma = \frac{|eE|}{ \pi} \exp \left(-\pi\frac{ m^2}{|eE|} \right). \\
\end{equation}
We will finish this section by computing the \gls{vacuum} decay rate. This rate was computed from the imaginary part of the one-loop effective action by Schwinger \cite{Schwinger1951}. It has been shown that Schwinger's result agrees with the canonical method for the case of pure electric field \cite{Nikishov1969,Narozhnyi:1970uv} as well as in \gls{dS} \cite{Mottola:1984ar}. In our case, the \gls{vacuum} decay rate is defined as
\begin{equation}
\Upsilon=\frac{1}{ V} \int \frac{dk}{2 \pi} \log(1-n_k).
\end{equation}
Expanding the logarithm and changing the $k$-integral to a time integral, as above, it is possible to find the following expression for the decay rate
\begin{equation}
\Upsilon= \sum_{r=\pm 1}r \sum_{j=1}^{\infty}\frac{ |\mu|  H^2}{2\pi j}  \e^{-\pi j( |\mu| \ - \lambda r) } \frac{\sinh^j(\pi (|\mu|+\lambda r)}{\sinh^j(2\pi |\mu|)}.
\end{equation}
Taking $H\rightarrow0$ gives the correct flat spacetime expression
\begin{equation}
 \lim_{H\rightarrow 0} \Upsilon = \sum_{j=1}^{\infty}\frac{|eE|}{ \pi j}  \exp \left(-j \pi \frac{m^2}{|eE|} \right).
\end{equation}

\subsection{Computation of the current}
\label{sec:current}
In this section we will compute the expectation value of the current in a locally inertial coordinate system given by \(\underline{\gamma}^\mu(x)=\gamma^\mu\). {These coordinates  were introduced to make the probability density positive semi-definite at each spacetime point \cite{Pollock2010,parker1980one,parker1980one2}. The expectation value of the current with respect to the \gls{vacuum} in the past infinity is then given by 
\begin{align}
 J^1&=-\frac{e}{2}\left\langle0\right|\left[\overline{\psi}(x),{\gamma}^1\psi(x)\right] \left|0\right\rangle\\
 &=-\frac{e}{2}\int_{-\infty}^{\infty}\frac{dk}{2\pi}\left[-\psi_\text{in}^+(\tau)^\dagger\gamma^0\gamma^1\psi_\text{in}^+(\tau)+\psi_\text{in}^-(\tau)^\dagger\gamma^0\gamma^1\psi_\text{in}^-(\tau)\right]\\
 &=-\frac{e}{2}\int_{-\infty}^{\infty}\frac{dk}{2\pi}\left[|\psi_1^+(\tau)|^2-|\psi_2^+(\tau)|^2-|\psi_1^-(\tau)|^2+|\psi_2^-(\tau)|^2\right]. \label{eq:current1}
\end{align}
where \(\psi_1^\pm(\tau)\) and \(\psi_2^\pm(\tau)\) are the first and second component of \(\psi_\text{in}^\pm\) respectively. 
Using the diagonal elements of the \gls{Wronskian condition} (\ref{eq:WronskianCondition}) we find a connection between the absolute square of the positive and negative frequency modes which can be used to simplify the current to
\begin{align}
 J^1=-e\int_{-\infty}^{\infty}\frac{dk}{2\pi}\left(|\psi_1^+(\tau)|^2-|\psi_2^+(\tau)|^2\right). \label{eq:current2}
\end{align}
Using the explicit form of the positive and negative frequency modes (\ref{eq:2DFermSol2p}) (\ref{eq:2DFermSol2m}) this can be computed as (see Appendix \ref{sec:Integral}) 
\begin{align}
 J^1=&\frac{eH}{\pi}i\left(\mu\frac{\sin(2\pi\kappa)}{\sin(2\pi\mu)}-\kappa\right)\label{eq:explicitcurrent}.
\end{align}
We will now regularize this current using \gls{adiabatic subtraction}. 
\subsubsection{Adiabatic Regularization}
As for the case of bosons, we will now renormalize our current to obtain a finite expression. A brief introduction to renormalization scheme is proposed in Sec.~\ref{sec:ren} where the corresponding problem for bosons is studied. We implement an \gls{adiabatic subtraction} for our problem as it has been done in order regularize the current in the bosonic case in $\ds_4$ \cite{Kobayashi:2014zza} as well as for fermions in flat Minkowski space in \cite{Kluger1992}.\\
To regularize the current using \gls{adiabatic subtraction} we compute the current for slow background variations and subtract it from our result (\ref{eq:explicitcurrent}). To quantify what is meant by slow varying background more precisely, we introduce a dimensionless slowness parameter T by replacing the \gls{scale factor} \(a(\tau)\) by a family of functions $a_T (\tau) := a(\tau/T )$. Observe that in the limit of infinitely slow backgrounds, $T \rightarrow \infty$, the derivatives of \(a(\tau)\) will tend to zero since for  $n \in \mathbf{N}$
\begin{align}\frac{d^n a(\tau/T)}{d\tau^n} \propto \frac{1}{T^n}.\end{align}
We define adiabatic orders as powers of $T^{-1}$. In our problem, it is equivalent to count time derivatives and adiabatic orders in a given expression. \\
For our purpose, we will expand our modes up to second adiabatic order and subtract it from the current to regularize it. To do so we start from the \acrshort{WKB}-like ansatz
\begin{align}
 \psi^+_1(\tau)=N_1\sqrt{\frac{1}{2\Omega(\tau)}}\exp\left(\int^\tau\left[-i \Omega(t)+\frac{p(t)}{2\Omega(t)}\left(\frac{p'(t)}{p(t)}-\frac{a'(t)}{a(t)}\right)\right]dt\right). \label{eq:adiabaticansatz}
\end{align}
Observe that in difference to the bosonic case of \cite{Kobayashi:2014zza} we are not using a pure \acrshort{WKB} ansatz. It was found that for fermions, a new kind of ansatz has to be proposed so that the imaginary part of the decoupled Dirac equation (\ref{eq:DecoupledDirac1}) is canceled. This has been used for fermions under the influence of an electric field in flat spacetime in \cite{Kluger1992} and in curved (\glslink{FLRW metric}{FLRW}) spacetimes without electric field \cite{Landete2014} (see also \cite{Ghosh:2015mva}). The ansatz (\ref{eq:adiabaticansatz}) is a combination of the two previous ones. By putting the ansatz in the decoupled Dirac equation (\ref{eq:DecoupledDirac1}) we find a reparametrization of it in terms of \(\Omega(\tau)\), namely
\begin{align}
\begin{split}
& \Omega(\tau)^2-\omega(\tau)^2=\left[\frac{a''(\tau)}{2 a(\tau)}\left(1+\frac{p(\tau)}{\Omega(\tau)}\right)-\frac{a'(\tau)^2}{a(\tau)^2}\left(\frac{3}{4}+\frac{p(\tau)}{2\Omega(\tau)}-\frac{p(\tau)^2}{4\Omega(\tau)^2}\right)\right.\\
 &\left.+\frac{a'(\tau)}{2 a(\tau)}\frac{p'(\tau)}{\Omega(\tau)}\left(1-\frac{p(\tau)}{\Omega(\tau)}\right)+\frac{\Omega'(t)p(t)}{\Omega(t)^2}\left(\frac{p'(t)}{p(t)}-\frac{a'(t)}{a(t)}\right)-\frac{p''(\tau)+\Omega''(\tau)}{2\Omega(\tau)}+\frac{3\Omega'(\tau)^2+p'(\tau)^2}{4\Omega(\tau)^2}\right]. \label{eq:Omega-omega}
\end{split}
 \end{align}
We can now expand (\ref{eq:Omega-omega}) to find \(\Omega(t)=\omega(t)+\O(T)^{-2}\). The \((n+1)\)-th order can be found by iteratively using the \(n\)-th order solution on the right hand side of (\ref{eq:Omega-omega}).\\ 
For the second component of the spinor we use the ansatz (\ref{eq:adiabaticansatz}) in the coupled Dirac equation (\ref{eq:CoupledDirac1}) to find
\begin{align}
 \frac{\psi^+_2(\tau)}{\psi_1^+(\tau)}=\frac{\Omega(\tau)+p(\tau)}{m a(\tau)}-\frac{i}{2m a(\tau)} \left[\frac{\Omega'(\tau)+p'(\tau)}{\Omega(\tau)}-\frac{a'(\tau)}{a(\tau)}\left(1+\frac{p(\tau)}{\Omega(\tau)}\right)\right]\label{eq:adiabaticansatz2}.
\end{align}
Solving the off-diagonal element of the \gls{Wronskian condition} (\ref{eq:WronskianCondition}) for \(\psi_1^-(\tau)\) and using it in one of the diagonal elements we can show that  
\begin{align}
 |\psi_1^+(\tau)|^2+|\psi_2^+(\tau)|^2=\frac{1}{a(\tau)}. \label{eq:normalization}
\end{align}
Using this normalization condition we can now write the current (\ref{eq:current2}) in terms of the fraction (\ref{eq:adiabaticansatz2})
\begin{align}
 J^x=-\frac{e}{a(\tau)}\int_{-\infty}^{\infty}\frac{dk}{2\pi}\frac{1- \frac{|\psi^+_2(\tau)|^2}{|\psi^+_1(\tau)|^2}}{ 1+\frac{|\psi^+_2(\tau)|^2}{|\psi^+_1(\tau)|^2}}. \label{eq:currentinbetween}
\end{align}
To perform the adiabatic expansion we can use the fact that \(\Omega(\tau)^2-\omega(\tau)^2\) is of second adiabatic order, which follows from (\ref{eq:Omega-omega}). It is possible to write \(\Omega(\tau)=\sqrt{\omega(\tau)^2+[\Omega(\tau)^2-\omega(\tau)^2]}\) and then expand the square root to find
\begin{align}
 \Omega(\tau)=\omega(\tau)+\frac{\Omega(\tau)^2-\omega(\tau)^2}{2\,\omega(\tau)}+\O(T)^{-4}.
\end{align}
Using this we can expand (\ref{eq:currentinbetween}) to second adiabatic order
\begin{align}
\begin{split}
 & J^1=\frac{e}{a(\tau)}\int_{-\infty}^{\infty}\frac{dk}{2\pi}\left(\frac{p(\tau)}{\omega(\tau)}+\frac{\omega(\tau)-p(\tau)}{2\omega(\tau)^3}\left[\Omega(\tau)^2-\omega(\tau)^2\right]  \right. \\
 &  \left. +\frac{1}{8}\frac{\left[ma'(\tau)p(\tau)-ma(\tau)p'(\tau)\right]^2}{\omega(\tau)^6}+\O\left(T\right)^{-4}\right),
 \end{split}
\end{align}
where for \(\Omega(\tau)^2-\omega(\tau)^2\) one can use (\ref{eq:Omega-omega}) and replace \(\Omega(\tau)\) by \(\omega(\tau)\) on the right hand side. We can now compute this for the constant electric field (\ref{vector}) and find 
\begin{align}
  J^1=\frac{e}{a(\tau)}\int_{-\infty}^{\infty}\frac{dk}{2\pi}\frac{p(\tau)}{\omega(\tau)}+\O\left(T\right)^{-4}=-\frac{eH}{\pi}\lambda+\O\left(T\right)^{-4}.
\end{align}
For the bosonic case, the same counter term was found using the Pauli-Villars regularization method in $\ds_2$ in \cite{Froeb2014}. The adiabatic regularization for bosons, in $\ds_2$, is very similar to the calculation presented above and the final result can be found to be the same.\\
Performing the \gls{adiabatic subtraction} for the current (\ref{eq:explicitcurrent}) we eventually find
\begin{align}
 J^1_\text{reg}=&\frac{eH}{\pi}\mu\frac{\sinh\left(2\pi\lambda\right)}{\sin(2\pi\mu)}.\label{eq:regcurrent}
\end{align}
In the next subsections, we will discuss the properties of this current in detail. Performing the limit of strong electrical and strong gravitational field we underline the effect of the respective contributions to the total \gls{Schwinger effect}. For a plot of the current (\ref{eq:regcurrent}) as a function of \(\lambda\) for different values of \(\gamma\) see Fig.~\ref{fig:current}. There we also show a comparison to the bosonic current of \cite{Froeb2014} which is plotted as dotted lines.
 \begin{figure}
 \center
 \includegraphics[width=0.7\textwidth]{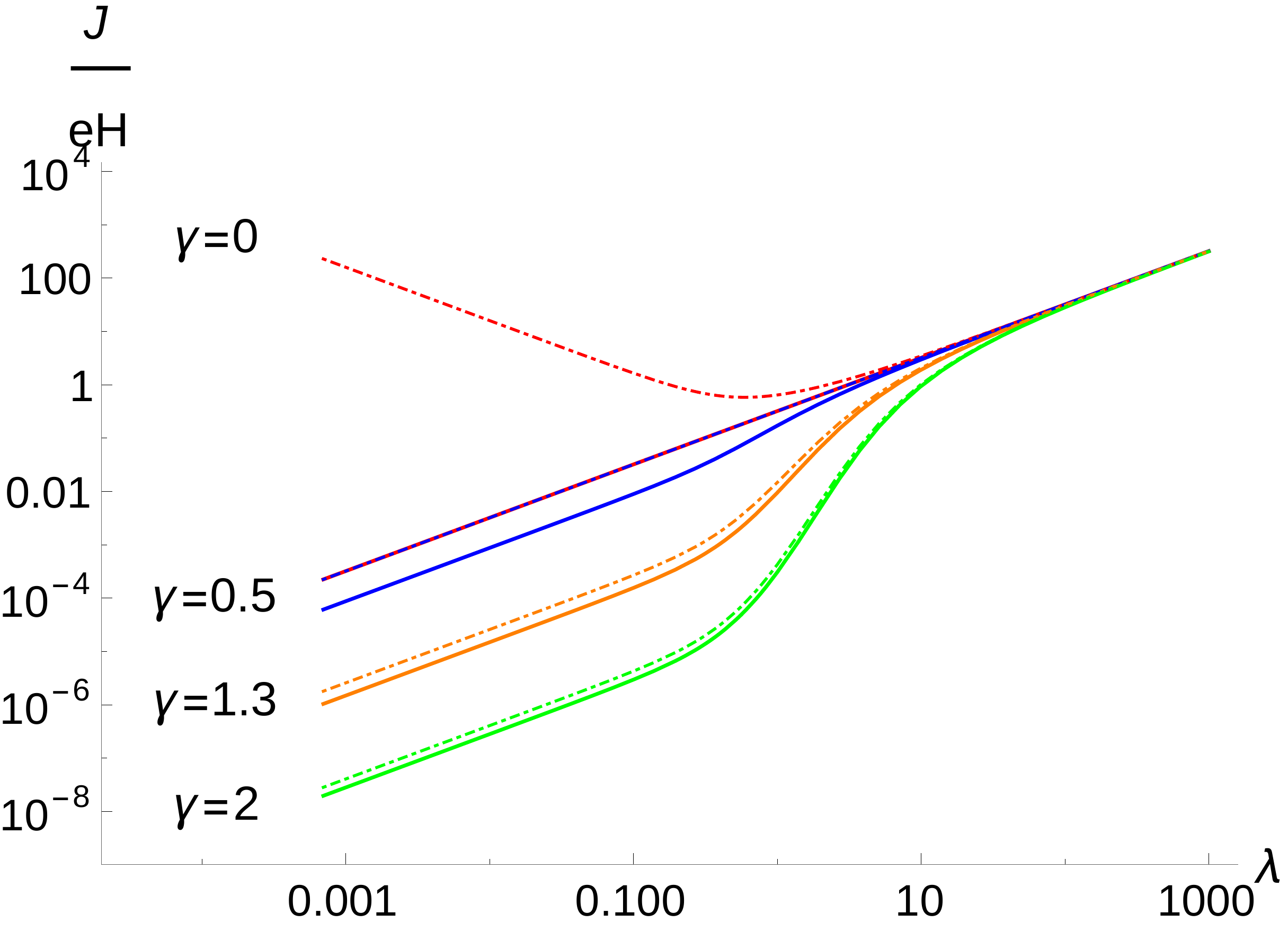} 
 \caption[The induced current in 1+1 D for bosons and fermions]{Regularized current for fermions and bosons (dotted) given by (\ref{eq:regcurrent}) and (\ref{eq:bosoncurrent}) respectively, as a function of $\lambda$ for different values of $\gamma$. One sees that for small electric fields there is a difference while all curves have the same asymptotic limit (\ref{eq:biglambda}) for negligible mass. Observe that only the bosonic case shows \acrshort{IRHC}, \textit{i.e.}~a large current for small electric field and mass. The curve with the linear response is given by $\gamma=0$ in the fermionic case while in the bosonic case it is found to be at $\gamma=0.5$.}
 \label{fig:current}
\end{figure}
 \subsubsection{Strong and weak electric field limit}
If we look at the limit \(\lambda\rightarrow \infty\) for fixed \(\gamma\), we find that the current is dominated by the electric field and has the asymptotic behavior
\begin{align}
 J^1_\text{reg}\sim \frac{eH}{\pi} \lambda, \label{eq:biglambda}
\end{align}
which is independent of the mass. This linear behavior can also be seen in Fig.~\ref{fig:current} where all the curves align for large \(\lambda\). In the bosonic case, in $\ds_2$, this linear behavior is also present \cite{Froeb2014} whereas, in $\ds_4$, a quadratic behavior is found which leads to a linear behavior for the conductivity (defined as ${J}/{\lambda}$). This can be used to impose strong constraints on \gls{magnetogenesis} scenarios \cite{Kobayashi:2014zza}. For fermions in $\ds_4$ the same might appear. The strong field limit is the same limit as the small mass limit \(\gamma\ll 1\) for which we find
\begin{align}
 J^1_\text{reg}=\frac{eH}{\pi}\left(\lambda+\left[\frac{1}{2\lambda}-\pi\coth\left(2\pi\lambda\right)\right]\gamma^2\right)+\O\left(\gamma\right)^3 \label{eq:smallmasslimit}.
\end{align}
For \(\gamma=0\) we find exactly the linear response which is analog to the asymptotic behavior (\ref{eq:biglambda}) since for strong enough fields the effect of the mass is negligible. The linear response is the usual flat spacetime response.} Indeed the \gls{pair production rate} for massless carriers in flat spacetime is given by
\begin{equation}
\Gamma=\frac{eE}{2\pi},
\end{equation} which leads to the \gls{induced current}
\begin{equation}
J=\frac{e^2 Et}{\pi}.
\end{equation} In an expanding spacetime the current can be computed in a comoving frame but will then be diluted in the physical frame. Naively making the substitution $t \rightarrow 1/H$, gives the linear response found in (\ref{eq:biglambda}). This substitution will be used again in the context of the flat spacetime limit in Sec.~\ref{sec:wog}. This shows that taking the limit $\gamma = m/H \rightarrow 0$ is equivalent to take the limit $H \rightarrow 0$ and $m=0$ which is the flat spacetime limit for massless particles. This illustrates the fact that a massless fermion is conformally invariant.

In the limit of strong electric fields, we can also allows compare to the result of the \gls{pair production rate} found in Sec.~\ref{sec:currentfromnk}. An approximation of the current for relativistic particles is given by
\begin{align}
 J\approx2en=2e\frac{\Gamma}{H}. \label{eq:approx_current}
\end{align}
Since the particle picture used to derive the \gls{pair production rate} (\ref{eq:paircreationrate}) only holds in the limit \(|\mu|\gg1\) which is equivalent to \(\sqrt{\gamma^2+\lambda^2}\gg 1\) and the assumption of relativistic particles only holds for small masses, the limit in which a comparison is possible is the strong field limit.\\
Using the \gls{pair production rate} (\ref{eq:paircreationrate}) to compute the estimated current (\ref{eq:approx_current}) in the limit \(|\mu|\gg1,\,\lambda\gg 1\) we find
\begin{align}
 J\approx\frac{e H}{\pi}|\mu|\exp\left[-2\pi\left(|\mu|-|\lambda|\right)\right],
\end{align}
which agrees with the regularized current (\ref{eq:regcurrent}) in this limit.\\
Expanding (\ref{eq:regcurrent}) for small electric fields \(\lambda\ll 1\) we find
\begin{align}
 J^1_\text{reg}=eH \frac{2\gamma \lambda}{\sinh\left(2\pi\gamma\right)}+\O\left(\lambda\right)^3.
\end{align}
As also visible in Fig.~\ref{fig:current} for small electric fields the mass begins to have an effect on the current. We find that for small electric fields the current gets strictly decreased by increasing mass. This is in contrast to what was found in the bosonic case of \cite{Froeb2014} and we will discuss this in the next section.
\subsubsection{Comparison to the bosonic case}
 \label{sec:comparison}
 We can compare the regularized current (\ref{eq:regcurrent}) to the scalar one in $\ds_2$ given by \cite{Froeb2014}
\begin{align}
 J^1_\text{boson,reg}=&\frac{eH}{\pi}\sigma\frac{\sinh\left(2\pi\lambda\right)}{\sin\left(2\pi\sigma\right)}, \label{eq:bosoncurrent}
\end{align}
where \(\sigma=\sqrt{\mu^2+\frac{1}{4}}\). \\
Observe that this is the same as the fermionic current (\ref{eq:regcurrent}) if we replace \(\sigma\rightarrow\mu\).
This means that the two currents agree in the limit \(|\mu|\gg1\) since there \(\sigma\approx\mu\) (see also Fig.~\ref{fig:current}), as we have already shown for the \gls{pair production rate} in Sec.~\ref{sec:currentfromnk}. However we find a different behavior for electric fields small with respect to the Hubble constant. This is due to the fact that \(\mu\ne\sigma\) and is in difference to the flat space case where the \gls{pair} production of a constant overcritical electric field is identical for scalar and spinor \acrshort{QED}.

The most striking result of the difference is, that the bosonic current is enhanced for small electric fields and values of the parameter \(\gamma<1/2\). In \cite{Froeb2014} this effect was given the name ``\acrlong{IRHC}''. We do not find this behavior for fermions. Mathematically this is due to the fact that \(\sigma\) in difference to the parameter \(\mu\) can become real for \(\gamma<1/2\). In the same way the linear response for the bosonic case is found for \(\gamma=1/2\) whereas it is found for \(\gamma=0\) for fermions. Since in flat space-time the linear response arises for massless particles the fermionic result is not as peculiar as the bosonic one.\\
The difference between \(\sigma\) and \(\mu\) comes from the last term in (\ref{eq:DecoupledDirac1})-(\ref{eq:DecoupledDirac2}). For the constant electric field (\ref{vector}) it evaluates to \(1/(4\tau^2)\) and accounts for the \(1/4\) term in the Whittaker equations (\ref{eq:Whittaker1})-(\ref{eq:Whittaker2}). These terms are absent in the Klein Gordon equation and thus an additional factor of \(1/4\) was introduced in the variable \(\sigma\).
\subsubsection{Vanishing gravitational field: flat spacetime limit}
  \label{sec:wog}
  For a comparison with the flat spacetime case the parameters \(\lambda\) and \(\gamma\) defined in (\ref{lambda}) are not convenient since they diverge in the flat space limit \(H\rightarrow0\). We therefore define the electric field \(\epsilon\) in units of the critical electric field as
  \begin{align}
   \epsilon \equiv \frac{eE}{m^2}=\frac{\lambda}{\gamma^2}. 
  \end{align}
  Performing the the flat spacetime limit \(H\rightarrow0\) of (\ref{eq:regcurrent}) we find
    \begin{align}
 \lim_{H\rightarrow0} J^1_\text{reg}&= \lim_{H\rightarrow0} \frac{e m^2}{\pi }\frac{\epsilon}{H} e^{-\frac{\pi}{|\epsilon|}}.
\end{align} 
 This current diverges, as it is expected for the current of an electric field in Minkowski spacetime, which was turned on at past infinity. This is due to missing Hubble dilution. One can however compare it to a current of an electric field which was turned on at finite time \(t\), as it was done in the bosonic case \cite{Kobayashi:2014zza}. Using the substitution \(1/H\rightarrow t\) introduced above, one finds agreement with the \glslink{induced current}{current induced} by the \gls{Schwinger effect} in flat Minkowski spacetime\footnote{See e.g.~Eq.~(5.22) of \cite{Anderson:2013ila} or Eq.~(2.14) of \cite{Anderson:2013zia} for the bosonic current in four dimensional flat spacetime and Eq.~(5.32) of \cite{Kluger:1998bm} for the one in two dimensional flat spacetime.}. 
\section{On infrared hyperconductivity}
\label{sec:IRHC} 
\acrfull{IRHC} is a regime where for a given interval of the electric field, a decreasing electric field gives an increasing conductivity.
\acrshort{IRHC} was first reported in \cite{Froeb2014} in the case of \gls{pair} production for bosons in 1+1 D but has been then showed to appear in 1+2 D \cite{Bavarsad:2016cxh} , 1+3 D for bosons \cite{Kobayashi:2014zza} and 1+3 D for fermions \cite{Hayashinaka:2016qqn}. 
In \cite{Froeb2014}, it was shown that the \gls{induced current} responds as $J\sim E^{-1}$ for small electric fields and \acrshort{IRHC} was present for $m/H<1/2$. In the case of $\ds_{4}$ the renormalization scheme introduces a term of the form $\log(m/H)$ \cite{Kobayashi:2014zza} in the
regularized current expression which arises from the second order adiabatic expansion.
Therefore, it signals that this renormalization method was not applicable for the case of exactly massless scalar field in $\ds_{4}$ \cite{Kobayashi:2014zza}; see also discussions in \cite{Parker1974,Fulling1974}.
Hence it was not possible to discuss \acrshort{IRHC} for the massless case but still \acrshort{IRHC} was present for $m/H<\sqrt{5/4}$.
In $\ds_{3}$, we have seen in Sec.~\ref{sec:result} that the current behaves as $J\sim E^{-1}$ in the regime of small electric field and massless minimally coupled charged particles and reported \acrshort{IRHC} for $\mds/H<\sqrt{3/4}$.
\par
These results lead us to propose a procedure to avoid an \acrshort{IRHC} regime by setting the value of the \gls{conformal coupling} to a specific range.
In $\dsd$, the \textit{nonrenormalized} in-vacuum state expectation of the spacelike component of the current operator is
\begin{align}\label{vevd}
\langle j^{1}\rangle_{in}&=\,_{in}\langle0|j^{1}|0\rangle_{in} \nn\\
&=2ea^{-D}(\tau)\int\frac{d^{d}\textbf{k}}{(2\pi)^{d}}\big(k_{x}+eA_{1}(\tau)\big)
\frac{e^{\kappa\pi i}}{2k}\big|\wwp(z_{+})\big|^{2}.
\end{align}
Generalizing to $D$ dimensions the step performed between Eq.~(\ref{vev}) and Eq.~(\ref{curent}), the integral~(\ref{vevd}) can be conveniently
rewritten as
\begin{equation}
\langle j^{1}\rangle_{in}=\frac{eH^{d}}{(2\pi)^{d}}a^{-1}(\tau)\int d\Sigma_{d-1}e^{\lambda r\pi}
\lim_{\Lambda\rightarrow\infty}
\int_{0}^{\Lambda}dp\,p^{d-2}\big(rp-\lambda\big)\big|\wwp(z_{+})\big|^{2}, \label{integrald}
\end{equation}
where $d\Sigma_{d-1}$ is given by Eq.~(\ref{aelement}).
As pointed out in \cite{Froeb2014,Kobayashi:2014zza}, in a \acrshort{IRHC} regime, the population of produced \glspl{pair} is dominated by IR contribution and
no longer by the \glspl{pair} created within a Hubble time.
Hence the asymptotic behavior of the wave function, in the limit $p\rightarrow 0$ will give the dominant term in a \acrshort{IRHC} regime. We recall here that in this regime $\mu$ is real and by convention is positive, so using (\ref{wout}), we find to the leading order that the integral~(\ref{integrald}) in the limit $p\rightarrow0$ behaves as
\begin{equation}\label{irint}
\lambda\int_{0}dp\,p^{D-2-2\mu}.
\end{equation}
Then power counting shows that in the regime
\begin{equation}\label{ircon}
\mu>\frac{D-2}{2},
\end{equation}
the current integrand diverges in the limit $p\rightarrow 0$.
However, since $\mu\leq\frac{D-1}{2}$, the total current integral remains finite.
From Eq.~(\ref{ircon}) and the definition of $\mu$, given by Eq.~(\ref{eq:mu}), we find first that
\begin{align}
\label{eq:approx}
\langle j^{1}\rangle_{in}\propto\frac{eH^{d}\lambda}{\rho^2}
=\frac{eH^{d}\lambda}{\lambda^{2}+\gamma^{2}}, && \rho\ll d.
\end{align}
Observe that setting $\gamma=0$, one recovers the behavior $J\propto E^{-1}$.
Fig.~\ref{fig4} shows a plot of the current in the \acrshort{IRHC} regime together with the analytical result of~(\ref{eq:approx}), and the curves agree
reasonably well. Similar plots could be produced for $D=2$ or $D=4$. \\
Second, again from Eqs.~(\ref{eq:mu}) and~(\ref{ircon}) it is also possible to deduce that \acrshort{IRHC} occurred when
\begin{equation}\label{iregime}
\gamma^{2}+\lambda^{2}<\frac{2D-3}{4}.
\end{equation}
Therefore, a sufficient condition to avoid \acrshort{IRHC} is
\begin{equation}\label{avoid}
\gamma^{2}\geq\frac{2D-3}{4},
\end{equation}
and we define thus $\lambda_{\text{m,min}}=\frac{\sqrt{2D-3}}{2}$. The previous condition implies also a minimal bound to avoid \acrshort{IRHC} for the conformal
coupling $\xi$ which in the case of a massless scalar field, reads
\begin{equation}\label{ximin}
\xi_{\mathrm{min}}=\frac{2D-3}{4D(D-1)}.
\end{equation}
We see that in nonconformally coupled theories, a \gls{conformal coupling} with values larger than $\xi_{\text{min}}$ can be used to avoid the
\acrshort{IRHC} regime. Conversely, $\forall\xi<\xi_{\mathrm{min}}$, \acrshort{IRHC} would appear for $m^{2}/H^{2}\in\mathcal{I}_{\text{IR-HC}}$, with
\begin{equation}\label{mathi}
\mathcal{I}_{\text{IR-HC}} \equiv \big(0,\frac{2D-3}{4}-D(D-1)\xi\big).
\end{equation}
These results are summarized in Table~\ref{tableIRHC}.
\begin{table}[h]
\begin{center}
\begin{tabular}{|c|c|c|c|}
  \hline
  & $\lambda_{\text{m,min}}^{2}$ & $\xi_{\mathrm{min}}$ & $\mathcal{I}_{\text{IR-HC}}$  \\
  \hline
  In $\ds_{2}$ & $1/4$    & $1/8$ &  $\big(0,\frac{1}{4}-2\xi\big)$ \\
  In $\ds_{3}$ & $3/4$    & $1/8$ &  $\big(0,\frac{3}{4}-6\xi\big)$  \\
   In $\ds_{4}$ & $5/4$    & $5/48$ & $\big(0,\frac{5}{4}-12\xi\big)$  \\
  In $\dsd$ & $\frac{2D-3}{4}$ & $\frac{2D-3}{4D(D-1)}$ & $\big(0,\frac{2D-3}{4}-D(D-1)\xi\big)$  \\
  \hline
\end{tabular}
\caption[Table summing up our suggestion to avoid infrared-hyperconductivity]{Minimum values of $\gamma^{2}$ and $\xi$ to avoid \acrshort{IRHC}.
The interval $\mathcal{I}_{\text{IR-HC}}$ is the range of $m^{2}/H^{2}$ for which \acrshort{IRHC} would appear after turning on the conformal coupling.} \label{tableIRHC}
\end{center}
\end{table}
In $D=2$ and $D=3$, Eq.~(\ref{avoid}) agrees very well with numerical investigations.
However, in the case of $D=4$, to avoid \acrshort{IRHC} one needs to have $\gamma\gtrsim 1.25$ \cite{Kobayashi:2014zza}; in this case, a small variation
from condition~(\ref{avoid}), comes from a term also dominant in the IR regime but not taken into account in the previous calculation: the one coming from the renormalization in ``$\log(m/H)$''.
\par
Table~\ref{tablelminlmax} presents the results of the numerical investigations for the value of $\lambda_{\text{min}}$ and $\lambda_{\text{max}}$
for dimensions $D=2,3,4$.
\begin{table}[h]
\begin{center}
\begin{tabular}{|c|c c c|c c c|c c c|}
  \hline
  $D$  &  & 2 & &  & 3 & & & 4 &  \\
  \hline
  $\gamma$ & 0.1,&0.01,&0.001 &0.1,&0.01,&0.001 & 0.1,&0.01,&0.001 \\
  \hline
  $\lambda_{\text{min}}$ & 0.54,&0.59,&0.59 &0.92,&0.95,&0.95 &1.27,&1.54,&1.68 \\
  \hline
  $\lambda_{\text{max}}$ &0.1,&0.01,&0.001 &0.1,&0.01,&0.001 &0.1,&0.01,&0.001 \\
  \hline
\end{tabular}
\caption[Numerical investigation of infrared hyperconductivity]{Numerically found values of $\lambda_{\text{min}}$ and $\lambda_{\text{max}}$ for different values of $\gamma$ and $D$.
\acrshort{IRHC} occurs for $\lambda\in(\lambda_{\text{max}},\lambda_{\text{min}})$.
Those results point toward the idea that $\lambda_{\text{max}}=\gamma$ and $\lambda_{\text{min}}=\lambda_{\text{m,min}}+\epsilon=\frac{\sqrt{2D-3}}{2}+\epsilon$, $\epsilon>0$.} \label{tablelminlmax}
\end{center}
\end{table}
Numerical investigations indicate that $\lambda_{\text{max}}=\gamma$ and $\lambda_{\text{min}}=\lambda_{\text{m,min}}+\epsilon$ with $\epsilon>0$.
Recall that $\lambda_{\text{m,min}}=0.5,0.87,1.12$ for $D=2,3,4$, respectively.
$\epsilon$ reaches an asymptotic value for $\gamma\rightarrow 0$ in $D=2,3$, whereas it is unbounded for $D=4$. This difference comes again from the renormalization term in ``$\log(m/H)$''.
\par
Looking at the fermionic \gls{induced current} in $\ds_{2}$ \cite{Stahl:2015gaa} and $\ds_{4}$ \cite{Hayashinaka:2016qqn}, no \acrshort{IRHC} was reported.
In $\ds_{2}$, the only difference between fermions and bosons was effectively a translation of the mass squared, \textit{i.e.}, $m^2_{\mathrm{fermion}}=m^2_{\mathrm{boson}}-H^2/4$.
It is furthermore known that a massless fermion is conformally invariant and gives, as in flat spacetime, a linear behavior for the current.
In the bosonic case this conformal behavior was found for $m^{2}/H^{2}=1/4$ and the \acrshort{IRHC} for $0\leq m^{2}/H^{2}<1/4$.
Hence, conformality plays an important role to understand \acrshort{IRHC}.
Note that for a fermionic particle in $D=2$, to have a regime of \acrshort{IRHC} one needs to let the mass parameters $m^{2}/H^{2}<0$, that is, to allow for
\glslink{tachyon}{tachyonic propagation}.
In parallel to \gls{tachyon}, \acrshort{IRHC} is a regime where decreasing one source (the electrical field) increases the consequence (the produced \glspl{pair}).
Therefore, it is against physical intuition and for massless cases leads even to a current unbounded from above.
The links between \glslink{tachyon}{tachyonic field}, conformality, and \acrshort{IRHC} remain to be explored.
\section{Concluding remarks and perspectives}
\label{sec:cclpairs}
We have seen different techniques to investigate \gls{pair} production in \gls{dS} under the influence of a constant electric field. After setting up the stage, we have studied \gls{semiclassical} estimates of the \gls{pair production rate} in Sec.~\ref{sec:smestim}. An extension of those calculations could be to explore more complicated electromagnetic field configurations in \gls{curved spacetime} and check if the equivalence of \gls{semiclassical} bosonic and fermionic \glspl{pair production rate} (see Sec.~\ref{sec:smestim}) still holds true. Indeed while for a constant electric field in flat spacetime this result is well known, it has been shown to be false for non constant fields with more than one component \cite{Strobel:2014tha}. Beyond \gls{semiclassical} results and the particle picture, we computed the \glspl{induced current}. It is a more relevant quantity than the usual \gls{pair production rate} per unit volume because it is not plagued by the absence of a clear definition of \gls{pair} production time. The final results were presented in (\ref{reg}) and (\ref{eq:regcurrent}). Finally in Sec.~\ref{sec:IRHC}, we investigated some aspects of \acrshort{IRHC} which is of course an effect to be clarified, maybe in connection with conformal field and \glspl{tachyon}. Another ingredient to understand \acrshort{IRHC} could be the link with the \gls{Breitenlohner-Freedman stability bound} \cite{Breitenlohner:1982jf} (see also \cite{McInnes:2001dq,Pioline:2005pf}) usually discussed in string theory in anti-de-Sitter space. In the case of $\ds$, it would be an upper bound on the mass of the field which could be violated as soon as \glslink{particle creation}{particles} begin to be created via the \gls{Schwinger effect}.

 As the gravitational and electrical field were taken to be external, a natural thing to check is to take their variation into account, \textit{i.e.}~use the currents in the generalized \gls{Maxwell equation} and the resulting number of \glspl{pair} in the \glspl{Einstein equation}. This will be done in the next chapter. Other open roads to continue this work are cosmological applications of our result. The produced \glspl{pair} might account for the asymmetry of matter/anti-matter in our universe with a modification of an Affleck-Dine mechanism or a specific model of \gls{reheating}. They could also give hints on the evolution of an accelerated period of expansion and how matter and gravitation interact in such periods.
 
 \subsection*{On baryogenesis}
 \Gls{baryogenesis} is a topic of cosmology not presented in chapter \ref{chap:introcosmo}, which aims at giving a mechanism for the matter content of the universe to overcome the anti-matter. This primordial asymmetry is required as the obervations today advocate for a universe mainly constituted of matter. \\
 The very idea to propose a \gls{baryogenesis} scenario with the \gls{pair} production mechanism studied in this chapter is to separate the \glspl{pair} particle/anti-particle one from each other with the help of strong gravitational and electrical fields present during \gls{inflation}. However one needs to define properly a charge operator $C$. The quantum fields theories manipulated in this chapter can be checked to be $C P T$ invariant. For the case of fermions in 1+1 D, the charge operator has been identified in equation (\ref{eq:charge}). Looking explicitly at the solutions for instance in equation (\ref{eq:2DFermSoloutm}), one recovers the Feynman physical picture of anti-particle traveling backwards in time. Those solutions for particles and anti-particles behave differently and it is indeed possible to observe a charge violation.

Hence the next steps would be to consider a quasi-de Sitter spacetime to allow a \gls{reheating} phase and to move to 3 spatial dimension. These technical and mathematical works being done, the physical picture would be that virtual \glspl{pair} could tunnel from the Dirac sea and become real and stretched to large scales because by the accelerated, non causal, expansion of the universe. Then the electrical field would sort them in a way that matter would aggregate in a given spatial region whereas an anti-matter would aggregate in another region.

The theory of inflationary multiverses is based on melting inflationary cosmology, anthropic considerations, and \gls{particle physics}. \cite{Vilenkin:1983xq,Linde:1986fd,Linde:2015edk} describes the growth of this theory from its infancy. The very idea of \gls{inflation} was to render our part of the Universe homogeneous by stretching any pre-existing inhomogeneities to scales inaccessible to us. If our universe consists in several parts, each of these parts after \gls{inflation} will become locally homogeneous and inhabitants of a given part will not be able to communicate with any other part and will conclude wrongly that the universe is homogeneous everywhere. These very large parts are sometimes called \emph{mini-universes}\cite{Linde:2015edk} or \emph{pocket universes}.\cite{carr2007universe} Finally the whole system consisting of many pocket universes is called \emph{multiverse} or \emph{inflationary multiverse}.

This picture of \gls{eternal inflation} may sound speculative but its ingredients are well rooted into theoretical physics' tools. It relies on three pillars: first, the non-uniqueness of \gls{vacuum} state. It is possible in the \gls{SMP} of \gls{particle physics}, common beyond standard model physics and inevitable in string theory. Second \gls{eternal inflation} is a theory based on a quantum field theory framework \cite{Peskin:1995ev} which has been shown, until now, to be an incredibly powerful tool to predict physical phenomenons. Third, \gls{eternal inflation} needs an accelerated phase of expansion. A late time expansion (\gls{dark energy}) was already observed and the early time \gls{inflation} is inferred from \acrshort{CMB} observations for instance. If one believes these three pillars, \gls{eternal inflation} is a direct natural consequence.

\glslink{particle creation}{Particles} created by the \gls{Schwinger effect} in \gls{dS} were already used in Ref.~\cite{Froeb2014} to model bubble nucleation in the context of inflationary multiverses. Within those models, it is possible that the transition via an Higgs mechanism from an inflationary universe to the standard \gls{reheating} phase does not occur simultaneously everywhere. One could imagine that the phase transition of the two zones created by the \gls{Schwinger effect} described in this chapter would not occur at the same time and hence form two causally disconnected patches: two pocket universes. One only filled with matter, the other only filled with anti-matter. This could be a proposal to explain the matter/anti-matter asymmetry, see also the discussion of Ref.~\cite{Linde:2015edk} about \gls{baryogenesis} proposals. This idea of using \glslink{particle creation}{particles} created by the gravitational and electrical \gls{Schwinger effect} to solve the \gls{baryogenesis} problem is a new one. See also Ref.~\cite{Goolsby-Cole:2015chd} for a topical discussion of the charge of the universe during \gls{inflation} and the connection to \gls{baryogenesis}.

 For our scenario to occur, one needs to assume a given number of premises. First that the decay of the \glslink{particle creation}{particles} created during \gls{reheating} preserves the \gls{baryon} number so that the already present asymmetry stays. Second that the expansion of the universe, which can still be accelerated in many models of \gls{reheating} or \gls{warm inflation} would not dilute the already present asymmetry. Third that all the other ways of creating asymmetry are negligible regarding the main one. Fourth that there is a range of parameter in this model which predicts the value of the parameter $\eta\equiv \frac{n_b}{n_{\gamma}}=10^{-10}$. If all these requirements appear to be satisfied then, the three Sakharov conditions would be fulfilled and this scenario would be a valuable attempt to solve the \gls{baryogenesis} problem. 

%% file: back/backelectro.tex
\epigraph{L'Italia è il giardino dell'europa.}{An old woman in a church in Penne, 2016}
\textit{In this chapter, we present two works on the possible backreaction of the pairs production mechanism studied in the previous chapter. How the produced pairs interact with the electric field which was assumed to be constant and how the pairs produced interact with the gravitational sector of the theory: the \gls{dS} are the two main focus. Those works require to solve the \glslink{Maxwell equation}{Maxwell} and \glspl{Einstein equation} with a new term on the right hand side. They are adapted from the following publications \cite{Stahl:2016geq,Bavarsad:2016cxh} }
\section{To the electromagnetic field}
\begin{figure}
    \centering
     \includegraphics[width=0.7\textwidth]{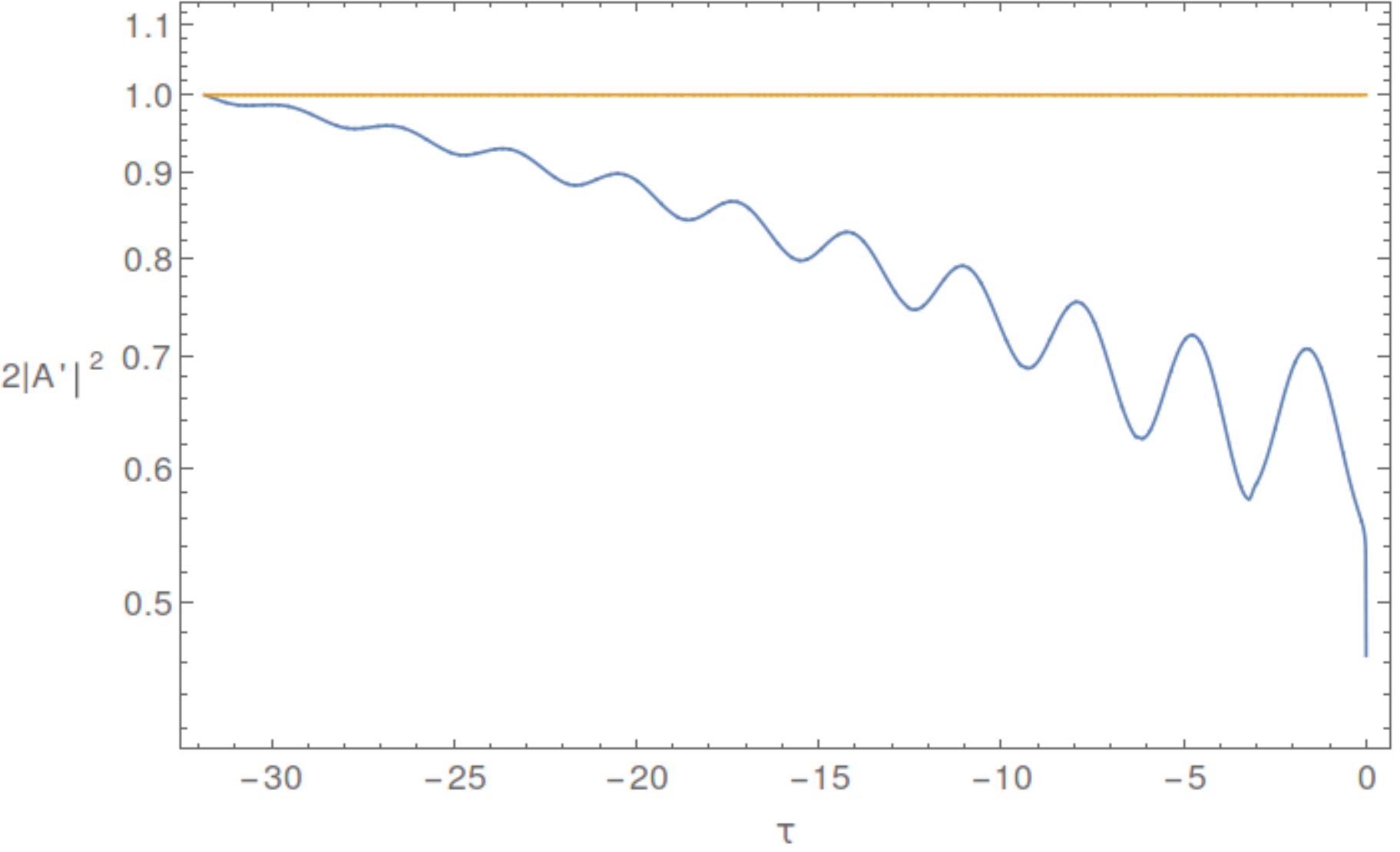} 
 \caption[Evolution of the electric field with backreaction]{We propose an example of the behavior of the mode function where $H=k=\epsilon=\mu=1$. The upper yellow constant line is the free case ($j=0$). The blue line is the mode function when the backreaction corrections are turn on. In the asymptotic past, the current vanishes and the two functions are equal. As one approaches the asymptotic future, the plasma oscillations appear.}
 \label{fig:mode function}
\end{figure}

\begin{figure}
    \centering
    \includegraphics[width=0.7\textwidth]{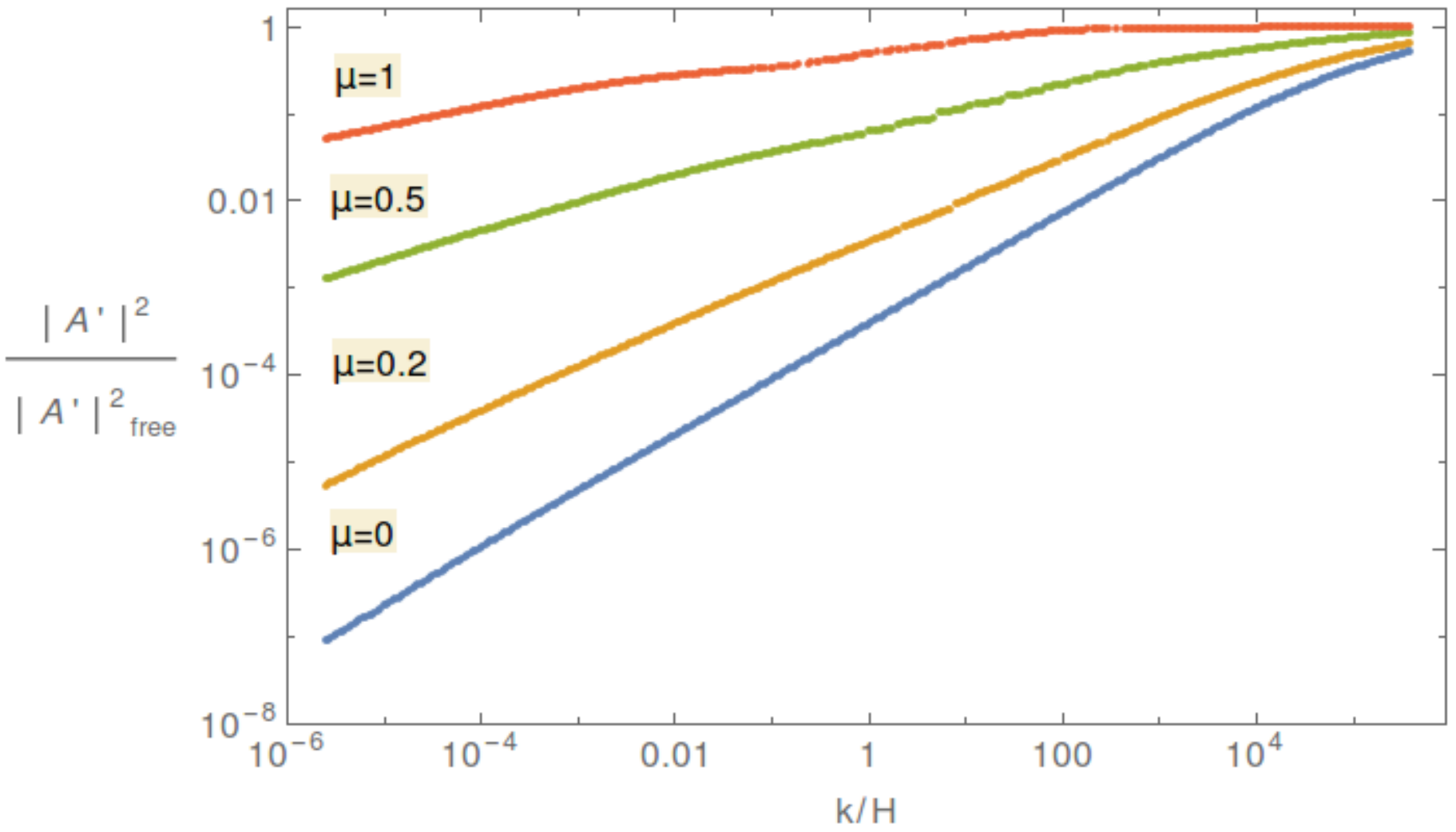} 
 \caption[Phenomenological analysis of the mass parameter]{Phenomenological analysis of the impact of the mass parameter to the gauge field evolution when backreaction is turned on. The selected parameters are: $H=\epsilon=1$. From bottom to top, the values of the mass are: $\mu=0,0.2,0.5,1$. All curves are plotted in the asymptotic future of the \gls{manifold}, that is $\tau \rightarrow 0$.}
 \label{fig:varym}
\end{figure}

\begin{figure}
        \centering
      \includegraphics[width=0.7\textwidth]{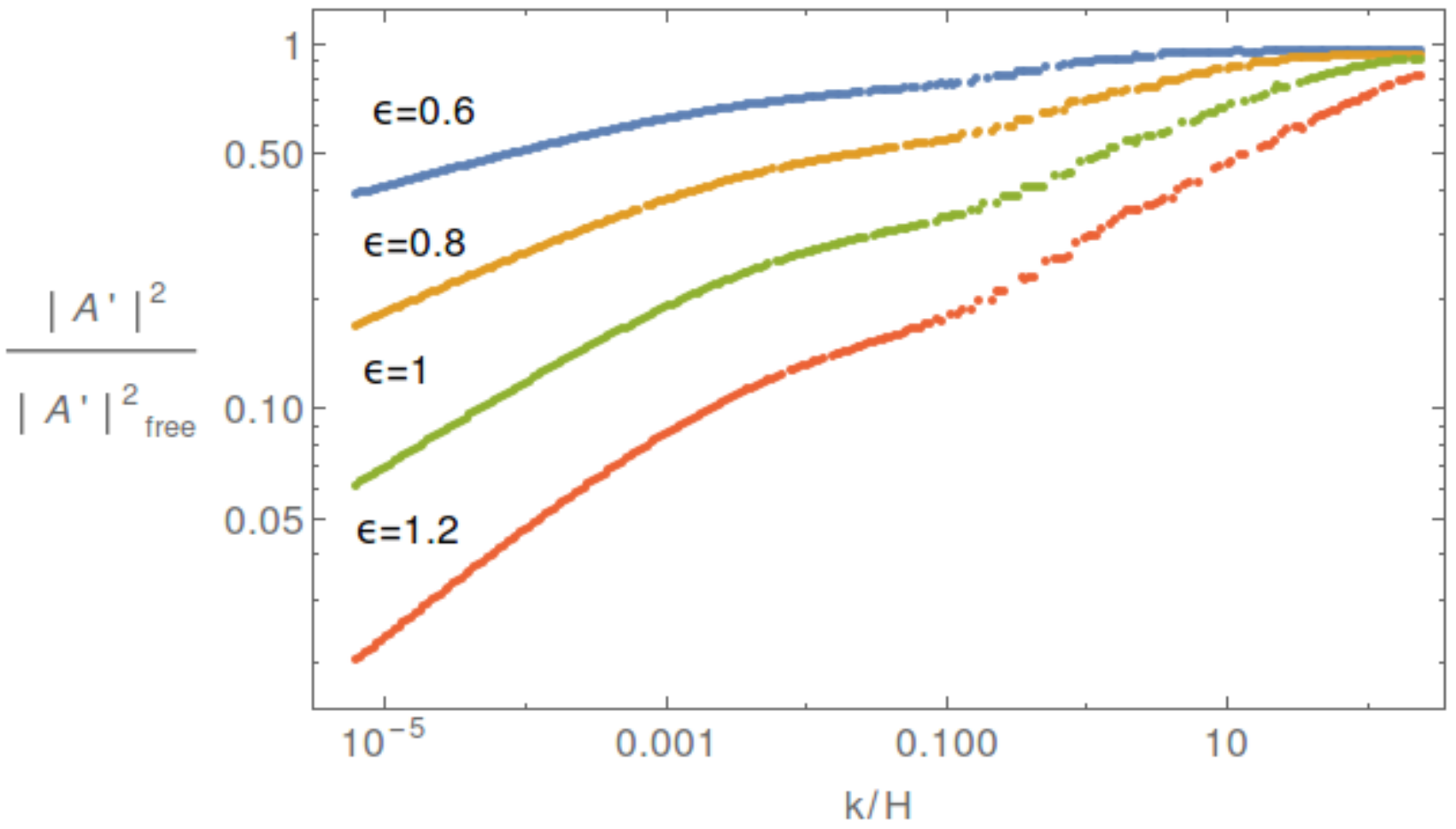} 
 \caption[Phenomenological analysis of the gauge coupling parameter]{Phenomenological analysis of the impact of the gauge coupling parameter to the gauge field evolution when backreaction is turned on. The selected parameters are: $H=\mu=1$. From top to bottom, the values of the gauge coupling are: $\epsilon=0.6,0.8,1,1.2$. For intermediate values of $k$, numerical instabilities do not allow us to solve (\ref{eq:tosolveback}) for all $k$. All curves are plotted in the asymptotic future of the \gls{manifold}, that is $\tau \rightarrow 0$.}
 \label{fig:varye}
\end{figure}

\paragraph*{}
\label{sec:backEM}
In the previous chapter, we assumed that the electric field was constant. We will investigate here some results on \gls{backreaction} to the electric field. In flat spacetime example, the \gls{backreaction} has been shown to be important: for instance, for 1 + 1 D fermions, the phenomenon of plasma oscillation has been discovered in \cite{Kluger:1991ib,Kluger:1992gb}, see also \cite{Ruffini:2003cr,Ruffini:2007jm}). In order to consider the \gls{backreaction} to the electromagnetic field, we consider that the currents (\ref{eq:bosoncurrent}) and (\ref{eq:regcurrent}) are coupled to a gauge field. To do so we add one term in the Lagrangians (\ref{action:sca})(\ref{action:fer}) in the form:
\begin{align}
\label{eq:int}
\mathcal{L}_I= -j^{\mu}(x)A_{\mu}(x),
\end{align}
where $j^{\mu}(x)$ is the current. We will not expand in this thesis the technical details, the reader is invited to follow \cite{Stahl:2016geq} where the main equations are displayed. The crux of the problem lies in the resolution of:
\begin{equation}
\label{eq:tosolveback}
A''(\tau,k)+\frac{k^2}{H^2}A(\tau,k)=\frac{\epsilon}{\pi \tau^2}\omega_F(\tau,k) \frac{\sinh(2\pi \epsilon A'(\tau,k) \tau)}{\sinh(2\pi\omega_F(\tau,k))},
\end{equation}
where the right hand side follows directly from (\ref{eq:regcurrent}), $\omega_F(\tau,k) \equiv \sqrt{\epsilon^2 A'(\tau,k)^2\tau^2+\mu^2}$. In this section, $\tau$ denotes the dimensionless conformal time which is related to the conformal time (\ref{metricds}) of the others of part of this thesis divided by $H$, the prime denotes then a partial derivative with respect to $\tau$, that is $A'(\tau,k) \equiv \partial_{\tau} A(\tau,k)$. The result for the resolution of (\ref{eq:tosolveback}) is displayed in figure \ref{fig:mode function} which shows a typical behavior for the mode function $|A'(\tau,k)|^2$. In the asymptotic past, the \gls{backreaction} term can be neglected. Then one observes plasma oscillations which resemble to the one found in flat spacetime \cite{Kluger:1992gb}. However these oscillations are damped by the dilution due to the expansion of the Universe. Qualitatively, the frequency of these plasma oscillations is inversely proportional to the comoving momentum $k$. Indeed, as we will also discuss later, in the ultraviolet regime: for $\frac{k}{H} \gg 1$, the electromagnetic field is not sensitive to gravity and is unaffected by the \gls{Schwinger effect}. Conversely, in the infrared regime, for $\frac{k}{H} \ll 1$, the plasma oscillations are the dominant physical phenomenon and their frequency is increased.
\paragraph{}
 Figures \ref{fig:varym} and \ref{fig:varye} show typical power spectra. In the ultraviolet regime ($\frac{k}{H} \gg 1$), the electric \gls{power spectrum} is unaffected by the \gls{Schwinger effect} because the \gls{backreaction} of the current is negligible. Indeed, in this regime, the electromagnetic oscillations are outside the de Sitter horizon and are unaffected by gravity effects. However in the infrared regime ($\frac{k}{H} \ll 1$), the \gls{power spectrum} gets significantly reduced due to the significant screening of the \gls{backreaction} of the Schwinger \glspl{pair}. Observe that varying the \gls{Hubble constant} $H$ just changes the window of the comoving momentum $k$ considered. Indeed the relevant quantity to understand the \gls{backreaction} effect is $\frac{k}{H}$.
The current is always positive in the direct space, as a consequence the \gls{power spectrum} is always reduced with respect to its flat spacetime value. As a result, the electromagnetic field is not significantly enhanced.
  \par
\begin{figure}
    \centering
     \includegraphics[width=0.7\textwidth]{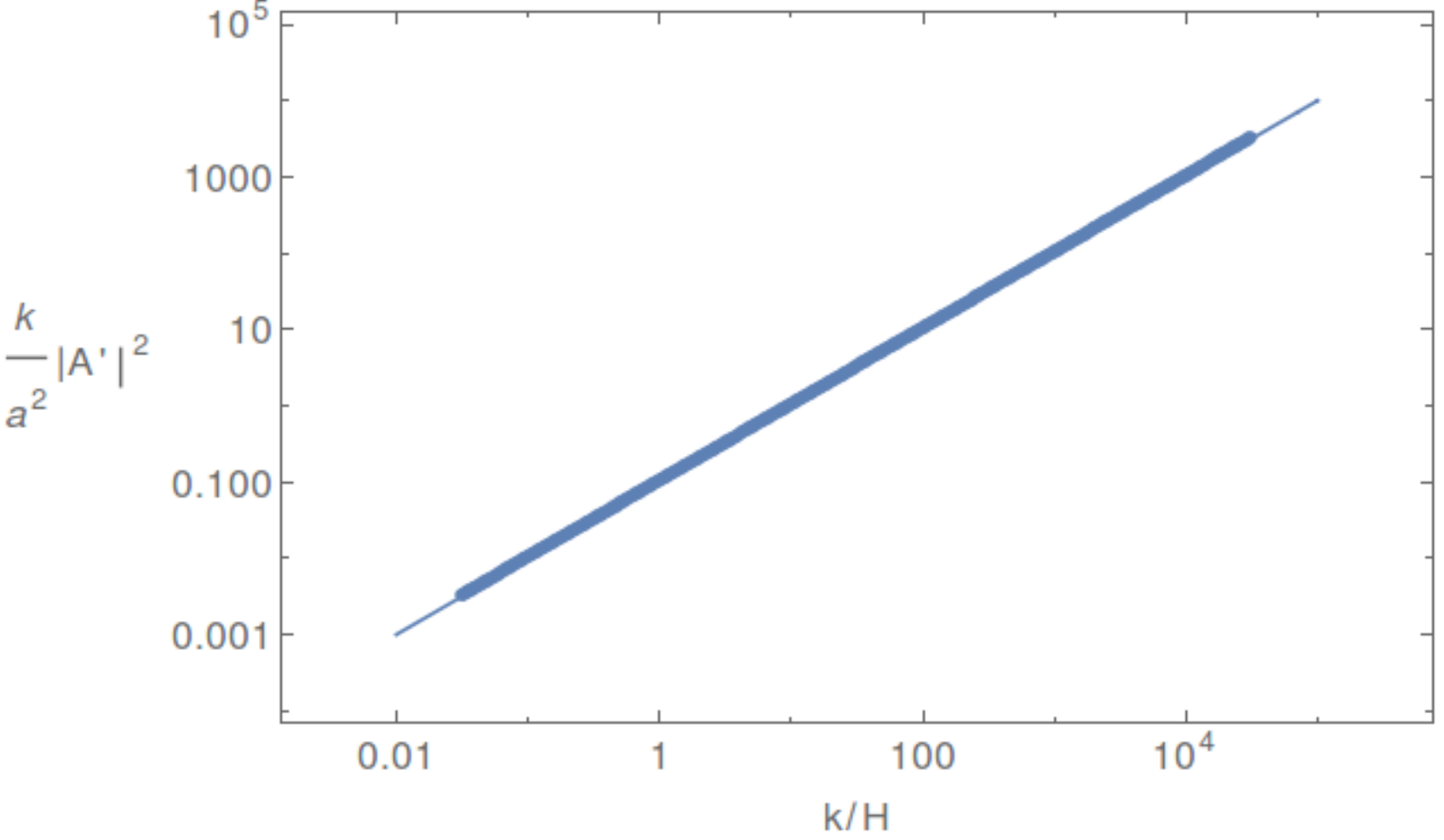} 
 \caption[A solution where the electric field is amplified]{We report a non-trivial solution where the electric field gets enhanced, in this case $\mu=0$ and $\epsilon=1$. The dots are the result of the numerical integration whereas the plain line is the analytic estimate $|c_{\text{sol}}|^2 k$.} 
 \label{fig:IRHCresul}
    \end{figure}

 We are now in the position to phenomenologically discuss the impact of the mass parameter $\mu$ (cf Fig.~\ref{fig:varym}). The more $\mu$ dwindles, the more the \gls{power spectrum} dwindles. Indeed the heavier the mass of the \glspl{pair} is, the smaller the current is (cf.~\cite{Stahl:2015gaa}). This implies that less \glspl{pair} are created so that the screening of the electric field is less important.
\paragraph{}
Now we turn to the phenomenological study of the gauge coupling $\epsilon$ (cf Fig.~\ref{fig:varye}). The more $\epsilon$ increases, the more the \gls{power spectrum} dwindles. The more the gauge coupling increases, the stronger the photon and \glspl{pair} are coupled together and the less important the amplification of the electric field is.
\paragraph{} In all the scenarii depicted so far, no enhancement of the electromagnetic field is reported and in agreement with \cite{Kobayashi:2014zza}. Namely via \gls{backreaction}, triggering the \gls{Schwinger effect} during \gls{inflation} has the only effect of constraining \gls{magnetogenesis} scenario.
\paragraph{Possible amplification of the electromagnetic field.} We now consider bosonic \glspl{pair} produced by the \gls{Schwinger effect}, in this case, following directly from (\ref{eq:bosoncurrent}) and the discussion around this equation, the the only change is:
\begin{equation}
\label{eq:presc}
\omega_B^2(\tau,k) \leftrightarrow \omega^2_F(\tau,k) -\frac{1}{4}.
\end{equation} Therefore the parameter space to be examined is $\mu \in (0,\frac{1}{4})$. We propose an ansatz:
\begin{equation}
\label{eq:anzzz}
A'(\tau)=\frac{c}{\tau},
\end{equation}
which is the minimal requirement for the solution in order to have enhancement of the electromagnetic field, where the parameter $c$ is to be determined. For infrared modes ($k/H \ll 1$), the current become the dominant term in the \gls{Maxwell equation}. Assuming it is the case and using (\ref{eq:anzzz}) in the equivalent equation of motion (\ref{eq:tosolveback}) modified by the bosonic prescription (\ref{eq:presc}), we find the following scalar equation which may or may not have solution depending on the parameters:
\begin{equation}
-c=\frac{\epsilon}{\pi} \frac{\sinh(2\pi \epsilon c) \omega_c}{\sinh(2\pi \omega_c)},
\label{eq:forc}
\end{equation}
where $ \omega_c \equiv \sqrt{\mu^2+\epsilon^2 c^2-1/4}$. Observe that in the fermionic case, the factor “$-1/4$” is absent, the only trivial solution is $c=0$, no enhancement solution is found. In the bosonic case, the factor “$-1/4$” is present and we consider the case of $\mu =0$. In this case but without assumption on $k$, a non-trivial solution of (\ref{eq:forc}) can be found numerically if $c$ is purely imaginary. In Fig.~\ref{fig:IRHCresul}, we plot a typical solution which arises for $\epsilon=1$, in this case $c_{\text{sol}} =  0.32 i $. We have numerically checked that it is the imaginary part of the electric field $A'(\tau)$ which is enhanced while the real part stays roughly constant. In the infrared regime ($k/H \ll 1$), we derived previously analytic estimates for the electric field. Both the numerical results and the analytic estimates are plotted in Fig.~\ref{fig:IRHCresul} and are shown there to agree, so that the estimates in the infrared regime are also valid in the ultraviolet regime.
\par To exist, this solution needs to be stable. We look at stability by considering a small \gls{perturbation} around the solution $A'(\tau)=\frac{c}{\tau}$. We introduce $\alpha(\tau)$ such that $A'(\tau)=\frac{c}{\tau}+\alpha(\tau)$, with $\alpha(\tau) \ll \frac{c}{\tau}$.
Using this ansatz together in the equation of motion (\ref{eq:tosolveback}) modified by the bosonic prescription (\ref{eq:presc}), to the leading order, we find that $\alpha(\tau)$ satisfies the following equation:
\begin{equation}
\alpha'(\tau)=\frac{\alpha(\tau)}{\tau} w,
\end{equation}
whose solution is $\alpha(\tau) \sim \tau^w$ and $w$ given by:
\begin{equation}
w= 2\epsilon^2  \omega_c \frac{\cosh(2\pi c \epsilon)}{\sinh(2 \pi  \omega_c)}+c \epsilon^3\frac{\sinh(2\pi c \epsilon)}{ \pi \omega_c \sinh(2\pi \omega_c)}-2 c \epsilon^3 \frac{\coth(2\pi \omega_c)\sinh(2\pi c \epsilon)}{\sinh(2 \pi \omega_c)}.
\end{equation} 
The solution is stable if $w>0$. In our specific case ($\mu=0$ and $\epsilon=1$), we find $w(c_{\text{sol}})= 2.2>0$ implying that our solution is stable.
\paragraph*{Conclusion and remarks} In this section, we completed the picture of the \gls{Schwinger effect} by studying the \gls{backreaction} of the Schwinger \glspl{pair} to an electromagnetic field. Our major conclusions are that the \gls{backreaction} of the fermion \glspl{pair} decreases or unaffects the electric field for all the parameters we considered. This corresponds to the screening of the fermions to the photon field. Conversely, for light bosons, we reported a solution where the electric field is enhanced. It corresponds to an anti-screening of the bosons to the photon field.
\subparagraph{} This regime where the amplification of the electromagnetic field occurs corresponds to a regime of \acrfull{IRHC} \cite{Froeb2014,Bavarsad:2016cxh,Kobayashi:2014zza}, where decreasing the electric field increases the resulting produced \glspl{pair}, see also Sec.~\ref{sec:IRHC} for a discussion about \acrshort{IRHC}. Our solution with $\mu=0$ corresponds to cases where \acrshort{IRHC} is maximal.
\subparagraph{} We considered the 1 + 1 D case so that the physics becomes apparent but the generalization to higher dimension is the straightforward next step. Observe that for the 4D problem, equation (3.4) of \cite{Kobayashi:2014zza}, without the kinetic coupling, corresponds exactly to the 2D. Furthermore, the \gls{induced current} has been found to have the same qualitative behavior up to corrections coming from dimension. Hence we expect our qualitative results: screening and anti-screening of the photon field to hold in higher dimensions.
\par Another direction worth to investigate would be to consider the full system $\{$Schwinger created \glslink{particle creation}{particles}, photon field$\}$ and deal with spatial variations of the electric field for the \gls{Schwinger effect}. We argue that to study such \gls{backreaction} problems, the approach presented in chapter \ref{chap:early} will not be sufficient and it will be required to study the \gls{Schwinger effect} in \gls{dS} with methods more suited for involved numerical investigation. For instance real-time lattice simulation techniques have been applied successfully in \cite{Kasper:2014uaa} to study the \gls{backreaction} problem in flat spacetime. More techniques are also described in \cite{Gelis:2015kya}.
\subparagraph{A connection to magnetogenesis} Magnetic fields are ubiquitous in the Universe but their origin is still a mystery \cite{Grasso:2000wj,Durrer:2013pga}. Their generation mechanism can be roughly divided into two categories: primordial ones (a review and a recent example: \cite{Subramanian:2015lua,Ferreira:2013sqa}) which happened before \gls{recombination} and astrophysical ones (a review and a recent example: \cite{Brandenburg:2004jv,Durrive:2015cja}) happening after \gls{recombination}. In this section, we investigated an alternative possibility of amplifying the electromagnetic field with the help of the \gls{backreaction} of the \glslink{Schwinger effect}{Schwinger} \glslink{particle creation}{particles} created during \gls{inflation}. For primordial scenario during \gls{inflation}, to enhance an electromagnetic field, one needs to break the conformal invariance of the Maxwell theory. Usually this is done by introducing a non-canonical kinetic term or by adding a mass to the photon field. Those \gls{magnetogenesis} scenarii are known to suffer problems such as ghosts\footnote{excitations with wrong sign kinetic term inducing pathologies, such as negative probabilities.}, the strong coupling problem and the \gls{backreaction} problem. However the \gls{backreaction} to the electric field was never investigated before and could change drastically the dynamics as we described already in 1 + 1 D (see also \cite{Yokoyama:2015wws} for a brief review of inflationary \gls{magnetogenesis} and its possible connection to the \gls{Schwinger effect} in \gls{curved spacetime}). Furthermore if the Schwinger \glspl{pair} are light enough, the electromagnetic field could then be enhanced without any other mechanism. The generalization to $4D$ is an exciting possibility, especially in the light of the result of \cite{Hayashinaka:2016qqn}.

%% file: back/backgrav.tex
\section{To the gravitational field}
Here we investigate some aspects of gravitational \gls{backreaction} to the \gls{pair} production mechanism presented in chapter \ref{chap:early}.
More specifically, our main goal is to focus on boson in $D$ dimensions and to naively estimate the variation of the \gls{Hubble constant} in the heavy scalar field regime.
So far, it was assumed that the produced pairs do not backreact to the background metric.
This assumption holds as far as the energy density of the pairs is much smaller than the background Hubble energy. For this chapter, we will focus on a \gls{semiclassical} computation of the \gls{energy momentum tensor}.
We assume that the effects of the \gls{pair} production to the \glspl{Einstein equation} are small; they give rise to an effective \gls{cosmological constant} $\Lambda_{\mathrm{eff}}$ in the \glspl{Einstein equation} (\ref{eq:einstein}) that we reproduce here:
\begin{equation}\label{einstein}
R^{\mu\nu}-\frac{1}{2}Rg^{\mu\nu}+\Lambda_{\mathrm{eff}}g^{\mu\nu}=-8\pi G_{D}T^{\mu\nu}_{\sem},
\end{equation}
where we take here $G_{D}=H^{4-D}M_{\mathrm{P}}^{-2}$ to be the gravitational constant in $D$ dimensions, with $M_{\mathrm{P}}$ being the Planck mass.
Now, we compute the \gls{semiclassical} \gls{energy momentum tensor} on the right-hand side of the \glspl{Einstein equation} (\ref{einstein}), it can be defined as
\begin{equation}\label{emtdef}
T^{\mu\nu}_{\sem} \equiv
|g|^{\frac{-1}{2}}\int\frac{d^{d}\textbf{k}}{(2\pi)^{d}}\frac{p^{\mu}_{\k}p^{\nu}_{\k}}{p^{0}_{\k}}|\beta_{\k}|^{2},
\end{equation}
where $|\beta_{\k}|^{2}$ is given by (\ref{betasq}) and $p^{\mu}_{\k}$ is the physical momentum vector of the created \glslink{particle creation}{particle}.
To perform the integral on the right-hand side of Eq.~(\ref{emtdef}), we follow the same integration procedure presented in Sec.~\ref{sec:smestim}:
impose the relation~(\ref{estimate}) to convert the $k$-integral into a $\tau$-integral.
In the heavy scalar field regime, $\gamma\gg\max(1,\lambda)$, the physical momentum takes the form $p^{\mu}_{\k} = \frac{H}{a} ( \gamma \delta^{\mu}_0-\lambda \delta^{\mu}_1)$, using this form together with (\ref{emtdef}) leads to
\begin{eqnarray}\label{emt}
T^{00}_{\sem}&\simeq&a^{-2}(\tau)\mathcal{E}, \hspace{1cm}
T^{01}_{\sem}\simeq-\frac{\lambda}{\gamma}T^{00}_{\sem}, \hspace{1cm}
T^{11}_{\sem}\simeq\frac{\lambda^{2}}{\gamma^{2}}T^{00}_{\sem}, \nn\\
T^{0i}_{\sem}&=&T^{ij}_{\sem}=0, \hspace{1cm} i=2,\cdots,d,
\end{eqnarray}
where $\mathcal{E}$ is given by
\begin{equation}\label{mathcale}
\mathcal{E} \equiv \frac{H^{D}}{(2\pi)^{D-2}(D-1)}\lambda^{\frac{3-D}{2}}\I_{\frac{D-3}{2}}(2\pi\lambda)\gamma^{D}e^{-2\pi\gamma}.
\end{equation}
\par
Considering the metric~(\ref{metricds}), in terms of the \gls{Hubble constant} $H(\tau)$, the components of the \gls{Ricci tensor} (\ref{ricci}) are obtained
\begin{eqnarray}\label{riccitensor}
R_{00}&=&(D-1)\Big(H^{2}(\tau)+a^{-1}(\tau)\dot{H}(\tau)\Big)a^{2}(\tau), \nn \\
R_{ij}&=&-\Big((D-1)H^{2}(\tau)+a^{-1}(\tau)\dot{H}(\tau)\Big)a^{2}(\tau)\delta_{ij}, \nn \\
R_{0i}&=&0, \hspace{1cm} i=1,\cdots,d,
\end{eqnarray}
and the \gls{Ricci scalar} (\ref{courbure}) is
\begin{equation}\label{ricciscalar}
R=(D-1)\Big(DH^{2}(\tau)+2a^{-1}(\tau)\dot{H}(\tau)\Big).
\end{equation}
The trace of the \glspl{Einstein equation} (\ref{einstein}) gives 
\begin{equation}\label{Lambda}
\Lambda_{\mathrm{eff}}=\frac{(D-2)R}{2D}-\frac{8\pi G_{D}\mathcal{E}}{D},
\end{equation}
and in the heavy scalar field regime, we find that the leading order terms for the \glspl{Einstein equation} (\ref{einstein}) involve $T_{\sem}^{00}$:
using Eqs.~(\ref{riccitensor}-\ref{Lambda}) it leads to
\begin{equation}\label{einsteineq}
a^{-1}(\tau)\frac{dH(\tau)}{d\tau}=-\frac{8\pi G_{D}\mathcal{E}}{(D-2)}.
\end{equation}
The above equation determines the evolution of the \gls{Hubble constant} with respect to the conformal time $\tau$.
In order to compare with the existing literature, we now work in cosmic time $t$: using Eqs.~(\ref{line}) and~(\ref{metricds}), it can be shown
that the evolution of the \gls{Hubble constant} with respect to the cosmic time $t$ is
\begin{equation}\label{hubbleq}
\frac{dH(t)}{dt}=-\frac{8\pi G_{D}\mathcal{E}}{(D-2)},
\end{equation}
which agrees with \cite{Mottola:1984ar,Cai:2005ra,Markkanen:2016aes}.
Thus, the \gls{Schwinger effect} leads to a decay of the \gls{Hubble constant} and as consequence of Eq.~(\ref{Lambda}), a decay of the cosmological
constant. This decay of the \gls{cosmological constant} begins with the \gls{pair} production and continues until $\Lambda_{\mathrm{eff}}=0$.
In this picture, as a classical black hole being evaporated into \gls{Hawking radiation} or the coherent energy of an electric field being dissipated
into $e^+$ $e^-$ \glspl{pair}, the coherent \gls{vacuum energy} is dissipated into a cloud of scalar \glspl{pair}.
The decay of the \gls{Hubble constant} affects $G_{D}$ for $D\neq 4$.
For $D<4$ the gravitational constant decays until it reaches zero and for $D>4$ the gravitational constant increases.
Similar to \cite{Mottola:1984ar}, the time scale for evolution of the \gls{Hubble constant} can be estimated by
\begin{equation}\label{timescale}
t_{B} \equiv -\frac{H}{\frac{dH(t)}{dt}}=\frac{(2\pi)^{D-3}(D-1)(D-2)M_{\mathrm{P}}^{2}}{4H^{3}}
\Big(\lambda^{\frac{3-D}{2}}\I_{\frac{D-3}{2}}(2\pi\lambda)\Big)^{-1}\gamma^{-D}e^{2\pi\gamma}.
\end{equation}
A series expansion of the time scale expression~(\ref{timescale}) around $\lambda=0$, with $\gamma$ fixed, leads to the leading order term
\begin{equation}\label{behavior}
t_{B}\simeq\frac{(4\pi)^{\frac{D-3}{2}}\Gamma\big(\frac{D-1}{2}\big)(D-1)(D-2)M_{\mathrm{P}}^{2}}{4H^{3}}\gamma^{-D}e^{2\pi\gamma},
\end{equation}
which is independent of $\lambda$.
In \cite{Mottola:1984ar}, the time scale has been computed in the global patch of $\ds_{4}$, without electric field, and the author showed there,
in the limit $m \gg H$, the time scale behaves as $Hm^{-4}\exp(\pi m/H)$.
Hence, in $D=4$ dimension, the result~(\ref{behavior}) agrees with the time scale obtained in the Ref.~\cite{Mottola:1984ar} up to a factor of 2 in
the exponent. This factor could come from the different definitions for the \gls{energy momentum tensor}.
\par
Observe that the calculation carried out in this section is not valid for $D =2$ as there is no Einstein gravity in 1+1 dimension.
Observe beside, that under our working assumption: heavy scalar field regime, $\lambdam\gg\max(1,\lambda)$, we find
$t_{B} \gg t_{\text{H}}=H^{-1}$ which still allow for a long \gls{inflation}.
Furthermore, we argue that this decay of the \gls{Hubble constant} presents similarities with generic models of \gls{slow roll} \gls{inflation} where a scalar field
sees its potential energy slowly decaying into kinetic energy to ultimately exhibit coherent oscillations around the minimum of its potential which
unleash a \gls{reheating} phase, \textit{cf.}~chapter \ref{chap:inflation}.
The next step is to consider the expectation value of the energy momentum operator, which as the current will present divergences. The computation of this \gls{tensor} is much more involving. We have seen in Sec.~\ref{sec:result} that the \gls{semiclassical} estimates agreed in the strong field regime, but were exponentially different in
the heavy scalar regime, so we argue that those results have to be checked by further study, mainly the exact computation of the \gls{energy momentum tensor} in order to see if those first estimates agree with the general case.
For instance, the authors of \cite{Markkanen:2016aes}, discovered an enhancement of the \gls{Hubble constant}, when considering $E=0$ and $D=4$, with a slightly different method,.
The same exponential behavior as in Eq.~(\ref{timescale}) was also found but with a different prefactor.
We argue that those changes are due to the renormalization procedure they carried out which gives different results than the replacement of the
$k$-integral into a $\tau$-integral we performed here.
\section{Conclusion}
In this chapter, we presented the first steps toward estimating \gls{backreaction} effects. Our main conclusions are that the electric field is damped by the \gls{pair} production, only in the \acrshort{IRHC} regime, it might be maintained to a constant value. The generalization to 4D in the fermionic case might be interesting as a negative current current has been found for some value of the parameter space \cite{Hayashinaka:2016qqn} which would imply an enhancement of the electric field due to \gls{pair} production. Regarding the gravitational field, we also found a decay of the \gls{Hubble constant}, implying that \glspl{pair} have taken energy both from the gravitational field and from the electromagnetic field. Again a generalization to 4D and beyond the \gls{semiclassical} regime would be a problem worth to investigate.

Interesting connections between the model for \gls{pair} production described in chapter \ref{chap:early}, its corresponding \gls{backreaction} and axion-like models could be build. In the \gls{axion inflation} scenario, one considers a pseudo scalar inflaton field coupled to a gauge field. This coupling was shown to lead to two phenomena: the production of gravitational waves (and scalar \glspl{perturbation}) but also a \gls{backreaction} of the \glspl{perturbation} to the background dynamics. This \gls{backreaction} of the gravity to the gauge sector has for consequence to slow down \gls{inflation} \cite{Barnaby:2011qe,Domcke:2016bkh}. In those models, it is hence the \glslink{particle creation}{particles} production which is the source of gravitational waves. The link with the model of chapter \ref{chap:early} remains to be explored. The physical application of axion models are both for \gls{dark matter} candidate, to drive an accelerated expansion and for primordial \gls{magnetogenesis}, which might also augment the range of application of the model of chapter \ref{chap:early}, together with the \gls{backreaction} effects studied in this chapter.

%% file: cours/DE.tex
 \part{Late Universe physics}
 \label{chap:late}
 \vspace{2cm}
Whenever life gets you down, Mrs. Brown,\\
And things seem hard or tough,\\
And people are stupid, obnoxious or daft,\\
\\
And you feel that you've had quite enough,\\
\\
Just remember that you're standing on a planet that's evolving\\
And revolving at 900 miles an hour.\\
It's orbiting at 19 miles a second, so it's reckoned,\\
The sun that is the source of all our power.\\
Now the sun, and you and me, and all the stars that we can see,\\
Are moving at a million miles a day,\\
In the outer spiral arm, at 40,000 miles an hour,\\
Of a galaxy we call the Milky Way.\\
\\
Our galaxy itself contains a hundred billion stars;\\
It's a hundred thousand light-years side to side;\\
It bulges in the middle sixteen thousand light-years thick,\\
But out by us it's just three thousand light-years wide.\\
We're thirty thousand light-years from Galactic Central Point,\\
We go 'round every two hundred million years;\\
And our galaxy itself is one of millions of billions\\
In this amazing and expanding universe.\\
  \vspace{1cm}
\begin{center}
\includegraphics[width=0.8\textwidth]{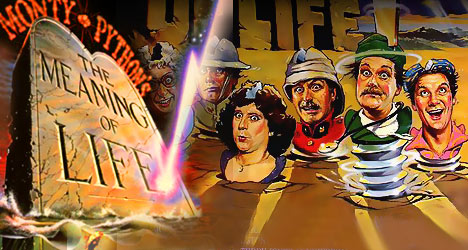} \\
The Monthy Python, The meaning of life, 1983
\end{center}
 \chapter{Dark energy}
 \label{chap:DE}
 \epigraph{Reports that say that something hasn't happened are always interesting to me, because as we know, there are known knowns; there are things we know we know. We also know there are known unknowns; that is to say we know there are some things we do not know. But there are also unknown unknowns – the ones we don't know we don't know.}{Secretary of Defense: D.~Rumsfeld, February 12, 2002 11:30 am EDT}
 
\textit{This chapter aims at giving more information about dark energy that was already introduced in the chapter \ref{chap:introcosmo}. Dark energy is a very generic name to describe all sort of theories to explain the observed acceleration of the expansion of the universe. Here we will discuss which problems dark energy raises and the main research lines to solve them.}
\paragraph*{} 
The existence of \gls{dark energy} relies historically on the discovery that the observed peak luminosity of the \glslink{SNe Ia}{supernovae Ia (SNe Ia)} was, after accounting for Phillips's relation, about 20\% fainter than the one expected for a $\Lambda=0$ \glslink{SM}{(FLRW) model} \cite{Perlmutter:1998np,Riess:1998cb,Perlmutter:1997zf,Schmidt:1998ys}. It leads most of the scientific community to believe that our Universe was expanding at an increasing rate. To drive this expansion, the old \gls{cosmological constant} introduced by \gls{Einstein} to render the universe static has been exhumed but with an opposite sign in order to accelerate the expansion. Putting the \gls{cosmological constant} to the right hand side of the \glspl{Einstein equation}, it is possible to interpret it as a \gls{perfect fluid} with \gls{vacuum} equation of state that has been dubbed \gls{dark energy}. The fitting procedure of the \glslink{SNe Ia}{SNe Ia} gives alone already $\Omega_M=0.27 \pm 0.04$, leading to \gls{dark energy} composing today 73 \% of our universe. The value $\Omega_{\Lambda} =0.683$ in the table \ref{table:cosmnum} is obtained when one takes into consideration other independent cosmological probes, among other, \acrshort{CMB} data. The two main puzzles regarding the \gls{cosmological constant} are the \gls{cosmological constant problem} and the \gls{coincidence problem}.
\section{The cosmological constant problem}
\label{sec:CCP}
The \gls{cosmological constant problem} arises when one tries to interpret the presence of a \gls{cosmological constant} in \glspl{Einstein equation} as a \gls{vacuum energy}. More precisely, various quantum mechanical effects should give corrections proportional to $g_{\mu \nu}$ in the \glspl{Einstein equation} and the sum of all those contributions should give rise to the \gls{cosmological constant} as in (\ref{eq:defCC}).
Converting a \gls{dark energy} contribution ($\Omega_{\Lambda} =0.683$) in Planck units gives a value of the \gls{cosmological constant} of:
\begin{equation}
\Lambda_{\text{obs}} = H_0^2 M_{\text{Pl}}^2=10^{-120} M_{\text{Pl}}^4.
\end{equation}
This value is a rather small one which could not be a tremendously worrying problem as small numbers (and big numbers!) appears often in physics. The electron mass $m_e = 10^{-7}$ TeV is a rather small number measured in units natural to the \gls{SMP} of \gls{particle physics} but its smallness is controlled as this parameter is stable under quantum correction. In the field theory language, the small parameter is said to be \emph{technically natural} \cite{tHooft:1979bh}. When the electron mass goes to zero, the theory enjoys a new \gls{symmetry}: the chiral symmetry which enforces that the quantum corrections to the electron mass are proportional to the mass itself. So if the electron mass is small, it stays small. No such \gls{symmetry} is known for the \gls{cosmological constant}, it leads to various big quantum corrections. The consequence is the \textit{cosmological constant problem}: why the observed value of the \gls{cosmological constant} is so small even though it is made of many big contributions.

\glslink{vacuum fluctuations}{Quantum fluctuations} from the known particles of the \gls{SMP} contribute largely to the value of $\Lambda$. The \gls{vacuum energy} has to be Lorentz-invariant as not preferred direction should be attributed to the \gls{vacuum energy}, using the \gls{equivalence principle}, the contributions to the \gls{vacuum energy} are of the form \cite{Zeldovich:1968,Weinberg:1988cp}:
\begin{equation}
\label{eq:contrCC}
<T_{\mu \nu}> = - <\rho> g_{\mu \nu},
\end{equation}
for quantum mechanical processes involving the fields of the \gls{SMP} of \gls{particle physics}. We model the fields of the \gls{SMP} of \gls{particle physics} as a collection of free scalar harmonic oscillators of pulsation $k$ which are known from quantum mechanics to have a zero point energy:
\begin{equation}
E= \frac{1}{2}\sqrt{k^2 +m^2}.
\end{equation}
The reasoning is here equivalent to the one done around equation (\ref{eq:Heinsenb}). The \gls{vacuum energy} density receives a contribution from these scalar fields:
\begin{equation}
 <\rho>= \int_0^{\infty} \frac{d^3 \textbf{k}}{(2\pi)^3} \frac{1}{2} \sqrt{k^2+m^2}.
\end{equation}
Those \gls{vacuum fluctuations} are divergent, as it is usually known from the first steps of a physicists into the vast field of quantum field theory \cite{Peskin:1995ev}. In flat spacetime, the fact that the \gls{vacuum energy} is infinite is usually ignored as one measures only differences of energy\footnote{To clarify, here the zoologies of quantum field infinities, continuing learning about quantum field theory, one will soon encounter new infinities that cannot be ignored in the same way as the \gls{vacuum} contributions. A whole branch of quantum field theory called renormalization deals with those infinities. We studied some ways of dealing with those infinities in Secs.~\ref{sec:curent} and \ref{sec:current}. Another point important is that, it should not be understood that the infinity arising from the \gls{vacuum fluctuations} cannot be renormalized but that its renormalization is unnecessary. Actually in \gls{curved spacetime}, many ways to renormalize it are possible all leading to the unsatisfactory subsequent result.}. But in \gls{curved spacetime}, we are precisely interested to know how the \gls{vacuum} gravitate, so we will assume that our picture hold until an energy scale $\Lambda_{UV}$ where new physics appear\footnote{Assuming a hard cutoff is one way of renormalizing the \gls{vacuum fluctuations}, it is not the most rigorous one as this regularization scheme do not respect the underlying \gls{symmetry} of the theory but we use it for illustrative purpose, a more rigorous treatment can be found in \cite{Martin:2012bt}.}. By conservatively assuming $\Lambda_{UV}= 1$ TeV where the \gls{SMP} of \gls{particle physics} is extremely well tested, we find:
\begin{equation}
 <\rho> \sim \Lambda_{UV}^4,
\end{equation} 
which is a terrible discrepancy between a theoretical prediction and its observed value:
\begin{equation}
\Lambda_{\text{theory}} = (TeV)^4=10^{-60} M_{\text{Pl}}^4.
\end{equation}
Such a failure by 60 orders of magnitude is annoying for a physicist (and not only), reinforced by the argument that within the \glslink{SM}{$\Lambda$CDM model} the observed value cannot be much different otherwise \glslink{S}{structures} would not form and stand the way we observe them nowadays. Of course it could be that the bare value of the \gls{cosmological constant} (the quantity in the Lagrangian) could cancel the 60 decimal discrepancy between the observed value and the theoretical prediction but this requires again a huge amount of \gls{fine-tuning}, which is again problematic\footnote{A brief discussion whether \gls{fine-tuning} is problematic or not is proposed in Sec.~\ref{sec:coincp}}. The simple calculation displayed in this section can be completed by reading \cite{Martin:2012bt} where many other contributions of the form of (\ref{eq:contrCC}) are computed in both from classical and quantum mechanical origin but the conclusion remains unchanged: the need of a massive cancellation to explain the observed value of the \gls{cosmological constant}. Beside, Weinberg's no-go theorem \cite{Weinberg:1988cp} showed that within \gls{general relativity} no dynamical solution to the \gls{cosmological constant problem} exists.
\section{The coincidence problem}
\label{sec:coincp}
\begin{figure}[h]
\begin{center}
\includegraphics[width=0.7\textwidth]{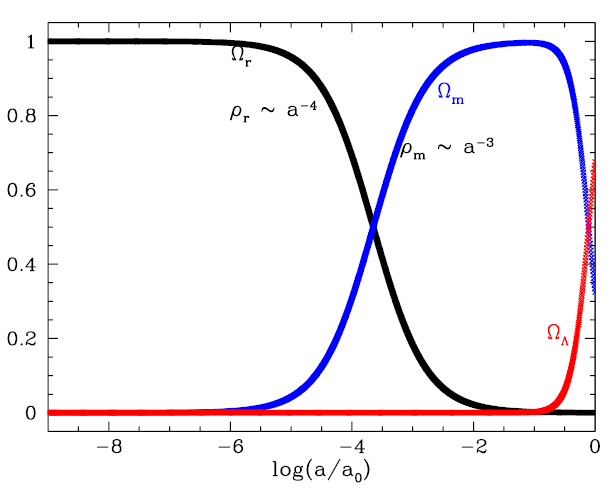}
\caption[Evolution of the cosmological fluids as an illustration of the coincidence problem]{Evolution of the different cosmological fluids as a function of the \gls{scale factor} in unit of critical density. We see that we are performing measure of our universe at the moment where the \gls{dark energy} and the \gls{dark matter} are of the same order of magnitude.}
\label{fig:coinc}
\end{center}
\end{figure}

The \gls{coincidence problem} is the fact that the contribution from \gls{dark energy} is found to be of the same order of magnitude than the contribution from \gls{dark matter} (both of order 1) in the present day even if their respective evolution through cosmic history is drastically different as it can be see from table \ref{tab:flu}. Figure \ref{fig:coinc} propose a graphical representation of the cosmological evolution of the different fluids composing the universe: matter, \gls{dark energy} and \gls{radiation} and we see obviously from this figure that the moment we are measuring the cosmological parameters is a special one. 

As for the \gls{cosmological constant problem} and for the horizon and flatness problems depicted in Sec.~\ref{sec:probinfl}, the \gls{coincidence problem} is a problem in the sense of the parameters of our universe are fine-tuned compared to typical parameters. It can always been argued what ``typical'' is for an observer. Again the fact that \gls{reproducibility} is a missing ingredient of the scientific method leads to this puzzle and hence depending on the definition of typicality, the \gls{fine-tuning} problems of the \glslink{SM}{standard model} could vanish if one defines different probability measure of typicality. The absence of a clear definition for a probability measure in cosmology is known as the measure problem. A famous argument related to the choose of a probability measure is the \gls{doomsday argument} \cite{Carter347,gott1993implications}. This argument gives a probabilistic guess of the life expectancy of the humanity by assuming a uniform distribution and by knowing our birth rank among the already born human. The statement is that there is 95 \% chance that within 9120 years, the human race will die out! This result gives clearly too much information regarding the inputs (birth rank and choose of a probability measure). Norton argues that, the problem lies in the fact that the uniform probability distribution does not reflect our ignorance of the question under consideration \cite{norton2010cosmic}. The use of imprecise probabilities in the Bayesian approach blocks also the conclusion of the \gls{doomsday argument} \cite{benetreau2013apocalypse}. This incursion in the domain of philosophy aimed at suggesting the fact that how \gls{fine-tuning} problems might lie in the existence of non-trivial probability measures rather than new physics. 
\section{Solutions to the problems and dark energy models}
\label{sec:DEsolution}
 \epigraph{A theory with mathematical beauty is more likely to be correct than an ugly one that fits some experimental data.}{P.~A.~M.~Dirac \cite{kragh1990dirac}}
A solution to the \gls{cosmological constant problem} would be to have as in the example of the the mass of the electron a \gls{symmetry} which would prevent the quantum corrections to be big. The perfect candidate is \gls{supersymmetry}: a postulated extension of the \gls{SMP} of \gls{particle physics}. However, \gls{supersymmetry}, if it exists\footnote{The \gls{supersymmetry} community has been pretty quiet in the past years, mainly because of the absence of detection of supersymmetric particles in the large hadron collider.}, is known to be broken at our energy scale, however the mechanism which allow the cancellation of the different contribution from bosons and fermions to the bare \gls{cosmological constant} could be maintained when gravity is turned on. This mechanism is known as adjustment mechanism \cite{Ford:1987de,Dolgov:2008rf} and is successful to solve the \gls{cosmological constant problem} but breaks the observational constraint on the variation of the gravitational constant $|\dot{G}/G|$ \cite{Williams:2004qba}. An overview of the different solutions to the solution to the \gls{cosmological constant problem} can be found in \cite{Nobbenhuis:2006yf}. We will now discuss some other solutions to the \gls{cosmological constant problem}: extensions of the \glslink{SM}{standard model} of cosmology, the anthropic perspective. We will finish by discussing \gls{Lovelock theorem} which gives hint on the structure of theories arising beyond \gls{general relativity}.
\subsection*{Modification of the standard model of cosmology}
As the \gls{cosmological constant problem} has been derived using the \glslink{SM}{standard model} of cosmology, it is a priori present only within the assumptions of the \glslink{SM}{standard model}. Looking at the hypothesis of the \glslink{SM}{standard model} of cosmology in Sec.~\ref{sec:cosmprin}, we see that two possible escapes to the \gls{dark energy} problems are to relax the hypothesis of homogenenity or to relax the assumption that it is \gls{general relativity} which governs the gravitational processes on large scales. In chapter \ref{chap:LTB}, we will investigate a simple inhomogeneous model and make a connection to \gls{fractal} matter distribution. Assuming that \gls{general relativity} is not the theory governing large scale gravitational processes leads to the research line entitled modify gravity \cite{Joyce:2014kja} even if gravity stays the same. Motivations to go in this direction are Weinberg's no-go theorem \cite{Weinberg:1988cp}, the fact that gravity is poorly constrained on larger-scales compared to the solar system tests and a fundamental motivation to understand alternatives to \glslink{general relativity}{GR} in echo to \gls{Lovelock theorem} discuss hereafter. The modifications to gravity as a way to explain the \gls{dark energy} are usually argued to be inelegant compared to \glslink{general relativity}{GR}.

Second, the cosmological problems happened when one tries to interpret the possible $\Lambda$ term in the \glspl{Einstein equation} (\ref{eq:Eisteq}) as the contribution from the \gls{vacuum energy}. It is sometimes argued that the \gls{dark energy} is dynamical, for instance, it is an unknown scalar field, this line of research is known as \gls{quintessence} \cite{Tsujikawa:2013fta}, in this case the formalism would be exactly the same discussed in chapter \ref{chap:inflation} making a bridge between early and late physics just as in this thesis. While the phenomenology for \gls{quintessence} models is rich, it is however difficult to understand why the \gls{cosmological constant problem} could be avoided. Finally, more recently also to solve the \gls{coincidence problem}, it has been argued that \gls{dark energy} and \gls{dark matter} may interact together, we will discuss this possibility in chapter \ref{chap:IDE}. Another way to explain the acceleration of the expansion of the universe is to consider that the small deviations from homogeneity and isotropy add up to induced an acceleration of the expansion of the universe, the field of research along this idea is usually refereed as to \gls{dynamical backreaction} \cite{Buchert:2007ik}. 

In most of those approaches, even if the \gls{dark energy} is not exactly a \gls{cosmological constant}, the \gls{cosmological constant problem} might still be present, a theorem known as \gls{Lovelock theorem} discuss in which cases such a cosmological problem could be present. 
\subsection*{Anthropic reasoning}
When one realizes that to solve both the \gls{cosmological constant problem} and the \gls{coincidence problem}, one needs just to give a natural explanation to the observed value of the \gls{cosmological constant}, a tempting possibility is to argue that any other value far away from the observed value is not allowed \textit{eg.}~for \glslink{S}{structures} to form the way we observe them or to permit us\footnote{Here one needs to be very cautious about the use of ``us'' as it is one of the time physics and religion may clash. The argument I am developing has nothing to do with spirituality, the nature of human being or why there is life as we know it but says only that we require the existence of physicists to do physics. Furthermore physicists, in this context, are trying to make statements about the formation of large scale \glslink{S}{structure} or of the atoms. Taking about life and ``us'' is catchier but should not be misunderstood: physicists don't try to build life. See as an example \cite{coucouppp} where the presence of intelligent civilization is proposed as an explanation for the coincidence problem, this idea is at the border of physics but leads to the beautiful metaphor that dark energy would \textit{somehow} fertilize our universe to allow life to start. Here in the main text, we just use ``us'', in the sense of having human being able to perform experiment on earth and observing the value of the \gls{cosmological constant}. Similar reasoning occurs very often in physics and are sometimes referred to as ``physical'' criteria. For instance, one discards Hamiltonians which are not bounded from below, as one could not talk about the \gls{vacuum} state of a theory, thus rending matter unstable, clearly leading to a non-physical theory. In the same sense, non-local theories, theories without a well defined initial value problem or even mathematically inconsistent theories are almost always discarded. See \cite{penrose2006road} on the relation between mathematics and physics and \cite{2016arXiv160808225L} for one possible reason why the physics world is much much smaller than the mathematical world, a complementary discussion was presented in Sec.~\ref{sec:cosmprin}.} to make observations. We define more generally the \gls{anthropic principle} as follow: our universe has some given properties as otherwise we would not be here to talk about those properties. Weinberg derived in \cite{Weinberg:1988cp,Weinberg:1987dv} an anthropic bound for the value of the \gls{cosmological constant}. An anthropic idea gets reinforced when the law of physics allows for multiple realization of our universe with different values of the \gls{cosmological constant}. In this sense, we would be naturally be in a universe with values of the fundamental constants hospitable for us. This line of idea has regain some interest in the context of the string theory landscape \cite{Kachru:2003aw,Bousso:2000xa}. It is the idea that the different way to compactify the extra-dimension gives rise to numerous de Sitter vaccua, each with a different values of the \gls{cosmological constant}. Combining the landscape together with the idea of \gls{eternal inflation} \cite{Vilenkin:1983xq,Linde:1986fd} discussed in Sec.~\ref{sec:cclpairs}, would give a logical consistant way of tackling the \glspl{cosmological constant problem} \cite{carr2007universe} but as it is based on speculative theories, it remains nothing but a possibility among others. More conservatively, anthropic reasoning could be seen as a guideline to complete other approaches to the \gls{cosmological constant problem}.
\subsection*{Lovelock theorem}
\gls{Lovelock theorem} \cite{Lovelock:1971yv,Lovelock:1972vz} characterize the type of theories one can construct from the \gls{metric tensor} alone. Precisely, it states that a gravitational theory, in a four-dimensional Riemannian \gls{manifold}, arising from an action principle, involving the \gls{metric tensor} and its first and second derivatives only, gives field equations of second order and less and are exactly \glspl{Einstein equation} with a \gls{cosmological constant}: \gls{general relativity}. In other words to go beyond \gls{general relativity}, one must do at least one of the following:
\begin{itemize}
\item consider other fields than the metric
\item allow for higher derivative than first derivative of the metric
\item assume $ D \neq 4$
\item consider other types of \glspl{tensor} that (2,0) or other \glslink{symmetry}{symmetries} for the the metric
\item consider non-local action
\end{itemize}
Appendix B of \cite{Joyce:2014kja} gives also a proof of \gls{Lovelock theorem}.
Any of this item will of course give new degree of freedom which could solve some of the problems of the \glslink{SM}{standard model} of cosmology but at the same time leads to a number of pathologies which have to be tackled, see \textit{eg.}~Appendix  D of \cite{Joyce:2014kja} for a non-exhaustive list of the diagnostics to make to obtain a healthy theory with the point of view of effective field theory. The other sections of \cite{Joyce:2014kja} review a number of alternatives to \gls{general relativity}.

\section*{Conclusion} To conclude on this introduction, on \gls{dark energy}, we introduced some riddles of the \glslink{SM}{standard model} of cosmology depicted in chapter \ref{chap:introcosmo}. They were related to the generic topic of \gls{dark energy}, we depicted then some proposals in order to solve those riddles in the chapters \ref{chap:LTB} and \ref{chap:IDE}, we will present two research projects belonging to two different research lines tackling the \glspl{cosmological constant problem}.

%% file: LTB/LTB.tex
\epigraph{When I entered graduate school I had carried out the instructions given to me by my father and had knocked on both Murray Gell-Mann's and Feynman's doors and asked them what they were currently doing. Murray wrote down the partition function for the three-dimensional Ising model and said it would be nice if I could solve it (at least that is how I remember the conversation). Feynman's answer was ``nothing''. }{Ken G. Wilson, 1966, quoted in \cite{yeomans1992statistical} p. 35}

\textit{In this chapter, we first motivate more the relaxation of the cosmological principle and detail the specific model we  focused on the \acrfull{LTB} cosmology. This chapter is adapted and augmented from its corresponding publication \cite{Stahl:2016vcl} }
\newcommand{\aj}{AJ}			
\newcommand{\sovast}{Soviet~Ast.}	
\newcommand{\jcap}{Journal of Cosmology and Astroparticle Physics}			
\newcommand{\mnras}{Monthly Notices of the Royal Astronomical Society}
 \newcommand{\apj}{ApJ} 
\section{Toward the cosmological principle}
We saw in chapter \ref{chap:introcosmo} that the \glslink{FLRW metric}{FLRW spacetime} exhibits spatial sections with constant curvature. That is possesses 3-surfaces which are homogeneous and isotropic. In differential geometry language, it means \glslink{FLRW metric}{FLRW spacetime} has 6 \glspl{Killing vector} which completely characterize its geometry up to a scalar: the curvature $k$. Going beyond \glslink{SM}{FLRW spacetimes} can be motivated by several points of view. On a theoretical point of view, it is interesting to go beyond homogeneity and isotropy in order to classify cosmological spacetimes. Indeed it might be that a result interpreted in a FLRW spacetime could also make sense in a more general spacetime. It is hence interesting to find counter-examples to some solution or to test their robustness in a more general framework. Furthermore, we will end up in a definition much more fundamental of homogeneity and isotropy that the one given in chapter \ref{chap:introcosmo}. We will also give some physical motivations why there is still some room for non-homogeneous or non-isotropic spacetimes.
\subsection{Classifying spacetimes, some bites of group theory}
\label{sec:groupth}
\par Of course most of the solutions to \glspl{Einstein equation} do not posses any sort of \glslink{symmetry}{symmetries} and are hard to exhibit because of the coupled non-linear nature of those equations. I like to say that \gls{Einstein} was also a genius in the sense that he was pioneer in the change of paradigm of physics at the beginning of the twentieth century. Indeed at that time, physicists started to consider the \glslink{symmetry}{symmetries} of a problem in order to solve a reduced and easier problem before writing down the full solution. This bright idea is ubiquitous in physics. However, to stay critics, it remains useful to have an idea of a non-approximate solution in order to strengthen or challenge some conclusions made with a more \glslink{symmetry}{symmetric} toy models. We will now overview some interesting group theory in order to classify the different cosmological spacetime, we want to emphasis that what we will do in the next paragraph is nothing by a classification of Lie group applied to cosmological purposes. The reader familiar with the topic can go straight to Table \ref{tab:iso} which will sum up the main results.  We will follow \cite{Peter}, more references and details can be found there.
\par For this section, we will assume we work on a D-dimensional euclidian \gls{manifold}, the connection to lorentzian \gls{manifold} is simply done by Wick rotating the time-like coordinate. We will say a diffeomorphism $\phi$ is a \gls{symmetry} of some tensor $T$, if the tensor is invariant after being pulled back into $\phi$:
\begin{equation}
\phi T = T.
\end{equation} A important special case of \gls{symmetry} is the \gls{isometry}. The \glslink{isometry}{isometries} of a given space are transformations of the space into itself that leave the metric tensor and all physical and geometrical properties invariant. For a Riemann space, a general theory of \glslink{isometry}{isometries} does not exists in the general case: one cannot describe discrete invariances (reflection, time reversal...). One usually focuses on continuously disformable transformations. They are usually characterized by a \gls{Killing vector} field $\xi$ obeying Killing's equation:
\begin{equation}
\mathfrak{L}_{\xi} g_{\mu \nu} \equiv \nabla_{\mu} \xi_{\nu} + \nabla_{\nu} \xi_{\mu} =0,
\end{equation}
where $\mathfrak{L}$ is the Lie derivative along $\xi$. If a spacetime possesses one \gls{Killing vector}, then there is a coordinate system where the metric is independent of one coordinate, called a cyclic coordinate. By virtue of Noether theorem, it implies that for the motion of a test particles (on a geodesic), there will be a conserved quantity. Restricting the space considered to Riemannian \glspl{manifold}, the group of the \gls{isometry} of a Riemannian \glspl{manifold} is a Lie group and the associated Lie algebra is formed by the set of \glspl{Killing vector}. Its dimension has an upper bound: $r \leq \frac{1}{2} D(D+1)$ reached for maximally \glslink{symmetry}{symmetric} space such as Minkowski or \gls{dS} which enjoys 10 degree of freedom in $D=4$. The product operation of the Lie algebra is the commutator of two \glspl{Killing vector}:
\begin{equation}
[\xi_a, \xi_b ] \equiv C_{ab}^{c} \xi_c,
\label{eq:conststr}
\end{equation}
where $C_{ab}^{c}$ are the \gls{structure constant} of the Lie algebra. Any \gls{structure constant} of a Lie algebra is antisymmetric that is $C_{ab}^{c}=-C_{ba}^{c}$ and satisfies the Jacobi identity: $C_{a[b}^{c} C_{cd]}^e=0 $, which follows from the definition of a commutator and ensure that the Lie algebra exists consistently. The \glslink{isometry}{isometries} of a given space are entirely characterized by the \gls{structure constant} plus two numbers roughly describing to degree of homogeneity and of isotropic of a given \gls{symmetry}. We will explain the origin of these numbers after a series of basic definitions of group theory.
\paragraph{}
The orbit of a point $P$ is the set of all points corresponding to the image of the point $P$ through all the \glslink{isometry}{isometries} of the space. By definition orbits are stable sets and are the smallest invariant set for all the \glslink{isometry}{isometries} of the space. If an orbit reduce to a point, it is called a fixed point and all the \glspl{Killing vector} vanish there. Generalizing this concept, it is possible to define the isotropy group at a point $P$ such that it is the subgroup of \glslink{isometry}{isometries} letting $P$ invariant. It is generated by the \glspl{Killing vector} vanishing at $P$ and its dimension $q$ is smaller than the subgroup of rotation (which is $\frac{1}{2}D(D-1)$). A group of \glslink{symmetry}{symmetries} is transitive on a space $S$ if it can move any point of $S$ to another point of $S$. Orbits are the largest space through each point on which the group is transitive and can be called surface of transitivity. The surface of transitivity have a finite dimension $s$ which has to be smaller than the dimension of the group of translation (which is $d$). To finish, the total dimension of the group $r$ is the sum of the dimension of the surface of transitivity and of the dimension of the isotropy group, giving $r=s+q \leq D + \frac{1}{2}D(D-1) = \frac{1}{2}D(D+1)$ and as announced the \gls{structure constant} in (\ref{eq:conststr}) and two of the three numbers $(r,s,q)$ will entirely describe the \glslink{isometry}{isometries} of a given space.
\par
For cosmological spacetimes, we have $D=4$, that is as discussed before $r \leq 10$. First let us emphasis that for our Universe, we have obviously $r=0$. $r$ is a fundamental property of the space considered so cannot change once the space is fixed, however $q$ can vary over space (but not over an orbit): it can be for instance greater at some point (\emph{eg.} an axis, center of \gls{symmetry}) where the dimension of the orbit is less. To continue, we will also assume an energy condition $\rho + P > 0$ which is an usual assumption of classical \glslink{general relativity}{GR} also needed for instance to derive the singularity theorems \cite{Hawking:1973uf}. It reveals once again that even if we were looking at deep geometrical properties of the space, \glspl{Einstein equation} always relate them to the matter content of our Universe. The only fluid not fulfilling this condition is \gls{dark energy} (for which $\rho+P=0$). Coming from the mathematical fact that $O(3)$ has not subgroup of dimension two, it can be shown that $q \in \{ 0,1,3 \}$. For $q=3$, a space is said to be isotropic. For $q=1$, it has a \acrfull{LRS}, in this case all kinematic quantities and observations at a general point are rotationally invariant under a specific space direction $q=1$. Finally a space is anisotropic for $q=0$. Table \ref{tab:iso} sums up the different possibilities for the $4D$ universe.
\begin{table}[h]
\begin{tabular}{|c||c|c|c|c|}
\hline
 & $s=4 $ & $s=3 $ & $s=2 $ & $s=0$ \\
 & spacetime homogeneous & space homogeneous & inhomogeneous & inhomogeneous \\
 & algebraic EFE & one non ignorable & two non ignorable &  \\
 & no \gls{redshift} & coordinate & coordinates &  \\
  \hline
  \hline
$q=6$  & Maximally \glslink{symmetry}{symmetric}:  &  none & none & none  \\
 isotropic & Minkowski, $\ds$ and AdS & & & \\
  \hline
$q=3$  & Einstein static  & FLRW & none & none \\
 isotropic & universe & & & \\
  \hline 
  $q=1$ & Gödel & \acrshort{LRS} Bianchi & \acrshort{LTB} & exists \\
  \acrshort{LRS} & & Kantowski-Sachs & & \\
  \hline
  $q=0$ & Osvath & Bianchi: & exists: spatially & Szekeres\\
  anisotropic &  & orthogonal \& tilted & self-similar & Our Universe ! \\
  \hline 
  \end{tabular}
\caption[Classification of the cosmological spacetimes according to their homogeneity and isotropy properties]{In $D=4$, classification of the cosmological spacetimes according to their homogeneity ($s)$ and isotropy properties ($q$). The total degree of \gls{symmetry} of the space is $r=q+s \leq 10$. See Ref.~\cite{Ellis:1998ct} p. 36 and references therein for details about the different models.} \label{tab:iso}
\end{table}

\paragraph{} The case $s=4$ and $q=3$ ($r=7$) is of interest for historical reasons with the \gls{Einstein universe} already mentioned in chapter \ref{chap:introcosmo}. It is the only non-expanding \glslink{SM}{FLRW model}, has hence no \gls{redshift} and is the first relativistic cosmological model discovered. The models with $s=3$ are of major interest for cosmological purpose, in the sense that they all give a mathematical realization of the \gls{cosmological principle} (see Sec.~\ref{sec:cosmprin}) For the Bianchi universes, to know exactly the group of \glslink{isometry}{isometries}, one needs to work out the \glspl{structure constant} which are 3x3 matrices, we will not detail the classification here, but the bottom line is that there exists 11 \gls{isometry} groups which give \gls{Bianchi spacetimes} numbered from I to IX. It is possible to work out the detailed observation properties of these models. For instance in the simplest Bianchi I universe, the strategy is usually to use \acrshort{CMB} or \glslink{BBN}{nucleosynthesis observations} to constrain the anisotropy parameter of this model. Via helium production, nucleosynthesis constraints are usually the strongest because it probes earlier time. The case $s=0$ and $q=0$ corresponds among others to our Universe. All the other models described here and in the references are only approximations of it. It is interesting to note that many universe with managable expressions have been exhibited: Szekeres' quasi-spherical model\cite{Szekeres:1974ct,Bolejko:2010eb}, Stephani's conformally flat models, Oleson's type $N$ solutions \cite{Ellis:1998ct} (and references therein). A last family is Swiss-Cheese models which is obtained by cutting and pasting spherically symetric models onto a \glslink{FLRW metric}{FLRW background}.
\paragraph{} The model, I mainly worked with is the \acrshort{LTB} model $(q,s,r)=(1,2,3)$. As described before, it is only one model among a lot and the job of the cosmologist is not only to find one model which fits the observational data but also to check if the others models don't fit the data equally better. In the next section, we will describe some physical and observation motivation to consider \acrshort{LTB} models.
\subsection{Inhomogeneous models: Why?}
\label{sec:whyinhomo}
\paragraph{} A direct response or as we say in french from the shepherd to the shepherdess is: why not? It is known that \glslink{SM}{FLRW universes} are a good first approximation of our Universe and in most of the cases inhomogeneous spacetimes are continuous deformations of FLRW ones meaning \glslink{SM}{FLRW} can be recovered by taking the correct limit. Curiosity to the unknown is a part of any person dedicating some of its time to academia and should be a motivation by itself. Beside that the opposite argument could also be put forward as a motivation to go for inhomogeneous cosmologies. 40 years ago, cosmology was not as popular as nowadays and most of the scientists were doing it as a ``part-time job'' while doing some more serious research such as \gls{particle physics} or \gls{general relativity}. When cosmology became a more respectable science, many scientists did cosmology full time but the distinction between cosmologists coming from \gls{particle physics} and cosmologists coming from \gls{general relativity} remained for some time. The way the people work in these two fields is sometimes drastically different and for instance, it is a successful tradition to postulate the existence of new particles to be discovered way later, see for instance the discovery of the Higgs boson. Some epidemiologists noted that the same has been done in cosmology: the \glslink{SM}{standard model} of cosmology has two big unknown namely two dark fluids: \gls{dark energy} and \gls{dark matter} which existence was postulated by cosmologists coming from \gls{particle physics}. The path to explore further \glslink{general relativity}{GR} and to look for generalization of \glslink{SM}{FLRW} is the one taken by the descendant of the \glslink{general relativity}{GR}-like cosmologists. Less sociologically orientated, one could just say that the questions answered by cosmology overlap with both \gls{particle physics} and \glslink{general relativity}{GR} questions.
\paragraph{} We already discuss in chapter \ref{chap:introcosmo}, a couple of caveats of the \glslink{SM}{FLRW model} which push to exploring different models. Statistical isotropy about our position has been established with a remarkable precision with the \acrshort{CMB} observations. Since most of the cosmological data are related to event happening on the past light cone, statistical homogeneity on a 3-surface is hard to probe because it is hard to distinguish a temporal from a spatial evolution on the past light cone (two good reviews: Refs.~\cite{Maartens:2011yx,Clarkson:2012bg}). In chapter \ref{chap:DE}, we already discussed a couple of ideas to explain the acceleration of the expansion of the Universe. In the framework of Einstein's \gls{general relativity}, a natural way to create an apparent acceleration is to introduce large-scale inhomogeneous matter distribution. This leads to many controversial philosophical debates (see \emph{eg}. Ref.~\cite{Ellis:2006fy}). It has been shown that the \acrshort{LTB} \cite{Lemaitre:1933gd,Tolman:1934za,Bondi:1947fta} metric can successfully fit the dimming of the \glslink{SNe Ia}{SNe Ia} by modeling radial inhomogeneities in a suitable way.\cite{Celerier:1999hp,Iguchi:2001sq} The fit of mass profiles to various datasets, including \glslink{SNe Ia}{SNe Ia}, was performed for instance in Ref.~\cite{Nadathur:2010zm}. The deduced acceleration of the expansion of the Universe induced by this unknown \gls{dark energy} would be nothing but a mirage due to light traveling through an inhomogeneous medium \cite{Mattsson:2007tj,Marra:2007pm,Celerier:2011zh}. 
\paragraph*{Dropping the Copernican principle?}  To account for the remarkable uniformity of the \acrshort{CMB} observation, our location in the Universe has to be fine-tuned at (or close up to 1\% to) the \gls{symmetry} center of the \acrshort{LTB} model \cite{Biswas:2010xm} clearly violating the \gls{Copernican principle}. Whether this \gls{fine-tuning} is problematic is still debated: \cite{Sundell:2013ova,Celerier:2011zh}, for instance the \glslink{SM}{FLRW model} is also non-copernican from a fully-relativistic point of view. Another direction would be not to take \acrshort{LTB} models too literally \cite{Celerier:2011zh,Celerier:2009sv}. As for the \glslink{SM}{FLRW model}, \acrshort{LTB} models should be considered as a an approximation of the reality. In some average sense, it could be that an inhomogeneous metric captures the real world better than a perfectly \glslink{symmetry}{symmetric} one.\cite{Bolejko:2010wc} Other exact non spherically \glslink{symmetry}{symmetric} solutions exist, like the Szekeres model \cite{Szekeres:1974ct,Bolejko:2010eb}, examples patching together \glslink{FLRW metric}{FLRW} and \acrshort{LTB} metrics in a kind of Swiss cheese model\cite{Biswas:2007gi}, meatball models \cite{Kainulainen:2009sx}. Thus those models should not be taken as way to describe exactly our reality but as attempt to investigate inhomogeneous metric and see how much of our reality it grasps. Most of those extensions are in agreement with the \gls{Copernican principle}.
\paragraph{} To finish this introduction on \acrshort{LTB} models, they are also used for non-cosmological purposes: modeling nucleus in nuclear physics, investigate the formation and the evolution of a black hole\footnote{This road leads, among other, to the concept of cosmological black holes which are black holes evolving within the \gls{Hubble flow} \textit{eg.}~\cite{Moradi:2015caa}}, investigate the influence of the electromagnetic field on a collapse of dust, investigate singularities theorem. \acrshort{LTB} models are also used to tackle various problems within cosmology: give simple \glslink{S}{structure formation} models, give universe models which are inhomogeneous on cosmological scale (without gluing a \glslink{FLRW metric}{FLRW model} at some cutoff scale), examine inhomogeneous \gls{big-bang} structures, look at how the expansion of the Universe impacts on planetary orbits, explore the geometrical dipole that would be seen in the \acrshort{CMB} by a off-center observer, investigate observational conditions for spatial homogeneity, trace the effect of averaging spatial inhomogeneities (see referencing in \cite{Ellis:1998ct} p. 40). 
\section{Lemaître-Tolman-Bondi model:}
\label{theory}
In this section, we will derive an expression for the \gls{luminosity distance} in a \acrshort{LTB} model. To do so, we will first present some generalities about \acrshort{LTB} model and then specify the class of models we will focus on. To connect to observations, the distances in \acrshort{LTB} spacetime are presented, there are simple generalization of the distances described in Sec.~\ref{sec:dista}. The final result for the \gls{luminosity distance} is shown in Eq.~\ref{eq:dl}. Then in Sec.~\ref{sec:data}, a data analysis is performed to determined if this model can also account for the observed data of the \glslink{SNe Ia}{SNe Ia}. A connection to \gls{fractal} cosmologies is proposed in Sec.~\ref{sec:frac}. Some concluding remarks and perspectives are drawn in Sec.~\ref{sec:ccl}. 
\subsection{Lemaître-Tolman-Bondi model: generalities}
The model describes a spherically \glslink{symmetry}{symmetric} dust distribution of matter. In coordinates comoving with the dust and in synchronous time gauge, the line element of the \acrshort{LTB} metric is given by:
\begin{equation}
\label{metricLTB}
ds^2=dt^2-\frac{R'^2(r,t)}{f^2(r)}dr^2-R^2(r,t) d\Omega^2. 
\end{equation}
$R(r,t)$ is the areal radius function, a generalization of the \gls{scale factor} in \glslink{FLRW metric}{FLRW spacetimes}. It also has a geometrical meaning as it can be interpreted as the \gls{angular distance}, as we will discuss in Sec.~\ref{distances}. $f(r)$ is the energy per unit mass in a comoving sphere and also represents a measure of the local curvature. 
\\
The time evolution of the areal radius function is given by the \glspl{Einstein equation} and reads (see also \ref{eq:LTBdyn}):
\begin{equation}
\label{EOM}
\frac{\dot{R^2}(r,t)}{2} - \frac{M(r)}{R(r,t)}= \frac{f(r)^2-1}{2},
\end{equation}
where $M(r)$ is a second free function. In Appendix \ref{ap:LTB}, we present a derivation of (\ref{metricLTB}) from the more general line element:
\begin{equation}
\label{generalmetric}
ds^2=dt^2-A^2(r,t)dr^2-B^2(r,t) d\Omega^2. 
\end{equation}
We present there also a derivation of the Einstein equations for the \acrshort{LTB} metric, together with a more detailed physical meaning of the free functions of the \acrshort{LTB} model. After integrating (\ref{EOM}), one will find the third free function of the \acrshort{LTB} model, namely $t_B(r)$ which is the bang time for worldlines at radius $r$. We will consider only a small class of \acrshort{LTB} solutions. First the equations of the \acrshort{LTB} model (\ref{metricLTB}, \ref{EOM}) are invariant under the change of radial coordinate $r \rightarrow r + g(r)$ where $g(r)$ is an arbitrary function. This gauge freedom gives the possibility to choose one of the free function of the \acrshort{LTB} model arbitrarily by picking a frame.

\paragraph{} However one caveat to these inhomogeneous \acrshort{LTB} models is that under very mild assumptions on regularity and asymptotic behavior of the free functions of the \acrshort{LTB} model, any pair of sets of conjugate observational data (for instance the pairs \{\gls{angular distance}, mass density in the redshift space\} or \{\gls{angular distance}, expansion rate\}) can be reproduced \cite{Mustapha:1998jb,Celerier:2009sv}. Those results were even shown through two theorems that we will reproduce here:
\subparagraph{Isotropic Observations Theorem (1)}
Any given isotropic set of source observations, together with any given source luminosity and number evolution functions, can be fitted by a spherically \glslink{symmetry}{symmetric} dust cosmology (a \acrshort{LTB} model) in which observations are spherically \glslink{symmetry}{symmetric} about us because we are located near the central worldline.
\subparagraph{Isotropic Observations Theorem (2)} Given any spherically \glslink{symmetry}{symmetric} geometry and any
spherically \glslink{symmetry}{symmetric} set of observations, we can find evolution functions that will make the
model compatible with the observations. This applies in particular if we want to fit observations to a \glslink{SM}{FLRW model}. 
\par For extensive discussion about these theorems, see \cite{Mustapha:1998jb,Ellis:1998ct,Celerier:2009sv}. For our purpose, what matters is that proposing an alternative to the \glslink{SM}{FLRW model} which has so much freedom that it can fit any data sets seems problematic. \emph{Hence physical criteria are required to constrain the \acrshort{LTB} models.} This is actually the main interrogation, our model aims at proposing a \gls{fractal} inspired input.
\par
The assumptions for the model are:
\begin{itemize}
\item The gauge freedom explained above allows us to choose one unique \gls{big-bang}: $t_B(r)$ constant.
\item We consider parabolic \acrshort{LTB} solutions so that the geometry is flat: $f(r)=1$
\item We choose the form of the free function $M(r)$ as following:
\begin{equation}
\label{condif}
M(r)=\mathcal{M}_g N(r)= \mathcal{M}_g \sigma r^{d},
\end{equation}
where $\mathcal{M}_g$ is the mass of a galaxy. $M(r)$ can be interpreted as the cumulative radius of matter inside a sphere of \textit{comoving} size $r$ \cite{Bondi:1947fta}, more details in Appendix \ref{physicalinterp}. At this point, it is worth noting that a \acrlong{EdS} universe (flat, matter only, \glslink{FLRW metric}{FLRW universe}) is recovered for $M(r)=M_0 r^3$ as shown in \ref{eq:limitFLRW}.
\end{itemize}
\paragraph*{}
Those assumptions are our prescriptions to restrain the free functions of the \acrshort{LTB} models. $(\sigma, d)$ are two new free parameters which will constrained by \glslink{SNe Ia}{supernovae} data. In Sec.~\ref{sec:frac}, a connection to \gls{fractal} cosmologies will be described.
\subsection{Observational distance}
\label{distances}
Ellis \cite{Ellis:1971pg} gave a definition of the \gls{angular distance} which applied to the \acrshort{LTB} metric gives $d_A=R$, one applies furthermore \glslink{distance duality relation}{Etherington reciprocity theorem} \cite{Ether}:
\begin{equation}
\label{Th}
d_L=(1+z)^2 d_A,
\end{equation}
which is true for general spacetime provided that source and observer are connected through null geodesics. Since, one considers a small class of parametric \acrshort{LTB} model, an analytic solution of (\ref{EOM}) is\footnote{Solutions for $f(r)>1$ and $f(r)<1$ also exists in the general case.}
\begin{equation}
\label{solu}
R(r,t)= \left(\frac{9M(r)}{2}\right)^{1/3} (t_B+t)^{2/3}.
\end{equation}
Assuming a single radial geodesic \cite{Mustapha:1998jb} and generalizing the calculation in Sec.~\ref{sec:redshift}, an analytical expression for the \gls{redshift} as a function of the radial coordinate has been found \cite{Nogueira:2013ypa}:
\begin{equation}
\label{SNR}
1+z(r) = \frac{t_B^{2/3}}{(t_B+t)^{2/3}}=\frac{t_B^{2/3}}{(t_B-r)^{2/3}}.
\end{equation}
Using Eqs.~(\ref{condif})-(\ref{SNR}), it is possible to propose an expression for the parametric \acrshort{LTB} \gls{luminosity distance}:
\begin{equation}
\label{eq:dl}
d_L=\left(\frac{9\sigma M_g}{2}\right)^{1/3} t_B^{\frac{d+2}{3}} \frac{\left((1+z)^{3/2}-1\right)^{d/3}}{(1+z)^{d/2-1}}.
\end{equation}
This will be the formula we will confront to its \glslink{SM}{FLRW counterpart}, it is only a function of the two parameters characterizing the matter distribution ($\sigma$,$d$). We will work with the following units: the unit of mass is $2.09\times 10^{22} M_{\odot}$, the time unit is $3.26\times 10^9$ years and the distances are given in Gpc. The \gls{big-bang} time will be taken for the data analysis to be equal to 4.3 corresponding to equation (\ref{eq:tbbt}).
\section{Supernovae data analysis}
\label{sec:data}
In this section, we will fit the Union2.1 compilation released by the Supernova Cosmology Project \cite{2012ApJ...746...85S}. It is composed of 580 uniformly analyzed \glslink{SNe Ia}{SNe Ia} and is currently the largest and most recent public available sample of standardized \glslink{SNe Ia}{SNe Ia}. The \glspl{redshift} range is up to $z=1.5$ \\
\begin{table}[h]
\center
\begin{tabular}{|c|c|c|c|}
  \hline
   & parameter 1 & parameter 2 &$\chi^2$ \\
  \hline
  flat FLRW & $\Omega_M=0.30\pm 0.03$ & $h=0.704 \pm 0.006$ & $538$\\
  parametric \acrshort{LTB} & $d=3.44 \pm 0.03$ & $\sigma^{1/3}=0.192 \pm 0.002$ & $973$ \\
  \hline
\end{tabular}
\caption[Result of the fit for the inhomogenenous model]{Quantitative results of the fitting procedure of the \glslink{SNe Ia}{SNe Ia} to the standard FLRW and our parametric \acrshort{LTB} model.}
\end{table}
A $\chi^2$ fit has been performed, it consists in minimizing the $\chi^2$ defined as:
\begin{equation}
\chi^2(\text{parameter 1},\text{parameter 2}) \equiv \sum_{i=1}^{580} \left[ \frac{d_L(i)-d_L(\text{parameter 1},\text{parameter 2})}{\Delta d_L(i) } \right]^2,
\label{eq:chisq}
\end{equation}
where $\Delta d_L(i)$ is the observational error bar for each data point indexed by $i$. The results are presented in the table together with the 95\% confidence interval  
\begin{figure}[h]
 \center
      \includegraphics[width=0.7\textwidth]{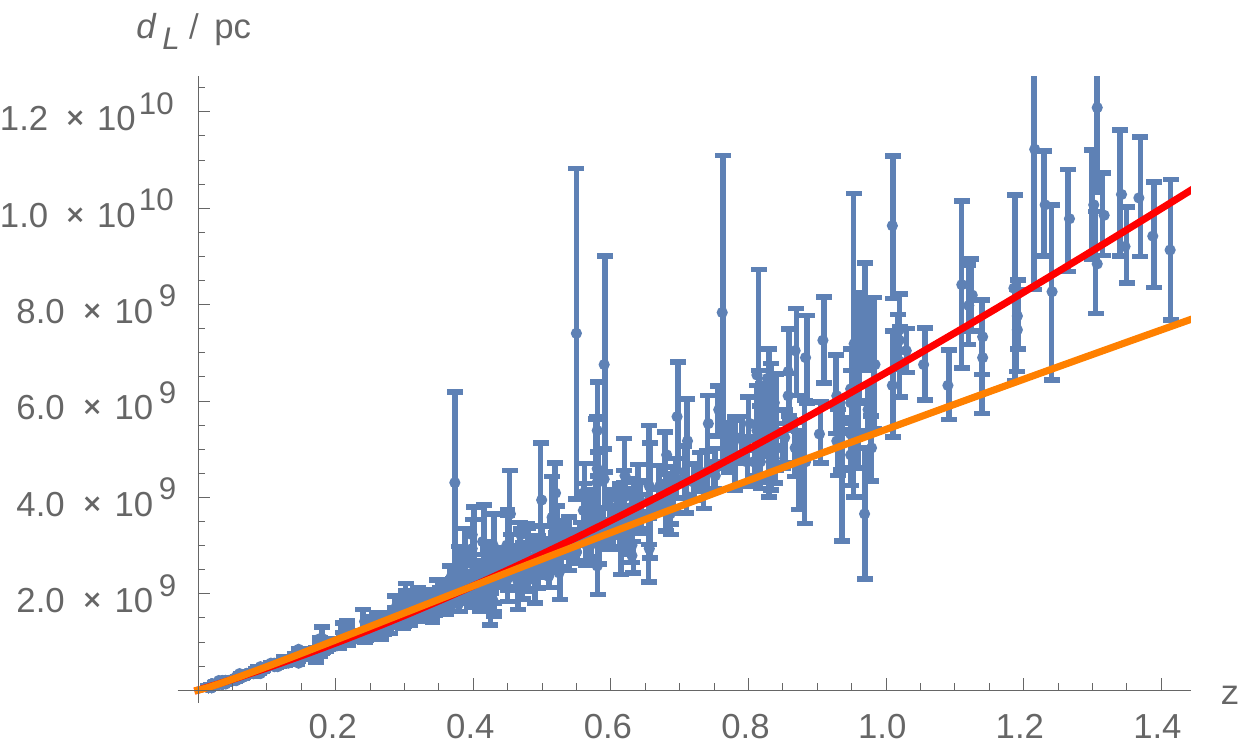}
\caption[Hubble diagram for the inhomogeneous model]{Best fitting line for the FLRW model (red, upper curve) and the parametric \acrshort{LTB} (orange, lower curve), the quantitative results are given in the table.}
        \label{fig:courbe}
        \end{figure}
        \begin{figure}[h] \center
         \includegraphics[width=0.7\textwidth]{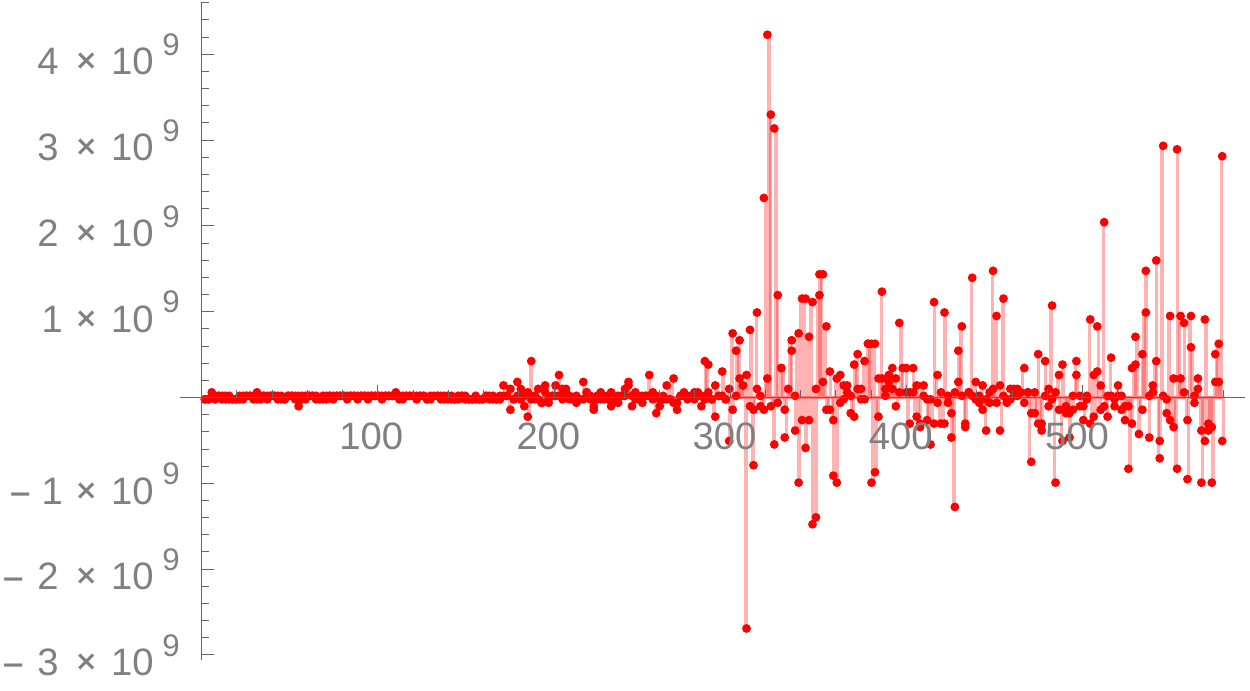}
\caption{Residual errors for the FLRW model}
\end{figure}
\begin{figure}[h] \center
         \includegraphics[width=0.7\textwidth]{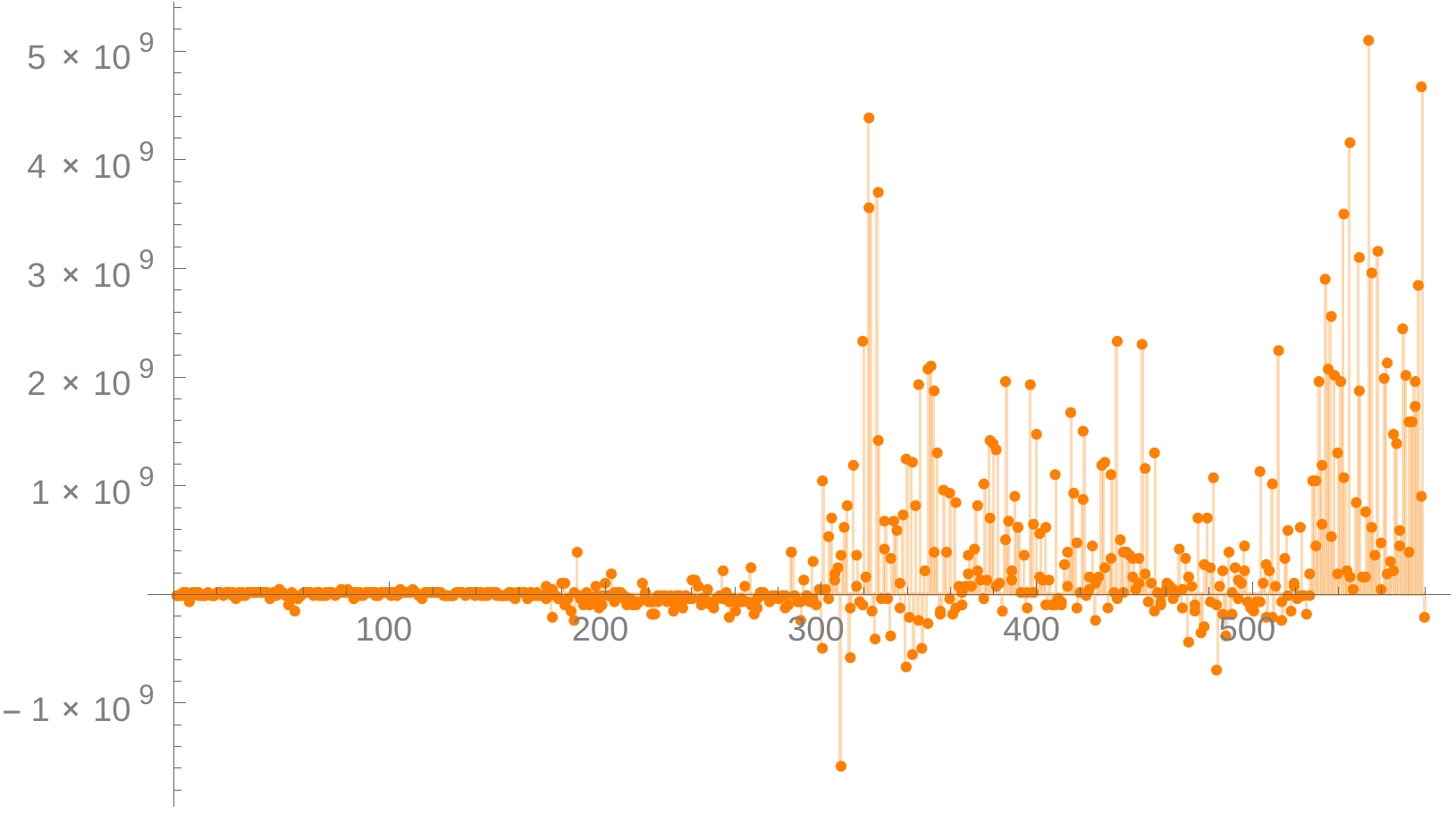}
\caption{Residual errors for the inhomogeneous model}
\end{figure}
$\Omega_M$ and $h$ are the two free parameters of the flat \glslink{SM}{FLRW model}. Recall that $\Omega_{\Lambda}$ is related to $\Omega_M$ with the relation $\Omega_{\Lambda}+\Omega_M=1$ ; h is related to the \gls{Hubble constant} via $H_0 \equiv 100 h \frac{\text{km}}{\text{Mpc s}}$. The \gls{Hubble diagram} and the associated residuals are plotted in Fig.~\ref{fig:courbe}. Our results for the \glslink{SM}{FLRW case} are in agreement with the current literature on the cosmological parameters \cite{Ade:2015xua} although the uncertainties are bigger. Interestingly the parametric \acrshort{LTB} models gives results with are not compatible with a FLRW universe in the sense that to recover the \glslink{SM}{FLRW model}, $d$ has to be exactly 3. \begin{figure}[h] \center
      \includegraphics[scale=0.7]{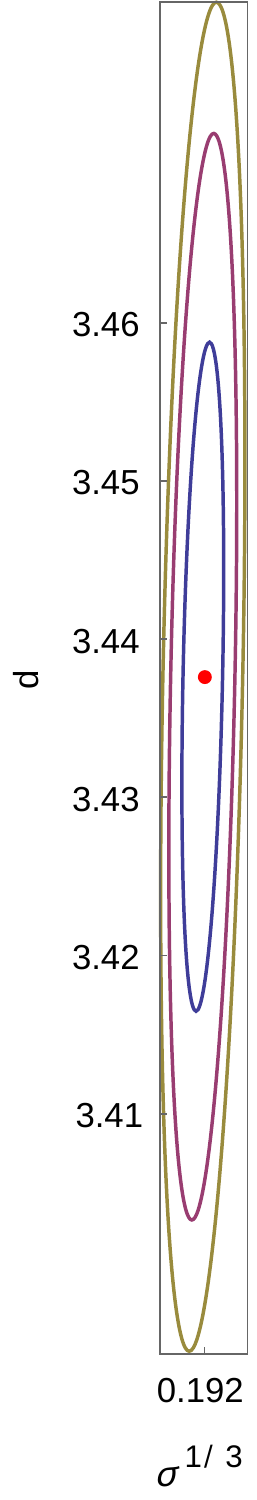}
        \caption{1, 3, 5 $\sigma$ confidence interval around the mean value given in the table for the parametric \acrshort{LTB} model.}
        \label{fig:prob1_6_1}
\end{figure}
 Looking at the confidence interval ellipses displayed in Fig.~(\ref{fig:prob1_6_1}) the value $d=3$ is ruled out by more than $5\sigma$. This illustrates that eventhough, the data are not better fitted by the parametric \acrshort{LTB} model, it still does better than an \acrshort{EdS} one ($\Omega_M=1$).
\paragraph*{} The shape of the \glslink{SNe Ia}{SNe Ia} luminosity curve is empirically well understood but their absolute magnitude is unknown and need to be calibrated. One need either to analytically marginalize the assumed \gls{Hubble constant} (or equivalently the absolute magnitude of the \glslink{SNe Ia}{supernovae}) \cite{Bridle:2001zv}, Appendix C.2 of Ref.~\cite{Biswas:2010xm}, or to use a weight matrix formalism (\emph{cf.} Ref.~\cite{Amanullah:2010vv}). Moreover, several authors pointed out the fact that the \glslink{SNe Ia}{supernovae} sample reduced with the SALT-II light-curve fitter from Ref.~\cite{Guy:2007dv} are systematically biased toward the \glslink{SM}{standard cosmological model} and are showing a tendency to disfavor alternative cosmologies \cite{Hicken:2009dk,Kessler:2009ys,Smale:2010vr}. This might be an reason why the parametric \acrshort{LTB} model fit has a bigger $\chi^2$.
\section{A connection to fractal cosmologies}
\label{sec:frac}
In this section, we will first review the use of \gls{fractal} and explain how our model could be related to a \gls{fractal} matter distribution. The idea of \glspl{fractal} relies on spatial power law scaling, self-similarity and structure recursiveness \cite{mandelbrot1983fractal}. These features are present in the formula:
\begin{equation}
N(r) \sim r^d,
\end{equation}
where $d$ is the \gls{fractal} dimension, $r$ the scale measure and $N(r)$ the distribution which manifests a \gls{fractal} behavior. If $d$ is an integer, it can be associated to the usual distributions (point-like for d=0, a line for d=1 and so on). The further from the spatial dimension, the more the \gls{fractal} structure is ``broken'' or irregular.  Note that topological arguments requires the \gls{fractal} dimension to be smaller than the spatial dimension in which the \gls{fractal} in embedded. This features of irregularity were useful to describe various structures from coastlines shapes to structure of clouds. In the context of cosmology, a \gls{fractal} distribution would simply describe how clumpy, inhomogeneous our Universe is. Historically, this idea was popular in the late 80's \cite{pietronero1987fractal,Coleman:1992cm,ruffini1988ino}, then some more modern models were developed \cite{Mureika:2006tz,Baryshev:2008nb,Grujic:2009hz} and the relation to cosmological observation was also was worked out (see Refs.~\cite{ChaconCardona:2012iu,Conde-Saavedra:2014dna,Bagla:2007tv} and references therein for an analysis with galaxy distribution).
\paragraph*{}
\begin{figure}
\center
\includegraphics[width=0.65\textwidth]{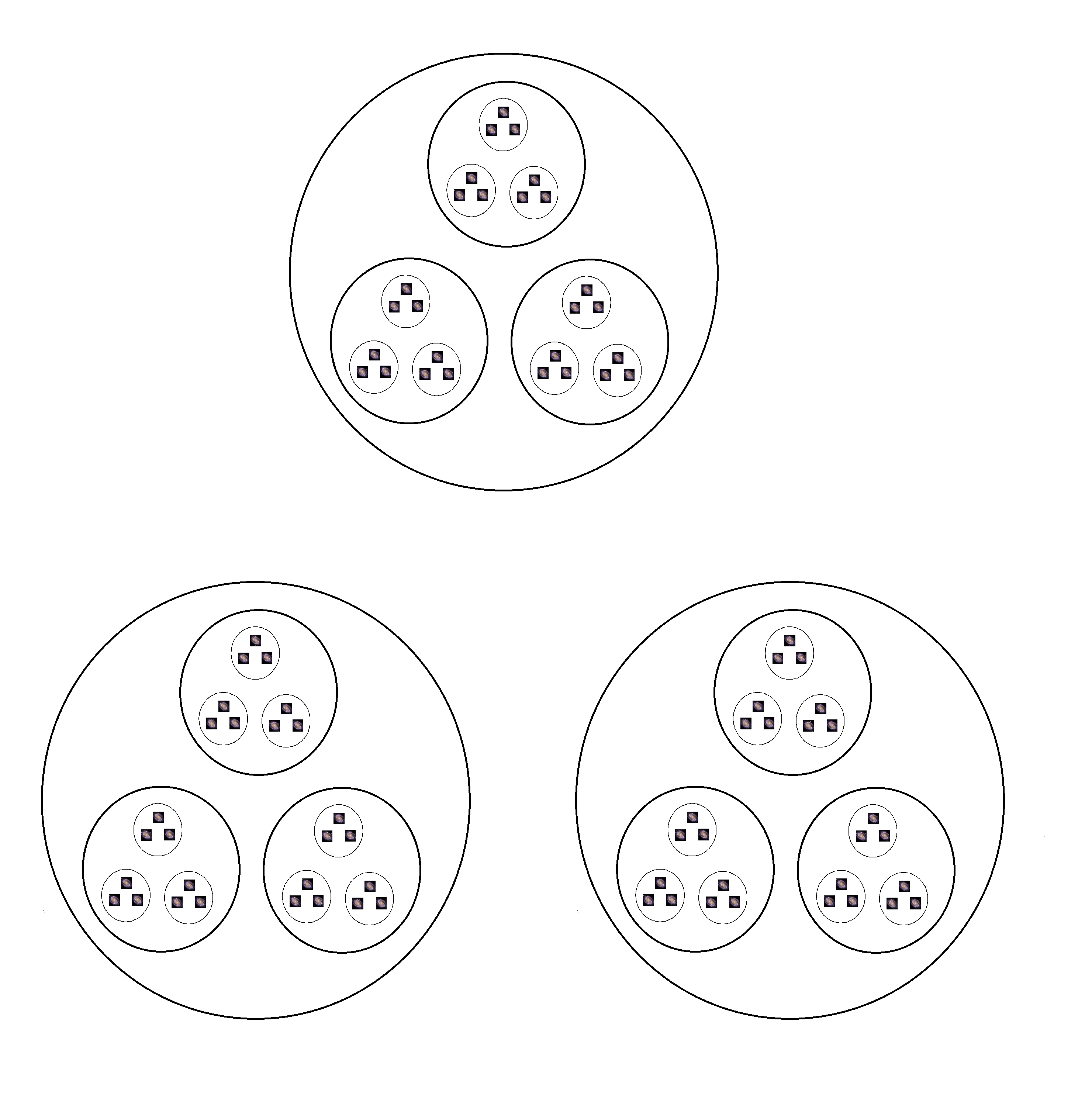}
\caption[An illustrative picture of a deterministic fractal.]{An illustrative picture of a deterministic \gls{fractal}. A specific pattern repeats itself in a self-similar way at different scales, in this case $\tilde{k}=3$, freely inspired from in \cite{pietronero1987fractal}.}
\label{fractal}
\end{figure}
To motivate the use of \glspl{fractal} to describe matter, I will review the model in \cite{pietronero1987fractal}. The goal is to get a number density of matter for a generic \gls{fractal} model. One possible illustration is given in figure \ref{fractal}. To do so, one starts from a point in the universe and counts the number of objects in a given 3-sphere of radius $r_0$. Then the number of objects within a sphere of radius $r_1=k r_0$ is $N_1 = \tilde{k} N_0$. Iterating this procedure for $n \in \mathbb{N}$, one finds:
\begin{align}
&N_n= \tilde{k}^n N_0, \\
&r_n=k^n r_0,
\end{align}
where $k$ is the scale of self-similarity and $\tilde{k}$ is the number of objects inside an iteration of the \gls{fractal} pattern. With some algebra, it is possible to propose an expression for $N_n$ as a function of $r_n$ ; it reads:
\begin{equation}
N_n=\sigma r_n^d,
\end{equation}
with $\sigma \equiv \frac{N_0}{r_0^d}$ is related to the lower cutoff of the \gls{fractal} object and $d \equiv \frac{\log(\tilde{k})}{\log(k)}$ is the \gls{fractal} dimension. This way of deriving the formula for the number of structure in a radius $r$ is given as an illustrative example as the \gls{fractal} under consideration here is deterministic and is embedded in a 2D plane. If a microphysical model is proposed (\emph{ie.} by presenting a model of matter which becomes \gls{fractal} at some moment of the structure formation), this model would in principle predict value for the parameters ($\sigma$,$d$) which would entirely characterize the \gls{fractal} distribution. 
\paragraph*{}
 Using \acrshort{LTB} models together with a \gls{fractal} matter distribution has been already considered in Refs.~\cite{Ribeiro:2008rs,Ribeiro:2008ru,Ribeiro:2008ug,SylosLabini:1997jg,Labini:2011dv,Labini:2011tj}. In \cite{Nogueira:2013ypa}, following \cite{pietronero1987fractal}, the authors proposed a \gls{fractal} inspired model by taking the \gls{luminosity distance} as the spatial separation of the \gls{fractal} ($N(r)=\sigma (d_L)^d$). This is motivated by the view of the \gls{fractal} structure as an observational feature of the galaxy distribution. Since astronomical observation are carried out on the past null cone, the underlying structure of galaxy may not be itself \gls{fractal} \cite{Ribeiro:2008rs,Ribeiro:2008ru}. The model proposed by Ref.~\cite{Nogueira:2013ypa} is interesting but a clear problem appears, if one uses the self-similarity condition for the \acrshort{LTB} model (Eq.~(3.17)), it is clear that, at $z=0$, the mass function is not zero which is unsatisfactory (\emph{cf.} also Fig.~3.5). In the same way, it can be shown that within the framework of Ref.~\cite{Nogueira:2013ypa}, the \gls{luminosity distance} is non zero at $z=0$.
 
 \paragraph*{} Our model is such that $N(r)=\sigma r^d$ and could correspond to a \gls{fractal} matter distribution if $0<d<3$ from topological considerations.  The \gls{fractal} is described by the coordinate $r$, that is the \gls{fractal} structure is a geometrical effect which does not necessarily translate itself in an astronomically observable quantity. Contemplating the best fit of $d=3.44$ and within our working hypothesis, we can state that \emph{\glslink{SNe Ia}{supernovae Ia} data do not support \gls{fractal} models}.
 
\section{Conclusion and Perspectives}
\label{sec:ccl}
\paragraph{} A parametric \acrshort{LTB} model has been introduced in this chapter. It is characterized by two parameters ($\sigma$,$d$) like the flat \glslink{SM}{FLRW model} ($H_0$,$\Omega_M$). The link to \glslink{SNe Ia}{SNe Ia} data was then worked out leading to a comparison between the flat \glslink{SM}{FLRW model} and the parametric \acrshort{LTB} model. The parametric \acrshort{LTB} model can fit the data reasonably but the standard \glslink{SM}{FLRW model} fits the data better. To keep testing such models, it is desirable to improve the data analysis with more elaborate techniques for \glslink{SNe Ia}{SNe Ia} but also with others data sets \emph{eg.}~\acrshort{CMB} anisotropies and polarization, Integrated Sachs Wolfe effect, \gls{BBN}{light element abundances}, \acrshort{BAO}, Galaxies Surveys... One of the motivation to build such a model was to propose a physical input to constrain the general \acrshort{LTB} where the two free functions that this model enjoys allow to fit any cosmological data under mild assumption on these free functions \cite{Mustapha:1998jb}. The parametric \acrshort{LTB} model of this paper can be generalized to cases where $t_B(r)$ is not constant \emph{or} (but not \emph{and} because of the gauge freedom of the free functions of the \acrshort{LTB} metric) by considering non uniformly flat geometries. It has been shown for instance in Ref.~\cite{krasinski2014accelerating} that a nonsimultaneous \gls{big-bang} can also account for the acceleration of the expansion of the Universe.

\paragraph{} Since \glslink{SM}{FLRW models} are nowadays the most popular ones, many caveats of them are known and investigated with special care (\emph{cf.}~Sec.~\ref{sec:cosmprin}, \ref{sec:status} and chapters \ref{chap:inflation} and \ref{chap:DE}). Even though, the \acrshort{LTB} metric is less popular, some problems have also been identified and investigated. When it comes to discuss \glslink{S}{structure formation} in the \glslink{SM}{FLRW model}, one still assumes spatial homogeneity and isotropy but considers the forming \glslink{S}{structures} as metric and matter \glspl{perturbation}. Performing the \gls{perturbation} theory in a \acrshort{LTB} metric is a really complex task, especially because a scalar-vector-tensor decomposition (see chapter \ref{chap:inflation}) does not allow anymore to study separately the scalar, vector and \gls{tensor} modes. Instead the ``natural'' variables to perform the \gls{perturbation} theory give in the FLRW limit a cumberstone combination of scalars, vectors and tensors. Efforts were done in this direction \cite{Clarkson:2009sc,Leithes:2014uda} but the \gls{perturbation} techniques are not yet advanced enough to be challenged with realistic numerical simulations and observations as the \glslink{SM}{FLRW model}. 
In addition, \gls{inflation} in \acrshort{LTB} would occur differently in separate space location. This is a bizarre feature on which one can only speculates the consequences. It might be one different way to touch the multiverse scenario \cite{carr2007universe}. With a loss of generality for the free functions, it is always possible to demand the \acrshort{LTB} model to approach the homogeneous limit in the early times, in which case the inflationary results apply \cite{dePutter:2012zx}. 

\paragraph{} To finish, all the models involving inhomogeneities do not solve the \gls{cosmological constant problem} but just shift it. From explaining a fairly unnatural value for the \gls{cosmological constant}, one just assumes it is zero without providing any explanation. \gls{Lovelock theorem} in this case ensure that a \gls{cosmological constant} should be present, see also the discussions in Sec.~\ref{sec:DEsolution} and in \cite{Martin:2012bt}.

%% file: IDE/IDE.tex
\epigraph{It is ... a good rule not to put overmuch confidence in observational results until they are confirmed by theory}{A.~Eddington, 1947, see \cite{Freeman:2013rz}}
\textit{In this chapter, we investigate a cosmological model of interacting dark matter and dark energy. An interaction between dark matter and dark energy is always possible as their nature is poorly constrained. Those models are known to alleviate the coincidence problem. Our model can be derived from high energy physics and has two extra parameters $\delta_G$ and $\delta_{\Lambda}$ which correspond to the interaction between dark matter and dark energy. We will study the consequences to observational cosmology with supernovae data. This chapter is adapted from the work to appear \cite{withD}}
\paragraph*{}
Interacting \gls{dark energy} models \cite{Friedman:1991dj,Gradwohl:1992ue,Wetterich:1994bg,Amendola:1999er,Amendola:2003wa,Lee:2006za,Pettorino:2008ez,Valiviita:2008iv,He:2008tn,Gavela:2009cy,Faraoni:2014vra,Salvatelli:2014zta,Xue:2014kna,Abdalla:2014cla,Koivisto:2015qua} rely on the idea that \gls{dark energy} and \gls{dark matter} do not evolve separately but interact with each other non-gravitationally (for a recent review, see \cite{Wang:2016lxa}). Most of the studies on interacting dark fluids focus on the relation to the cosmological data introducing an \textit{ad hoc} coupling between \gls{dark matter} and \gls{dark energy} \mbox{\cite{Salvatelli:2014zta,Murgia:2016ccp}}. A classification of those models was given in \mbox{\cite{Koyama:2009gd}}. In this work, we adopt a less phenomenological approach as we test a model arising directly from quantum field theory and perform the link to a set of observational data: the \glslink{SNe Ia}{SNe Ia}. Interacting \gls{dark energy} models are known to alleviate the \gls{coincidence problem} depicted in Sec.~\ref{sec:coincp}. 

The model considered in this chapter relies on the Einstein-Cartan gravitational theory where the universe would be in the scale invariant ultra-violet fixed point of the theory. It has been shown in \cite{Xue:2014kna} that it is possible from this model to propose an expression for the \gls{luminosity distance}, and hence compare it to observational data. This chapter aims at realizing this possibility and investigating whether this model, as other interacting \gls{dark energy} models, offers a viable alternative to the \glslink{SM}{$\Lambda$CDM model}. Our work shares the same spirit than \cite{Sola:2016jky}: not only it provides a phenomenological interacting \gls{dark energy} framework, but it also presents theoretical motivations arising from quantum field theory. This chapter is divided as follow: in Secs.~\ref{sec:SFC} and \ref{sec:latea}, we present the basic equations of the model and calculate an expression for the \gls{luminosity distance}. In Sec.~\ref{sec:fit}, we perform a $\chi^2$ fit to the \glslink{SNe Ia}{SNe Ia} data. Motivated by the encouraging results, we investigate in Sec.~\ref{sec:paramstudy} more systematically the parameter space of our model by asking the following question: which range of the parameter space of our model offers a better alternative than the \glslink{SM}{$\Lambda$CDM model}. Finally, we give conclusions and perspectives in Sec.~\ref{sec:ccl}.

\section{Quantum Field Cosmology}
\label{sec:SFC}
As one of the fundamental theories for interactions in Nature, \gls{general relativity}, which plays an essential role in the \glslink{SM}{standard model} of cosmology, should be realized in the scaling-invariant domain of a fixed point of its quantum field theory\footnote{It was suggested by Weinberg \cite{zichichi2012understanding} that the quantum field theory of gravity regularized with an ultraviolet (UV) cutoff might have a non-trivial UV-stable fixed point and asymptotic safety, namely the renormalization group (RG) flows are attracted into the UV-stable fixed point with a finite number of physically renormalizable operators for the gravitational field.}. 
It was proposed \cite{Xue:2014kna,Xue:2015tmw} that the present (low-\gls{redshift} $z<1$) cosmology is realized in the scaling-invariant domain of an ultraviolet-stable fixed point ($\sim G_0$) of the quantum field theory of Einstein gravity\footnote{Instead, the inflationary cosmology is realized in the scaling-invariant domain of an ultraviolet-unstable fixed point $\tilde{G}_0 \neq G_0$.}, and is described by
\begin{eqnarray}
H^2
&=& H_0^2
\Big[\Omega^0_M a^{-3+\delta_{G}} + 
\Omega^0_{_\Lambda}a^{-\delta_{\Lambda}}\Big],
\label{re3}\\
a\frac{dH^2}{da}\! +\! 2H^2 \!&= &\!H^2_0
\Big[2\Omega^0_{_\Lambda}a^{-\delta_{\Lambda}}\!-\!(1\!+\!3\omega_M )\Omega^0_M a^{-3+\delta_{G}}\Big].
\label{e2}
\end{eqnarray}
Notation are introduced in chapter \ref{chap:introcosmo}. In deriving Equations (\ref{re3}) and (\ref{e2}), motivated by observations, it was assumed that the curvature is null $k=0$, implying $\Omega_\Lambda^0 + \Omega_ M^0 = 1$. In addition, in the framework of our model, both the gravitation constant and $\Omega_\Lambda/\Omega_\Lambda^0$ can vary, following the scaling evolutions $G/G_0\approx a^{\delta_G}$ and $\Omega_\Lambda/\Omega^0_{\Lambda}\approx a^{\delta_\Lambda}$. In other words, the evolution of the two dark sectors is a slight deformation of (\ref{eq:Fried1}) and (\ref{eq:Fried2}), described by the two critical indexes $\delta_{G}$ and $\delta_\Lambda$. 

The \gls{dark energy} and matter interact and can be converted from one to another. They obey the generalized \gls{Bianchi identity} (total energy conservation), 
\begin{eqnarray}
a\frac{d}{da}\left[(G/G_0)(\Omega_{_\Lambda}+\Omega_{M})\right]
&=&-3(G/G_0)(1+\omega_{M})\Omega_{M},
\label{cgeqi20}
\end{eqnarray}
where effective variations of the gravitational coupling constant and of the \gls{cosmological constant} generalize the standard \gls{Bianchi identity}. Of course those variations are only effective as a direct time dependence of constants in \glspl{Einstein equation} (\ref{eq:Eisteq}) would simply break general covariance. For small redshift, assuming $\delta_\Lambda <\delta_G\ll 1$, Equation (\ref{cgeqi20}) leads to the relation
\begin{equation}
\label{eq:rela}
\delta_{\Lambda} = \left(\frac{\Omega_M^0}{\Omega_{\Lambda}^{0}}\right)\delta_G>0.
\end{equation}
In this work, such a \acrfull{QFC} model with theoretical parameters $\Omega_M^0$, $\Omega_\Lambda^0$, $\delta_G$ and $\delta_\Lambda$ is compared with the observational cosmology, the case of 
$\delta_{_G} =\delta_{_\Lambda}=0$ reducing to the the \glslink{SM}{$\Lambda$CDM model} of chapter \ref{chap:introcosmo}.

\section{Effective Equation of state and interaction of dark energy and matter}
\label{sec:latea}
Equations (\ref{re3}) and (\ref{e2}) in the \acrshort{QFC} model can be obtained by phenomenologically introducing a slight deformation of the evolution of the dark sector of our Universe, obtained from (\ref{eq:Fried1}) and (\ref{eq:Fried2}):
\begin{align}
& \rho_M = \rho_M^0 a^{-3+\delta_G}, \label{eq:8}\\
& \rho_{\Lambda} = \rho_{\Lambda}^0 a^{-\delta_{\Lambda}}.\label{eq:9}
\end{align}
Although the parameters $\delta_G$ and $\delta_{\Lambda}$ are motivated by the \acrshort{QFC} model, these parameters can also be explored on phenomenological grounds. It leads to another interpretation of parameters $\delta_G$ and $\delta_\Lambda$.

Using the individual conservation of the \glspl{energy momentum tensor} for the \gls{dark matter} and \gls{dark energy} sectors
\begin{align}
\dot{\rho}_{M,\Lambda} + 3 H (1+ \omega_{M,\Lambda}) \rho_{M,\Lambda} =0,
\end{align}
 
and the \gls{Friedmann equation} (\ref{eq:Fried1}), we find that the parameters  $\delta_G$ and $\delta_\Lambda$ in Eq.~(\ref{re3}) can be interpreted as effective modifications of the equation of state with
\begin{align}
& \omega_M = -\frac{\delta_G}{3}, \\
& \omega_{\Lambda} = -1+\frac{\delta_{\Lambda}}{3}.
\end{align}
Beside, assuming the standard equation of state ($\omega_M=0, \omega_{\Lambda}=-1$), it is also possible to relate the parameters $\delta_G$ and $\delta_{\Lambda}$ to an interaction between \gls{dark matter} and \gls{dark energy} by introducing an interaction term $Q$

\begin{align}
&\dot{\rho}_M + 3 H \rho_M =+Q, \\
& \dot{\rho}_{\Lambda}  =-Q.
\end{align}
$Q$ is assumed to be zero in the \glslink{SM}{standard model} of cosmology, see discussion around equation (\ref{eq:conservatoten}). In our case, $Q$ reads:
\begin{align}
Q = H \delta_G \rho_M = H \delta_{\Lambda} \rho_{\Lambda},
\label{eq:Q}
\end{align}
which leads also to the relation (\ref{eq:rela}). Such interaction terms have been shown to alleviate the \gls{coincidence problem} \cite{Abdalla:2014cla}. We stress that this last interpretation in term of $Q$ of the deformation the dark sector (\ref{eq:8}) (\ref{eq:9}) is valid only for small redshifts ($z \ll 1$), as otherwise the different evolutions in \gls{redshift} for $\rho_M$ and $\rho_{\Lambda}$ invalidate (\ref{eq:Q}). To obtain the general interaction term, one needs to consider the general evolution of the effective gravitational constant in equation (36) of \cite{Xue:2014kna}.

\paragraph*{} In this chapter, we will treat the parameters $\delta_G$ and $\delta_\Lambda$ in Eq.~(\ref{re3}) as free parameters determined by the observational cosmology. Observe that in order to have a coherent model of \gls{dark energy}, the constraint $\delta_G \delta_{\Lambda} >0$ must be fulfilled. In this case, the parameters $\delta_G$ and $\delta_{\Lambda}$ can be interpreted as the rate of conversion of \gls{dark matter} into \gls{dark energy} and of \gls{dark energy} into \gls{dark matter}. A phenomenological investigation shows that increasing $\delta_G$ or decreasing $\delta_{\Lambda}$ induces an acceleration of the expansion of the universe: \gls{dark matter} is converted into \gls{dark energy}. Conversely, decreasing $\delta_G$ or increasing $\delta_{\Lambda}$ induces a deceleration of the expansion of the universe: \gls{dark energy} is converted into \gls{dark matter}.

Now, the \gls{luminosity distance} can be obtained
\begin{align}
 d_L (z) = \frac{c}{H_0} (1+z) \int_1^{\frac{1}{1+z}} \frac{da}{a^2 \sqrt{\Omega^0_{\Lambda} a^{-\delta_{\Lambda}}+\Omega_M^0 a^{\delta_G -3} }},
 \end{align}
where the speed of light $c$ is included for clarity. An analytic representation of the integral exists. It reads
 \begin{align*}
d_L(z)= &\frac{2c}{H_0}\frac{1+z}{\Omega_M^0 (1-\delta_G)} \times \Bigg[\text{ }_2F_1\bigg(1,\frac{4-\delta_{\Lambda}-2 \delta_G}{2(3-\delta_{\Lambda}-\delta_G)},\frac{7-2\delta_{\Lambda}-3 \delta_G}{2(3-\delta_{\Lambda}-\delta_G	)};-\frac{\Omega^0_{\Lambda}}{\Omega_M^0} \bigg) \\
& -\frac{\sqrt{(1+z)^{3-\delta_G}\Omega_M^0+(1+z)^{\delta_{\Lambda}}\Omega_{\Lambda}^0}}{(1+z)^{2-\delta_G}} \\
& \times 2F_1\bigg(1,\frac{4-\delta_{\Lambda}-2 \delta_G}{2(3-\delta_{\Lambda}-\delta_G)},\frac{7-2\delta_{\Lambda}-3 \delta_G}{2(3-\delta_{\Lambda}-\delta_G	)};-\frac{\Omega^0_{\Lambda}}{\Omega_M^0} (1+z)^{-3+\delta_G +\delta_{\Lambda}}\bigg)\Bigg] ,
 \end{align*}
 where $\text{}_2F_1$ is the hypergeometric function defined as: $\text{}_2F_1(a,b,c;z) \equiv \sum_{n=0}^{\infty}\frac{(a)_n (b)_n}{(c)_n} \frac{z^n}{n!}$, with the Pochhammer symbol $(x)_n$ given by: $(x)_n \equiv \frac{\Gamma(x+n)}{\Gamma(x)}$. In deriving this equation, the assumption of a flat universe was used, that is to say $\Omega_M^0 + \Omega_{\Lambda}^0 =1$.
 \section{Fit to supernovae data}
 \label{sec:fit}
\begin{figure}[h]
        \centering
     \includegraphics[width=0.7\textwidth]{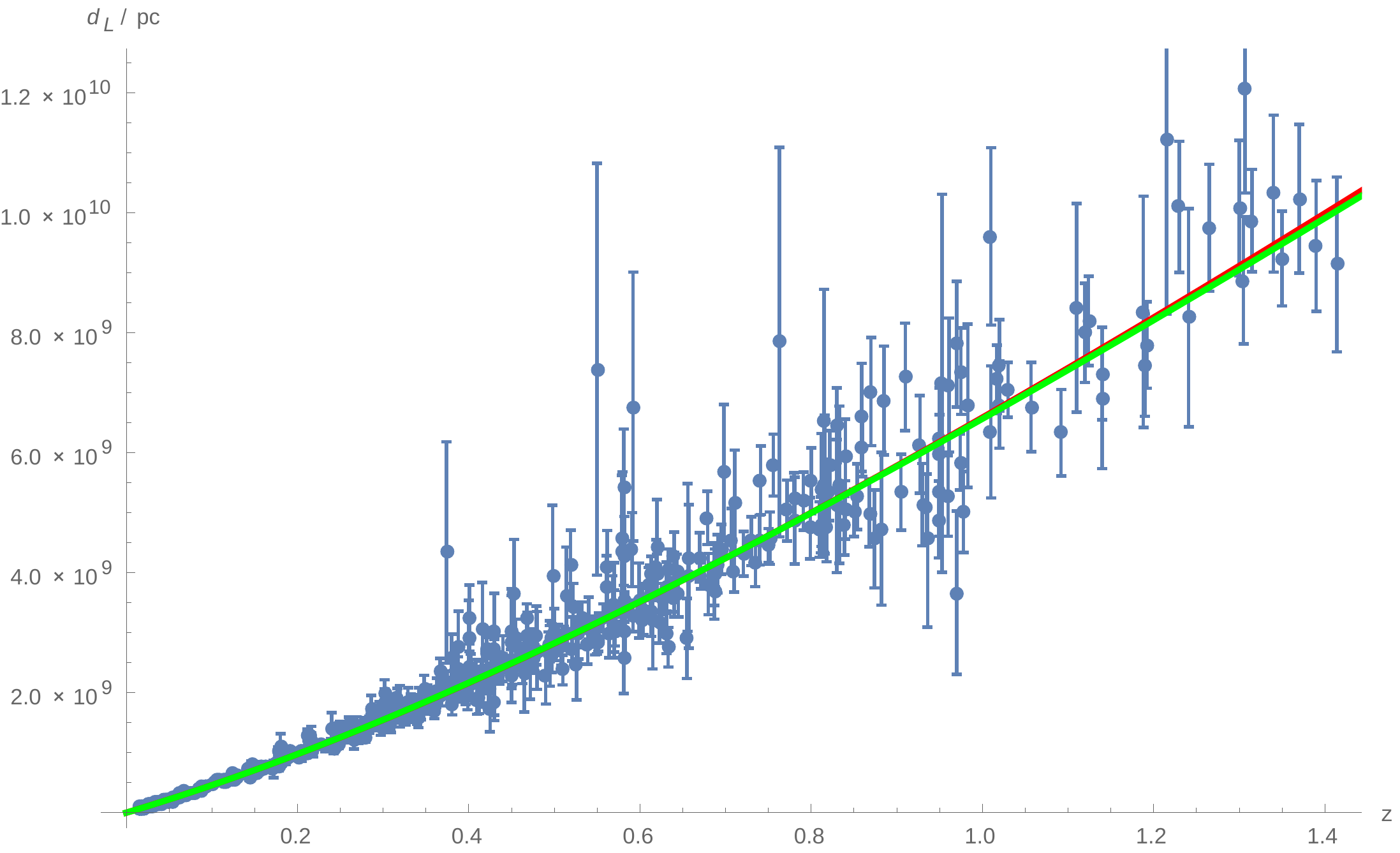}
\caption[Hubble diagram for the interacting dark energy model]{Best fitting line for the \glslink{SM}{FLRW models} (red, upper curve) and the \acrshort{QFC} model (green, lower curve), the quantitative results are given in table \ref{table}.}
       \label{fig:1} 
\end{figure}

\begin{figure}[h]
    \centering
        \includegraphics[width=0.7\textwidth]{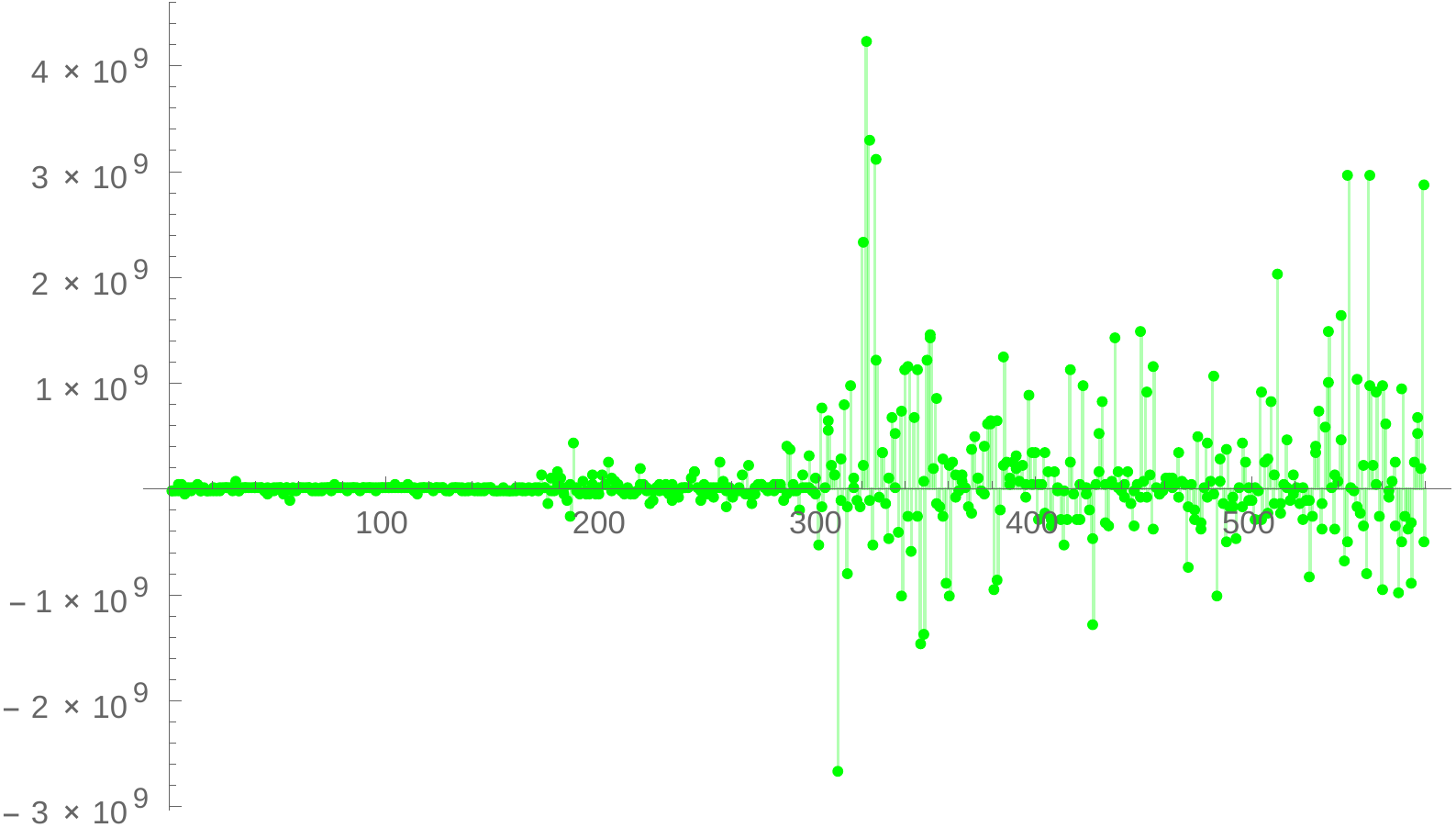}
\caption{Residuals errors of the the interacting dark energy model.}
\label{fig:2}
\end{figure}

In this section, the relation given by equation (\ref{eq:rela}) will be enforced. It gives a constraint between the two new parameters of the model. In Sec.~\ref{sec:paramstudy}, this constraint will be relaxed in order to study a broader range of the parameter space. We are now in position to perform a $\chi^2$ fit with the Union 2.1 compilation released by the Supernova Cosmology Project \cite{2012ApJ...746...85S}. We follow the same procedure as in Sec.~\ref{sec:data}, in particular, we minimize again the $\chi^2$ given by equation (\ref{eq:chisq}). A comparison of the result for our model and the standard flat \glslink{SM}{FLRW models} ($\delta_G=\delta_\Lambda = 0$) is presented in the table \ref{table} together with the 95\% confidence interval. Figure \ref{fig:1} displays the resulting \gls{Hubble diagram} for both models and the residual of the fit for the \acrshort{QFC} model is shown on Figure \ref{fig:2}.

\begin{table}[h]
\centering
\begin{tabular}{|c|c|c|c|c|}
  \hline
   & $\Omega_M^0$ & $h$ & $\delta_G$ &$\chi^2$ \\
  \hline
  flat FLRW & $0.30\pm 0.03$ & $0.704 \pm 0.006$ & $0$ & $538.754$\\
  \acrshort{QFC} model & $0.27 \pm 0.11$ & $0.705 \pm 0.008$ & $-0.19 \pm 0.9$ & $538.596$ \\
  \hline
\end{tabular}
\caption[Results of the fit for the interacting dark energy model]{The best fit parameters for the two models under consideration and the associated $\chi^2$, together with the 95 \% confidence intervals.  $h$ is related to the \gls{Hubble constant} via $H_0 \equiv 100 h \frac{\text{km}}{\text{Mpc.s}}$.}
\label{table}
\end{table}


From table \ref{table}, we note that the parameters of the \glslink{SM}{FLRW model} are recovered with reasonable precision. However, there is a strong degeneracy between $\delta_G$ and $\Omega_M^0$ for the \acrshort{QFC} model. To investigate this issue, in the next section we inspect in more details the parameter space of the model without imposing any constraints.

Only with \glslink{SNe Ia}{SNe Ia} the value $\Omega_M$ is poorly constrained (see table \ref{table}) but with joined data sets, the current best value is $\Omega_M=0.308 \pm 0.012$ \textit{cf.}~table \ref{table:cosmnum}. An important point is that in most of interacting \gls{dark energy} models both the \gls{dark matter} and the \gls{baryon}ic matter fluids are assumed to interact with the \gls{dark energy}, but as the nature of \gls{baryon}ic matter is known from ground experiment, the type of interaction allowed between this matter and the hypothetical \gls{dark energy} is hugely constrained. So a more conservative approach would be to allow only \gls{dark matter} (accounting for $\Omega_{DM} = 0.268 \pm 0.013$ \textit{cf.}~table \ref{table:cosmnum}) to interact with \gls{dark energy} and account for this interaction with a new term in the bias quantity describing the different behavior of \gls{baryon}ic and \gls{dark matter} in the cosmic history. The fact that the best fit from the \acrshort{QFC} model is $\Omega_M = 0.27$ might be a hint pointing to this requirement of having non-interacting \gls{baryon}ic matter \cite{Amendola:2001rc}.
\section{Parameter space study}
\label{sec:paramstudy}
In this section, we drop the relation (\ref{eq:rela}) and explore the parameter space $\delta_G$-$\delta_\Lambda$ for set values of the parameters $H_0$, $\Omega_\Lambda^0$ and $\Omega_M^0$ compatible with constraints from cosmological observations. $H_0$ is fixed to it standard value in the \glslink{SM}{FLRW model} from \glslink{SNe Ia}{SNe Ia}: $H_0 = 70.4 \frac{\text{km}}{\text{Mpc.s}}$ \textit{cf.}~table \ref{table:cosmnum}. The purpose of this section is to identify regions for which the $\chi^2$ of the \acrshort{QFC} model is smaller than that of the \glslink{SM}{$\Lambda$CDM model}.

\begin{figure}[h]
\centering
\includegraphics[width=0.9\textwidth]{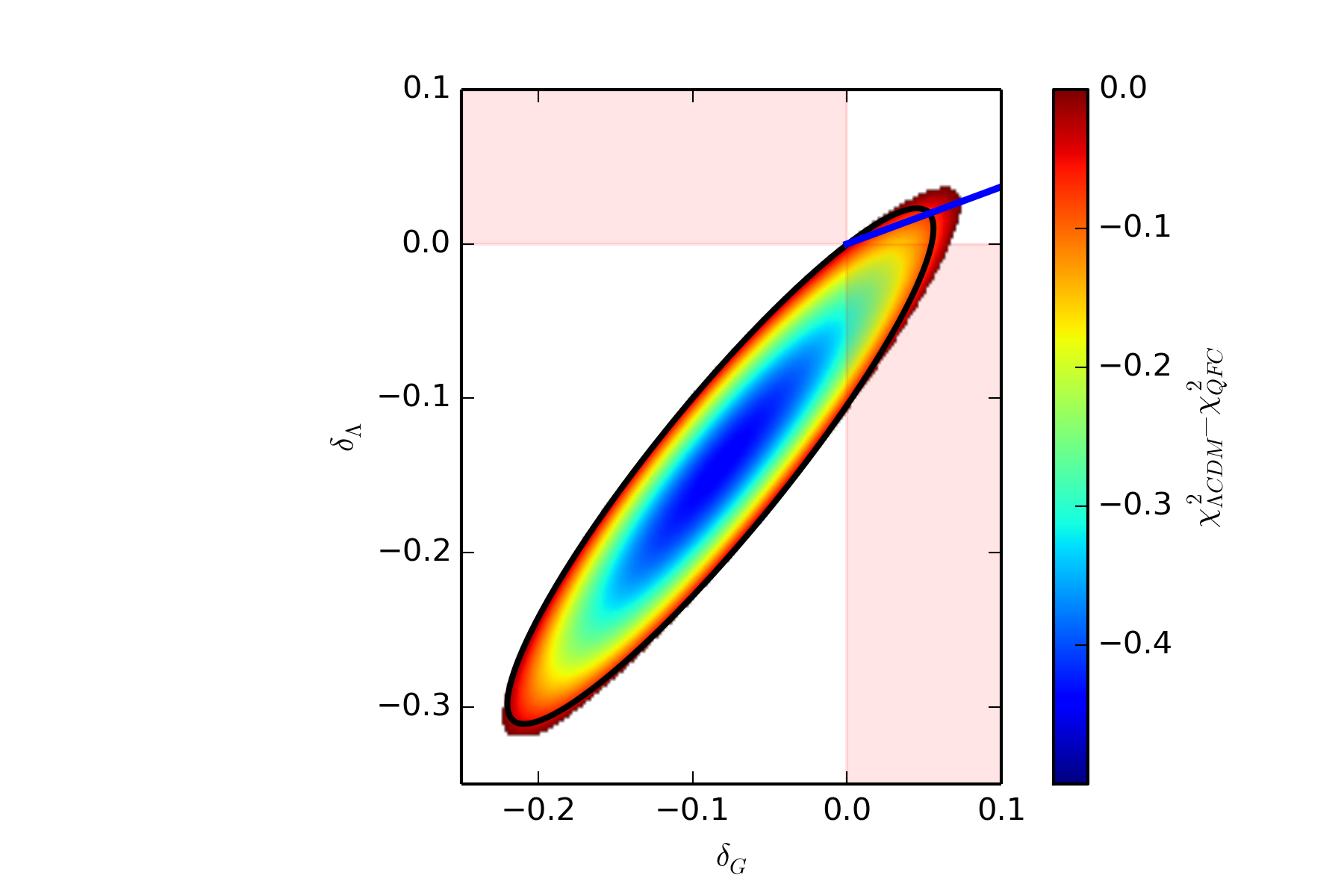}
\caption[Parameter space study of the interacting dark energy model (1/2)]{Contour which obtains a better $\chi^2$ than the \glslink{SM}{FLRW model} as a function of the two new parameters introduced. The parameters $\Omega_M^0$ and $\Omega_\Lambda^0$ were set to the values 0.3 and 0.7 respectively. The red regions are non-physical with $\delta_G \delta_\Lambda <0$. Equation (\ref{eq:rela} has been also represented in blue.}
\label{fig:3D} 
\end{figure}

\begin{figure}[h]
\centering
\includegraphics[width=0.9\textwidth]{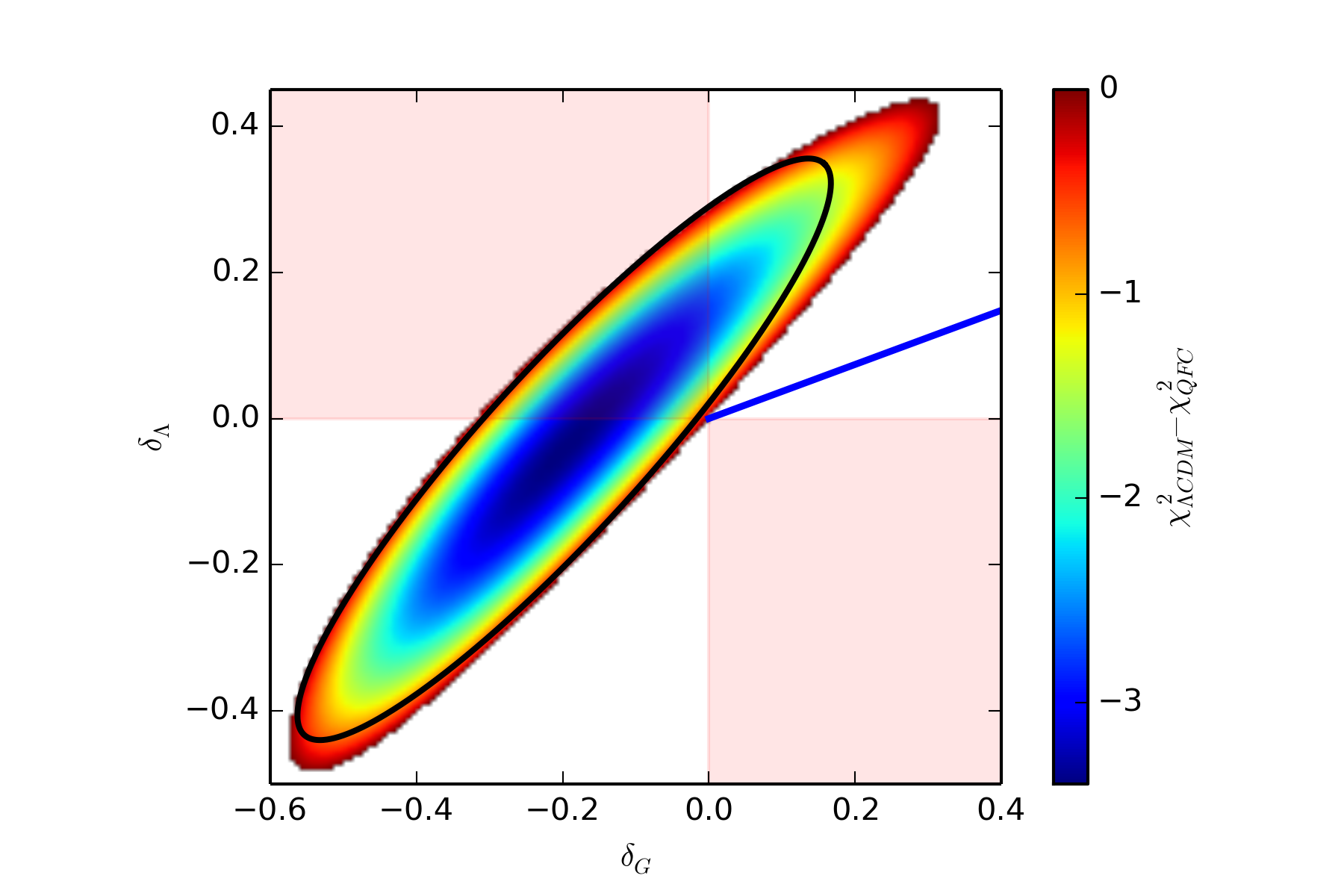}
\caption[Parameter space study of the interacting dark energy model (2/2)]{Contour which obtains a better $\chi^2$ than the \glslink{SM}{FLRW model} as a function of the two new parameters introduced. The parameters $\Omega_M^0$ and $\Omega_\Lambda^0$ were set to the values 0.27 and 0.73 respectively. The red regions are non-physical with $\delta_G \delta_\Lambda <0$. The linear relation (\ref{eq:rela}) is also displayed in blue}
\label{fig:4} 
\end{figure}

On figure \ref{fig:3D}, we plotted the difference of $\chi^2$ between the \acrshort{QFC} and the \glslink{SM}{FLRW model}. It is found that a large zone of the parameter space allows for the \acrshort{QFC} model to have a smaller $\chi^2$ than the one of the \glslink{SM}{FLRW model}. Figure \ref{fig:3D} also shows the different quadrants allowing a physical \acrshort{QFC} models satisfying the constraint $ \delta_G \delta_{\Lambda} >0$. The linear relation (\ref{eq:rela}) is also displayed in blue but we see that no value for $\delta_g$ and $\delta_{\Lambda}$ allows for a better $\chi^2$ than the \glslink{SM}{FLRW model} together with the relation (\ref{eq:rela}).
Continuing investigating different values of the matter content of the universe, we ran a similar type of program when $\Omega_M$ is fixed to the best fit value for the \acrshort{QFC} model: $\Omega_M=0.27$, the result is displayed in figure \ref{fig:4}. In this case, we find an intersection between the values of the parameters giving a smaller $\chi^2$ than the \glslink{SM}{FLRW model} and the relation (\ref{eq:rela}).

\section{Discussion and conclusion}
\label{sec:ccl}
We presented a new model for interacting \gls{dark energy} arising from high energy physics in Sec.~\ref{sec:SFC} and explored its observational consequences on\glslink{SNe Ia}{SNe Ia}. We find that it obtains a better $\chi^2$ than the standard \glslink{SM}{$\Lambda$CDM model}. Motivated by this result, we explored the parameter space for this model to know for which value of the parameters, the model has a better $\chi ^2$ than its \glslink{SM}{FLRW} counterpart. We found that if one decreases slightly the matter content of the universe, a substantial zone of the parameter space offers a smaller $\chi^2$ than the \glslink{SM}{FLRW model}. It could be a possible hint pointing to the need to consider a correction due to interacting \gls{dark energy} to the halo bias for \glslink{S}{structure formation}. It is not the first time than decent alternative based on time variation of the \gls{cosmological constant} and interaction between \gls{dark matter} and \gls{dark energy} are explored. Our model provide one more example of such an alternative. 

The model of \cite{Xue:2014kna} is also constrained by other independent experiments which constrain the effective variation of the gravitational constant: by Lunar Laser Ranging experiment  \cite{Williams:2004qba}, we find a bound for $\delta_g$:
\begin{equation}
|\delta_g| < 0.02.
\end{equation}
Furthermore the \glslink{BBN}{big-bang nucleosynthesis} gives an even stringent constraint on the variation of the gravitational constant which translates for our parameter into:
\begin{equation}
|\delta_g|< \mathcal{O}(10^{-3}),
\end{equation}
depending on the model used for nucleosynthesis \cite{Iocco:2008va}.
\par This study is a first step towards better exploring the parameter space and setting constraints on the parameters. As already done for other interacting \gls{dark energy} models, it needs to be challenged with different independent data set such as \acrshort{BAO}. An important extension of this work would be to follow the research plan carried for instance in \cite{Sola:2016ecz} where similar models were confronted to 5 different data sets. We stress that to do so for this model, one needs not only to consider the dynamics explored in equations (\ref{re3}) and (\ref{e2}) which is only valid for small redshift but a more general one given by equation (36) of \cite{Xue:2014kna}.

%% file: concl.tex
\phantomsection
\addcontentsline{toc}{chapter}{Conclusion}
\chapter*{Conclusion}
\epigraph{It is difficult to make predictions, especially about the future}{Anonymous danish (sometimes attributed to N.~Bohr)}
This last chapter finished the presentation of the different results obtained in this thesis. As a physicist studying theoretical problems, it is sometimes hard to see, as we say in french, the end of tunnel. That is to know whether the results derived have any chance to be observed one day as being part of our physical world. In this sense some days, I felt during those past three years as a science-fiction writer studying other realities. No matter how hard you work how rigorous and imaginative you are, sometimes the world is just not the way you believed or the observational confirmation of your theories will come much later. In this sense asking for a general conclusion needs time.

I believe quantum processes in accelerated period of expansion will be of great importance in order to understand our universe near the initial singularity. The careful generalization to the presence of an electric field could be of importance for the generation of primordial electromagnetic field, the asymmetry matter/antimatter and for the propagation of little perturbations in the early universe together with the understanding of the general dynamics of our newly born universe.

The presence of inhomogeneities on the larger scales is a fact, their impact on cosmic measured quantities is still under debate. This path to the understanding of the cosmos is difficult as physicists prefer usually to consider easy problems with a limited number of particle or particles of all the same type. When the simplest try does not work, the physicist tries then to perturb a little bit the simplest one to make it work. It is this very idea that we illustrated by studying an interacting dark energy model but for these ideas to hold true a definite microscopical justification of the type of interaction has to be provided.

 Harder problems require tools which might not be in the physicist palet but which are more familiar to chemist, geologist or sociologist. Or tools, coming back to the specific example of large scale inhomogeneities, which require advanced computational techniques. Hence this path through the complexity is sometimes ignored or underrepresented because it does not fit the physicist mind which prefers simpler problems. I would hence not be surprised that the existance of dark energy or dark matter will be questioned in the next decades toward an explanation requiring more complexity than physicist usually deals with. Finishing with this aspect, all the physics actually function in this way and all (most) of the linear problems have been solved already: the ones which remain are the hard ones: NP-problem such as the N-body problem, the physics of piles of sand, biological systems, networks, human beings...

%% file: cours/appendix.tex
\epigraph{Never! Never, Marge. I can't live the button-down life like you. I want it all: the terrifying lows, the dizzying highs, the creamy middles. Sure, I might offend a few of the bluenoses with my cocky stride and musky odors -- oh, I'll never be the darling of the so-called ``City Fathers'' who cluck their tongues, stroke their beards, and talk about What's to be done with this Homer Simpson?}{Homer Simpson, in Lisa's Rival (1994), Matt Groening}

\section{Classification of the topologies of cosmological spaces}
\label{ap:topo}
\begin{figure}[h]
 \center
      \includegraphics[scale=0.8]{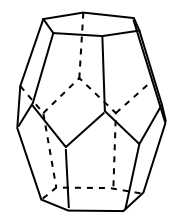}
\caption[Example of a fundamental polyhedron]{For Löbell topology \cite{lobell1931beispiele,gott1980chaotic} the fundamental cell is given is by the 14-hedra displayed in this figure. All the angle in the figure should be right angles in $\mathbb{H}^3$. To obtain the full fundamental polyhedron of \gls{lb}, one should glue together 8 of those 14-hedra. In fact it is even possible to construct an infinite number of compact hyperbolic 3-space by gluing together various number of these 14-hedra and identifying the unattached faces.}
        \label{fig:ploy}
        \end{figure}

\paragraph{}  Studying \gls{topology} consists, crudely speaking, in counting the number of holes in a given space. More precisely, the job of the topologist is to characterize properties of a given \gls{manifold} which are invariant by a \textit{continuous} transformation such as stretching or bending but not cutting or gluing. \Gls{topology} is hence not at all interested in distances which we see already might be a problem in cosmology as the universe might be so big that its topological properties might be so far away that we will never observe them. As a matter of examples, a topologist cannot differentiate a square from a star or a triangle, but knows well the difference between a beer glass and a bowl. In order to characterize the number of holes, one needs to define invariant quantites, which can be just numbers such as dimension of a \gls{manifold}, the degree of connectedness (aka. the Poincaré-Euler characteristic), or whole mathematical objects such as the homology group, the homotopy group, the \gls{holonomy group} that we will discuss later. The whole job of the topologist is to fully characterize different possible topologies and to place any given \gls{manifold} in its given class. So far this has only been done for $2D$ and $3D$ closed and flat surfaces.
\paragraph{} In order to count the number of holes, we introduce loops, which are paths going from one point to another on a \gls{manifold}. Two loops are said to be homotopic if one can continuously deform one to another. A \gls{manifold} is said to be simply connected if all loops on a \gls{manifold} are homotopic (to a point) and multi-connected otherwise. The group of equivalence classes of homotopic loops is the fundamental group $\pi_1(\mathcal{M})$. It is a topological invariant and is particularly well suited for 2-surfaces as loops are 1$D$ structures, for instance the fundamental group is trivial for simply connected space in 2$D$. For higher dimensional space, the fundamental group is not enough to determine the connectedness of a \gls{manifold} and a whole branch of mathematics called \gls{algebric topology} investigates the possibility of higher dimensional groups than the fundamental group. In 1904, Poincaré asked whether it is possible to have a 3$D$ compact \gls{manifold} without boundary with a trivial fundamental group not topologically equivalent to the sphere $\mathbb{S}^3$. This question kept mathematician busy for a little while and the answer has been shown to be negative in 2002 by Perelman \cite{Perelman02theentropy}. The analogous higher dimensional problems were already proved at that time. It gave a new characterization of a sphere.
\paragraph{}
Our spacetime is described in the \glslink{SM}{standard model} of cosmology by a 4$D$ lorentzian \gls{manifold}: $ \mathcal{M} = \mathbb{R} \times \Sigma_3$, where the spatial section $\Sigma_3$ have constant curvature as a direct consequence of the \gls{cosmological principle}. \textit{Locally}, $\Sigma_3$ can have three different structures: a 3-sphere $\mathbb{S}^3$, the euclidian space $\mathbb{E}^3$ or a 3-Hyperboloid $\mathbb{H}^3$. Those three spaces are simply connected and are called covering space. They have the same geometry as $\Sigma_3$. However other possibilities of covering space that we will denote $\mathbb{X}$ exist. To study them, it is convenient to define the \gls{holonomy group} $\Gamma$ such that:
\begin{equation}
\Sigma_3 = \mathbb{X} / \Gamma.
\end{equation}
The \gls{holonomy group} is a discrete sub-group of the \gls{isometry} group\footnote{Some other properties of the \gls{isometry} group are also discussed in Sec.~\ref{sec:groupth}.} $G$, if $\Gamma$ is trivial, the space is simply connected. The \gls{holonomy group} acts discretely and has no fixed points, so no rotations can be element of $\Gamma$. Since its elements are \glslink{isometry}{isometries}, they satisfy by definition:
\begin{equation}
\forall x,y \in \Sigma_3, \forall g \in \Gamma, \text{d}(x,y) = \text{d}(g(x),g(y)),
\end{equation}
where $d$ is the distance measure on the \gls{manifold} usually defined as the $\inf$ of the size of all curves between two points. Classifying the different topological spaces of $\Sigma_3$, is equivalent to classify the sub-groups $\Gamma$ as $\pi_1(\Sigma_3$) is isomorphic to $\Gamma$. Furthermore it simplifies the problem to consider the fundamental polyhedrons for which the fundamental group is isomorphic to $\Gamma$, those polyhedrons are convex and have a even number of faces related by pairs by the generators of the \gls{holonomy group}. An example of fundamental polyhedron is given in figure \ref{fig:ploy}.
        The \gls{holonomy group} can be thought of a set of instructions for identifying the faces of the fundamental polyhedron from which the \gls{manifold} can be reconstructed. The total classification of polyhedrons and hence spaces depends on the curvature: in 3$D$ flat spaces, there exist 17 different topological spaces which cosmological application were studied in \cite{Riazuelo:2003ud}, for closed spaces, the topologies are also all known and investigated for cosmology in \cite{Gausmann:2001aa}, however in hyperbolic spaces, the question of the classification of the topological space is open.
        \newpage
        \section{Numerical value of the cosmological parameters}
        \begin{table}[!h] 
        \begin{tabular}{|c|c|c|c|}
\hline
         Physical parameter & Numerical value & Reference & Section encountered \\
\hline      
\hline
 $H_0$ & $67.8 \pm 0.9$ km.s$^{-1}$.Mpc$^{-1}$ & \cite{Ade:2015xua} & \ref{sec:expofu} \\
\hline
$\Omega_{\text{rad}}^0$ & $ (9.16 \pm 0.24) \times 10^{-5}$ & with \cite{Ade:2015xua} & \ref{sec:compoo} \\
\hline
$\Omega_{\text{b}}^0$ & $ 0.049 \pm 0.005 $ & with \cite{Ade:2015xua} & \ref{sec:compoo} \\
\hline
$\Omega_{\text{CDM}}^0$ & $ 0.268 \pm 0.013 $ & with \cite{Ade:2015xua} & \ref{sec:compoo} \\
\hline
$\Omega_{\text{K}}^0$ & $ -0.005^{+0.016}_{-0.017} $ & \cite{Ade:2015xua} & \ref{sec:compoo} \\
\hline
$\Omega_{\text{DE}}^0$ & $ 0.683 \pm 0.013 $ & \cite{Ade:2015xua} & \ref{sec:compoo} \\
\hline
$t_{\text{BB}}$ & $13.799 \pm 0.021 \text{ Gyr}$ & \cite{Ade:2015xua} & \ref{sec:compoo} \\
\hline
$n_S$ & $0.968 \pm 0.006 $ & \cite{Ade:2015xua} & \ref{sec:quasi} \\
\hline
        \end{tabular}
        \caption{Numerical value of some useful cosmological parameters}
        \label{table:cosmnum}
        \end{table}
\section{Derivation of the Friedmann equation}
\label{app:fried}
In this appendix, we provide a derivation of (\ref{eq:Fried1}) and (\ref{eq:Fried2}) from \glspl{Einstein equation}.
Using (\ref{FLRW1}) and (\ref{eq:defCHR}), the \glspl{Christoffel symbol} are:
\newcommand\Gamzero{\Gamma^{0}_{\mu\nu}=\left(
\begin{array}{cccc}
0 & 0 & 0 & 0 \\
0 & \frac{a\dot{a}}{1-kr^2} & 0 & 0 \\
0 & 0 & a\dot{a}r^2 & 0 \\
0 & 0 & 0 & a\dot{a}r^2 \sin^2\theta \\
\end{array}\right)}

\newcommand\Gamun{\Gamma^{1}_{\mu\nu}=\left(
\begin{array}{cccc}
0 & \frac{\dot{a}}{a} & 0 & 0 \\
\frac{\dot{a}}{a} & \frac{\dot{kr}}{1-kr^2} & 0 & 0 \\
0 & 0 & r(kr^2-1) & 0 \\
0 & 0 & 0 & (r\sin^2\theta)(kr^2-1) \\
\end{array}\right)}

\newcommand\Gamdeux{\Gamma^{2}_{\mu\nu}=\left(
\begin{array}{cccc}
0 & 0 & \frac{\dot{a}}{a} & 0 \\
0 & 0 & \frac{1}{r} & 0 \\
\frac{\dot{a}}{a} & \frac{1}{r} & 0 & 0 \\
0 & 0 & 0 & -\sin\theta\cos\theta\\
\end{array}\right)}

\newcommand\Gamtrois{\Gamma^{3}_{\mu\nu}=\left(
\begin{array}{cccc}
0 & 0 & 0 & \frac{\dot{a}}{a} \\
0 & 0 & 0 & \frac{1}{r} \\
0 & 0 & 0 & \cot\theta \\
\frac{\dot{a}}{a} & \frac{1}{r} & \cot\theta & 0 \\
\end{array}\right)} 
\begin{equation}
\left\{
\begin{array}{llll}
 \Gamzero \\
 \Gamun \\
 \Gamdeux \\
 \Gamtrois \\
\end{array}
\right.
\end{equation}

Hence with the \glspl{Christoffel symbol} together with (\ref{riemann}) and (\ref{ricci}), we deduce the \gls{Ricci tensor} which is diagonal:
\begin{equation}
\left\{
\begin{array}{l}
 R_{00}=\frac{3\ddot{a}}{a} \\
 R_{11}=\frac{a\ddot{a}+2\dot{a}^2+2k}{kr^2-1} \\
 R_{22}=-r^2(a\ddot{a}+2\dot{a}^2+2k) \\
 R_{33}=R_{22}\sin^2\theta \\
\end{array}
\label{eq:rigds}
\right.
\end{equation}
We can also deduce the curvature scalar $R$ given by (\ref{courbure}):
\begin{equation}
R=R_{00}-\frac{1-kr^2}{a^2}R_{11}-\frac{1}{a^2r^2}R_{22}-\frac{1}{a^2r^2\sin^2\theta}R_{33}.
\label{eq:expressricc}
\end{equation}
Using (\ref{eq:rigds}) in (\ref{eq:expressricc}) gives:
\begin{equation}
\label{Rf}
R=6\left(\frac{\ddot{a}}{a}+\left(\frac{\dot{a}}{a}\right)^2+\frac{k}{a^2}\right).
\end{equation}
Finally we can build the \gls{Einstein tensor} given by (\ref{einsteinTenseur}):
\begin{equation}
\label{eq:einstt}
\left\{
\begin{array}{l}
G_{00}=-3((\frac{\dot{a}}{a})^2+\frac{k}{a^2})\\
G_{11}=\frac{a^2}{1-kr^2}(2\frac{\ddot{a}}{a}+(\frac{\dot{a}}{a})^2+\frac{k}{a^2})\\
G_{22}=a^2r^2(2\frac{\ddot{a}}{a}+(\frac{\dot{a}}{a})^2+\frac{k}{a^2})\\
G_{33}=\sin^2\theta G_{22}\\
\end{array}
\right.
\end{equation}
the non diagonal terms vanishing. Now that we have the left hand side of the \glspl{Einstein equation} (\ref{eq:Eisteq}), we assume that the universe is filled with various \glspl{perfect fluid}:
\begin{equation}
T_{\mu\nu}=(\rho +P)\delta^{0}_{\mu}\delta^{0}_{\nu}-Pg_{\mu\nu}.
\label{eq:perfectf}
\end{equation}
The assumption of idealized \glspl{perfect fluid} is common in cosmology. It implies that that on larger scales, the effect of viscosity, anisotropic stress or shear, or diffusive heat transport are negligible. Using the (\ref{eq:einstt}) and (\ref{eq:perfectf}), the \glspl{Einstein equation} (\ref{eq:Eisteq}) becomes:
\begin{align}
& H^2 \equiv (\frac{\dot{a}}{a})^2=\frac{8\pi G}{3}\rho-\frac{k}{a^2}+\frac{\Lambda}{3}, \label{equafried1}\\
& \frac{\ddot{a}}{a}=-\frac{4\pi G}{3}(\rho+3P)+\frac{\Lambda}{3},
\label{equafried2}
\end{align}
where we also incorporated a \gls{cosmological constant} term of the form (\ref{eq:defCC}) following the remark that it is compatible with the \glspl{Einstein equation} (\ref{eq:Eisteq}). By taking the derivative of (\ref{equafried1}) and using (\ref{equafried2}), we get the equation of continuity:
\begin{equation}
\dot{\rho}+\frac{3\dot{a}}{a}\rho=0
\label{equarho}.
\end{equation}

%% file: cours/appendixinfl.tex
\epigraph{This party's over}{Mace Windu in Star Wars: Episode II Attack of the Clones, 22 BBY}
In this appendix, we will fill some details which were eluded in Sec.~\ref{sec:quasi}. We will present two computations to generalize (\ref{eq:psmassles}), one including a general potential together with a quasi-\gls{dS}. A second where we will consider a even more general scalar field which could have a non minimal kinetic term. The strategy will be similar to the one applied in Sec.~\ref{sec:massless}:
\begin{itemize}
\item Establish the equation of motion
\item Solve the equation of motion in the regime of short wave length
\item Determine the initial conditions: the \gls{Bunch-Davies vacuum}
\item Solve the equation in the long wave length regime, the mode will be typically frozen
\item Calculate the power spectrum at the end of \gls{inflation}
\end{itemize}
To avoid heavy expressions, we adopt only for this appendix a new convention for derivative: $\frac{\partial f}{\partial X} \equiv f,_X$.
\section{Inflation with a general potential}
\label{sec:IWAGP}
\subsection*{Establishing the set of differential equations generalizing (\ref{eq:KGmassless})}
We consider a scalar field $\varphi$ with potential $V(\varphi)$. The action is: 
\begin{equation}
\label{eq:actmore}
S=\int \mathrm{d}^{4}x \sqrt{-g} \left[\frac{1}{2}\partial_{\mu}\varphi \partial^{\mu}\varphi-V(\varphi)\right].
\end{equation}
Varying the action with respect to $\varphi$, we get the Klein-Gordon equation: 
\begin{equation}
\label{KG}
\frac{1}{\sqrt{-g}}\frac{\partial}{\partial x^{\alpha}}\left(\sqrt{-g} g^{\alpha\beta} \frac{\partial \varphi}{\partial x^{\beta}} \right) + \frac{\partial V}{\partial \varphi} = 0.
\end{equation}
Considering the SVT decomposition in (\ref{g}), we will work in the Newton or longitudinal gauge where $\phi=\phi_l, \psi=\psi_l, B=E=0,$ so the metric is diagonal. Hence the Klein-Gordon equation to linear order in the \gls{perturbation} becomes:
\begin{equation}
\label{perturbation}
\delta \varphi ''+ 2\mathcal{H} \delta \varphi' -\Delta \delta \varphi - \varphi_{0}'(3\psi_{l}+\phi_{l})'+\delta \varphi V_{,\varphi \varphi} a^{2} + 2a^{2} \phi_{l} V_{,\varphi}=0.
\end{equation}
This equation describes the evolution of the \glspl{perturbation} in an expanding universe in the longitudinal gauge. The 3 unknowns are: $\delta \varphi, \phi_l, \psi_l$. To get gauge invariant equations, we promote the variables to their corresponding gauge invariant quantities: \begin{equation} \label{eq:gaugeINV} \begin{split} & \delta \varphi \rightarrow \overline{\delta \varphi} \equiv \delta \varphi - \varphi_{0}'(B-E'),\\ 
& \phi_{l} \rightarrow \Phi \equiv \phi_{l}-\frac{1}{a}\left[a(B-E')\right]', \\ 
&\psi_{l} \rightarrow \Psi \equiv \psi_{l} + \frac{a'}{a} (B-E').\\ \end{split} \end{equation} We still need to find 2 others equations to be able to solve the problem. We will use \glspl{Einstein equation} (\ref{eq:Eisteq}) and for this we will have to compute the \gls{energy momentum tensor}. Varying (\ref{eq:actmore}) with respect to the metric $g_{\mu \nu}$ gives: \begin{equation}
T_{\alpha}^{\beta}=\partial_{\alpha}\varphi \partial^{\beta} \varphi - \delta_{\alpha}^{\beta} \mathcal{L}.
\end{equation}
Aware of this, it is possible to compute the variation of the \gls{energy momentum tensor}. Computing $\delta T_{i}^{0}$ and $\delta T_{i}^{j}$ and using the \glspl{Einstein equation} (\ref{eq:Eisteq}) give a system of three independent equations:
\begin{equation}
\label{syst}
\left\{ \begin{split} &
\overline{\delta \varphi }''+ 2\mathcal{H} \overline{\delta \varphi}' -\Delta \overline{\delta \varphi} - \varphi_{0}'(3\Psi+\Phi)'+\overline{\delta \varphi} V_{,\varphi \varphi} a^{2} + 2a^{2} \Phi V_{,\varphi}=0 \\
&\Psi' + \mathcal{H} \Phi = 4\pi \varphi_{0}' \delta \overline{\varphi} \\
&\Psi=\Phi \\
\end{split} \right.
\end{equation}
Using (\ref{eq:gaugeINV}) and (\ref{syst}), we find the gauge invariant \gls{Bardeen potential} (\ref{eq:Bardeen}) which describe the gravity sector. Observe that the results of Sec.~\ref{sec:massless} are shed with a new light by remarking that the de Sitter phase is recovered for $\varphi_{0}'=0$ which gives zero gravitational potentials: $\Phi=\Psi=0$. This does not mean that the scalar doesn't fluctuate but just that its fluctuations don't couple to the geometry and don't generate any metric fluctuations. This also explains why the de Sitter limit will be often singular in the equation.
\subsection*{Solving (\ref{syst})}
\subsubsection{For sub-Hubble modes}
We assume that the wavelength of the \glspl{perturbation} is small compared to the \gls{Hubble radius} ($k|\tau|\gg 1$) and we look for oscillatory solutions so that $\Phi' \sim k \Phi$. hence using the second equation of (\ref{syst}) we get:
\begin{equation}
\Phi= \frac{\varphi_{0}' \overline{\delta \varphi}}{k}.
\end{equation} 
We neglect then, for large $k$, in the first equation of (\ref{syst}), the last three terms. Assuming the condition $V,_{\varphi \varphi} \ll V $, we finally have: 
\begin{equation}
\overline{\delta \varphi}_{k}''+ 2\mathcal{H} \overline{\delta \varphi}_{k}' +k^{2} \overline{\delta \varphi}_{k}=0.
\end{equation}
To solve this equation, we introduce the auxiliary field $u_{k}= a\overline{ \delta \varphi}_{k}$: 
\begin{equation}
\label{eq:geneapp}
u_{k}''+ \left(k^{2}-\frac{a''}{a}\right)u_{k}=0.
\end{equation}
This equation is a generalization of (\ref{eq:KGmassless}) but the approximation $k|\tau|\gg 1$ allows us to neglect the mass term in $\frac{a''}{a}$, we then get an harmonic oscillator so the different modes don't interact one to each others. This is coherent with the physical assumption, we made, which is that we begin on scale smaller than the curvature scale. The solution is:
\begin{equation}
\label{solution}
\overline{\delta \varphi}_{k}=\frac{c_{k}}{a}\exp(\pm i k \tau).
\end{equation}
We will now determine the integration constants $c_{k}$ as the minimum allowed by the \glslink{Heisenberg uncertainty principle}{Heisenberg principle}. The physical ingredient for that is \gls{vacuum fluctuations} of the inflaton.
\subsubsection{Vacuum fluctuations} We assume that the field is nearly homogeneous in a volume $V \sim L^{3}$. The action can then be written $S=\int dt(\frac{\dot{X}}{2}+...)$, where $X \equiv \varphi L^{\frac{3}{2}}$. We will use $X$ as the canonical quantization variable with conjugate momentum $P=\dot{X}=\frac{X}{L}$. Hence using Heisenberg uncertainty relation $\Delta X \Delta P \sim 1$, we get the minimal amplitude $\delta \phi \sim L^{-1}$. With (\ref{solution}) and the definition of the comoving wavenumber $k \equiv \frac{a}{L}$, we get:
\begin{equation}
\overline{\delta \varphi}_{k}=\frac{1}{a\sqrt{k}}\exp(\pm i k \tau).
\end{equation}
Since the modes are independent, the \gls{vacuum} is preserved. 
\subsubsection{Long wavelength}
In this section, we assume that one can neglect\footnote{Once the solution is found, it can be checked that those terms are negligible.} terms proportional to $\ddot{\delta \varphi}$ and $\dot{\Phi}$. The equations for \glspl{perturbation}  (\ref{syst}) become in cosmic time (\ref{time}):
\begin{equation}
\left\{ \begin{split} &
3H \dot{\overline{\delta \varphi}} +\overline{\delta \varphi} V_{,\varphi \varphi}  + 2 \Phi V_{,\varphi}=0 \\
&H \Phi = 4\pi \dot{\varphi_{0}} \delta \overline{\varphi} \\
\end{split} \right.
\end{equation}
We introduce $y \equiv \frac{\overline{\delta \varphi}}{V,_{\varphi}}$ and remembering the fact that \gls{Friedmann equation} for \gls{inflation} is $H^{2} \sim 8 \pi V$, we get $\frac{\mathrm{d}(yV)}{\mathrm{d}t}=0$. After integrating we get, in term of $\Phi$ and $\overline{\delta \varphi}$.
\begin{equation}
\label{eq:finnnn}
\left\{ \begin{split}
&\delta \varphi _{k}=A_{k}\frac{V,_{\varphi}}{V} \\
&\Phi _{k} = -\frac{1}{2} A_{k} \left(\frac{V,_{\varphi}}{V}\right)^{2} \\
\end{split} \right.
\end{equation}
To determine the integration constant, we use the previous part with the short wavelength solution and get $A_{k} \sim \left[\frac{1}{a(\tau) \sqrt{k}} \left(\frac{V}{V,_{\varphi}}\right) \right]_{k \sim Ha}$, where $k \sim Ha$ means that the quantity is evaluated at the horizon crossing, it is the equivalent of $k \tau$ in the case of de Sitter of Sec.~\ref{sec:massless} but as $H=H(\tau)$ it is reformulated in $k \sim a H$. In the limit $k \ll aH$ the solution $\Phi_k(\tau)$ is constant; the mode are frozen and $\frac{V,_{\varphi}}{V}$ is constant and of order unity. In this case, $\Phi(\tau) \sim A_k$ and the \gls{power spectrum} reads:
\begin{equation}
\label{eq:PSAP}
P_{\Phi}(k)=\left(\frac{H}{2 \pi}\right)^2  \frac{1}{2}\left(\frac{V}{V,_{\varphi}} \right)^2_{k \sim Ha}.
\end{equation}
As said below equation (\ref{syst}), the limit to \gls{dS} is singular. To make the connection with (\ref{Final power spectrum}), it can be shown that $\left(\frac{V,_{\varphi}}{V}\right)^2= 2\epsilon$. We will now turn to a even more general calculation as we do not assume anymore that there is a canonical kinetic term in the action.
\section{Inflation with non minimal kinetic term:}
We start with a flat universe filled with a scalar field condensate described by the action: 
\begin{equation}
\label{eq:actflui}
S=\int \mathrm{d}^{4}x \sqrt{-g} \text{ }p(X,\varphi),
\end{equation}
where $X \equiv \frac{1}{2}\partial_{\mu}\varphi \partial^{\mu}\varphi$ and the Lagrangian plays the role of the pressure. The \gls{energy momentum tensor} is then written in the form of an ideal fluid:
\begin{equation}
T_{\mu \nu} \equiv \frac{\partial S}{\partial g^{\mu \nu}}=(\rho+p)u_{\mu}u_{\nu}-g_{\mu \nu} p,
\end{equation}
where the energy density is given by $\rho \equiv 2X p,_{X} - p $ and the velocity of the fluid is $u_{\mu} \equiv \frac{\varphi,_{\mu}}{\sqrt{2X}}$. The action (\ref{eq:actflui}) can describe many different fluids, the canonical scalar field is described if $P = X- V(\varphi)$ and thus $\rho = X+ V$. If $P=X^{n}$, we find $P=\rho/(2n-1)$ recovering many cases studied in table \ref{tab:flu}.
\subsection*{Establishing the set of differential equations generalizing (\ref{syst})}
The steps to derive the set of differential equations that govern the dynamics of the \glspl{perturbation} are the same: work in the Newton gauge, calculate $\delta X$, $\delta T^{0}_{i}$, $\delta T^{0}_{0}$, $\delta T^{i}_{j}$ \footnote{$\delta T^{i}_{j}$ will be found to be still equal to zero, we get the same conclusion that the two \glspl{Bardeen potential} are equal $\Phi=\Psi$. It is only beyond \gls{general relativity} that the two gravitational are different.}. The final answer corresponding to (\ref{syst}) in the previous part is:
\begin{equation}
\label{motion}
\left\{ \begin{split}
&c_{s} \Delta u = z \left(\frac{v}{z}\right)' \\
&c_{s} v= \theta \left(\frac{u}{\theta}\right)' \\
\end{split}
\right.
\end{equation} 
where we define:
\begin{align}
& c_{s}^{2} \equiv \frac{P,_{X}}{\rho,_{X}},  \label{eq:cs} \\
& u \equiv \frac{\Psi}{4\pi\sqrt{\rho +p}}, \\
& v \equiv \sqrt{\rho,_{X}} a(\tau) \left( \overline{\delta \varphi} + \frac{\varphi_{0}'}{\mathcal{H} \Psi}\right), \\
& z \equiv \frac{a(\tau)^{2} \sqrt{\rho+p}}{c_{s} \mathcal{H}}, \label{eq:defz}\\
&\theta \equiv \frac{1}{c_{s}z}.
\end{align} $c_s$ is the speed of sound, if it is equal to $c=1$ when $p=X$ which correspond to the case of Sec.~\ref{sec:massless}. By isolating $u$, we get the equivalent of (\ref{eq:KGmassless}):
\begin{equation}
u''-c_{s}^{2} \Delta u -\frac{\theta''}{\theta}u=0.
\end{equation}
\subsection*{Solving (\ref{motion})}
We will solve the equation in two regimes: short wavelength which is now defined as $c_{s}^{2}k^{2} \gg \left|\frac{\theta''}{\theta}\right|$ and the long wavelength: $c_{s}^{2}k^{2} \ll \left|\frac{\theta''}{\theta}\right|$ . We sum up the result in table \ref{tab:sol}. 
\begin{table}[h]
\begin{tabular}{|l|l|l|}
  \hline
   & short wavelength & long wavelength \\
  \hline
  $\Phi$ & $4\pi c \dot{\varphi_{0}}\sqrt{\frac{p,_{X}}{c_{s}}}\exp\left(\pm ik\int \mathrm{d}t \frac{c_{s}}{a}\right)$ & $A \frac{\mathrm{d}}{\mathrm{d}t}\left(\frac{1}{a} \int a \mathrm{d}t \right) =A\left(1- \frac{H}{a}\int a \mathrm{d}t \right) $\\
  $\overline{\delta \varphi}$ & $c\sqrt{\frac{1}{c_{s}} P,_{X}}\left(\pm \frac{i c_{s} k}{a} + H+.. \right) \exp\left( \pm ik \int \frac{c_{s}}{a} \mathrm{d}t \right)$ & $A \dot{\varphi_{0}} \frac{1}{a}\int a \mathrm{d}t$ \\
  \hline
\end{tabular}
\caption{General solutions for the vacuum fluctuations in quasi de Sitter spacetime.}
\label{tab:sol}
\end{table}

$c$ and $A$ are integration constant to be determined. Discarding again the higher time derivative and integrating by part in the long wavelength regime. We get to the leading order: 
\begin{equation}
\label{sloroll}
\left\{ \begin{split}
& \Phi=- \frac{A \dot{H}}{H^{2}} \\
& \overline{\delta \varphi}= \frac{A \dot{\varphi_{0}}}{H} \\
\end{split}
\right.
\end{equation}
Finally we assume at the end of \gls{inflation} a period dominated by \gls{radiation}, that is $a \sim t^{1/2}$, using table \ref{tab:sol} and neglecting the decaying mode, we get that $\Phi= \frac{2A}{3}, \overline{\delta \varphi}=\frac{2A t \dot{\varphi_{0}}}{3}$ where $A$ is to be determined. We see that in this gauge, the gravitational potential freezes out after \gls{inflation}. Determining $A$ with (\ref{sloroll}) and assuming \glspl{perturbation} leave the horizon during \gls{inflation}, we get the final result:
\begin{equation}
\Phi= \frac{2}{3} \left( H \frac{\overline{\delta \varphi}}{\dot{\varphi_{0}}} \right)_{c_{s}k \sim Ha}.
\end{equation} 
This is consistent with (\ref{eq:finnnn}) and is a little bit more general since we can use it to calculate \glspl{perturbation} in theories with non-minimal kinetic term.

The equation of motion derived from (\ref{eq:actmoremore}) namely (\ref{eq:eomag}), can be obtain also by isolating $v$ in  (\ref{motion}) showing that the approach followed in this appendix and in the main text are equivalent.
\subsection*{Power spectrum} Using (\ref{motion}) and similar steps as in Sec.~\ref{sec:IWAGP}, one can find an expression for the \gls{power spectrum}:
\begin{equation}
P_{\Phi}(k)=4 (\rho + p) |u_{k}|^{2} k^{3}.
\end{equation}
Using the result for long wavelength solution for $\Phi$ in table \ref{tab:sol}, integrating it by part and neglecting higher order terms we get: $u_{k}(\tau)=\frac{A_{k}}{4 \pi \sqrt{\rho+p}}\frac{\dot{H}}{H^{2}}$. With the \glspl{Friedmann equation} (\ref{eq:F1infl}) and (\ref{eq:R2infl}), we get:
\begin{equation}
u_{k}(\tau)=\frac{A_{k} \sqrt{\rho + p}}{H^{2}}.
\end{equation}
To determine $A_{k}$, we use the result for the short wavelength solution for $\Phi$ in table \ref{tab:sol}, (\ref{motion}) taken at $\eta_{i}$, the initial time of \gls{inflation}, together with (\ref{CI}), we get $u_{k} \cong -\frac{i}{\sqrt{c_{s}} k^{3/2}}\exp \left(i k \int^{\tau}_{\tau_{i}}c_{s} \mathrm{d} \tilde{\tau}\right)$. By a direct comparison, this gives the value of $A_{k}=-\frac{i}{k^{3/2}}\left( \frac{H^{2}}{\sqrt{c_{s}(\rho + p)}} \right)$ and the \gls{power spectrum}:
\begin{equation}
\label{power spectrum}
\delta \Phi^{2} (k,t)=
\left\{ \begin{split}
& \frac{4(\rho +p)}{c_{s}},\text{ short wavelength}\\
& \frac{16}{9}\left(\frac{\rho}{c_{s} (1+\frac{p}{\rho})}\right)_{c_{s}k=Ha}\left(1-\frac{H}{a}\int_{t_{i}}^{t} a \mathrm{d}\tilde{t}\right)^{2},\text{ long wavelength}\\
\end{split}
\right.
\end{equation}
Calculating the integral in (\ref{power spectrum}) for a post inflationary epoch where the mode freezes out, we get for a general inflationary scenario:
\begin{equation}
\label{eq:fpsa}
P_{\Phi}(k)=\frac{64}{81}\left( \frac{\rho}{ c_{s}(1+P/\rho}\right)_{c_{s}k \sim H a}.
\end{equation}
This equation can be found with more details for the derivation in \cite{Mukhanov} p.345. Using (\ref{eq:F1infl}) and (\ref{eq:R2infl}), we find (\ref{Final power spectrum}). 

%% file: current/appendixbess.tex
\epigraph{Mathematics is a collection of cheap tricks and dirty jokes.}{L.~Bers}
\section{\label{app:math}Useful mathematical functions}
In this appendix, we have gathered some useful relations and properties of mathematical functions needed in this thesis.
More can be found in, e.g. \cite{Olver2010}.
\subsection{\label{app:aw}Whittaker functions}
The Whittaker differential equation is
\begin{equation}\label{whittaker}
\frac{d^{2}}{dz^{2}}F(z)+\Big(-\frac{1}{4}+\frac{\kappa}{z}+\frac{1/4-\mu^{2}}{z^{2}}\Big)F(z)=0.
\end{equation}
It has the two linearly independent solutions: $\wwp(z)$ and $\wmp(z)$. Under conjugation they behave as
\begin{align}
 \left[W_{\kappa,\mu}(z)\right]^*=W_{\kappa^*,\mu^*}(z^*),&& \left[M_{\kappa,\mu}(z)\right]^*=M_{\kappa^*,\mu^*}(z^*).\label{eq:Wconjugated}
\end{align}
And they have the following properties known as connection formulas:
\begin{align}
\label{conectionw}
& \wwp(z)=\wwm(z), \\
\label{conectionm}
& \wmp(e^{\pm i\pi}z)=\pm ie^{\pm\mu i\pi}\m_{-\kappa,\mu}(z), \\
& W_{\kappa,\mu}(z)=\frac{\Gamma(-2\mu)}{\Gamma(\frac{1}{2}-\mu-\kappa)}M_{\kappa,\mu}(z)+\frac{\Gamma(2\mu)}{\Gamma(\frac{1}{2}+\mu-\kappa)}M_{\kappa,-\mu}(z), \label{eq:WtoM} \\
& \forall \mu \notin \frac{\mathbb{N}}{2}, \frac{1}{\Gamma(1+2 \mu)}\wmp(z)=\frac{\e^{\pm(\kappa-\mu-1/2)\pi i}}{\Gamma(\frac{1}{2} +\mu+\kappa)}W_{\kappa,\mu}(z) +\frac{\e^{\pm \kappa \pi i}}{\Gamma(\frac{1}{2}+\mu-\kappa)}W_{-\kappa,\mu}(\e^{\pm \pi i}z).
\end{align}
The function \(W_{\kappa,\mu}(z)\) can be expressed with an integral in the complex plane known as the Mellin–Barnes integral representation (valid for $ \frac{1}{2} \pm \mu - \kappa \notin -\mathbb{N}$ and $|\arg z| < \frac{3\pi}{2}$):
\begin{align}
W_{\kappa,\mu}\left(z\right)=\e^{-\frac{z}{2}}\int_{-i\infty}^{i\infty}\frac{ds}{2i\pi}\frac{\Gamma\left(\frac12+\mu+s\right)\Gamma\left(\frac12-\mu+s\right)\Gamma\left(-\kappa-s\right)}{\Gamma\left(\frac12+\mu-\kappa\right)\Gamma\left(\frac12-\mu-\kappa\right)}z^{-s}. \label{eq:MellinBarnes}
\end{align}
By using \(\Gamma(n+1)=\Gamma(n)n\) in (\ref{eq:MellinBarnes}) one can find
\begin{align}
 W_{\mu,\kappa-\frac{1}{2}}(z)=\frac{2\kappa+1-z}{2(\mu^2-\kappa^2)} W_{\mu,\kappa+\frac{1}{2}}(z)-\frac{z}{(\mu^2-\kappa^2)}\frac{dW_{\mu,\kappa+\frac{1}{2}}(z)}{dz}. \label{eq:pmonehalfIdentity}
\end{align}
The asymptotic expansions of the Whittaker functions as $|z|\rightarrow\infty$ are given by
\begin{align}
\label{win}
\wwp(z)& \underset{|z|\rightarrow\infty}{\sim} e^{-\frac{z}{2}}z^{\kappa}, \\
\wmp(z)&\underset{|z|\rightarrow\infty}{\sim}\frac{\Gamma(1+2\mu)}{\Gamma(\frac{1}{2}+\mu-\kappa)}
\,e^{\frac{z}{2}}z^{-\kappa} +\frac{\Gamma(1+2\mu)}{\Gamma(\frac{1}{2}+\mu+\kappa)} \,e^{-\frac{z}{2}\pm(\frac{1}{2}+\mu-\kappa)\pi i}z^{\kappa}, \nn\\
\label{min}
-\frac{1}{2}\pi+\delta&\leq\pm \arg(z)\leq\frac{3}{2}\pi-\delta,
\end{align}
here $\delta$ is an arbitrary small positive constant. In the limit $|z|\rightarrow0$, we have
\begin{align}
\label{mout}
&\wmp(z)\underset{|z|\rightarrow0}{\sim} z^{\frac{1}{2}+\mu}, \\
\label{wout}
&\wwp(z)\underset{|z|\rightarrow0}{\sim}\frac{\Gamma(2\mu)}{\Gamma(\frac{1}{2}+\mu-\kappa)}
z^{\frac{1}{2}-\mu}+\frac{\Gamma(-2\mu)}{\Gamma(\frac{1}{2}-\mu-\kappa)}
z^{\frac{1}{2}+\mu}, & 0\leq\Re(\mu)<\frac{1}{2}, &&\mu\neq0.
\end{align}
Finally, some useful \glslink{Wronskian condition}{Wronskian}\footnote{The Wronskian between two functions $f$ and $g$ is defined as ${\cal{W}}(f,g)=f g'-gf'$, where a prime is the standard derivation.} are
\begin{eqnarray}
\label{wrww}
{\cal{W}}\Big\{\wwp(z),\w_{-\kappa,\mu}(e^{\pm i\pi}z)\Big\}&=&e^{\mp i\pi\kappa}, \\ \label{wrmm}
{\cal{W}}\Big\{\wmp(z),\m_{\kappa,-\mu}(z)\Big\}&=&-2\mu, \\ \label{wrmw}
{\cal{W}}\Big\{\wwp(z),\wmp(z)\Big\}&=&\frac{\Gamma(1+2\mu)}{\Gamma(\frac{1}{2}+\mu-\kappa)}.
\end{eqnarray}
\subsection{\label{app:bessel}Modified Bessel functions}
The modified Bessel function has integral representation along the real line
\begin{equation}\label{bessel}
\I_{\nu}(z)=\frac{z^{\nu}}{2^{\nu}\pi^{\frac{1}{2}}\Gamma(\nu+\frac{1}{2})}
\int_{0}^{\pi}(\sin\theta)^{2\nu}e^{\pm z\cos\theta}d\theta.
\end{equation}
For $n \in \mathbb{N}$,
\begin{equation}\label{parity}
\I_{\nu}(e^{n\pi i}z)=e^{n\nu\pi i}\I_{\nu}(z).
\end{equation}
For $\nu \notin \mathbb{N}^-$, fixed, and $z\rightarrow 0$
\begin{align}\label{origin}
\I_{\nu}(z) \underset{|z|\rightarrow 0}{\sim} \frac{z^{\nu}}{2^{\nu}\Gamma(1+\nu)}.
\end{align}
For $\nu$ fixed and $z\rightarrow\infty$
\begin{align}\label{infty}
\I_{\nu}(z) \underset{|z|\rightarrow\infty}{\sim} \frac{e^{z}}{\sqrt{2\pi z}},
\end{align}
with $ |\arg(z)| \leq\frac{\pi}{2}-\delta$. For $\nu=-\frac{1}{2}$ and $\nu=\frac{1}{2}$, the relations
\begin{eqnarray}
\I_{-\frac{1}{2}}(z)&=&\sqrt{\frac{2}{\pi z}}\cosh(z), \label{cosh} \\
\I_{\frac{1}{2}}(z)&=&\sqrt{\frac{2}{\pi z}}\sinh(z) \label{sinh}
\end{eqnarray}
are satisfied. The following mathematical formulae can be shown
\begin{eqnarray}
\int_{0}^{\pi}(\sin\theta)^{2\nu}d\theta&=&\frac{\sqrt{\pi}}{\Gamma\big(1+\nu\big)}
\Gamma\big(\frac{1}{2}+\nu\big), \label{formul} \\
\lim_{|\lambda|\rightarrow\infty}\int_{\frac{\pi}{2}}^{\pi}(\sin\theta)^{\nu}
e^{-2\pi|\lambda|\cos\theta}d\theta&=&
\frac{\Gamma\big(\frac{\nu+1}{2}\big)}{2(\pi\lambda)^{\frac{\nu}{2}}}e^{2\pi|\lambda|}. \label{formula}
\end{eqnarray}
\subsection{\label{app:spher}Spherical coordinates}
In order to evaluate the integrals (\ref{semidif}) and (\ref{intk}), we use of the spherical coordinates to decompose the momentum vector $\k$ in the flat $d$-dimensional Euclidean space. Hence in this space, the volume element is
\begin{equation}\label{velement}
d^{d}\k=d\Sigma_{d-1}\k^{d-1}dk,
\end{equation}
where $d\Sigma_{d-1}$ is the area element of the unit sphere in the $d$-dimensional Euclidean space $\mathbb{S}^d$. Convenient coordinates
on this sphere are specified by
\begin{eqnarray}\label{cordinat}
\omega^{1}&=&\cos\theta_{1}, \nn\\
\omega^{2}&=&\sin\theta_{1}\cos\theta_{2}, \nn\\
&\vdots& \nn\\
\omega^{d-1}&=&\sin\theta_{1}\cdots\sin\theta_{d-2}\cos\theta_{d-1}, \nn\\
\omega^{d}&=&\sin\theta_{1}\cdots\sin\theta_{d-2}\sin\theta_{d-1},
\end{eqnarray}
where $0\leq\theta_{i}<\pi$ for $1\leq i\leq d-2$ and $0\leq\theta_{d-1}<2\pi$. Then, the metric on the sphere is
\begin{equation}\label{sphere}
d\varpi_{d-1}=\sum_{i=1}^{d}(d\omega^{i})^{2}=d\theta_{1}^{2}+\sin^{2}\theta_{1}d\theta_{2}^{2}+\cdots
+\sin^{2}\theta_{1}\cdots\sin^{2}\theta_{d-2}d\theta_{d-1}^{2},
\end{equation}
and consequently, the area element is
\begin{equation}\label{aelement}
d\Sigma_{d-1}=(\sin\theta_{1})^{d-2}\cdots\sin\theta_{d-2}d\theta_{1}\cdots d\theta_{d-1}.
\end{equation}
Therefore the area of the sphere is
\begin{equation}\label{area}
\Sigma_{d-1}=\int d\Sigma_{d-1}=\frac{2\pi^{\frac{d}{2}}}{\Gamma(\frac{d}{2})}.
\end{equation}
Using Eqs.~(\ref{bessel}) (\ref{formul}) (\ref{aelement}) (\ref{area}) the following formula can be shown
\begin{equation}\label{areaint}
\int d\Sigma_{d-1}e^{2\pi\lambda\cos\theta_{1}}=2\pi^{\frac{d}{2}}(\pi\lambda)^{1-\frac{d}{2}}
\I_{\frac{d}{2}-1}(2\pi\lambda).
\end{equation}
\section{\label{app:int}Computation of the integral for the current}
\epigraph{Nature laughs at the difficulties of integration.}{P.~Laplace (1749-1827) quoted in I.~Gordon and S.~Sorkin, The Armchair Science Reader, 1959.}
In (\ref{curent}), we encounted the following integral
\begin{equation}\label{integral}
\mathcal{J}=\lim_{\Lambda\rightarrow\infty}\int_{-1}^{1}\frac{dr}{\sqrt{1-r^{2}}}\int_{0}^{\Lambda}dp
\big(rp-\lambda\big)e^{\lambda r\pi}\big|\w_{-i\lambda r,\gamma}(-2ip)\big|^{2}.
\end{equation}
We work out this integral by using the Mellin-Barnes representation of the Whittaker function given in (\ref{eq:MellinBarnes}) and together with (\ref{conectionw}) we find
\begin{eqnarray}\label{contourint}
\mathcal{J}&=&\lim_{\Lambda\rightarrow\infty}\int_{-1}^{1}\frac{dr}{\sqrt{1-r^{2}}}c_{r}
\int_{-i\infty}^{+i\infty}\frac{ds}{2\pi i}\Gamma(\frac{1}{2}+\gamma+s)\Gamma(\frac{1}{2}-\gamma+s)
\Gamma(i\lambda r-s) \nn\\
&\times&\int_{-i\infty}^{+i\infty}\frac{dt}{2\pi i}\Gamma(\frac{1}{2}+\gamma+t)\Gamma(\frac{1}{2}-\gamma+t)
\Gamma(-i\lambda r-t)e^{\frac{i\pi}{2}(s-t)}2^{-s-t} \nn\\
&\times&\int_{0}^{\Lambda}dp\big(rp-\lambda\big)p^{-s-t},
\end{eqnarray}
where $c_{r}$ is defined as
\begin{equation}\label{cr}
c_{r} \equiv e^{\pi\lambda r}\Big(\Gamma\big(\frac{1}{2}+\gamma+i\lambda r\big)
\Gamma\big(\frac{1}{2}-\gamma+i\lambda r\big)\Gamma\big(\frac{1}{2}+\gamma-i\lambda r\big)
\Gamma\big(\frac{1}{2}-\gamma-i\lambda r\big)\Big)^{-1}.
\end{equation}
If we choose both $s$ and $t$ integration contours to run in a similar way as Ref.~\cite{Kobayashi:2014zza}, then we obtain the final result
\begin{eqnarray}\label{computed}
\mathcal{J}&=&-\frac{\pi}{2}\lambda\lim_{\Lambda\rightarrow\infty}\Lambda
+\frac{\pi}{4}\lambda\gamma\cot(2\pi\gamma)
+\frac{\gamma}{4\sin(2\pi\gamma)}\Big(3\I_{1}(2\pi\lambda)-2\pi\lambda\I_{0}(2\pi\lambda)\Big) \nn\\
&+&\frac{i}{2\sin(2\pi\gamma)}\int_{-1}^{1}\frac{dr}{\sqrt{1-r^{2}}}b_{r}
\Big\{\big(e^{2\pi\lambda r}+e^{-2\pi i\gamma}\big)\psi\big(\frac{1}{2}+i\lambda r-\gamma\big) \nn\\
&-&\big(e^{2\pi\lambda r}+e^{2\pi i\gamma}\big)\psi\big(\frac{1}{2}+i\lambda r+\gamma\big)\Big\},
\end{eqnarray}
where $b_{r}$ is given by Eq.~(\ref{br}).
\section{Wronskian condition for (\ref{eq:2DFermSol})}
\label{sec:WronskianCondition}
In this appendix we show how to derive the value of parameters (\ref{eq:C_14}) of the positive and negative frequency solutions at asymptotic past (\ref{eq:2DFermSol}) so that the \gls{Wronskian condition} (\ref{eq:WronskianCondition}) holds. We therefore define 
\begin{align}
M \equiv \psi^+(\tau)\psi^+(\tau)^\dagger+\psi^-(\tau)\psi^-(\tau)^\dagger.
\end{align}
Now we can use the behavior of the Whittaker function under conjugation given by (\ref{eq:Wconjugated}) and the specific form of the solutions (\ref{eq:2DFermSol}) given in (\ref{eq:Csolutions}) and (\ref{eq:C_23}) to find
\begin{align}
 \psi_1^+(\tau)^*=\psi_2^-(\tau)\cdot\begin{cases}
                  -\frac{C_2^*}{C_3} &\text{ for } k>0\\
                  +\frac{C_3^*}{C_2} &\text{ for } k<0
                 \end{cases},&&
 \psi_1^-(\tau)^*=\psi_2^+(\tau)\cdot\begin{cases}
                  +\frac{C_3^*}{C_2} &\text{ for } k>0\\
                  -\frac{C_2^*}{C_3} &\text{ for } k<0
                 \end{cases},  \label{eq:psiconjugated}
\end{align}
where \(\psi_1^\pm(\tau)\) and \(\psi_2^\pm(\tau)\) are the first and second component of \(\psi_\text{in}^\pm\) respectively. 
This can be used to find
\begin{align}
 {M_{12}}^*=M_{21}
 =&\psi_2^+(\tau)\psi_2^-(\tau)\left[\frac{C_3^*}{C_2}-\frac{C_2^*}{C_3}\right].
\end{align}
Now requiring the \gls{Wronskian condition} (\ref{eq:WronskianCondition}), i.e. \(M_{12}=M_{21}=0\), we find
\begin{align}
 |C_2|^2=|C_3|^2. \label{eq:C_14^2}
\end{align}
Using (\ref{eq:psiconjugated}) for the diagonal elements of \(M\) we find
\begin{align}
 M_{11}={M_{22}}^*=-\frac{C_3^*}{C_2}\sgn(k)\left[\psi_1^+(\tau)\psi_2^-(\tau)-\psi_1^-(\tau)\psi_2^+(\tau)\right].
\end{align}
We can now use the Dirac equation (\ref{eq:CoupledDirac1}) to bring this into the form
\begin{align}
 M_{11}={M_{22}}^*=-\frac{i}{m a(\tau)}\frac{C_3^*}{C_2}\sgn(k)\left[\psi_1^+(\tau){\psi_1^-}'(\tau)-\psi_1^-(\tau){\psi_1^+}'(\tau)\right].
\end{align}
Using the solutions (\ref{eq:2DFermSol}) we find 
\begin{align}
 M_{11}={M_{22}}^*=-\frac{i}{m a(\tau)}\frac{C_3^*}{C_2}\left[\psi_1^a(z)\frac{d}{d\tau}\psi_1^b(z)-\psi_1^b(z)\frac{d}{d\tau}\psi_1^a(z)\right],\end{align}
using the explicit form of the solutions (\ref{eq:Csolutions}) and (\ref{eq:C_23}) this becomes
\begin{align}
 M_{11}={M_{22}}^*=-i\frac{|C_3|^2}{ a(\tau)}\frac{2k}{H}\left[W_{\kappa-\frac{1}{2},\mu}(z)\frac{dW_{-\kappa+\frac{1}{2},-\mu}(-z)}{dz}-\frac{dW_{\kappa-\frac{1}{2},\mu}(z)}{dz}W_{-\kappa+\frac{1}{2},-\mu}(-z)\right].
 \end{align}
 Using a \glslink{Wronskian condition}{Wronskian} (\ref{wrmm}) we find
 \begin{align}
 M_{11}={M_{22}}^*=-i\frac{|C_3|^2}{ a(\tau)}\frac{2k}{H}\e^{-i\pi\sgn(k)\left(\kappa-\frac{1}{2}\right)}=\frac{|C_3|^2}{ a(\tau)}\frac{2|k|}{H}\e^{-i\pi\sgn(k)\kappa}.
 \end{align}
Now requiring the \gls{Wronskian condition} (\ref{eq:WronskianCondition}), i.e.~\(M_{11}=M_{22}=a(\eta)^{-1}\), we find
\begin{align}
 |C_3|^2=\frac{H}{2|k|}\e^{i\pi\kappa\sgn(k)}.
\end{align}
Choosing a physically irrelevant phase this leads to (\ref{eq:C_14}) using (\ref{eq:C_14^2}). \\
Observe that from (\ref{eq:psiconjugated}) and (\ref{eq:C_14^2}) we find
\begin{align}
 |\psi_1^+(\tau)|^2=|\psi_2^-(\tau)|^2,&&|\psi_2^+(\tau)|^2=|\psi_1^-(\tau)|^2. \label{eq:psi+psi-}
\end{align}
\section{Integration in the complex plane of (\ref{eq:current2})}
\label{sec:Integral}
If we change the variable in (\ref{eq:current2}) to \(v:=|k|\tau \) we find
\begin{align}
 J^x&=-\frac{e }{\tau}\int_0^{\infty}\frac{dv}{2\pi}\Big(|\psi_1^+(\tau)|^2\Big|_{k>0}-|\psi_2^+(\tau)|^2\Big|_{k>0}+|\psi_1^+(\tau)|^2\Big|_{k<0}-|\psi_2^+(\tau)|^2\Big|_{k<0}\Big),\\
     &=-\frac{2 e }{\tau}\int_0^{\infty}\frac{dv}{2\pi}\Big(|\psi_1^+(\tau)|^2\Big|_{k>0}-|\psi_2^+(\tau)|^2\Big|_{k<0}\Big),
\end{align}
where we used the normalization (\ref{eq:normalization}) of the modes. Using the specific form of the solutions (\ref{eq:2DFermSol2p}) (\ref{eq:2DFermSol2m}) this takes the form
\begin{align}
 J^x&=-\gamma^2\frac{He}{2\pi}\int_0^\infty\frac{dv}{v}\sum_{r=-1,1}r\,\e^{-r\pi i\kappa}\left|W_{ r \kappa-\frac12,r\mu}(2i v)\right|^2.
\end{align}
Using the Mellin-Barnes form of the Whittaker function (\ref{eq:MellinBarnes}) we can write the current as
\begin{align}
 \begin{split}
 J^x=-&\gamma^2\frac{He}{2\pi}\lim_{\xi\rightarrow\infty}\int_0^\xi\frac{dv}{v}\int_{-i\infty}^{i\infty}\frac{ds}{2i\pi}\int_{-i\infty}^{i\infty}\frac{dt}{2i\pi}(2i v)^{-s}(-2i v)^{-t}\sum_{r=-1,1}r\,e^{-r\pi i \kappa}\\
 &\times\frac{\Gamma(1/2+\mu+s)\Gamma(1/2-\mu+s)\Gamma(1/2-r\kappa-s)}{\Gamma(1+\mu-r\kappa)\Gamma(1-\mu-r\kappa)} \\
&\times \frac{\Gamma(1/2-\mu+t)\Gamma(1/2+\mu+t)\Gamma(1/2+r\kappa-t)}{\Gamma(1-\mu+r\kappa)\Gamma(1+\mu+r\kappa)}.
 \end{split}
 \end{align}
 We now perform the integral over \(v\) 
 \begin{align}
 \begin{split}
 J^x=&-\gamma^2\frac{He}{(2\pi)^3}\lim_{\xi\rightarrow\infty}\frac{1}{\Gamma(1+\mu-\kappa)\Gamma(1-\mu-\kappa)\Gamma(1-\mu+\kappa)\Gamma(1+\mu+\kappa)}\\
 &\times\int_{-i\infty}^{i\infty}{ds}\,\sum_{r=-1,1}r\,e^{-r\pi i\kappa}e^{-i\frac{\pi}{2}s}\,\Gamma(1/2+\mu+s)\Gamma(1/2-\mu+s)\Gamma(1/2-r\kappa-s)\\
 &\times\int_{-i\infty}^{i\infty}{dt}\,\e^{i\frac{\pi}{2}t}\,{\Gamma(1/2-\mu+t)\Gamma(1/2+\mu+t)\Gamma(1/2+r\kappa-t)}\frac{(2\xi)^{-s-t}}{s+t}.
 \end{split}
 \end{align}
We can close the contour of the integral over \(t\) in the \(t>0\) -plane in order to use the residue theorem. The contributing poles thus are \(t=-s\) and \(t=-r\kappa+1/2+n\).  Due to the \(\xi\rightarrow\infty\) limit the integral is only non-zero for \(t+s\le0\). Since we could close the integral over \(s\) in a similar way in the \(s>0\) -plane for \(t\ne-s\) the only pole which gives non-zero contribution is \(t=-s\). Using the residue theorem we thus find
 \begin{align}
 \begin{split}
 J^x=-&\gamma^2\frac{i eH}{(2\pi)^2}\int_{-i\infty}^{i\infty}{ds}\,\,\frac{\Gamma(1/2+\mu+s)\Gamma(1/2-\mu+s)
 \Gamma(1/2-\mu-s)\Gamma(1/2+\mu-s)}{\Gamma(1+\mu-\kappa)\Gamma(1-\mu-\kappa)\Gamma(1-\mu+\kappa)\Gamma(1+\mu+\kappa)}\\
 &\hspace{3cm}\times\sum_{r=-1,1}r\,\e^{-i{\pi}(r\kappa+s)}\Gamma(1/2-r\kappa-s)\Gamma(1/2+r\kappa+s)
 \end{split},\\
 =&-\gamma^2\frac{1}{\mu^2-\kappa^2}\frac{eH}{4\pi}i\int_{-i\infty}^{i\infty}{ds}\,\frac{\sin[(\mu+\kappa)\pi]\sin[(\mu-\kappa)\pi]}{\cos[(\mu+s)\pi]\cos[(\mu-s)\pi]}\sum_{r=-1,1}r\,\frac{\e^{-i{\pi}(r\kappa+s)}}{\cos[(r\kappa+s)\pi]}.
 \end{align}
 To solve this integral we now write trigonometric functions as exponentials and change the integration variable to \(X:=\exp(i\pi s)\). This leads to a integral which can be solved using the standard decomposition theorems for rational fractions
 \begin{align}
 J^x=-&\gamma^2\frac{4}{\mu^2-\kappa^2}\frac{eH}{\pi^2}i\int_{\infty}^{0}{dX}\,\frac{\sin[2\pi\kappa]\sin[(\mu+\kappa)\pi]\sin[(\mu-\kappa)\pi]X^4}{(X^2+\e^{2i\pi\kappa})(X^2+\e^{-2i\pi\kappa})(X^2+\e^{2i\pi\mu})(X^2+\e^{-2i\pi\mu})}\\
 =&\frac{eH}{\pi}i\left(\mu\frac{\sin(2\pi\kappa)}{\sin(2\pi\mu)}-\kappa\right).
 \end{align}

%% file: LTB/LTBappendix.tex
\epigraph{Taci. Su le soglie \\
    del bosco non odo \\
    parole che dici \\
    umane; ma odo \\
    parole più nuove \\
    che parlano gocciole e foglie \\
    lontane.\\
    Ascolta.}{La pioggia nel pineto, G.~D’Annunzio, 1902}
As Appendix \ref{app:fried}, we will use the \glspl{Einstein equation} discussed in Sec.~\ref{sec:Einst} to obtain the cosmological evolution of the the metric functions $A(r,t)$ and $B(r,t)$. The notations used in this appendix differ a bit from the body text in order to save space and time regarding the calculations. In order to remain clear, we sum up in the next equations the quantities which will appear though the appendix and will be different from the body text:
\begin{align}
& B(r,t) \equiv R(r,t), \\
& A(r,t) = \frac{B'(r,t)}{f(r)} \equiv \frac{R'(r,t)}{f(r)} \text{, see also equation \ref{eq:AtoB},} \\
& 1 + 2 E(r) \equiv f^2(r) \text{, see also equation \ref{eq:Etof}.}
\end{align}
\section{Deriving Einstein equation for the Lemaître-Tolman-Bondi metric}
\par We work in comoving coordinates which exist as long as there exist a time-like vector field $u^{\alpha}$ in the spacetime. Furthermore, if $u^{\alpha}$ has no rotation, then the comoving coordinate can be chosen so that they are also synchronous. A proof of those two statement together with the definitions is given in \cite{krasinski2007introduction}. For our purpose, we will assume that we have such a vector, which can be the 4-velocity of matter for instance and we will furthermore assume that we have only pressureless matter. A consequence will be that up to a redefinition of time $g_{tt}=1$. Working with another equation of state would be somehow odd in the sense that the entropy per particle will be a universal constant. We see already that \gls{radiation} can be hardly described by an inhomogeneous universe, that is this model will be more suited after decoupling, for late universe physics.
\par Our starting point will then be a spherically \glslink{symmetry}{symmetric} metric with two free functions: $A(r,t)$ and $B(r,t)$. The line element reads:
\begin{equation}
\label{generalmetric}
ds^2=dt^2-A^2(r,t)dr^2-B^2(r,t) d\Omega^2. 
\end{equation}
$B(r,t)$ has a geometrical interpretation as the \gls{angular distance}. The ratio $\frac{\dot{A}}{A}$ and $\frac{\dot{B}}{B}$ are the radial and transverse expansion rate respectively. The \gls{FLRW metric} can be recovered for:
\begin{align}
\label{eq:Flimit1}
& A(r,t)=\frac{a(t)}{\sqrt{1-kr^2}}, \\
& B(r,t) = a(t) r.
\end{align}
Using (\ref{eq:defCHR}) together with (\ref{generalmetric}), the Christoffel symbols are:
\begin{equation}
\left\{\begin{split} &\Gamma^{0}_{\mu\beta}=\left(
\begin{array}{cccc}
0 & 0 & 0 & 0 \\
0 & A\dot{A} & 0 & 0  \\
0 & 0 & B\dot{B} & 0 \\
0 & 0 & 0 & B\dot{B}\sin^2\theta A\theta  \\
\end{array}\right)
\\
&\Gamma^{1}_{\mu\beta}=\left(
\begin{array}{cccc}
0 & \frac{\dot{A}}{A} & 0 & 0 \\
\frac{\dot{A}}{A} & \frac{A'}{A} & 0 & 0  \\
0 & 0 & -\frac{BB'}{A^2} & 0 \\
0 & 0 & 0 & -\frac{BB'\sin^2\theta}{A^2}  \\
\end{array}\right)
\\
&\Gamma^{2}_{\mu\beta}=\left(
\begin{array}{cccc}
0 & 0 & \frac{\dot{B}}{B} & 0 \\
0 & 0 & \frac{B'}{B} & 0  \\
\frac{\dot{B}}{B} & \frac{B'}{B} & 0 & 0 \\
0 & 0 & 0 & -\sin \theta \cos \theta \\
\end{array}\right)
\\
&\Gamma^{3}_{\mu\beta}=\left(
\begin{array}{cccc}
0 & 0 & 0 & \frac{\dot{B}}{B} \\
0 & 0 & 0 & \frac{B'}{B}  \\
0 & 0 & 0 & \cot\theta \\
\frac{\dot{B}}{B} & \frac{B'}{B} & \cot\theta & 0  \\
\end{array}\right)
\end{split}
\right.
\end{equation}
Hence with the Christoffel symbols together with (\ref{ricci}) and (\ref{riemann}), we deduce the \gls{Ricci tensor}:
\begin{equation}
\left\{\begin{split} 
&R_{00}=\frac{\ddot{A}}{A}+2{\ddot{B}}{B}  \\
&R_{11}=\frac{2B''}{B}-A\ddot{A}-2A\dot{A}\dot{B}-2\frac{A'B'}{AB}   \\
&R_{22}=\dot{B}^2-B\ddot{B}+\frac{B'^2}{A^2}+\frac{BB''}{A^2}-\frac{BA'B'}{A^3}-\frac{B\dot{A}\dot{B}}{A} \\
&R_{33}=  \sin^2\theta R_{22}   \\
\end{split}
\right.
\end{equation}
We can also deduce the curvature scalar $R$ given by (\ref{courbure}):
\begin{equation}
\frac{R}{2}=\frac{\ddot{A}}{A}+2{\ddot{B}}{B}-2\frac{B''}{A^2B}+2\frac{\dot{A}\dot{B}}{AB}+2\frac{A'B'}{A^3B}+\frac{1}{B^2}+\frac{\dot{B}^2}{B^2}-\frac{B'}{A^2B^2}.
\end{equation}
We build now the \gls{Einstein tensor} given by (\ref{einsteinTenseur}):
\begin{equation}
\left\{\begin{split}
&G_{00}=2\frac{B''}{AÂ²B}-2\frac{\dot{A}\dot{B}}{AB}-2\frac{A'B'}{A^3B}-\frac{1}{B^2}-\frac{\dot{B}^2}{B^2}+\frac{B'}{A^2B^2} \\
&G_{11}=\frac{1}{B^2}(2A^2B\ddot{B}+A^2+A^2\dot{B}^2-B'^2) \\
&G_{22}=\frac{B}{A^3}(A^3\ddot{B}-AB''+A'B'+A^2\dot{A}\dot{B}+A^2\ddot{A}B) \\
&G_{33}= \sin^2\theta\ G_{22}    \\
&G_{01}=\dot{B'}-\frac{\dot{A}B'}{A} \\
\end{split}\right.
\end{equation}
And we easily check that we once again find the correct limit by using (\ref{eq:Flimit1}).
Assuming a dust \gls{perfect fluid} (pressureless: taking $P=0$ in (\ref{eq:perfectf})), we get:
\begin{align}
&B'^2=2A^2(B\ddot{B}+1+\dot{B}^2), \label{eq:ET1} \\
&AB''-A'B'=A^2(A\ddot{B}+\dot{A}\dot{B}+B\ddot{A}), \label{eq:ET2} \\
&\frac{2}{AB} \left(\frac{B''}{A}-\dot{A}\dot{B}-\frac{A'B'}{A^2} \right)=\frac{A}{B^2} \left(1+\dot{B}^2-\frac{B'^2}{A^2} \right)-8\pi G\rho,  \label{eq:ET3} \\
& \dot{B'}=\frac{\dot{A}B'}{A}. \label{eq:ET4}
\end{align}
\section{Simplifying the equations}
Putting (\ref{eq:ET1}) and (\ref{eq:ET2}) into (\ref{eq:ET3}) gives:
\begin{equation}
\label{bol}
\frac{\ddot{A}}{A}+2\frac{\ddot{B}}{B}=-4\pi G\rho.
\end{equation}
We may use (\ref{bol}) instead of \ref{eq:ET3}. The conservation of the \gls{energy momentum tensor} $\nabla_{\mu} T^{\mu\nu}=0. $ gives:
\begin{equation}
\dot{\rho}+\left(\frac{\dot{A}}{A}+2\frac{\dot{B}}{B} \right)\rho=0.
\end{equation}
Gathering these equations we get the following set of equations:
\begin{equation}
\label{eq:setLTB}
\left\{\begin{split}
&B'^2=2A^2(B\ddot{B}+1+\dot{B}^2)  \\
&AB''-A'B'=A^2(A\ddot{B}+\dot{A}\dot{B}+B\ddot{A})  \\
&\frac{\ddot{A}}{A}+2\frac{\ddot{B}}{B}=-4\pi G\rho \\
&\dot{\rho}+\left(\frac{\dot{A}}{A}+2\frac{\dot{B}}{B} \right)\rho=0\\
& \dot{B'}=\frac{\dot{A}B'}{A} \\
\end{split}
\right.
\end{equation}
Integrating (\ref{eq:ET4}) gives:
\begin{equation}
\label{eq:AtoB}
B'=\frac{A}{f(r)},
\end{equation}
with $f(r)$ being an arbitrary function of $r$. Observe that if $B'(r,t)=0$, the last integration step is not possible. In that case, we talk about Nariai solution. If one consider charged dust, Nariai solution become the so-called Datt-Ruban solution which interestingly enough has no corresponding solution neither in Newtonian approximation of GR, nor in linearised GR. 

Now, for physical reasons we will explain afterwards (see \ref{physicalinterp}), we define two functions $E(r)$ and $M(r,t)$ as follows:
\begin{equation}
\label{eq:Etof}
\left\{\begin{split}
&f(r)\equiv\frac{1}{\sqrt{1+2E(r)}}\ \text{where}\ \forall r, E(r)>-\frac{1}{2}\\
&M(r,t)\equiv-B^2\ddot{B}\\
\end{split}
\right.
\end{equation}

This way, it is possible to simplify (\ref{eq:setLTB})
\begin{align}
\label{set2}
& B'=A\sqrt{1+2E},\\
& 4\pi G\rho=\frac{M'}{B'B^2}, \label{eq:mattereq} \\
& \frac{1}{2}\dot{B}^2-\frac{GM}{B}=E.  \label{eq:LTBdyn}
\end{align}
Inspecting equation (\ref{eq:mattereq}), we see that the matter density can become infinite in two cases: $M'(r) \neq 0$ and $B=0$ or $M'(r) \neq 0$ and $B'=0$. The first case corresponds to the usual \gls{big-bang} singularity but the second cases defines shell-crossing singularities where the mass density goes to infinity and change sign to become negative. It separates two regions of spacetime with different velocities of the matter fluid and could indicate a breakdown of the assumption of the models or new physics such as the so-called wormhole or neck. Those singularities would reduce to \gls{big-bang} singularities in the FLRW limit but can have much more wilder behavior in the general case.
\section{Physical interpretation of these equations}
\label{physicalinterp}
\subsection{The mass function $M(r)$:} The $M$ function has two important properties. First, a priori $M$ depends on $r$ and $t$: $M=M(r,t)$. But in fact a direct computation $\dot{M}(r,t)$ gives 0. Thus we have: $M=M(r)$.
Second, noticing that the mass inside a sphere of radius $r$ is given in term of the trace of the \gls{energy momentum tensor} $T$ by:
\begin{equation}
\frac{\tilde M(r)}{G}=\int_{0}^{r}dr \int_{0}^{\pi}d\theta \int_{0}^{2\pi}d\phi T\sqrt{-g}=4\pi\int^{r}_{0}dr\rho AB^2,
\end{equation}
we have the following relation:
\begin{equation}
\tilde M'(r)=4\pi G\rho AB^2,
\end{equation}
which is strikingly similar to the second equation of \eqref{set2}. Thus $M(r)$ and $\tilde M$(r) are linked by:
\begin{equation}
\tilde M'\sqrt{1+2E}=M'.
\end{equation}
For this reason we call $\tilde M$(r) the effective mass for the gravitational effect of this sphere of radius $r$, it is the sum of the masses of the particles composing the total gravitating body. $M(r)$ is the active gravitational mass which generates the gravitational field.
In a bound system, ($E<0$ or $\tilde M > M$), $\tilde M - M$ is called the \emph{relativistic mass defect} which is analogue of the mass defect known from nuclear and elementary \gls{particle physics}. Indeed, for bound system, some of the energy contained in the particle is shed in order to bind the system together, hence a mass defect \cite{Bondi:1947fta}.
\subsection{The energy function $E(r)$:} Since $M(r)$ is a mass and $B(r,t)$ a distance, the third equation of \eqref{set2} can be read as an analogue of the energy conservation law for the dust particles at comoving coordinate radius $r$:
\begin{equation}
\tag{\ref{set2}.c}
\frac{1}{2}\dot{B}^2-\frac{GM}{B}=E,
\end{equation}
where:
\begin{itemize}
\item $\frac{1}{2}\dot{B}^2$ is the kinetic energy per unit mass
\item $-\frac{GM(r)}{B}$ is the gravitational potential energy per unit mass
\item $E(r)$ is the total energy per unit mass
\end{itemize}
This interpretation of $E(r)$, analogue of the Newtonian approach of gravity, was first given by Bondi in 1947 \cite{Bondi:1947fta}.

So $E(r)$ is an energy. But what is striking is that if we compare the two metrics (taking (\ref{set2}.a) into account in the \acrshort{LTB} metric):
\begin{equation}
\tag{FLRW metric}
ds^2=dt^2-\frac{a^2}{1-kr^2}dr^2-a^2r^2[d\theta^2+\sin^2\theta d\phi^2],
\end{equation}
\begin{equation}
\tag{LTB metric}
ds^2=dt^2-\frac{B'^2}{1+2E}dr^2-B^2[d\theta^2+\sin^2\theta d\phi^2].
\end{equation}
Comparing the two metrics and (\ref{eq:LTBdyn}) to (\ref{EnergyConservation}), we see that to recover \glslink{FLRW metric}{FLRW} from \acrshort{LTB} we just have to take:
\begin{align}
& B(r,t)=a(t).r, \\
& E(r)=-\frac{1}{2}kr^2, \\
& M(r) = M_0 r^3. \label{eq:limitFLRW}
\end{align}  
Thus $E(r)$ is also the analogue of the curvature $k$ and their relation is summed up in table \ref{tab:curv}.
\begin{table}[h]
   \centering
\begin{tabular}{|c|c|c|c|c|}
\hline  & geometry & spherical & euclidean & hyperbolic \\ 
\hline FLRW & curvature $k$ & positive & nul & negative \\
\hline \acrshort{LTB} & function $E(r)$ & negative & nul & positive \\
\hline
\end{tabular}
\caption{A comparison of the curvature function in \acrshort{LTB} and FLRW model.}
\label{tab:curv}
\end{table}
It is important to notice that since $k$ is a constant and $E(r)$ is a function, the geometry of the Universe is unique with an \gls{FLRW metric}, while it depends on the distance to the center with the \acrshort{LTB} metric. It can be understand as the \glslink{FLRW metric}{FLRW geometry} has more \glspl{Killing vector} than the \acrshort{LTB} one.

%% file: post.tex
\phantomsection
\addcontentsline{toc}{chapter}{Postface}
\chapter*{Postface}
\epigraph{Mischief managed.}{Harry Potter in the  Prisoner of Azkaban, 1993 \cite{rowling1999harry}}
It is with great satisfaction that I put a final point to the most massive, tiring but also very rewarding piece of work I ever carried out. For the readers who made it to the end: first I hope you had a pleasant journey into my every-day scientific problems (and solutions) of the past three years. Second and last: my more sincere congratulations!
\paragraph*{}
Clément Stahl, Pescara, 30 october 2016

%% file: PhD_Template.bbl
\begin{thebibliography}{100}

\bibitem{smith1863dreamthorp}
A.~Smith, {\em Dreamthorp: A Book of Essays Written in the Country}.
\newblock 1863.

\bibitem{Bavarsad:2016cxh}
E.~Bavarsad, C.~Stahl, and S.-S. Xue, ``{Scalar current of created pairs by
  Schwinger mechanism in de Sitter spacetime},'' {\em Phys. Rev.}, vol.~D94,
  no.~10, p.~104011, 2016.

\bibitem{Stahl:2015cra}
C.~Stahl and S.~Eckhard, ``{Semiclassical fermion pair creation in de Sitter
  spacetime},'' {\em AIP Conf. Proc.}, vol.~1693, p.~050005, 2015.

\bibitem{Stahl:2015gaa}
C.~Stahl, E.~Strobel, and S.-S. Xue, ``{Fermionic current and Schwinger effect
  in de Sitter spacetime},'' {\em Phys. Rev.}, vol.~D93, no.~2, p.~025004,
  2016.

\bibitem{Stahl:2016qjs}
C.~Stahl, E.~Strobel, and S.-S. Xue, ``{Pair creation in the early universe},''
  in {\em {14th Marcel Grossmann Meeting on Recent Developments in Theoretical
  and Experimental General Relativity, Astrophysics, and Relativistic Field
  Theories (MG14) Rome, Italy, July 12-18, 2015}}, 2016.

\bibitem{Stahl:2016geq}
C.~Stahl and S.-S. Xue, ``{Schwinger effect and backreaction in de Sitter
  spacetime},'' {\em Phys. Lett.}, vol.~B760, pp.~288--292, 2016.

\bibitem{Stahl:2016vcl}
C.~Stahl, ``{Inhomogeneous matter distribution and supernovae},'' {\em Int. J.
  Mod. Phys.}, vol.~D25, no.~06, p.~1650066, 2016.

\bibitem{withD}
D.~Bégué, C.~Stahl, and S.-S. Xue, ``{A model of interacting dark fluids
  tested with supernovae data},'' 2017.

\bibitem{Serre}
``\url{https://people.math.osu.edu/goss.3/hint.pdf},''

\bibitem{daudet1869lettres}
A.~Daudet, {\em Lettres de mon Moulin}.
\newblock 1869.

\bibitem{zweig2013schachnovelle}
S.~Zweig, {\em Schachnovelle}.
\newblock 1943.

\bibitem{peebles1993principles}
P.~J.~E. Peebles, {\em Principles of physical cosmology}.
\newblock Princeton University Press, 1993.

\bibitem{Dodelson:2003ft}
S.~Dodelson, {\em {Modern Cosmology}}.
\newblock Amsterdam: Academic Press, 2003.

\bibitem{Mukhanov}
V.~Mukhanov, {\em {Physical Foundations of Cosmology}}.
\newblock Oxford: Cambridge University Press, 2005.

\bibitem{Peter}
P.~Peter and J.~Uzan, {\em Primordial cosmology}.
\newblock Oxford graduate texts, Oxford University Press, 2009.

\bibitem{lyth2009primordial}
D.~H. Lyth and A.~R. Liddle, {\em The primordial density perturbation:
  Cosmology, inflation and the origin of structure}.
\newblock Cambridge University Press, 2009.

\bibitem{Slipher:1917zz}
V.~M. Slipher, ``{Nebulae},'' {\em Proc. Am. Phil. Soc.}, vol.~56,
  pp.~403--409, 1917.

\bibitem{Einstein:1915by}
A.~Einstein, ``{On the General Theory of Relativity},'' {\em Sitzungsber.
  Preuss. Akad. Wiss. Berlin (Math. Phys.)}, vol.~1915, pp.~778--786, 1915.
\newblock [Addendum: Sitzungsber. Preuss. Akad. Wiss. Berlin (Math.
  Phys.)1915,799(1915)].

\bibitem{Einstein:1915ca}
A.~Einstein, ``{The Field Equations of Gravitation},'' {\em Sitzungsber.
  Preuss. Akad. Wiss. Berlin (Math. Phys.)}, vol.~1915, pp.~844--847, 1915.

\bibitem{Einstein:1916vd}
A.~Einstein, ``{The Foundation of the General Theory of Relativity},'' {\em
  Annalen Phys.}, vol.~49, pp.~769--822, 1916.
\newblock [Annalen Phys.14,517(2005)].

\bibitem{Wagner:2012ui}
T.~A. Wagner, S.~Schlamminger, J.~H. Gundlach, and E.~G. Adelberger,
  ``{Torsion-balance tests of the weak equivalence principle},'' {\em Class.
  Quant. Grav.}, vol.~29, p.~184002, 2012.

\bibitem{Joyce:2014kja}
A.~Joyce, B.~Jain, J.~Khoury, and M.~Trodden, ``{Beyond the Cosmological
  Standard Model},'' {\em Phys. Rept.}, vol.~568, pp.~1--98, 2015.

\bibitem{Einstein:1917ce}
A.~Einstein, ``{Cosmological Considerations in the General Theory of
  Relativity},'' {\em Sitzungsber. Preuss. Akad. Wiss. Berlin (Math. Phys.)},
  vol.~1917, pp.~142--152, 1917.

\bibitem{deSitter:1917zz}
W.~de~Sitter, ``{Einstein's theory of gravitation and its astronomical
  consequences, Third Paper},'' {\em Mon. Not. Roy. Astron. Soc.}, vol.~78,
  pp.~3--28, 1917.

\bibitem{friedmann1922125}
A.~Friedmann, ``125. on the curvature of space,'' {\em Zeitschrift f{\"u}r
  Physik}, vol.~10, pp.~377--386, 1922.

\bibitem{Friedmann:1924bb}
A.~Friedmann, ``{On the Possibility of a world with constant negative curvature
  of space},'' {\em Z. Phys.}, vol.~21, pp.~326--332, 1924.
\newblock [Gen. Rel. Grav.31,2001(1999)].

\bibitem{Lemaitre:1927zz}
G.~Lemaitre, ``{A Homogeneous Universe of Constant Mass and Growing Radius
  Accounting for the Radial Velocity of Extragalactic Nebulae},'' {\em Annales
  Soc. Sci. Brux. Ser. I Sci. Math. Astron. Phys.}, vol.~A47, pp.~49--59, 1927.

\bibitem{magueijo2004faster}
J.~Magueijo, {\em Faster Than the Speed of Light: The Story of a Scientific
  Speculation}.
\newblock Penguin Books, 2004.

\bibitem{biofried}
V.~Y. Frenkel', ``Aleksandr aleksandrovich fridman (friedmann): a biographical
  essay,'' {\em Soviet Physics Uspekhi}, vol.~31, no.~7, p.~645, 1988.

\bibitem{LachiezeRey:1995kj}
M.~Lachieze-Rey and J.-P. Luminet, ``{Cosmic topology},'' {\em Phys. Rept.},
  vol.~254, pp.~135--214, 1995.

\bibitem{Levin:2001fg}
J.~J. Levin, ``{Topology and the cosmic microwave background},'' {\em Phys.
  Rept.}, vol.~365, pp.~251--333, 2002.

\bibitem{Hawking:1973uf}
S.~W. Hawking and G.~F.~R. Ellis, {\em {The Large Scale Structure of
  Space-Time}}.
\newblock Cambridge Monographs on Mathematical Physics, Cambridge University
  Press, 2011.

\bibitem{Zeldovich:1984vk}
{\relax Ya}.~B. Zeldovich and A.~A. Starobinsky, ``{Quantum creation of a
  universe in a nontrivial topology},'' {\em Sov. Astron. Lett.}, vol.~10,
  p.~135, 1984.

\bibitem{Ade:2013vbw}
P.~A.~R. Ade {\em et~al.}, ``{Planck 2013 results. XXVI. Background geometry
  and topology of the Universe},'' {\em Astron. Astrophys.}, vol.~571, p.~A26,
  2014.

\bibitem{Ade:2015xua}
P.~A.~R. Ade {\em et~al.}, ``{Planck 2015 results. XIII. Cosmological
  parameters},'' 2015.

\bibitem{Tegmark:2003ud}
M.~Tegmark {\em et~al.}, ``{Cosmological parameters from SDSS and WMAP},'' {\em
  Phys. Rev.}, vol.~D69, p.~103501, 2004.

\bibitem{Seljak:2004xh}
U.~Seljak {\em et~al.}, ``{Cosmological parameter analysis including SDSS
  Ly-alpha forest and galaxy bias: Constraints on the primordial spectrum of
  fluctuations, neutrino mass, and dark energy},'' {\em Phys. Rev.}, vol.~D71,
  p.~103515, 2005.

\bibitem{Abbott:2005bi}
T.~Abbott {\em et~al.}, ``{The dark energy survey},'' 2005.

\bibitem{tamburini2011no}
F.~Tamburini, C.~Cuofano, M.~Della~Valle, and R.~Gilmozzi, ``No quantum gravity
  signature from the farthest quasars,'' {\em Astronomy \& Astrophysics},
  vol.~533, p.~A71, 2011.

\bibitem{Celerier:2011zh}
M.-N. Celerier, ``{Some clarifications about Lema\^itre-Tolman models of the
  Universe used to deal with the dark energy problem},'' {\em Astron.
  Astrophys.}, vol.~543, p.~A71, 2012.

\bibitem{Cyburt:2015mya}
R.~H. Cyburt, B.~D. Fields, K.~A. Olive, and T.-H. Yeh, ``{Big Bang
  Nucleosynthesis: 2015},'' {\em Rev. Mod. Phys.}, vol.~88, p.~015004, 2016.

\bibitem{Kamionkowski:2015yta}
M.~Kamionkowski and E.~D. Kovetz, ``{The Quest for B Modes from Inflationary
  Gravitational Waves},'' 2015.

\bibitem{Guzzetti:2016mkm}
C.~Guzzetti, M., N.~Bartolo, M.~Liguori, and S.~Matarrese, ``{Gravitational
  waves from inflation},'' {\em Riv. Nuovo Cim.}, vol.~39, no.~9, pp.~399--495,
  2016.

\bibitem{Hubble}
E.~{Hubble}, ``{A Relation between Distance and Radial Velocity among
  Extra-Galactic Nebulae},'' {\em Proceedings of the National Academy of
  Science}, vol.~15, pp.~168--173, Mar. 1929.

\bibitem{kant}
I.~Kant, {\em Allgemeine Naturgeschichte und Theorie des Himmels}.
\newblock APS, 1755.

\bibitem{hubleserisou}
A.~{Sandage}, ``{Current Problems in the Extragalactic Distance Scale.},'' {\em
  \apj}, vol.~127, p.~513, May 1958.

\bibitem{Verde:2013wza}
L.~Verde, P.~Protopapas, and R.~Jimenez, ``{Planck and the local Universe:
  Quantifying the tension},'' {\em Phys. Dark Univ.}, vol.~2, pp.~166--175,
  2013.

\bibitem{Bernal:2016gxb}
J.~L. Bernal, L.~Verde, and A.~G. Riess, ``{The trouble with $H_0$},'' {\em
  JCAP}, vol.~1610, no.~10, p.~019, 2016.

\bibitem{blogpot}
``\url{https://aeon.co/ideas/what-i-learned-as-a-hired-consultant-for-autodidact-physicists},''

\bibitem{christianson1996edwin}
G.~Christianson, {\em Edwin Hubble: Mariner of the Nebulae}.
\newblock University of Chicago Press, 1996.

\bibitem{teresi}
D.~Teresi, ``{The Cosmic Egoist},'' {\em The New York Times}, 03/09/1995.

\bibitem{fitzgerald2003great}
F.~Fitzgerald, {\em The Great Gatsby}.
\newblock 1925.

\bibitem{Wells:2014fia}
C.~G. Wells, ``{On The Equivalence of the FRW Field Equations and those of
  Newtonian Cosmology},'' {\em Eur. Phys. J. Plus}, vol.~129, p.~168, 2014.

\bibitem{Einstein:1932zz}
A.~Einstein and W.~de~Sitter, ``{On the Relation between the Expansion and the
  Mean Density of the Universe},'' {\em Proc. Nat. Acad. Sci.}, vol.~18,
  pp.~213--214, 1932.

\bibitem{krasinski2007introduction}
J.~Plebanski and A.~Krasinski, {\em {An introduction to general relativity and
  cosmology}}.
\newblock 2006.

\bibitem{Balazs:2015xsa}
L.~G. Balazs, Z.~Bagoly, J.~E. Hakkila, I.~Horvath, J.~Kobori, I.~Racz, and
  L.~V. Toth, ``{A giant ring-like structure at 0.78<z<0.86 displayed by
  GRBs},'' {\em Mon. Not. Roy. Astron. Soc.}, vol.~452, p.~2236, 2015.

\bibitem{1963SvA.....6..699S}
M.~F. {Shirokov} and I.~Z. {Fisher}, ``{Isotropic Space with Discrete
  Gravitational-Field Sources. On the Theory of a Nonhomogeneous Isotropic
  Universe},'' {\em \sovast}, vol.~6, p.~699, Apr. 1963.

\bibitem{Buchert:1999er}
T.~Buchert, ``{On average properties of inhomogeneous fluids in general
  relativity. 1. Dust cosmologies},'' {\em Gen. Rel. Grav.}, vol.~32,
  pp.~105--125, 2000.

\bibitem{Zalaletdinov:2008ts}
R.~Zalaletdinov, ``{The Averaging Problem in Cosmology and Macroscopic
  Gravity},'' {\em Int. J. Mod. Phys.}, vol.~A23, pp.~1173--1181, 2008.

\bibitem{Fleury:2013sna}
P.~Fleury, H.~Dupuy, and J.-P. Uzan, ``{Interpretation of the Hubble diagram in
  a nonhomogeneous universe},'' {\em Phys. Rev.}, vol.~D87, no.~12, p.~123526,
  2013.

\bibitem{Hogg:1999ad}
D.~W. Hogg, ``{Distance measures in cosmology},'' 1999.

\bibitem{Lilly:2006va}
S.~J. Lilly {\em et~al.}, ``{zCOSMOS: A Large VLT/VIMOS redshift survey
  covering 0<z<3 in the COSMOS field},'' {\em Astrophys. J. Suppl.}, vol.~172,
  pp.~70--85, 2007.

\bibitem{2016arXiv160505503S}
M.~{Siudek}, K.~{Ma{\l}ek}, M.~{Scodeggio}, B.~{Garilli}, A.~{Pollo}, C.~P.
  {Haines}, A.~{Fritz}, M.~{Bolzonella}, S.~{de la Torre}, B.~R. {Granett},
  L.~{Guzzo}, U.~{Abbas}, C.~{Adami}, D.~{Bottini}, A.~{Cappi}, O.~{Cucciati},
  G.~{De Lucia}, I.~{Davidzon}, P.~{Franzetti}, A.~{Iovino}, J.~{Krywult},
  V.~{Le Brun}, O.~{Le F{\`e}vre}, D.~{Maccagni}, A.~{Marchetti}, F.~{Marulli},
  M.~{Polletta}, L.~A.~M. {Tasca}, R.~{Tojeiro}, D.~{Vergani}, A.~{Zanichelli},
  S.~{Arnouts}, J.~{Bel}, E.~{Branchini}, O.~{Ilbert}, A.~{Gargiulo},
  L.~{Moscardini}, T.~T. {Takeuchi}, and G.~{Zamorani}, ``{The VIMOS Public
  Extragalactic Redshift Survey (VIPERS). Star formation history of passive
  galaxies},'' {\em ArXiv e-prints}, May 2016.

\bibitem{Wisnioski:2014xwa}
E.~Wisnioski {\em et~al.}, ``{The KMOS$^{3D}$ Survey: design, first results,
  and the evolution of galaxy kinematics from $0.7 \leq z \leq 2.7$},'' {\em
  Astrophys. J.}, vol.~799, no.~2, p.~209, 2015.

\bibitem{Kriek:2014ita}
M.~Kriek {\em et~al.}, ``{The MOSFIRE Deep Evolution Field (MOSDEF) Survey:
  Rest-Frame Optical Spectroscopy for ~1500 H-Selected Galaxies at 1.37 $\le z
  \le$ 3.8},'' {\em Astrophys. J. Suppl.}, vol.~218, no.~2, p.~15, 2015.

\bibitem{Kessler:2009ys}
R.~Kessler {\em et~al.}, ``{First-year Sloan Digital Sky Survey-II (SDSS-II)
  Supernova Results: Hubble Diagram and Cosmological Parameters},'' {\em
  Astrophys. J. Suppl.}, vol.~185, pp.~32--84, 2009.

\bibitem{Karman:2014sqa}
W.~Karman {\em et~al.}, ``{MUSE integral-field spectroscopy towards the
  Frontier Fields Cluster Abell S1063},'' {\em Astron. Astrophys.}, vol.~574,
  p.~A11, 2015.

\bibitem{2012A&A...539A..91C}
T.~{Contini}, B.~{Garilli}, O.~{Le F{\`e}vre}, M.~{Kissler-Patig}, P.~{Amram},
  B.~{Epinat}, J.~{Moultaka}, L.~{Paioro}, J.~{Queyrel}, L.~{Tasca},
  L.~{Tresse}, D.~{Vergani}, C.~{L{\'o}pez-Sanjuan}, and E.~{Perez-Montero},
  ``{MASSIV: Mass Assemby Survey with SINFONI in VVDS. I. Survey description
  and global properties of the 0.9 < z < 1.8 galaxy sample},'' {\em Astron.
  Astrophys.}, vol.~539, p.~A91, Mar. 2012.

\bibitem{2003MNRAS.344.1000B}
G.~{Bruzual} and S.~{Charlot}, ``{Stellar population synthesis at the
  resolution of 2003},'' {\em Mon. Not. Roy. Astron. Soc.}, vol.~344,
  pp.~1000--1028, Oct. 2003.

\bibitem{Sandage}
A.~{Sandage}, ``{The Change of Redshift and Apparent Luminosity of Galaxies due
  to the Deceleration of Selected Expanding Universes.},'' {\em \apj},
  vol.~136, p.~319, Sept. 1962.

\bibitem{Steinmetz:2008gp}
T.~Steinmetz {\em et~al.}, ``{Laser frequency combs for astronomical
  observations},'' {\em Science}, vol.~321, pp.~1335--1337, 2008.

\bibitem{Ribeiro:2004nk}
M.~B. Ribeiro, ``{Cosmological distances and fractal statistics of the galaxy
  distribution},'' {\em Astron. Astrophys.}, vol.~429, pp.~65--74, 2005.

\bibitem{2012NewAR..56..122W}
B.~{Wang} and Z.~{Han}, ``{Progenitors of type Ia supernovae},'' {\em New
  Astronomy Reviews}, vol.~56, pp.~122--141, June 2012.

\bibitem{1993ApJ...413L.105P}
M.~M. {Phillips}, ``{The absolute magnitudes of Type IA supernovae},'' {\em The
  Astrophysical Journal, Letters}, vol.~413, pp.~L105--L108, Aug. 1993.

\bibitem{2007ecf..book...95D}
M.~{Della Valle}, R.~{Gilmozzi}, N.~{Panagia}, J.~{Bergeron}, P.~{Madau},
  J.~{Spyromilio}, and P.~{Dierickx}, {\em {ELT Observations of Supernovae at
  the Edge of the Universe}}, p.~95.
\newblock 2007.

\bibitem{2015MNRAS.448.2312B}
C.~{Barbarino}, M.~{Dall'Ora}, M.~T. {Botticella}, M.~{Della Valle},
  L.~{Zampieri}, J.~R. {Maund}, M.~L. {Pumo}, A.~{Jerkstrand}, S.~{Benetti},
  N.~{Elias-Rosa}, M.~{Fraser}, A.~{Gal-Yam}, M.~{Hamuy}, C.~{Inserra},
  C.~{Knapic}, A.~P. {LaCluyze}, M.~{Molinaro}, P.~{Ochner}, A.~{Pastorello},
  G.~{Pignata}, D.~E. {Reichart}, C.~{Ries}, A.~{Riffeser}, B.~{Schmidt},
  M.~{Schmidt}, R.~{Smareglia}, S.~J. {Smartt}, K.~{Smith}, J.~{Sollerman},
  M.~{Sullivan}, L.~{Tomasella}, M.~{Turatto}, S.~{Valenti}, O.~{Yaron}, and
  D.~{Young}, ``{SN 2012ec: mass of the progenitor from PESSTO follow-up of the
  photospheric phase},'' {\em \mnras}, vol.~448, pp.~2312--2331, Apr. 2015.

\bibitem{CristinaPhD}
C.~Barbarino, {\em {The fickle death of massive stars: from Hydrogen-rich to
  He-poor Supernova explosions.}}
\newblock PhD thesis, Sapienza U., 2016.

\bibitem{2013IJMPD..2230028A}
L.~{Amati} and M.~{Della Valle}, ``{Measuring Cosmological Parameters with
  Gamma Ray Bursts},'' {\em International Journal of Modern Physics D},
  vol.~22, p.~1330028, Dec. 2013.

\bibitem{izzo2015new}
L.~Izzo, M.~Muccino, E.~Zaninoni, L.~Amati, and M.~Della~Valle, ``New
  measurements of $\omega$m from gamma-ray bursts,'' {\em Astronomy \&
  Astrophysics}, vol.~582, p.~A115, 2015.

\bibitem{Ether}
I.~M.~H. Etherington {\em Phil.Mag.}, vol.~15, p.~761, 1933.

\bibitem{Uzan:2006mf}
J.-P. Uzan, ``{The acceleration of the universe and the physics behind it},''
  {\em Gen. Rel. Grav.}, vol.~39, pp.~307--342, 2007.

\bibitem{1996PASP..108..190K}
A.~{Kim}, A.~{Goobar}, and S.~{Perlmutter}, ``{A Generalized K Correction for
  Type IA Supernovae: Comparing R-band Photometry beyond z=0.2 with B, V, and
  R-band Nearby Photometry},'' {\em Publ. Astron. Soc. Pac.}, vol.~108, p.~190,
  Feb. 1996.

\bibitem{Nugent:2002si}
P.~Nugent, A.~Kim, and S.~Perlmutter, ``{K-Corrections and Extinction
  Corrections for Type Ia Supernovae},'' {\em Publ. Astron. Soc. Pac.},
  vol.~114, pp.~803--819, 2002.

\bibitem{d1992introducing}
R.~D'Inverno, {\em Introducing Einstein's Relativity}.
\newblock Clarendon Press, 1992.

\bibitem{dirac1996general}
P.~Dirac, {\em General Theory of Relativity}.
\newblock Physics notes, Princeton University Press, 1996.

\bibitem{lichnerowicz1955elements}
A.~Lichnerowicz, {\em Elements de calcul tensoriel}.
\newblock Armand Colin, A. Colin, 1955.

\bibitem{Narimani:2014zha}
A.~Narimani, D.~Scott, and N.~Afshordi, ``{How does pressure gravitate?
  Cosmological constant problem confronts observational cosmology},'' {\em
  JCAP}, vol.~1408, p.~049, 2014.

\bibitem{Rappaport:2007ct}
S.~Rappaport, J.~Schwab, S.~Burles, and G.~Steigman, ``{Big Bang
  Nucleosynthesis Constraints on the Self-Gravity of Pressure},'' {\em Phys.
  Rev.}, vol.~D77, p.~023515, 2008.

\bibitem{Schwab:2008ce}
J.~Schwab, S.~A. Hughes, and S.~Rappaport, ``{The Self-Gravity of Pressure in
  Neutron Stars},'' 2008.

\bibitem{Kamiab:2011am}
F.~Kamiab and N.~Afshordi, ``{Neutron Stars and the Cosmological Constant
  Problem},'' {\em Phys. Rev.}, vol.~D84, p.~063011, 2011.

\bibitem{Bertone:2004pz}
G.~Bertone, D.~Hooper, and J.~Silk, ``{Particle dark matter: Evidence,
  candidates and constraints},'' {\em Phys. Rept.}, vol.~405, pp.~279--390,
  2005.

\bibitem{Bertone:2016nfn}
G.~Bertone and D.~Hooper, ``{A History of Dark Matter},'' {\em Submitted to:
  Rev. Mod. Phys.}, 2016.

\bibitem{Lemaitre:1931zzb}
G.~Lemaitre, ``{Republication of: The beginning of the world from the point of
  view of quantum theory},'' {\em Nature}, vol.~127, p.~706, 1931.
\newblock [Gen. Rel. Grav.43,2929(2011)].

\bibitem{Lemaitre:1933gd}
G.~Lemaitre, ``{The expanding universe},'' {\em Gen. Rel. Grav.}, vol.~29,
  pp.~641--680, 1997.
\newblock [Annales Soc. Sci. Brux. Ser. I Sci. Math. Astron.
  Phys.A53,51(1933)].

\bibitem{Martin:2014nya}
J.~Martin, C.~Ringeval, and V.~Vennin, ``{Observing Inflationary Reheating},''
  {\em Phys. Rev. Lett.}, vol.~114, no.~8, p.~081303, 2015.

\bibitem{Gamow:1946eb}
G.~Gamow, ``{Expanding universe and the origin of elements},'' {\em Phys.
  Rev.}, vol.~70, pp.~572--573, 1946.

\bibitem{Alpher:1948ve}
R.~A. Alpher, H.~Bethe, and G.~Gamow, ``{The origin of chemical elements},''
  {\em Phys. Rev.}, vol.~73, pp.~803--804, 1948.

\bibitem{meneguzzi1971production}
M.~Meneguzzi, J.~Audouze, and H.~Reeves, ``The production of the elements li,
  be, b by galactic cosmic rays in space and its relation with stellar
  observations.,'' {\em Astronomy and Astrophysics}, vol.~15, pp.~337--359,
  1971.

\bibitem{Burbidge:1957vc}
M.~E. Burbidge, G.~R. Burbidge, W.~A. Fowler, and F.~Hoyle, ``{Synthesis of the
  elements in stars},'' {\em Rev. Mod. Phys.}, vol.~29, pp.~547--650, 1957.

\bibitem{2013ApJS..208...19H}
G.~{Hinshaw}, D.~{Larson}, E.~{Komatsu}, D.~N. {Spergel}, C.~L. {Bennett},
  J.~{Dunkley}, M.~R. {Nolta}, M.~{Halpern}, R.~S. {Hill}, N.~{Odegard},
  L.~{Page}, K.~M. {Smith}, J.~L. {Weiland}, B.~{Gold}, N.~{Jarosik},
  A.~{Kogut}, M.~{Limon}, S.~S. {Meyer}, G.~S. {Tucker}, E.~{Wollack}, and
  E.~L. {Wright}, ``{Nine-year Wilkinson Microwave Anisotropy Probe (WMAP)
  Observations: Cosmological Parameter Results},'' {\em The Astrophysical
  Journal, Supplement}, vol.~208, p.~19, Oct. 2013.

\bibitem{gamow1948origin}
G.~Gamow, ``The origin of elements and the separation of galaxies,'' {\em
  Physical Review}, vol.~74, no.~4, p.~505, 1948.

\bibitem{Penzias:1965wn}
A.~A. Penzias and R.~W. Wilson, ``{A Measurement of excess antenna temperature
  at 4080-Mc/s},'' {\em Astrophys. J.}, vol.~142, pp.~419--421, 1965.

\bibitem{Dicke:1965zz}
R.~H. Dicke, P.~J.~E. Peebles, P.~G. Roll, and D.~T. Wilkinson, ``{Cosmic
  Black-Body Radiation},'' {\em Astrophys. J.}, vol.~142, pp.~414--419, 1965.

\bibitem{Ade:2013sjv}
P.~A.~R. Ade {\em et~al.}, ``{Planck 2013 results. I. Overview of products and
  scientific results},'' {\em Astron. Astrophys.}, vol.~571, p.~A1, 2014.

\bibitem{Ma:1995ey}
C.-P. Ma and E.~Bertschinger, ``{Cosmological perturbation theory in the
  synchronous and conformal Newtonian gauges},'' {\em Astrophys. J.}, vol.~455,
  pp.~7--25, 1995.

\bibitem{Amendola:2012ys}
L.~Amendola {\em et~al.}, ``{Cosmology and fundamental physics with the Euclid
  satellite},'' {\em Living Rev. Rel.}, vol.~16, p.~6, 2013.

\bibitem{flatness}
R.~H. {Dicke} and P.~J.~E. {Peebles}, ``{The big bang cosmology - enigmas and
  nostrums.},'' in {\em General Relativity: An Einstein centenary survey}
  (S.~W. {Hawking} and W.~{Israel}, eds.), pp.~504--517, 1979.

\bibitem{Rindler:1956yx}
W.~Rindler, ``{Visual horizons in world-models},'' {\em Gen. Rel. Grav.},
  vol.~34, pp.~133--153, 2002.
\newblock [Mon. Not. Roy. Astron. Soc.116,662(1956)].

\bibitem{Misner:1967uu}
C.~W. Misner, ``{The Isotropy of the universe},'' {\em Astrophys. J.},
  vol.~151, pp.~431--457, 1968.

\bibitem{Baumann:2014nda}
D.~Baumann and L.~McAllister, {\em {Inflation and String Theory}}.
\newblock Cambridge University Press, 2015.

\bibitem{Guth:1979bh}
A.~H. Guth and S.~H.~H. Tye, ``{Phase Transitions and Magnetic Monopole
  Production in the Very Early Universe},'' {\em Phys. Rev. Lett.}, vol.~44,
  p.~631, 1980.
\newblock [Erratum: Phys. Rev. Lett.44,963(1980)].

\bibitem{Zeldovich:1978wj}
{\relax Ya}.~B. Zeldovich and M.~{\relax Yu}. Khlopov, ``{On the Concentration
  of Relic Magnetic Monopoles in the Universe},'' {\em Phys. Lett.}, vol.~B79,
  pp.~239--241, 1978.

\bibitem{Preskill:1979zi}
J.~Preskill, ``{Cosmological Production of Superheavy Magnetic Monopoles},''
  {\em Phys. Rev. Lett.}, vol.~43, p.~1365, 1979.

\bibitem{Guth:1980zm}
A.~H. Guth, ``{The Inflationary Universe: A Possible Solution to the Horizon
  and Flatness Problems},'' {\em Phys. Rev.}, vol.~D23, pp.~347--356, 1981.

\bibitem{Martin:2007bw}
J.~Martin, ``{Inflationary perturbations: The Cosmological Schwinger effect},''
  {\em Lect. Notes Phys.}, vol.~738, pp.~193--241, 2008.

\bibitem{Bunch:1978yq}
T.~S. Bunch and P.~C.~W. Davies, ``{Quantum Field Theory in de Sitter Space:
  Renormalization by Point Splitting},'' {\em Proc. Roy. Soc. Lond.},
  vol.~A360, pp.~117--134, 1978.

\bibitem{book:Parker}
L.~Parker and D.~Toms, {\em Quantum field theory in curved spacetime: quantized
  fields and gravity}.
\newblock Cambridge University Press, 2009.

\bibitem{Bardeen:1980kt}
J.~M. Bardeen, ``{Gauge Invariant Cosmological Perturbations},'' {\em Phys.
  Rev.}, vol.~D22, pp.~1882--1905, 1980.

\bibitem{Mukhanov:1990me}
V.~F. Mukhanov, H.~A. Feldman, and R.~H. Brandenberger, ``{Theory of
  cosmological perturbations. Part 1. Classical perturbations. Part 2. Quantum
  theory of perturbations. Part 3. Extensions},'' {\em Phys. Rept.}, vol.~215,
  pp.~203--333, 1992.

\bibitem{Gomez:2013xza}
L.~G. Gomez and Y.~Rodriguez, ``{Statistical Anisotropy in Inflationary Models
  with Many Vector Fields and/or Prolonged Anisotropic Expansion},'' {\em AIP
  Conf. Proc.}, vol.~1548, pp.~270--276, 2013.

\bibitem{Sasaki:1986hm}
M.~Sasaki, ``{Large Scale Quantum Fluctuations in the Inflationary Universe},''
  {\em Prog. Theor. Phys.}, vol.~76, p.~1036, 1986.

\bibitem{Mukhanov:1988jd}
V.~F. Mukhanov, ``{Quantum Theory of Gauge Invariant Cosmological
  Perturbations},'' {\em Sov. Phys. JETP}, vol.~67, pp.~1297--1302, 1988.
\newblock [Zh. Eksp. Teor. Fiz.94N7,1(1988)].

\bibitem{Martin:2013tda}
J.~Martin, C.~Ringeval, and V.~Vennin, ``{Encyclopædia Inflationaris},'' {\em
  Phys. Dark Univ.}, vol.~5-6, pp.~75--235, 2014.

\bibitem{dyson1997innovation}
F.~J. Dyson, ``Innovation in physics,'' in {\em JingShin Theoretical Physics
  Symposium in Honor of Professor Ta-You Wu}, pp.~73--90, World Scientific,
  1997.

\bibitem{Sauter1931}
F.~Sauter, ``{\"U}ber das verhalten eines elektrons im homogenen elektrischen
  feld nach der relativistischen theorie diracs,'' {\em Z. Phys.}, vol.~69,
  no.~11-12, pp.~742--764, 1931.

\bibitem{Heisenberg1936}
W.~Heisenberg and H.~Euler, ``Folgerungen aus der diracschen theorie des
  positrons (consequences of dirac theory of the positron),'' {\em Z. Phys.},
  vol.~98, pp.~714--732, 1936.

\bibitem{Schwinger1951}
J.~Schwinger, ``On gauge invariance and vacuum polarization,'' {\em Phys.
  Rev.}, vol.~82, no.~5, p.~664, 1951.

\bibitem{Gelis:2015kya}
F.~Gelis and N.~Tanji, ``{Schwinger mechanism revisited},'' {\em Prog. Part.
  Nucl. Phys.}, vol.~87, pp.~1--49, 2016.

\bibitem{XFEL}
``\url{http://www.xfel.eu/},''

\bibitem{HIBEF}
``\url{http://www.hzdr.de/hibef},''

\bibitem{ELIA}
``\url{http://www.eli-beams.eu/},''

\bibitem{HiPer}
``\url{http://www.hiperlaser.org},''

\bibitem{XCELS}
``\url{http://www.xcels.iapras.ru/img/XCELS-Project-english-version.pdf},''

\bibitem{Peskin:1995ev}
M.~E. Peskin and D.~V. Schroeder, {\em {An Introduction to quantum field
  theory}}.
\newblock 1995.

\bibitem{greiner2008quantum}
W.~Greiner and J.~Reinhardt, {\em Quantum Electrodynamics}.
\newblock Physics and Astronomy, Springer Berlin Heidelberg, 2008.

\bibitem{itzykson2012quantum}
C.~Itzykson and J.~Zuber, {\em Quantum Field Theory}.
\newblock Dover Books on Physics, Dover Publications, 2012.

\bibitem{Toms:2012bra}
D.~J. Toms, {\em {The Schwinger Action Principle and Effective Action}}.
\newblock Cambridge University Press, 2012.

\bibitem{Durrer:2013pga}
R.~Durrer and A.~Neronov, ``{Cosmological Magnetic Fields: Their Generation,
  Evolution and Observation},'' {\em Astron. Astrophys. Rev.}, vol.~21, p.~62,
  2013.

\bibitem{Garriga1994}
J.~Garriga, ``Nucleation rates in flat and curved space,'' {\em Physical Review
  D}, vol.~49, no.~12, p.~6327, 1994.

\bibitem{Froeb2014}
M.~B. Fr{\"o}b, J.~Garriga, S.~Kanno, M.~Sasaki, J.~Soda, T.~Tanaka, and
  A.~Vilenkin, ``Schwinger effect in de sitter space,'' {\em Journal of
  Cosmology and Astroparticle Physics}, vol.~2014, no.~04, p.~009, 2014.

\bibitem{Kobayashi:2014zza}
T.~Kobayashi and N.~Afshordi, ``Schwinger effect in 4d de sitter space and
  constraints on magnetogenesis in the early universe,'' {\em Journal of High
  Energy Physics}, vol.~2014, no.~10, pp.~1--36, 2014.

\bibitem{Fischler2014}
W.~Fischler, P.~H. Nguyen, J.~F. Pedraza, and W.~Tangarife, ``{Holographic
  Schwinger effect in de Sitter space},'' {\em Phys. Rev.}, vol.~D91, no.~8,
  p.~086015, 2015.

\bibitem{Landete2014}
A.~Landete, J.~Navarro-Salas, and F.~Torrent{\'\i}, ``Adiabatic regularization
  and particle creation for spin one-half fields,'' {\em Physical Review D},
  vol.~89, no.~4, p.~044030, 2014.

\bibitem{kim2015}
S.~P. Kim, H.~K. Lee, and Y.~Yoon, ``Thermal interpretation of schwinger effect
  in near-extremal rn black hole,'' {\em arXiv preprint arXiv:1503.00218},
  2015.

\bibitem{Hayashinaka:2016qqn}
T.~Hayashinaka, T.~Fujita, and J.~Yokoyama, ``{Fermionic Schwinger effect and
  induced current in de Sitter space},'' {\em JCAP}, vol.~1607, no.~07, p.~010,
  2016.

\bibitem{Hayashinaka:2016dnt}
T.~Hayashinaka and J.~Yokoyama, ``{Point splitting renormalization of Schwinger
  induced current in de Sitter spacetime},'' {\em JCAP}, vol.~1607, no.~07,
  p.~012, 2016.

\bibitem{Yokoyama:2015wws}
J.~Yokoyama, ``{Issues on the inflationary magnetogenesis},'' {\em Comptes
  Rendus Physique}, vol.~16, no.~10, pp.~1018--1026, 2015.

\bibitem{Birrell1984}
N.~D. Birrell and P.~C.~W. Davies, {\em Quantum fields in curved space}.
\newblock No.~7, Cambridge university press, 1984.

\bibitem{Barata:2016ves}
J.~C.~A. Barata, C.~D. Jäkel, and J.~Mund, ``{Interacting Quantum Fields on de
  Sitter Space},'' 2016.

\bibitem{Kachru:2003sx}
S.~Kachru, R.~Kallosh, A.~D. Linde, J.~M. Maldacena, L.~P. McAllister, and
  S.~P. Trivedi, ``{Towards inflation in string theory},'' {\em JCAP},
  vol.~0310, p.~013, 2003.

\bibitem{Parker1980}
L.~Parker, ``One-electron atom in curved space-time,'' {\em Physical Review
  Letters}, vol.~44, no.~23, p.~1559, 1980.

\bibitem{Pollock2010}
M.~Pollock, ``On the dirac equation in curved space-time,'' {\em Acta Physica
  Polonica B}, vol.~41, no.~8, p.~1827, 2010.

\bibitem{Olver2010}
F.~Olver, D.~Lozier, R.~Boisvert, and C.~Clark, {\em NIST Handbook of
  Mathematical Functions}.
\newblock Cambridge University Press, Cambridge, London and New York, 2010.

\bibitem{Kim:2016dmm}
S.~P. Kim, ``{Schwinger Effect, Hawking Radiation, and Unruh Effect},'' in {\em
  {2nd LeCosPA Symposium: Everything about Gravity, Celebrating the Centenary
  of Einstein's General Relativity (LeCosPA2015) Taipei, Taiwan, December
  14-18, 2015}}, 2016.

\bibitem{Strobel:2014tha}
E.~Strobel and S.-S. Xue, ``{Semiclassical pair production rate for rotating
  electric fields},'' {\em Phys. Rev.}, vol.~D91, p.~045016, 2015.

\bibitem{Dumlu2011}
C.~K. Dumlu and G.~V. Dunne, ``Interference effects in schwinger vacuum pair
  production for time-dependent laser pulses,'' {\em Phys. Rev. D}, vol.~83,
  no.~6, p.~065028, 2011.

\bibitem{Kluger1992}
Y.~Kluger, J.~Eisenberg, B.~Svetitsky, F.~Cooper, and E.~Mottola, ``Fermion
  pair production in a strong electric field,'' {\em Physical Review D},
  vol.~45, no.~12, p.~4659, 1992.

\bibitem{Brezin1970}
E.~Brezin and C.~Itzykson, ``Pair production in vacuum by an alternating
  field,'' {\em Phys. Rev. D}, vol.~2, pp.~1191--1199, 1970.

\bibitem{Popov1971}
V.~S. {Popov}, ``{Production of $e^{+}e^{-}$ Pairs in an Alternating External
  Field},'' {\em ZhETF Pis ma Redaktsiiu}, vol.~13, pp.~261--+, Mar. 1971.

\bibitem{Popov1972}
V.~S. {Popov}, ``{Pair Production in a Variable External Field (Quasiclassical
  Approximation)},'' {\em J. Exp. Theor. Phys.}, vol.~34, p.~709, 1972.

\bibitem{Popov1973}
V.~S. Popov and M.~S. Marinov, ``\(e^+e^-\) pair production in an alternating
  electric field,'' {\em Sov. J. Nucl. Phys.}, vol.~16, no.~4, pp.~449--456,
  1973.

\bibitem{Popov2001}
V.~S. {Popov}, ``{Schwinger Mechanism of Electron-Positron Pair Production by
  the Field of Optical and X-ray Lasers in Vacuum},'' {\em J. Exp. Theor. Phys.
  Lett.}, vol.~74, pp.~133--138, Aug. 2001.

\bibitem{Kleinert2008}
H.~Kleinert, R.~Ruffini, and S.-S. Xue, ``{Electron-Positron Pair Production in
  Space- or Time-Dependent Electric Fields},'' {\em Phys. Rev.}, vol.~D78,
  p.~025011, 2008.

\bibitem{Kleinert2013}
H.~Kleinert and S.-S. Xue, ``Vacuum pair-production in classical electric field
  and electromagnetic wave,'' {\em Annals of Physics}, vol.~333, pp.~104--126,
  2013.

\bibitem{Strobel:2013vza}
E.~Strobel and S.-S. Xue, ``{Semiclassical pair production rate for
  time-dependent electrical fields with more than one component: WKB-approach
  and world-line instantons},'' {\em Nucl. Phys.}, vol.~B886, pp.~1153--1176,
  2014.

\bibitem{StrobelPhD}
E.~Strobel, {\em {Critical and overcritical Electromagnetic Fields}}.
\newblock PhD thesis, Sapienza U., 2015.

\bibitem{Blinne:2015zpa}
A.~Blinne and E.~Strobel, ``{Comparison of semiclassical and Wigner function
  methods in pair production in rotating fields},'' {\em Phys. Rev.}, vol.~D93,
  no.~2, p.~025014, 2016.

\bibitem{Gavrilov:1996pz}
S.~P. Gavrilov and D.~M. Gitman, ``{Vacuum instability in external fields},''
  {\em Phys. Rev.}, vol.~D53, pp.~7162--7175, 1996.

\bibitem{Kluger:1998bm}
Y.~Kluger, E.~Mottola, and J.~M. Eisenberg, ``{The Quantum Vlasov equation and
  its Markov limit},'' {\em Phys. Rev.}, vol.~D58, p.~125015, 1998.

\bibitem{Anderson:2013ila}
P.~R. Anderson and E.~Mottola, ``{Instability of global de Sitter space to
  particle creation},'' {\em Phys. Rev.}, vol.~D89, p.~104038, 2014.

\bibitem{Berry1982}
M.~Berry, ``Semiclassically weak reflections above analytic and non-analytic
  potential barriers,'' {\em Journal of Physics A: Mathematical and General},
  vol.~15, no.~12, p.~3693, 1982.

\bibitem{Narozhnyi:1970uv}
N.~B. Narozhnyi and A.~I. Nikishov, ``{The Simplist processes in the pair
  creating electric field},'' {\em Yad. Fiz.}, vol.~11, p.~1072, 1970.
\newblock [Sov. J. Nucl. Phys.11,596(1970)].

\bibitem{Nikishov:1970br}
A.~I. Nikishov, ``{Barrier scattering in field theory removal of klein
  paradox},'' {\em Nucl. Phys.}, vol.~B21, pp.~346--358, 1970.

\bibitem{Haouat2013}
S.~Haouat and R.~Chekireb, ``Comment on “creation of spin 1/2 particles by an
  electric field in de sitter space”,'' {\em Physical Review D}, vol.~87,
  no.~8, p.~088501, 2013.

\bibitem{Garriga:1994bm}
J.~Garriga, ``{Pair production by an electric field in (1+1)-dimensional de
  Sitter space},'' {\em Phys. Rev.}, vol.~D49, pp.~6343--6346, 1994.

\bibitem{Parker1966}
L.~Parker, {\em The Creation of Particles in an Expanding Universe, Ph.D.
  thesis}.
\newblock Harvard University, 1966.

\bibitem{Parker1974}
L.~Parker and S.~Fulling, ``Adiabatic regularization of the energy-momentum
  tensor of a quantized field in homogeneous spaces,'' {\em Physical Review D},
  vol.~9, no.~2, p.~341, 1974.

\bibitem{Fulling1974}
S.~Fulling and L.~Parker, ``Renormalization in the theory of a quantized scalar
  field interacting with a robertson-walker spacetime,'' {\em Annals of
  Physics}, vol.~87, no.~1, pp.~176--204, 1974.

\bibitem{Fulling1974B}
S.~Fulling, L.~Parker, and B.~Hu, ``Conformal energy-momentum tensor in curved
  spacetime: Adiabatic regularization and renormalization,'' {\em Physical
  Review D}, vol.~10, no.~12, p.~3905, 1974.

\bibitem{Feynman:1949hz}
R.~P. Feynman, ``{The Theory of positrons},'' {\em Phys. Rev.}, vol.~76,
  pp.~749--759, 1949.

\bibitem{Nikishov1969}
A.~I. {Nikishov}, ``Pair production by a constant electric field,'' {\em J.
  Exp. Theor. Phys.}, vol.~30, pp.~660--662, 1970.

\bibitem{Mottola:1984ar}
E.~Mottola, ``{Particle Creation in de Sitter Space},'' {\em Phys. Rev.},
  vol.~D31, p.~754, 1985.

\bibitem{parker1980one}
L.~Parker, ``One-electron atom in curved space-time,'' {\em Physical Review
  Letters}, vol.~44, no.~23, p.~1559, 1980.

\bibitem{parker1980one2}
L.~Parker, ``One-electron atom as a probe of spacetime curvature,'' {\em
  Physical Review D}, vol.~22, no.~8, p.~1922, 1980.

\bibitem{Ghosh:2015mva}
S.~Ghosh, ``{Creation of spin $1/2$ particles and renormalization in FLRW
  spacetime},'' 2015.

\bibitem{Anderson:2013zia}
P.~R. Anderson and E.~Mottola, ``{Quantum vacuum instability of “eternal”
  de Sitter space},'' {\em Phys. Rev.}, vol.~D89, p.~104039, 2014.

\bibitem{Breitenlohner:1982jf}
P.~Breitenlohner and D.~Z. Freedman, ``{Stability in Gauged Extended
  Supergravity},'' {\em Annals Phys.}, vol.~144, p.~249, 1982.

\bibitem{McInnes:2001dq}
B.~McInnes, ``{Exploring the similarities of the dS / CFT and AdS / CFT
  correspondences},'' {\em Nucl. Phys.}, vol.~B627, pp.~311--329, 2002.

\bibitem{Pioline:2005pf}
B.~Pioline and J.~Troost, ``{Schwinger pair production in AdS(2)},'' {\em
  JHEP}, vol.~03, p.~043, 2005.

\bibitem{Vilenkin:1983xq}
A.~Vilenkin, ``{The Birth of Inflationary Universes},'' {\em Phys. Rev.},
  vol.~D27, p.~2848, 1983.

\bibitem{Linde:1986fd}
A.~D. Linde, ``{Eternally Existing Selfreproducing Chaotic Inflationary
  Universe},'' {\em Phys. Lett.}, vol.~B175, pp.~395--400, 1986.

\bibitem{Linde:2015edk}
A.~Linde, ``{A brief history of the multiverse},'' 2015.

\bibitem{carr2007universe}
B.~Carr, {\em Universe or multiverse?}
\newblock Cambridge University Press, 2007.

\bibitem{Goolsby-Cole:2015chd}
C.~Goolsby-Cole and L.~Sorbo, ``{On the electric charge of the observable
  Universe},'' 2015.

\bibitem{Kluger:1991ib}
Y.~Kluger, J.~M. Eisenberg, B.~Svetitsky, F.~Cooper, and E.~Mottola, ``{Pair
  production in a strong electric field},'' {\em Phys. Rev. Lett.}, vol.~67,
  pp.~2427--2430, 1991.

\bibitem{Kluger:1992gb}
Y.~Kluger, J.~M. Eisenberg, B.~Svetitsky, F.~Cooper, and E.~Mottola, ``{Fermion
  pair production in a strong electric field},'' {\em Phys. Rev.}, vol.~D45,
  pp.~4659--4671, 1992.

\bibitem{Ruffini:2003cr}
R.~Ruffini, L.~Vitagliano, and S.~S. Xue, ``{On plasma oscillations in strong
  electric fields},'' {\em Phys. Lett.}, vol.~B559, pp.~12--19, 2003.

\bibitem{Ruffini:2007jm}
R.~Ruffini, G.~V. Vereshchagin, and S.~S. Xue, ``{Vacuum polarization and
  plasma oscillations},'' {\em Phys. Lett.}, vol.~A371, pp.~399--405, 2007.

\bibitem{Kasper:2014uaa}
V.~Kasper, F.~Hebenstreit, and J.~Berges, ``{Fermion production from real-time
  lattice gauge theory in the classical-statistical regime},'' {\em Phys.
  Rev.}, vol.~D90, no.~2, p.~025016, 2014.

\bibitem{Grasso:2000wj}
D.~Grasso and H.~R. Rubinstein, ``{Magnetic fields in the early universe},''
  {\em Phys. Rept.}, vol.~348, pp.~163--266, 2001.

\bibitem{Subramanian:2015lua}
K.~Subramanian, ``{The origin, evolution and signatures of primordial magnetic
  fields},'' {\em Rept. Prog. Phys.}, vol.~79, no.~7, p.~076901, 2016.

\bibitem{Ferreira:2013sqa}
R.~J.~Z. Ferreira, R.~K. Jain, and M.~S. Sloth, ``{Inflationary magnetogenesis
  without the strong coupling problem},'' {\em JCAP}, vol.~1310, p.~004, 2013.

\bibitem{Brandenburg:2004jv}
A.~Brandenburg and K.~Subramanian, ``{Astrophysical magnetic fields and
  nonlinear dynamo theory},'' {\em Phys. Rept.}, vol.~417, pp.~1--209, 2005.

\bibitem{Durrive:2015cja}
J.-B. Durrive and M.~Langer, ``{Intergalactic Magnetogenesis at Cosmic Dawn by
  Photoionization},'' {\em Mon. Not. Roy. Astron. Soc.}, vol.~453, no.~1,
  pp.~345--356, 2015.

\bibitem{Cai:2005ra}
R.-G. Cai and S.~P. Kim, ``{First law of thermodynamics and Friedmann equations
  of Friedmann-Robertson-Walker universe},'' {\em JHEP}, vol.~02, p.~050, 2005.

\bibitem{Markkanen:2016aes}
T.~Markkanen and A.~Rajantie, ``{Massive scalar field evolution in de
  Sitter},'' 2016.

\bibitem{Barnaby:2011qe}
N.~Barnaby, E.~Pajer, and M.~Peloso, ``{Gauge Field Production in Axion
  Inflation: Consequences for Monodromy, non-Gaussianity in the CMB, and
  Gravitational Waves at Interferometers},'' {\em Phys. Rev.}, vol.~D85,
  p.~023525, 2012.

\bibitem{Domcke:2016bkh}
V.~Domcke, M.~Pieroni, and P.~Binétruy, ``{Primordial gravitational waves for
  universality classes of pseudoscalar inflation},'' {\em JCAP}, vol.~1606,
  p.~031, 2016.

\bibitem{Perlmutter:1998np}
S.~Perlmutter {\em et~al.}, ``{Measurements of Omega and Lambda from 42 high
  redshift supernovae},'' {\em Astrophys. J.}, vol.~517, pp.~565--586, 1999.

\bibitem{Riess:1998cb}
A.~G. Riess {\em et~al.}, ``{Observational evidence from supernovae for an
  accelerating universe and a cosmological constant},'' {\em Astron. J.},
  vol.~116, pp.~1009--1038, 1998.

\bibitem{Perlmutter:1997zf}
S.~Perlmutter {\em et~al.}, ``{Discovery of a supernova explosion at half the
  age of the Universe and its cosmological implications},'' {\em Nature},
  vol.~391, pp.~51--54, 1998.

\bibitem{Schmidt:1998ys}
B.~P. Schmidt {\em et~al.}, ``{The High Z supernova search: Measuring cosmic
  deceleration and global curvature of the universe using type Ia
  supernovae},'' {\em Astrophys. J.}, vol.~507, pp.~46--63, 1998.

\bibitem{tHooft:1979bh}
G.~'t~Hooft, ``{Naturalness, chiral symmetry, and spontaneous chiral symmetry
  breaking},'' {\em NATO Sci. Ser. B}, vol.~59, p.~135, 1980.

\bibitem{Zeldovich:1968}
Y.~B. Zel’dovich, ``The cosmological constant and the theory of elementary
  particles,'' {\em Phys. Usp.}, vol.~11, no.~3, pp.~381--393, 1968.

\bibitem{Weinberg:1988cp}
S.~Weinberg, ``{The Cosmological Constant Problem},'' {\em Rev. Mod. Phys.},
  vol.~61, pp.~1--23, 1989.

\bibitem{Martin:2012bt}
J.~Martin, ``{Everything You Always Wanted To Know About The Cosmological
  Constant Problem (But Were Afraid To Ask)},'' {\em Comptes Rendus Physique},
  vol.~13, pp.~566--665, 2012.

\bibitem{Carter347}
B.~Carter and W.~H. McCrea, ``The anthropic principle and its implications for
  biological evolution [and discussion],'' {\em Philosophical Transactions of
  the Royal Society of London A: Mathematical, Physical and Engineering
  Sciences}, vol.~310, no.~1512, pp.~347--363, 1983.

\bibitem{gott1993implications}
J.~R. Gott, ``Implications of the copernican principle for our future
  prospects,'' {\em Nature}, vol.~363, no.~6427, pp.~315--319, 1993.

\bibitem{norton2010cosmic}
J.~D. Norton, ``Cosmic confusions: Not supporting versus supporting not,'' {\em
  Philosophy of Science}, vol.~77, no.~4, pp.~501--523, 2010.

\bibitem{benetreau2013apocalypse}
Y.~Ben{\'e}treau-Dupin, ``Apocalypse not now,'' 2013.

\bibitem{kragh1990dirac}
H.~Kragh, {\em Dirac: A Scientific Biography}.
\newblock Cambridge University Press, 1990.

\bibitem{Ford:1987de}
L.~H. Ford, ``{Cosmological constant damping by unstable scalar fields},'' {\em
  Phys. Rev.}, vol.~D35, p.~2339, 1987.

\bibitem{Dolgov:2008rf}
A.~D. Dolgov and F.~R. Urban, ``{Dynamical vacuum energy via adjustment
  mechanism},'' {\em Phys. Rev.}, vol.~D77, p.~083503, 2008.

\bibitem{Williams:2004qba}
J.~G. Williams, S.~G. Turyshev, and D.~H. Boggs, ``{Progress in lunar laser
  ranging tests of relativistic gravity},'' {\em Phys. Rev. Lett.}, vol.~93,
  p.~261101, 2004.

\bibitem{Nobbenhuis:2006yf}
S.~Nobbenhuis, {\em {The Cosmological Constant Problem, an Inspiration for New
  Physics}}.
\newblock PhD thesis, Utrecht U., 2006.

\bibitem{Tsujikawa:2013fta}
S.~Tsujikawa, ``{Quintessence: A Review},'' {\em Class. Quant. Grav.}, vol.~30,
  p.~214003, 2013.

\bibitem{Buchert:2007ik}
T.~Buchert, ``{Dark Energy from Structure: A Status Report},'' {\em Gen. Rel.
  Grav.}, vol.~40, pp.~467--527, 2008.

\bibitem{coucouppp}
M.~Livio, ``How rare are extraterrestrial civilizations, and when did they
  emerge?,'' {\em The Astrophysical Journal}, vol.~511, no.~1, p.~429, 1999.

\bibitem{penrose2006road}
R.~Penrose and P.~E. Jorgensen, ``The road to reality: A complete guide to the
  laws of the universe,'' {\em The Mathematical Intelligencer}, vol.~28, no.~3,
  pp.~59--61, 2006.

\bibitem{2016arXiv160808225L}
H.~W. {Lin} and M.~{Tegmark}, ``{Why does deep and cheap learning work so
  well?},'' {\em ArXiv e-prints}, Aug. 2016.

\bibitem{Weinberg:1987dv}
S.~Weinberg, ``{Anthropic Bound on the Cosmological Constant},'' {\em Phys.
  Rev. Lett.}, vol.~59, p.~2607, 1987.

\bibitem{Kachru:2003aw}
S.~Kachru, R.~Kallosh, A.~D. Linde, and S.~P. Trivedi, ``{De Sitter vacua in
  string theory},'' {\em Phys. Rev.}, vol.~D68, p.~046005, 2003.

\bibitem{Bousso:2000xa}
R.~Bousso and J.~Polchinski, ``{Quantization of four form fluxes and dynamical
  neutralization of the cosmological constant},'' {\em JHEP}, vol.~06, p.~006,
  2000.

\bibitem{Lovelock:1971yv}
D.~Lovelock, ``{The Einstein tensor and its generalizations},'' {\em J. Math.
  Phys.}, vol.~12, pp.~498--501, 1971.

\bibitem{Lovelock:1972vz}
D.~Lovelock, ``{The four-dimensionality of space and the einstein tensor},''
  {\em J. Math. Phys.}, vol.~13, pp.~874--876, 1972.

\bibitem{yeomans1992statistical}
J.~Yeomans, {\em Statistical Mechanics of Phase Transitions}.
\newblock Clarendon Press, 1992.

\bibitem{Ellis:1998ct}
G.~F.~R. Ellis and H.~van Elst, ``{Cosmological models: Cargese lectures
  1998},'' {\em NATO Sci. Ser. C}, vol.~541, pp.~1--116, 1999.

\bibitem{Szekeres:1974ct}
P.~Szekeres, ``{A Class of Inhomogeneous Cosmological Models},'' {\em Commun.
  Math. Phys.}, vol.~41, p.~55, 1975.

\bibitem{Bolejko:2010eb}
K.~Bolejko and M.-N. Celerier, ``{Szekeres Swiss-Cheese model and supernova
  observations},'' {\em Phys. Rev.}, vol.~D82, p.~103510, 2010.

\bibitem{Maartens:2011yx}
R.~Maartens, ``{Is the Universe homogeneous?},'' {\em Phil. Trans. Roy. Soc.
  Lond.}, vol.~A369, pp.~5115--5137, 2011.

\bibitem{Clarkson:2012bg}
C.~Clarkson, ``{Establishing homogeneity of the universe in the shadow of dark
  energy},'' {\em Comptes Rendus Physique}, vol.~13, pp.~682--718, 2012.

\bibitem{Ellis:2006fy}
G.~F.~R. Ellis, ``{Issues in the philosophy of cosmology},'' 2006.

\bibitem{Tolman:1934za}
R.~C. Tolman, ``{Effect of imhomogeneity on cosmological models},'' {\em Proc.
  Nat. Acad. Sci.}, vol.~20, pp.~169--176, 1934.
\newblock [Gen. Rel. Grav.29,935(1997)].

\bibitem{Bondi:1947fta}
H.~Bondi, ``{Spherically symmetrical models in general relativity},'' {\em Mon.
  Not. Roy. Astron. Soc.}, vol.~107, pp.~410--425, 1947.

\bibitem{Celerier:1999hp}
M.-N. Celerier, ``{Do we really see a cosmological constant in the supernovae
  data?},'' {\em Astron. Astrophys.}, vol.~353, pp.~63--71, 2000.

\bibitem{Iguchi:2001sq}
H.~Iguchi, T.~Nakamura, and K.-i. Nakao, ``{Is dark energy the only solution to
  the apparent acceleration of the present universe?},'' {\em Prog. Theor.
  Phys.}, vol.~108, pp.~809--818, 2002.

\bibitem{Nadathur:2010zm}
S.~Nadathur and S.~Sarkar, ``{Reconciling the local void with the CMB},'' {\em
  Phys. Rev.}, vol.~D83, p.~063506, 2011.

\bibitem{Mattsson:2007tj}
T.~Mattsson, ``{Dark energy as a mirage},'' {\em Gen. Rel. Grav.}, vol.~42,
  pp.~567--599, 2010.

\bibitem{Marra:2007pm}
V.~Marra, E.~W. Kolb, S.~Matarrese, and A.~Riotto, ``{On cosmological
  observables in a swiss-cheese universe},'' {\em Phys. Rev.}, vol.~D76,
  p.~123004, 2007.

\bibitem{Biswas:2010xm}
T.~Biswas, A.~Notari, and W.~Valkenburg, ``{Testing the Void against
  Cosmological data: fitting CMB, BAO, SN and H0},'' {\em JCAP}, vol.~1011,
  p.~030, 2010.

\bibitem{Sundell:2013ova}
P.~Sundell and I.~Vilja, ``{Inhomogeneous cosmological models and fine-tuning
  of the initial state},'' {\em Mod. Phys. Lett.}, vol.~A29, no.~10,
  p.~1450053, 2014.

\bibitem{Celerier:2009sv}
M.-N. Celerier, K.~Bolejko, and A.~Krasinski, ``{A (giant) void is not
  mandatory to explain away dark energy with a Lemaitre -- Tolman model},''
  {\em Astron. Astrophys.}, vol.~518, p.~A21, 2010.

\bibitem{Bolejko:2010wc}
K.~Bolejko and R.~A. Sussman, ``{Cosmic spherical void via coarse-graining and
  averaging non-spherical structures},'' {\em Phys. Lett.}, vol.~B697,
  pp.~265--270, 2011.

\bibitem{Biswas:2007gi}
T.~Biswas and A.~Notari, ``{Swiss-Cheese Inhomogeneous Cosmology and the Dark
  Energy Problem},'' {\em JCAP}, vol.~0806, p.~021, 2008.

\bibitem{Kainulainen:2009sx}
K.~Kainulainen and V.~Marra, ``{SNe observations in a meatball universe with a
  local void},'' {\em Phys. Rev.}, vol.~D80, p.~127301, 2009.

\bibitem{Moradi:2015caa}
R.~Moradi, J.~T. Firouzjaee, and R.~Mansouri, ``{Cosmological black holes: the
  spherical perfect fluid collapse with pressure in a FRW background},'' {\em
  Class. Quant. Grav.}, vol.~32, no.~21, p.~215001, 2015.

\bibitem{Mustapha:1998jb}
N.~Mustapha, C.~Hellaby, and G.~F.~R. Ellis, ``{Large scale inhomogeneity
  versus source evolution: Can we distinguish them observationally?},'' {\em
  Mon. Not. Roy. Astron. Soc.}, vol.~292, pp.~817--830, 1997.

\bibitem{Ellis:1971pg}
G.~F.~R. Ellis, ``{Relativistic cosmology},'' {\em Gen. Rel. Grav.}, vol.~41,
  pp.~581--660, 2009.
\newblock [Proc. Int. Sch. Phys. Fermi47,104(1971)].

\bibitem{Nogueira:2013ypa}
F.~A. M.~G. Nogueira, {\em {Single Past Null Geodesic in the
  Lemaitre-Tolman-Bondi Cosmology}}.
\newblock PhD thesis, Observatorio do Valongo, Universidade Federal do Rio de
  Janeiro, 2013.

\bibitem{2012ApJ...746...85S}
N.~{Suzuki}, D.~{Rubin}, C.~{Lidman}, G.~{Aldering}, R.~{Amanullah},
  K.~{Barbary}, L.~F. {Barrientos}, J.~{Botyanszki}, M.~{Brodwin},
  N.~{Connolly}, K.~S. {Dawson}, A.~{Dey}, M.~{Doi}, M.~{Donahue},
  S.~{Deustua}, P.~{Eisenhardt}, E.~{Ellingson}, L.~{Faccioli}, V.~{Fadeyev},
  H.~K. {Fakhouri}, A.~S. {Fruchter}, D.~G. {Gilbank}, M.~D. {Gladders},
  G.~{Goldhaber}, A.~H. {Gonzalez}, A.~{Goobar}, A.~{Gude}, T.~{Hattori},
  H.~{Hoekstra}, E.~{Hsiao}, X.~{Huang}, Y.~{Ihara}, M.~J. {Jee},
  D.~{Johnston}, N.~{Kashikawa}, B.~{Koester}, K.~{Konishi}, M.~{Kowalski},
  E.~V. {Linder}, L.~{Lubin}, J.~{Melbourne}, J.~{Meyers}, T.~{Morokuma},
  F.~{Munshi}, C.~{Mullis}, T.~{Oda}, N.~{Panagia}, S.~{Perlmutter},
  M.~{Postman}, T.~{Pritchard}, J.~{Rhodes}, P.~{Ripoche}, P.~{Rosati}, D.~J.
  {Schlegel}, A.~{Spadafora}, S.~A. {Stanford}, V.~{Stanishev}, D.~{Stern},
  M.~{Strovink}, N.~{Takanashi}, K.~{Tokita}, M.~{Wagner}, L.~{Wang},
  N.~{Yasuda}, H.~K.~C. {Yee}, and T.~{Supernova Cosmology Project}, ``{The
  Hubble Space Telescope Cluster Supernova Survey. V. Improving the Dark-energy
  Constraints above z > 1 and Building an Early-type-hosted Supernova
  Sample},'' {\em \apj}, vol.~746, p.~85, Feb. 2012.

\bibitem{Bridle:2001zv}
S.~L. Bridle, R.~Crittenden, A.~Melchiorri, M.~P. Hobson, R.~Kneissl, and A.~N.
  Lasenby, ``{Analytic marginalization over CMB calibration and beam
  uncertainty},'' {\em Mon. Not. Roy. Astron. Soc.}, vol.~335, p.~1193, 2002.

\bibitem{Amanullah:2010vv}
R.~Amanullah {\em et~al.}, ``{Spectra and Light Curves of Six Type Ia
  Supernovae at 0.511 < z < 1.12 and the Union2 Compilation},'' {\em Astrophys.
  J.}, vol.~716, pp.~712--738, 2010.

\bibitem{Guy:2007dv}
J.~Guy {\em et~al.}, ``{SALT2: Using distant supernovae to improve the use of
  Type Ia supernovae as distance indicators},'' {\em Astron. Astrophys.},
  vol.~466, pp.~11--21, 2007.

\bibitem{Hicken:2009dk}
M.~Hicken, W.~M. Wood-Vasey, S.~Blondin, P.~Challis, S.~Jha, P.~L. Kelly,
  A.~Rest, and R.~P. Kirshner, ``{Improved Dark Energy Constraints from ~100
  New CfA Supernova Type Ia Light Curves},'' {\em Astrophys. J.}, vol.~700,
  pp.~1097--1140, 2009.

\bibitem{Smale:2010vr}
P.~R. Smale and D.~L. Wiltshire, ``{Supernova tests of the timescape
  cosmology},'' {\em Mon. Not. Roy. Astron. Soc.}, vol.~413, pp.~367--385,
  2011.

\bibitem{mandelbrot1983fractal}
B.~B. Mandelbrot, {\em The fractal geometry of nature}, vol.~173.
\newblock Macmillan, 1983.

\bibitem{pietronero1987fractal}
L.~Pietronero, ``The fractal structure of the universe: correlations of
  galaxies and clusters and the average mass density,'' {\em Physica A:
  Statistical Mechanics and its Applications}, vol.~144, no.~2-3, pp.~257--284,
  1987.

\bibitem{Coleman:1992cm}
P.~H. Coleman and L.~Pietronero, ``{The fractal structure of the universe},''
  {\em Phys. Rept.}, vol.~213, pp.~311--389, 1992.

\bibitem{ruffini1988ino}
R.~Ruffini, D.~Song, and S.~Taraglio, ``The'ino'mass and the cellular
  large-scale structure of the universe,'' {\em Astronomy and Astrophysics},
  vol.~190, pp.~1--9, 1988.

\bibitem{Mureika:2006tz}
J.~R. Mureika, ``{Fractal holography: A Geometric re-interpretation of
  cosmological large scale structure},'' {\em JCAP}, vol.~0705, p.~021, 2007.

\bibitem{Baryshev:2008nb}
{\relax Yu}.~V. Baryshev, ``{Field Fractal Cosmological Model As an Example of
  Practical Cosmology Approach},'' in {\em {International Conference on
  Problems of Practical Cosmology St. Petersburg, Russia, June 23-27, 2008}},
  p.~56, 2008.

\bibitem{Grujic:2009hz}
P.~V. Grujic and V.~D. Pankovic, ``{On the Fractal Structure of the
  Universe},'' 2009.

\bibitem{ChaconCardona:2012iu}
C.~A. Chacon-Cardona and R.~A. Casas-Miranda, ``{Millennium Simulation Dark
  Matter Haloes: Multi-fractal and Lacunarity Analysis with Homogeneity
  Transition},'' {\em Mon. Not. Roy. Astron. Soc.}, vol.~427, p.~2613, 2012.

\bibitem{Conde-Saavedra:2014dna}
G.~Conde-Saavedra, A.~Iribarrem, and M.~B. Ribeiro, ``{Fractal analysis of the
  galaxy distribution in the redshift range 0.45 < z < 5.0},'' {\em Physica},
  vol.~A417, pp.~332--344, 2015.

\bibitem{Bagla:2007tv}
J.~S. Bagla, J.~Yadav, and T.~R. Seshadri, ``{Fractal Dimensions of a Weakly
  Clustered Distribution and the Scale of Homogeneity},'' {\em Mon. Not. Roy.
  Astron. Soc.}, vol.~390, p.~829, 2007.

\bibitem{Ribeiro:2008rs}
M.~B. Ribeiro, ``{On Modelling a Relativistic Hierarchical (Fractal) Cosmology
  by Tolman's Spacetime. I. Theory},'' {\em Astrophys. J.}, vol.~388, pp.~1--8,
  1992.

\bibitem{Ribeiro:2008ru}
M.~B. Ribeiro, ``{On Modelling a Relativistic Hierarchical (Fractal) Cosmology
  by Tolman's Spacetime. II. Analysis of the Einstein-de Sitter Model},'' {\em
  Astrophys. J.}, vol.~395, pp.~29--33, 1992.

\bibitem{Ribeiro:2008ug}
M.~B. Ribeiro, ``{On Modelling a Relativistic Hierarchical (Fractal) Cosmology
  by Tolman's Spacetime. III. Numerical Results},'' {\em Astrophys. J.},
  vol.~415, pp.~469--485, 1993.

\bibitem{SylosLabini:1997jg}
F.~Sylos~Labini, M.~Montuori, and L.~Pietronero, ``{Scale invariance of galaxy
  clustering},'' {\em Phys. Rept.}, vol.~293, pp.~61--226, 1998.

\bibitem{Labini:2011dv}
F.~S. Labini, ``{Very large scale correlations in the galaxy distribution},''
  {\em Europhys. Lett.}, vol.~96, p.~59001, 2011.

\bibitem{Labini:2011tj}
F.~S. Labini, ``{Inhomogeneities in the universe},'' {\em Class. Quant. Grav.},
  vol.~28, p.~164003, 2011.

\bibitem{krasinski2014accelerating}
A.~Krasi{\'n}ski, ``Accelerating expansion or inhomogeneity? a comparison of
  the $\lambda$ cdm and lema{\^\i}tre-tolman models,'' {\em Physical Review D},
  vol.~89, no.~2, p.~023520, 2014.

\bibitem{Clarkson:2009sc}
C.~Clarkson, T.~Clifton, and S.~February, ``{Perturbation Theory in
  Lemaitre-Tolman-Bondi Cosmology},'' {\em JCAP}, vol.~0906, p.~025, 2009.

\bibitem{Leithes:2014uda}
A.~Leithes and K.~A. Malik, ``{Conserved Quantities in Lemaitre-Tolman-Bondi
  Cosmology},'' {\em Class. Quant. Grav.}, vol.~32, no.~1, p.~015010, 2015.

\bibitem{dePutter:2012zx}
R.~de~Putter, L.~Verde, and R.~Jimenez, ``{Testing LTB Void Models Without the
  Cosmic Microwave Background or Large Scale Structure: New Constraints from
  Galaxy Ages},'' {\em JCAP}, vol.~1302, p.~047, 2013.

\bibitem{Freeman:2013rz}
K.~Freeman, ``{Slipher and the Nature of the Nebulae},'' {\em ASP Conf. Ser.},
  vol.~471, p.~63, 2013.

\bibitem{Friedman:1991dj}
J.~A. Frieman and B.-A. Gradwohl, ``{Dark matter and the equivalence
  principle},'' {\em Phys. Rev. Lett.}, vol.~67, pp.~2926--2929, 1991.

\bibitem{Gradwohl:1992ue}
B.-A. Gradwohl and J.~A. Frieman, ``{Dark matter, long range forces, and large
  scale structure},'' {\em Astrophys. J.}, vol.~398, pp.~407--424, 1992.

\bibitem{Wetterich:1994bg}
C.~Wetterich, ``{The Cosmon model for an asymptotically vanishing time
  dependent cosmological 'constant'},'' {\em Astron. Astrophys.}, vol.~301,
  pp.~321--328, 1995.

\bibitem{Amendola:1999er}
L.~Amendola, ``{Coupled quintessence},'' {\em Phys. Rev.}, vol.~D62, p.~043511,
  2000.

\bibitem{Amendola:2003wa}
L.~Amendola, ``{Linear and non-linear perturbations in dark energy models},''
  {\em Phys. Rev.}, vol.~D69, p.~103524, 2004.

\bibitem{Lee:2006za}
S.~Lee, G.-C. Liu, and K.-W. Ng, ``{Constraints on the coupled quintessence
  from cosmic microwave background anisotropy and matter power spectrum},''
  {\em Phys. Rev.}, vol.~D73, p.~083516, 2006.

\bibitem{Pettorino:2008ez}
V.~Pettorino and C.~Baccigalupi, ``{Coupled and Extended Quintessence:
  theoretical differences and structure formation},'' {\em Phys. Rev.},
  vol.~D77, p.~103003, 2008.

\bibitem{Valiviita:2008iv}
J.~Valiviita, E.~Majerotto, and R.~Maartens, ``{Instability in interacting dark
  energy and dark matter fluids},'' {\em JCAP}, vol.~0807, p.~020, 2008.

\bibitem{He:2008tn}
J.-H. He and B.~Wang, ``{Effects of the interaction between dark energy and
  dark matter on cosmological parameters},'' {\em JCAP}, vol.~0806, p.~010,
  2008.

\bibitem{Gavela:2009cy}
M.~B. Gavela, D.~Hernandez, L.~Lopez~Honorez, O.~Mena, and S.~Rigolin, ``{Dark
  coupling},'' {\em JCAP}, vol.~0907, p.~034, 2009.
\newblock [Erratum: JCAP1005,E01(2010)].

\bibitem{Faraoni:2014vra}
V.~Faraoni, J.~B. Dent, and E.~N. Saridakis, ``{Covariantizing the interaction
  between dark energy and dark matter},'' {\em Phys. Rev.}, vol.~D90, no.~6,
  p.~063510, 2014.

\bibitem{Salvatelli:2014zta}
V.~Salvatelli, N.~Said, M.~Bruni, A.~Melchiorri, and D.~Wands, ``{Indications
  of a late-time interaction in the dark sector},'' {\em Phys. Rev. Lett.},
  vol.~113, no.~18, p.~181301, 2014.

\bibitem{Xue:2014kna}
S.-S. Xue, ``{How universe evolves with cosmological and gravitational
  constants},'' {\em Nucl. Phys.}, vol.~B897, pp.~326--345, 2015.

\bibitem{Abdalla:2014cla}
E.~Abdalla, E.~G.~M. Ferreira, J.~Quintin, and B.~Wang, ``{New evidence for
  interacting dark energy from BOSS},'' 2014.

\bibitem{Koivisto:2015qua}
T.~S. Koivisto, E.~N. Saridakis, and N.~Tamanini, ``{Scalar-Fluid theories:
  cosmological perturbations and large-scale structure},'' {\em JCAP},
  vol.~1509, p.~047, 2015.

\bibitem{Wang:2016lxa}
B.~Wang, E.~Abdalla, F.~Atrio-Barandela, and D.~Pavon, ``{Dark Matter and Dark
  Energy Interactions: Theoretical Challenges, Cosmological Implications and
  Observational Signatures},'' {\em Rept. Prog. Phys.}, vol.~79, no.~9,
  p.~096901, 2016.

\bibitem{Murgia:2016ccp}
R.~Murgia, S.~Gariazzo, and N.~Fornengo, ``{Constraints on the Coupling between
  Dark Energy and Dark Matter from CMB data},'' {\em JCAP}, vol.~1604, no.~04,
  p.~014, 2016.

\bibitem{Koyama:2009gd}
K.~Koyama, R.~Maartens, and Y.-S. Song, ``{Velocities as a probe of dark sector
  interactions},'' {\em JCAP}, vol.~0910, p.~017, 2009.

\bibitem{Sola:2016jky}
J.~Sola, A.~Gomez-Valent, and J.~d.~C. Perez, ``{First evidence of running
  cosmic vacuum: challenging the concordance model},'' 2016.

\bibitem{zichichi2012understanding}
A.~Zichichi, {\em Understanding the fundamental constituents of matter},
  vol.~14.
\newblock Springer Science \& Business Media, 2012.

\bibitem{Xue:2015tmw}
S.-S. Xue, ``{The quantum gravitational field theory and the domains of its fix
  points for inflationary or low-redshift universe},'' {\em Int. J. Mod.
  Phys.}, vol.~A30, no.~28, p.~1545003, 2015.

\bibitem{Amendola:2001rc}
L.~Amendola and D.~Tocchini-Valentini, ``{Baryon bias and structure formation
  in an accelerating universe},'' {\em Phys. Rev.}, vol.~D66, p.~043528, 2002.

\bibitem{Iocco:2008va}
F.~Iocco, G.~Mangano, G.~Miele, O.~Pisanti, and P.~D. Serpico, ``{Primordial
  Nucleosynthesis: from precision cosmology to fundamental physics},'' {\em
  Phys. Rept.}, vol.~472, pp.~1--76, 2009.

\bibitem{Sola:2016ecz}
J.~Sola, J.~d.~C. Perez, A.~Gomez-Valent, and R.~C. Nunes, ``{Dynamical Vacuum
  against a rigid Cosmological Constant},'' 2016.

\bibitem{Perelman02theentropy}
G.~Perelman, ``The entropy formula for the ricci flow and its geometric
  applications,'' 2002.

\bibitem{lobell1931beispiele}
F.~L{\"o}bell, ``Beispiele geschlossener dreidimensionaler clifford-kleinscher
  r{\"a}ume negativer kr{\"u}mmung,'' {\em Ber. Sächs. Akad. Wiss. Leipzig},
  vol.~83, pp.~167--171, 1931.

\bibitem{gott1980chaotic}
J.~R. Gott, ``Chaotic cosmologies and the topology of the universe,'' {\em
  Monthly Notices of the Royal Astronomical Society}, vol.~193, no.~1,
  pp.~153--169, 1980.

\bibitem{Riazuelo:2003ud}
A.~Riazuelo, J.~Weeks, J.-P. Uzan, R.~Lehoucq, and J.-P. Luminet, ``{Cosmic
  microwave background anisotropies in multi-connected flat spaces},'' {\em
  Phys. Rev.}, vol.~D69, p.~103518, 2004.

\bibitem{Gausmann:2001aa}
E.~Gausmann, R.~Lehoucq, J.-P. Luminet, J.-P. Uzan, and J.~Weeks,
  ``{Topological lensing in spherical spaces},'' {\em Class. Quant. Grav.},
  vol.~18, pp.~5155--5186, 2001.

\bibitem{rowling1999harry}
J.~Rowling, {\em Harry Potter and the Prisoner of Azkaban}.
\newblock 1999.

\end{thebibliography}


\nocite{*}
